\documentclass[11pt,oneside]{article}
\DeclareUnicodeCharacter{2605}{\ensuremath{\star}}
\usepackage{epsfig,lscape}
\usepackage{rotating}
\usepackage{graphicx} 
\usepackage{amsmath}
\usepackage{amsthm}
\usepackage{amssymb}
\usepackage{subcaption}
\usepackage{bbm}
\usepackage{dsfont}
\usepackage{float}
\usepackage{hyperref}
\usepackage{enumitem}
\usepackage{algpseudocode}
\usepackage{natbib} 
\usepackage{multibib}
\newcites{Supp}{Supplement References}

\usepackage[ruled]{algorithm}
\usepackage{lipsum} 
\usepackage{titling} 

\providecommand{\keywords}[1]{\bigskip\noindent\textbf{Keywords: }#1}

\usepackage{color}
\usepackage[dvipsnames]{xcolor}

\usepackage{float}

\usepackage[dvipsnames]{xcolor}
\newcommand{\zt}{\textcolor{blue}}
\newcommand{\yz}{\textcolor{red}}

\def\E{\mathrm E}
\def\var{\mathrm{Var}}

\def\Softmax{\mathrm{Softmax}}
\def\TV{\mathrm{TV}}
\def\log{\mathrm{log}}
\def\ESS{\mathrm{ESS}}

\def\Unif{\mathrm{Unif}}
\def\exp{\mathrm{exp}}
\def\DiGamma{\mathrm{DiGamma}}
\def\InvGamma{\mathrm{InvGamma}}

\def\diag{\mathrm{diag}}
\def\tr{\mathrm{tr}}
\def\T{ {\mathrm{\scriptscriptstyle T}} }

\makeatother

\newtheorem{proposition}{Proposition}

\allowdisplaybreaks

\begin{document}

\setlength{\abovedisplayskip}{5pt}
\setlength{\belowdisplayskip}{5pt}


\begin{titlepage}

\begin{center}
{\Large Discrete Hamiltonian-Assisted Metropolis Sampling}

\vspace{.1in}Yuze Zhou\footnotemark[1] \& Zhiqiang Tan \footnotemark[1]

\vspace{.1in}
\today
\end{center}

\footnotetext[1]{Department of Statistics, Rutgers University. E-mails: yz909@scarletmail.rutgers.edu, ztan@stat.rutgers.edu.}

\begin{abstract}
Gradient-based Markov Chain Monte Carlo methods have recently received much attention for sampling discrete distributions,
with interesting connections to their continuous counterparts. For examples, there are two discrete analogues to the Metropolis-adjusted Langevin Algorithm (MALA).
As motivated by Hamiltonian-Assisted Metropolis Sampling (HAMS),
we propose Discrete HAMS (DHAMS), a discrete sampler which, for the first time, not only exploits gradient information
but also incorporates a Gaussian momentum variable and samples a Hamiltonian as an augmented distribution.
DHAMS is derived through several steps,
including an auxiliary-variable proposal scheme, negation and gradient correction for the momentum variable,
and over-relaxation for the state variable. Two distinctive properties are achieved simultaneously.
One is generalized detailed balance, which enables irreversible exploration of the target distribution.
The other is a rejection-free property for a target distribution with a linear potential function.
In experiments involving both ordinal and binary distributions, DHAMS algorithms consistently yield superior performance compared with existing algorithms.
\end{abstract}


\end{titlepage}

\section{Introduction}
In numerous statistical applications, the primary objective is to sample from a discrete probability distribution, defined by
a probability mass function $\pi(s)$, over a discrete domain $\mathcal{S} \subset \mathbb{R}^{d}$. The support $\mathcal{S}$ is typically assumed to be a $d$-dimensional discrete lattice expressed as $\mathcal{S} = \mathcal{A}^{d}$ with $\mathcal{A} = \{a_{1}, a_{2}, \cdots , a_{K}\}$, where each coordinate can take $K$ different real values $a_{1}, a_{2}, \cdots, a_{K}$. By introducing a negative potential function $f(s)$, the target distribution can be written as
\begin{equation}
\log (\pi(s)) = f(s) -\log Z, \nonumber
\end{equation}
where $Z=\sum_{s \in \mathcal{S}}\exp(f(s))$ is the normalizing constant. We further assume $f$ is a restriction of a differentiable function originally defined over $\mathbb{R}^{d}$ onto $\mathcal{S}$. Consequently we use $\nabla f$ to denote its gradients.
Sampling from such distributions is often highly challenging, primarily due to the intractability of computing the normalizing constant $Z$. This intractability arises for several reasons. High dimensionality ($d$) and a large lattice size $(k)$ can render normalization infeasible. Additionally, the complexity of the probability mass function itself can pose difficulties.

For sampling continuous distributions, gradient-based Markov Chain Monte Carlo (MCMC) methods have been extensively studied. Such methods are often grounded in Langevin dynamics and Hamiltonian mechanics. Examples include Metropolis-adjusted Langevin Algorithm (MALA) \citep{Besag1994mala, Roberts1996mala}, Hamiltonian Monte Carlo (HMC) \citep{duane1987HMC, neal2011mcmc}, Underdamped Langevin Sampling \citep{Bussi2207UDL},
and Hamiltonian-assisted Metropolis Sampling (HAMS) \citep{Song2023hams} among others.
In contrast, gradient-based samplers have only recently received attention for discrete distributions.
\cite{Grathwohl2021gwg} demonstrated that discrete distributions can be interpreted as restrictions of continuous distributions to a discrete domain, thereby allowing gradients to inform more effective transition kernels. They introduced Gibbs-with-Gradients (GWG),
by employing gradient-informed proposals via the approach of \cite{Zanella2020lbp} while using a Hamming ball as a neighborhood.
By taking gradient-informed proposals but using a Gaussian kernel,
\cite{Rhodes2022GradientMC} and \cite{Zhang2022dmala} proposed the Norm-Constrained Gradient (NCG) method, also referred to as Discrete MALA (DMALA), which represents a discrete analogue to the continuous MALA sampler. Furthermore, \cite{Rhodes2022GradientMC} also introduced Auxiliary Variable Gradient (AVG),
which provides another discrete analogue to continuous MALA in a different approach using auxiliary variables \citep{Titsias2018auxavg}.
See Section \ref{sec:related-methods} for further information.

In this work, we propose Discrete-HAMS (DHAMS), a novel discrete sampler which not only exploits gradient information
but also incorporates a Gaussian momentum variable and samples a Hamiltonian as an augmented distribution.
DHAMS appears to be the first of such discrete samplers.
We conduct several numerical experiments and find that DHAMS algorithms consistently yield superior results over existing algorithms,
as measured by the total variation (TV) distances over iterations
and effective sample size (ESS) estimated from multiple chains.

At a high level, DHAMS is a generalized Metropolis--Hastings sampler in employing a proposal scheme
and then performing acceptance-rejection based on a generalized Metropolis--Hastings probability
such that generalized detailed balance is achieved  \citep{Song2023hams}.
More concretely, DHAMS is derived as a discrete analogue to continuous HAMS through several steps
including an auxiliary-variable proposal scheme, negation and gradient correction for updating the momentum variable,
and over-relaxation for updating the state variable.
While these steps can be connected to similar steps underlying continuous HAMS,
we introduce new techniques to tackle the discreteness of the target distribution,
notably, incorporating an auto-regression step for the momentum to overcome non-erdogicity and
designing an over-relaxation scheme for discrete distributions to mimic Gaussian over-relaxation \citep{Adler1981overrelaxation}.
This seems to be the first time that over-relaxation is developed for discrete distributions in a proposal scheme subject to
acceptance-rejection, as opposed to Gibbs sampling without any rejection \citep{Neal1998overrelax}.

Compared with existing discrete samplers, DHAMS achieves two distinctive properties simultaneously.
First, as mentioned above,
DHAMS satisfies a generalized detailed balance condition, where the backward transition is related to the forward transition by negating the momentum.
This condition is known for various sampling algorithms with continuous state variables in physics
\citep{Scemama2006irreversible, Bussi2207UDL, Fang2014compressible}.
Due to the momentum negation, generalized detailed balance enables irreversible exploration of the target distribution,
similarly to a stochastic version of each trajectory in HMC before resetting the momentum.
Second, DHAMS is rejection-free when the target distribution is of the product form, $\pi(s) \propto \exp(a^{\T}s)$,
or equivalently the negative potential $f(s)$ is linear in $s$. An example of such product distributions is a product of Bernoulli distributions (possibly non-identical).
This property mirrors the property that HAMS is rejection-free when the target distribution is standard Gaussian
(or Gaussian with a pre-specified variance matrix).
It can be shown that AVG satisfies this rejection-free property, but achieves standard detailed balance (hence reversible sampling).

For our augmented distribution, the momentum variable $u$ is Gaussian, independently of the state variable $s$ as in a standard Hamiltonian.
In contrast, \cite{pakman2014auxiliary} and \cite{Zhang2012MRF} proposed augmenting a discrete distribution with additional variables which are Gaussian
conditionally on the state variable.
Furthermore, the state variable in \cite{pakman2014auxiliary} is restricted to be binary componentwise, whereas the target distribution in \cite{Zhang2012MRF}
is limited to a quadratic potential function.

\paragraph{Notation.}
For a (vector-valued) state variable $s$, we denote as $s_i$ the $i$-th coordinate of $s$.
When a sequence of draws is discussed, $s_{t}$ refers to the $t$-th draw, while $s_{t,i}$ denotes the $i$-th coordinate of $s_{t}$.
We denote as $\Unif([a,b))$ a uniform distribution with support $[a,b)$. We use $\mathcal{N}(\mu, V)$ to represent a Gaussian distribution with mean $\mu$ and variance matrix $V$, and $\mathcal{N}(\cdot|\mu, V)$ to represent the corresponding density function. We use $\Softmax$ to denote
the probabilities in a discrete distribution with negative potential $g(\cdot)$ over the discrete set $\mathcal{A} = \{a_1, \ldots, a_K\}$:
\begin{align}
   \Softmax( g(a_k))  & = \frac{\exp(g(a_k))}{\sum\limits_{j=1}^{K} \exp(g(a_j))}, \quad k= 1, \ldots, K.\nonumber
\end{align}
where the dependency on $\mathcal{A}$ is suppressed.

\section{Related Methods} \label{sec:related-methods}
We describe several discrete samplers that are related to our work. For a target distribution $\pi(s) \propto \exp(f(s))$ We write the current state as $s_{t}$, a proposal as $s^{*}$ and the next state as $s_{t+1}$.
\subsection{Metropolis--Hastings}
A prototypical approach for sampling from target distributions with intractable normalizing constants is the Metropolis-Hastings algorithm \citep{Metropolis1953MH, Hastings1970mh}.  This method constructs a Markov chain by iteratively proposing a new state $s^{*}$ from a transition kernel $Q(\cdot|s_{t})$ and accepting $s_{t+1}=s^*$ with probability
\begin{align}
     \min\{1, \frac{\exp(f(s^{*}))Q(s_{t}|s^{*})}{\exp(f(s_{t}))Q(s^{*}|s_{t})}\},
     \label{eq:metropolis_prob}
\end{align}
or rejecting $s^*$ and taking $s_{t+1}=s_t$. A key challenge in Metropolis--Hastings lies in selecting an effective proposal distribution $Q(\cdot | \cdot)$. A commonly used choice is a uniform proposal over a neighbor of the current state $s_{t}$. However, such a strategy may often result in slow mixing, particularly when the target distribution is complex.

\subsection{Informed Proposals and Variations}\label{sec:inform_proposal}

\cite{Zanella2020lbp} proposed an interesting approach called informed proposals to design proposal distributions. Given a symmetric kernel $K_{\delta}(\cdot,\cdot)$ and a continuous function $g$, the next state $s^{*}$ is proposed according to
\begin{equation}
    s^{*} \sim Q(s|s_{t}) \propto g(\frac{\pi(s)}{\pi(s_{t})})K_{\delta}(s, s_{t}),
\label{eq:pointwise}
\end{equation}
where $K_\delta$ is indexed by a scale parameter $\delta$ such that $K_\delta(x, \cdot)$
converge weakly to a point mass at $x$ as $\delta\to 0$ for any $x$.
The proposal $s^*$ is then accepted using Metropolis--Hastings probability \eqref{eq:metropolis_prob} or otherwise rejected.
A family of proposal distributions $Q$ in the form \eqref{eq:pointwise} is said to be locally balanced with respect to $\pi$ if $Q$ with kernel $K_\delta$ is reversible with respect to some distribution $\pi_\delta$ such that
$\pi_\delta$ converges weakly to $\pi$ as $\delta \to 0$.
\cite{Zanella2020lbp} showed that $Q$ is locally balanced if and only if  $g(t) = t \cdot g\left(\frac{1}{t}\right)$, for example $g(t) =\sqrt{t}$.
Recently, informed proposals with $g(t) = t^{\gamma}$ for general $\gamma\ge\frac{1}{2}$  (hence possibly non-locally balanced
for $\gamma\not=\frac{1}{2}$) are studied by \cite{sun2023anyscale} and \cite{pynadath2024scale}.

To facilitate drawing from the proposal distribution \eqref{eq:pointwise}, consider a first-order Taylor expansion of $f(s)$ about $s_t$:
\begin{align}
    \frac{\pi(s)}{\pi(s_{t})} \approx \exp(\nabla f(s_{t})^{\T}(s-s_{t}) ) . \label{eq:first-taylor}
\end{align}
For $g(t) =\sqrt{t}$, using the proceeding approximation and taking $K_{\delta}$ to be a Gaussian kernel, $K_{\delta}(x,y)
\propto \exp(-\frac{1}{2\delta}\|x-y\|_{2}^{2})$, in \eqref{eq:pointwise} leads to
\begin{align}
   s^{*} \sim
   Q (s|s_{t}) \propto \exp(\frac{1}{2}\nabla f(s_{t})^{\T}(s-s_{t}))\exp(-\frac{1}{2\delta}\|s-s_{t}\|_{2}^{2}).
\label{eq:NCG1-new}
\end{align}
For a continuous distribution, \eqref{eq:NCG1-new} recovers the MALA proposal, $s^*= s_t + \frac{\delta}{2} \nabla f(s_{t}) + \sqrt{\delta}Z$ with $Z \in \mathcal{N}(0,I)$  \citep{Besag1994mala, Roberts1996mala}. For completeness, see Supplement \ref{NCG_cont} for a direct demonstration.
For a discrete distribution which is of interest in our paper,
\eqref{eq:NCG1-new} can be put in a product form as
\begin{equation}
    s^{*} \sim Q (s|s_{t}) = \prod\limits_{i=1}^{d} \Softmax ([\frac{1}{2}\nabla f(s_{t})_{i}+\frac{1}{\delta}s_{t,i}]s_{i}-\frac{1}{2\delta}s_{i}^{2}).
\label{eq:NCG2-new}
\end{equation}
By the product structure, the proposal distribution \eqref{eq:NCG2-new} can be easily sampled from by drawing the components of $s^*$ independently, hence amenable to parallel computing. The proposal $s^*$ can be accepted with the usual Metropolis--Hastings probability \eqref{eq:metropolis_prob}, i.e.,
\begin{equation}
    \min \{1, \exp(f(s^{*})-f(s_{t}))\frac{Q (s_{t}|s^{*})}{Q (s^{*}|s_{t})}\}, \nonumber
\end{equation}
or otherwise rejected. The sampling method based on \eqref{eq:NCG1-new} or \eqref{eq:NCG2-new} is a discrete analogue of MALA, and in fact is proposed independently by \cite{Rhodes2022GradientMC} and \cite{Zhang2022dmala}. For convenience, this method is called as Norm Constrained Gradient (NCG) algorithm as in \cite{Rhodes2022GradientMC}.

For $g(t) =\sqrt{t}$,
\cite{Grathwohl2021gwg} proposed the Gibbs with Gradients (GWG) algorithm, by using the first-order approximation \eqref{eq:first-taylor} but
taking $K_{\delta}(x, y) =\mathds{1}\{d_{H}(x,y)\leq \delta\}$, where $d_{H}$ is the Hamming distance. The proposal in \eqref{eq:pointwise} leads to:
\begin{align}
s^{*} \sim Q(s|s_{t}) \propto \exp(\frac{1}{2}\nabla f(s_{t})^{\T}(s-s_{t})) \mathds{1}
\{d_{H}(s,s_{t})\leq \delta\}.
\label{eq:GWG-new}
\end{align}
The GWG algorithm can be connected to the Hamming ball sampler \citep{Titsias2017Hamming},
which draws an auxiliary variable $z_t$ uniformly over a Hamming ball centered at $s_t$
and then generates the next state with probabilities proportional to $f$ over the Hamming ball centered at $z_t$.
This method requires evaluating $f$ over all states in the latter Hamming ball.
By comparison, the proposal \eqref{eq:GWG-new} of GWG avoids this computational cost by relying solely on a single gradient computation, $\nabla f(s_{t})$. However, due to the use of the Hamming ball in \eqref{eq:GWG-new}, GWG only allows updates in a few dimensions of $s_t$, making it less efficient for high-dimensional distributions.

\subsection{Auxiliary Variable Gradient} \label{sec:AVG}

\cite{Rhodes2022GradientMC} also proposed another discrete sampler, called Auxiliary Variable Gradient (AVG), by exploiting an auxiliary variable technique analogous to that used by \cite{Titsias2018auxavg} to derive MALA as a marginalized auxiliary variable sampler for a continuous distribution. For completeness, see Supplement Section \ref{AVG_cont} for a description.
For AVG, an auxiliary variable $ z$ is introduced such that $\pi(z|s) \propto \mathcal{N}(z; s,\delta^{2} I)$.
Similarly to \eqref{eq:first-taylor}, consider a first-order approximation to $\pi(s)$ at the current state $s_t$:
$\tilde{\pi}(s; s_{t}) \propto \exp(\nabla f(s_{t})^{\T}(s-s_{t}))$. The joint probability $\pi(s,z) \propto \pi(s) \mathcal{N}(z;s,\delta^{2}I) $ is then approximated as
\begin{align}
\tilde{\pi}  (s,z;s_{t}) & \propto \tilde{\pi}(s;s_{t})\mathcal{N}(z;s,\delta^{2}I) \nonumber \\
& \propto \exp(\nabla f(s_{t})^{\T}(s-s_{t})-\frac{1}{2\delta^{2}}\| z -s\|_{2}^{2}).
\label{AVG-1}
\end{align}
The proposal $s^*$ is obtained by applying Metropolis within Gibbs to sample
$\tilde{\pi}(s,z ;s_{t})$ in \eqref{AVG-1}:
\begin{itemize}
        \item[(i)] Sample
        \begin{align}
            z_{t} \sim \mathcal{N}(z;s_{t}, \delta^{2}I).
        \label{eq:AVG-2z}
        \end{align}
        \item[(ii)]  Propose
        \begin{align}
        s^{*} \sim Q(s|z_{t} ; s_{t})
        \propto \exp(\nabla f(s_{t})^{\T}(s-s_{t})-\frac{1}{2\delta^{2}}\|z_{t}-s\|_{2}^{2}),
        \label{eq:AVG-2}
        \end{align}
where $Q(s|z_{t} \zt{;} s_{t}) = \tilde\pi (s|z_t; s_t)$, the conditional distribution of $s$ given $z=z_{t}$ determined from the joint distribution \eqref{AVG-1}.
\end{itemize}
Similarly to \eqref{eq:NCG1-new} and \eqref{eq:NCG2-new}, the proposal distribution \eqref{eq:AVG-2} can be written in a product form as
\begin{equation}
Q(s|z_{t};s_{t}) = \prod\limits_{i=1}^{d}\Softmax(([\nabla f(s_{t})_{i}+\frac{1}{\delta^{2}}z_{t,i})]s_{i}-\frac{1}{2\delta^{2}}s_{i}^{2}).
\label{eq:AVG-prod}
\end{equation}
The proposal is either accepted ($s_{t+1} = s^{*}$) with the Metropolis within Gibbs probability below, or rejected ($s_{t+1}=s_t$):
\begin{equation}
\min\{1, \exp(f(s^{*})-f(s_{t}))\frac{\mathcal{N}(z_{t}| s^{*},\delta^{2}I)Q(s_{t}|z_{t};s^{*})}{\mathcal{N}(z_{t} | s_{t}, \delta^{2}I)Q(s^{*}|z_{t}; s_{t})}\}.
\label{eq:AVG_acc}
\end{equation}
By the product form, sampling from the proposal distribution \eqref{eq:AVG-prod} is computationally easy.

We make two remarks, associated with two notable differences between NCG and AVG.
First, as mentioned earlier, NCG is a discrete analogue of MALA in the approach of informed proposals. MALA can also be derived by marginalizing an auxiliary variable sampler for a continuous distribution (Supplement Section \ref{AVG_cont}). From the preceding derivation, AVG is a discrete analogue of the auxiliary variable sampler before applying the marginalization to obtain MALA. However, in contrast with the continuous case, marginalizing out the auxiliary variable $z_t$ above does not yield a tractable algorithm and, in particular, does not lead to NCG. Hence both NCG and AVG can be seen as discrete relatives of MALA, but in two different manners. Second, as shown in Supplement Section \ref{sec:AVGrejectfree}, AVG is rejection-free (i.e., always accepting the proposal)
if the target distribution is of the product form, $\pi(s) \propto \exp(a^{\T}s)$,
i.e., the negative potential function (or log of the probability function) $f(s)$ is linear in $s$.
An example of such product distributions is a product of Bernoulli distributions. On the other hand, NCG does not satisfy such a rejection-free property.

\section{Vanilla Discrete-HAMS (V-DHAMS)}

We develop our method, called the Vanilla Discrete-HAMS, through the following steps. First, we augment the target distribution with a momentum variable and construct an auxiliary variable scheme for generating a proposal of both the state and momentum. Then we introduce negation and gradient correction in updating the momentum, and define a generalized acceptance probability to achieve a generalized detailed balance condition.
While these steps can be seen to, in spirit, mirror similar steps underlying
a specific case of HAMS, HAMS-A, for a continuous distribution, our development needs to overcome various
challenges due to the discreteness of the target distribution.
For comparison, we present in Supplement Section~\ref{sec:HAMS-A} a related derivation of HAMS\yz{-A} proposal using similar steps.

\subsection{Auxiliary Variable Scheme} \label{sec:auxiliary_scheme}

We introduce a momentum variable $u  \in \mathbb{R}^{d}$
with a standard normal distribution, $u \sim \mathcal{N}(0, I)$. The augmented target distribution takes the Hamiltonian form:
\begin{equation}
    \pi(s,u) \propto \exp(f(s) -\frac{1}{2}\|u\|_{2}^{2}).
    \label{eq:V-DHAMS-prob}
\end{equation}
To start, consider a first-order approximation to the original target distribution given the current state $s_t$:
$\tilde{\pi}(s; s_{t}) \propto \exp(\nabla f(s_{t})^{\T}(s-s_{t}))$, similarly as in AVG.\
However, with the momentum $u$, we introduce an auxiliary variable,
$z = s+\delta u$, as a linear combination of $s$ and $u$. The conditional distribution of $(s,u)$ given $z$ is approximated as
\begin{align}
        \tilde{\pi}(s, u|z; s_{t} ) \propto \exp(\nabla f(s_{t})^{\T}(s-s_{t})-\frac{1}{2}\|u\|_{2}^{2})
        \mathds{1}\{s+\delta u = z \}.
\label{eq:V-DHAMS1}
\end{align}
Given the current pair $(s_t,u_t)$ and $z_t = s_{t} + \delta u_{t}$, we generate the proposal by sampling from the conditional distribution \eqref{eq:V-DHAMS1} for $z=z_t$ in two steps as follows:
\begin{itemize}
\item[(i)] Propose
\begin{align}
 s^* & \sim Q(s | z_{t} = s_t+\delta u_t; s_t )\nonumber \\
& \propto \exp(\nabla f(s_{t})^{\T}(s-s_{t})-\frac{1}{2}\|\frac{ s_t+\delta u_t-s}{\delta}\|_{2}^{2}),
\label{eq:V-DHAMS-s}
\end{align}
or equivalently
\begin{align}
 s^* \sim \prod_{i=1}^{d} \Softmax( [\nabla f(s_{t})_{i}+\frac{1}{\delta^{2}}(s_{t,i}+\delta u_{t,i})]s_{i}-\frac{1}{2\delta^{2}}s_{i}^{2}), \label{eq:V-DHAMS-prod}
\end{align}
where $Q(s | z_{t}; s_t ) = \tilde\pi(s | z_t; s_t)$, the conditional distribution of $s$ given $z=z_t$ induced from $\tilde{\pi}(s, u|z_t; s_{t} ) $
by setting $u=\frac{z_t-s}{\delta}$.
\item[(ii)] Compute
\begin{align}
        u^{*} &= u_{t} + \frac{s_{t}-s^{*}}{\delta}.  \label{eq:V-DHAMS-u}
\end{align}
\end{itemize}
Compared with AVG, the main difference lies in the maintenance of a momentum variable $u_t$. In fact, if $u_t$ were drawn independently from $\mathcal{N}(0, I)$, the sampling of $s^*$ in \eqref{eq:V-DHAMS-s} and \eqref{eq:V-DHAMS-prod}
would be the same as in \eqref{eq:AVG-2} for AVG.

The proposal scheme above, however, suffers a degeneracy issue.
As seen from the expression \eqref{eq:V-DHAMS-u} for $u^*$, the constraint $ s^* + \delta u^* = s_t + \delta u_t$ is enforced, and hence $s_{t+1} + \delta u_{t+1} = s_t + \delta u_t$ always holds no matter whether $(s^*,u^*)$ is accepted or not.
Because $s$ is a discrete variable with a countable support set $\mathcal{S}$, this relation implies that the sequence of $u_{t}$ is also restricted to a countable set, depending on initial value $u_0$, and hence fails to be ergodic with respect to $u \sim \mathcal{N}(0, I)$.

To address the degeneracy (or non-ergodicity) issue, we introduce an auto-regression step for the momentum $u$. Specifically, we define $z_{t} = s_{t}+\delta u_{t+1/2}$ after generating an intermediate momentum $u_{t+1/2}$ as
\begin{equation}
    u_{t+1/2} = \epsilon u_{t} +\sqrt{1-\epsilon^{2}}Z , \quad Z \sim \mathcal{N}(0, I),
    \label{eq:auto-regressive}
\end{equation}
where $\epsilon\in [0,1)$ and $Z$ is drawn independently.
The auto-regression step \eqref{eq:auto-regressive} is always accepted, being reversible to the momentum distribution $u \sim \mathcal{N}(0, I)$ as well as the augmented target distribution $\pi(s,u)$. With the auto-regression step, the proposal pair $(s^{*}, u^{*})$ is defined by \eqref{eq:V-DHAMS-s}--\eqref{eq:V-DHAMS-u} with $u_t$ replaced by $u_{t+1/2}$ (hence $z_t = s_t+\delta u_{t+1/2}$), i.e.,
\begin{align}
 s^* & \sim Q( s | z_{t}=s_t+ \delta u_{t+1/2}; s_t ),  \label{eq:auto-regressive-s} \\
        u^{*} &= u_{t+1/2} + \frac{s_{t} -s^{*}}{\delta}.  \label{eq:auto-regressive-u}
\end{align}
Then we either accept $(s_{t+1}, u_{t+1}) = (s^{*}, u^{*})$ with the Metropolis--Hastings
probability
\begin{equation}
\min\{1, \frac{\pi(s^{*}, u^{*}) Q(s_{t}, u_{t+1/2}|s^{*}, u^{*} )}{\pi(s_{t}, u_{t+1/2}) Q(s^{*}, u^{*}|s_{t}, u_{t+1/2} )}\},
\label{eq:auto-regressive-acc}
\end{equation}
or reject the proposal and set $(s_{t+1}, u_{t+1}) =(s_{t}, u_{t+1/2})$,
where $Q (s^*,u^*|s_t, u_{t+1/2} )$, the transition probability from
$(s_t,u_{t+1/2})$ to $(s^*,u^*)$, can be evaluated as $Q( s^* | z_t=s_t+\delta u_{t+1/2}; s_t) $ in \eqref{eq:auto-regressive-s}, and $Q(s_{t}, u_{t+1/2}|s^*, u^*)$ is evaluated as $Q(s_{t}|z_t = s^*+\delta u^*; s^*)$.

We make three additional remarks. First,
in the special case of $\epsilon=0$,
the sampling scheme from $(s_t,u_t)$ to $(s_{t+1},u_{t+1})$,
defined by \eqref{eq:auto-regressive}--\eqref{eq:auto-regressive-acc}, reduces to AVG, with $u_{t+1/2}\sim \mathcal{N}(0,I)$ drawn independently. See Supplement Section \ref{limitations}.
Second, we show in Supplement Section~\ref{rejection_free} that for any $\epsilon\in [0,1)$, the sampling scheme from $(s_t,u_t)$ to $(s_{t+1},u_{t+1})$  becomes rejection-free, i.e., the acceptance probability \eqref{eq:auto-regressive-acc} is always 1, for a product target distribution $\pi(s) \propto \exp(a^{\T}s)$. This extends a similar rejection-free property of AVG mentioned in Section~\ref{sec:AVG}. Finally, for a continuous distribution, while the general HAMS proposal scheme corresponds to marginalizing an auxiliary variable scheme as noted in \cite{Song2023hams},
the HAMS-A proposal as a degenerate case can be shown to be similar to \eqref{eq:V-DHAMS-s}--\eqref{eq:V-DHAMS-u} without marginalization.
An auto-regression step is not needed to overcome non-ergodicity. See Supplement Section \ref{sec:HAMS-A} for further information.
For completeness, we also discuss in Supplement Sections~\ref{sec:auto_regressiveHAMS} and \ref{sec:insymmetricHAMS}, two possible ways of incorporating auto-regression in HAMS and their consequences.

\subsection{Momentum Negation} \label{sec:negation}
To induce irreversible exploration as in HAMS for continuous distributions, we modify
the proposal \eqref{eq:auto-regressive}--\eqref{eq:auto-regressive-acc} with auto-regression
by reversing the intermediate momentum from $u_{t+1/2}$ to $-u_{t+1/2}$ and then applying the proposal scheme. The modified proposal from $(s_t, u_{t+1/2})$ to $(s^*, u^*)$ is
\begin{align}
 s^* & \sim Q( s | z_{t}=s_t- \delta u_{t+1/2}; s_t) \nonumber \\
 & \quad \propto
        \exp(\nabla f(s_{t})^{\T}(s-s_{t})-\frac{1}{2}\|\frac{ s_t-\delta u_{t+1/2} -s}{\delta}\|_{2}^{2}),  \label{eq:negation-s} \\
        u^{*} &= - u_{t+1/2} + \frac{s_{t} -s^{*}}{\delta},  \label{eq:negation-u}
\end{align}
where as in \eqref{eq:V-DHAMS-s},  $Q(s|z_{t};s_t)$ is determined from $\tilde{\pi}(s, u|z_t; s_{t})$. The proposal \eqref{eq:negation-s} can also be written in a product form as
\begin{align}
 s^* \sim \prod_{i=1}^{d} \Softmax( [\nabla f(s_{t})_{i}+\frac{1}{\delta^{2}}(s_{t,i}-\delta u_{t+1/2,i})]s_{i}-\frac{1}{2\delta^{2}}s_{i}^{2}).
 \label{eq:negation-prod}
\end{align}
The new proposal $(s^*,u^*)$ from $(s_t, u_{t+1/2})$ is accepted or rejected, but with an acceptance probability
different from the standard Metropolis-Hastings probability to account for the momentum reversal.
Specifically, we either accept $(s_{t+1},u_{u+1}) = (s^*,u^*)$ with the following probability
\begin{equation}
 \min\{1, \frac{\pi(s^{*}, -u^{*}) Q (s_{t}, -u_{t+1/2}|s^{*}, -u^{*})}{\pi(s_{t}, u_{t+1/2}) Q(s^{*}, u^{*}|s_{t}, u_{t+1/2})}\},
\label{eq:negation-acc}
\end{equation}
or reject the proposal and set $(s_{t+1},u_{t+1}) = (s_t, -u_{t+1/2})$,
where $Q(s^{*}, u^{*}|s_{t}, u_{t+1/2})$ is the transition probability under \eqref{eq:negation-s}--\eqref{eq:negation-prod}, and can be evaluated as $Q( s^* | z_t=s_t -\delta u_{t+1/2}; s_t) $ in \eqref{eq:auto-regressive-s}. Note that $u_{t+1} = u^*$ upon acceptance, but $u_{t+1} = -u_{t+1/2}$ in case of rejection.
The resulting transition can be shown to satisfy generalized detailed balance as follows.

\begin{proposition}\label{prop:1}
Let $K_0 (s_{t+1}, u_{t+1}|s_{t}, u_{t+1/2})$ be the transition kernel from $(s_t,  u_{t+1/2} )$ to $(s_{t+1},u_{t+1})$ defined by
the proposal scheme \eqref{eq:negation-s}--\eqref{eq:negation-prod} and acceptance probability \eqref{eq:negation-acc}.
Then the generalized detailed balance condition \eqref{eq:negation_dbc} holds:
\begin{equation}
    \pi(s_{t}, u_{t+1/2}) K_0 (s_{t+1}, u_{t+1}|s_{t}, u_{t+1/2}) = \pi(s_{t+1}, -u_{t+1}) K_0 (s_{t}, -u_{t+1/2}|s_{t+1}, -u_{t+1}).
    \label{eq:negation_dbc}
\end{equation}
Moreover, the augmented target distribution $\pi(s,u)$ is a stationary distribution of the resulting Markov chain.
\end{proposition}

Our algorithm described so far falls in the class of generalized Metropolis--Hastings algorithms from \cite{Song2023hams}, while using an augmented distribution with a momentum variable. See Supplement Section \ref{sec:GMH} for more information.
The generalized detailed balance above differs from the standard detailed balance in that the backward transition is defined with a reversed momentum.
Such a generalized condition is often used in the derivation of continuous samplers, such as Underdamped Langevin Sampling \citep{Bussi2207UDL} and HAMS \citep{Song2023hams}. Our work appears to be the first in exploiting generalized detailed balance together with gradient information for discrete distributions.

The transition from $(s_t,u_t)$ to $(s_t, u_{t+1/2})$ by \eqref{eq:auto-regressive} with $\epsilon\in [0,1)$ and then
to $(s_{t+1}, u_{t+1})$ with momentum negation by \eqref{eq:negation-s}--\eqref{eq:negation-prod} and acceptance probability \eqref{eq:negation-acc}
can be equivalently presented as a transition which applies an auto-regression step \eqref{eq:auto-regressive} with $\epsilon\in (-1,0]$
and then the proposal scheme \eqref{eq:V-DHAMS-s}--\eqref{eq:V-DHAMS-u} with $u_t$ replaced by $u_{t+1/2}$ and subsequent acceptance-rejection
with the standard Metropolis--Hastings probability \eqref{eq:auto-regressive-acc}. See Supplement Section \ref{eq:negation_equi} for details.
In spite of this equivalence, our presentation here separates the auto-regression step to address the non-ergodicity issue and the momentum negation to achieve irreversibility.

When the target distribution is of the product form $\pi(s) \propto \exp(a^{\T}s)$,
our algorithm with momentum negation can be shown to remain rejection-free as noted in Section~\ref{sec:auxiliary_scheme},
provided that the generalized acceptance probability \eqref{eq:negation-acc} is used
(Supplement Section \ref{rejection_free}).

\subsection{Gradient Correction for Updating Momentum}\label{sec:modification}

A potential limitation of the proposal \eqref{eq:negation-s}--\eqref{eq:negation-acc} is that, when updating the momentum $u^{*}$, the gradient $\nabla f(s^{*})$ at the newly generated $s^*$ is ignored. Such gradient information from a proposal state $s^*$ (not the current state $s_t$) has been introduced for Hamiltonian-based continuous samplers, such as HMC \citep{duane1987HMC, neal2011mcmc} through the leap-frog scheme
and HAMS \citep{Song2023hams} through a direct gradient correction.
Motivated by this consideration, we keep \eqref{eq:negation-s} for $s^*$ and modify \eqref{eq:negation-u} for $u^*$ to
\begin{align}
        u^{*} &= -u_{t+1/2}+\frac{s_{t}-s^{*}}{\delta} + \phi (\nabla f(s^{*})-\nabla f(s_{t})),
    \label{eq:mod1}
\end{align}
where $\phi \geq 0$ is a tuning parameter. The update \eqref{eq:mod1} can be rearranged to
\begin{align}
        -u_{t+1/2} &= u^{*} +\frac{s^{*}-s_{t}}{\delta} + \phi (\nabla f(s_{t})-\nabla f(s^{*})).
    \label{eq:mod1-rev}
\end{align}
Importantly, the $u$-updates $(s_t, u_{t+1/2}, s^*) \mapsto u^*$ in \eqref{eq:mod1} and  $(s^*, -u^*, s_t) \mapsto - u_{t+1/2}$ in \eqref{eq:mod1-rev} satisfy the \textit{same} mapping deterministically.
Hence the forward transition $(s_t, u_{t+1/2}) \mapsto (s^*,u^*)$ under \eqref{eq:negation-s} and \eqref{eq:mod1}
can be reduced to only the $s$-update $(s_t,u_{t+1/2}) \mapsto s^*$,
and the backward transition $(s^*,-u^*) \mapsto (s_t, -u_{t+1/2})$ can be reduced to only the $s$-update $(s^*,-u^*) \mapsto s_t$. To distinguish the joint proposal of $(s,u)$ using \eqref{eq:negation-s} and \eqref{eq:mod1} from that using \eqref{eq:negation-s} and \eqref{eq:negation-u}, we denote the new proposal with gradient correction as $Q_{\phi}(s^{*}, u^{*}|s_{t}, u_{t+1/2})$.
Then we either accept $(s_{t+1},u_{u+1}) = (s^*,u^*)$ with the following probability
\begin{equation}
   \min\{1, \frac{\pi(s^{*}, -u^{*}) Q_{\phi}(s_{t}, -u_{t+1/2}|s^{*}, -u^{*})}{\pi(s_{t}, u_{t+1/2}) Q_{\phi} (s^{*}, u^{*}|s_{t}, u_{t+1/2})}\},
    \label{eq:mod-acc}
\end{equation}
or reject the proposal and set $(s_{t+1},u_{t+1}) = (s_t, -u_{t+1/2})$. The forward transition probability $Q_{\phi}(s^{*}, u^{*}|s_{t}, u_{t+1/2})$
can be evaluated as $Q( s^* | z_t=s_t -\delta u_{t+1/2}; s_t) $,
and backward transition probability $Q_{\phi}(s_{t}, -u_{t+1/2}|s^{*}, -u^{*})$ can be evaluated as $Q (s_t | z_t = s^* + \delta u^*; s^*)$, as in \eqref{eq:negation-s}. The acceptance probability \eqref{eq:mod-acc} appear the same as \eqref{eq:negation-acc}, independently of $\phi$, but the $u$-update is modified, depending on $\phi$.
The resulting transition can be shown to also satisfy generalized detailed balance.
\begin{proposition}\label{prop:2}
Let $K_\phi (s_{t+1}, u_{t+1}|s_{t}, u_{t+1/2})$ be the transition kernel from $(s_t, u_{t+1/2})$ to $(s_{t+1},u_{t+1})$ defined by
the proposal scheme \eqref{eq:negation-s} and \eqref{eq:mod1} and acceptance probability \eqref{eq:mod-acc}.
Then the generalized detailed balance holds:
\begin{align}
    \pi(s_{t}, u_{t+1/2}) K_\phi (s_{t+1}, u_{t+1}|s_{t}, u_{t+1/2}) = \pi(s_{t+1}, -u_{t+1}) K_\phi (s_{t}, -u_{t+1/2}|s_{t+1}, -u_{t+1}),
    \label{eq:mod-dbc}
\end{align}
Moreover, the augmented target distribution $\pi(s,u)$ is a stationary distribution of the resulting Markov chain.
\end{proposition}
When the target distribution is of the product form $\pi(s) \propto \exp(a^{\T}s)$, the correction term with $\phi$ vanishes in \eqref{eq:mod1},
thus preserving the rejection-free property as noted in Section \ref{sec:negation}.
We summarize all the operations from Sections  \ref{sec:auxiliary_scheme}--\ref{sec:modification} in Algorithm \ref{algo:V-DHAMS}, which we call the Vanilla Discrete-HAMS (V-DHAMS), prior to introducing over-relaxation in the next section.
\begin{algorithm}[tbp]
With current state and momentum $(s_{t}, u_{t})$
\begin{itemize}
\item Generate $u_{t+1/2} = \epsilon u_{t} +\sqrt{1-\epsilon^{2}}Z$, where $Z \sim \mathcal{N}(0,I)$.
\item Compute $z_{t} = s_{t} -\delta u_{t+1/2}$.
\item Propose $s^*$ by \eqref{eq:negation-s} or equivalently \eqref{eq:negation-prod},
and compute $u^{*}$ by \eqref{eq:mod1}.
\item Accept $(s_{t+1}, u_{t+1}) = (s^{*}, u^{*})$ with probability \eqref{eq:mod-acc}, or otherwise set $(s_{t+1}, u_{t+1}) = (s_{t}, -u_{t+1/2})$.
\end{itemize}
\caption{Vanilla Discrete-HAMS}
\label{algo:V-DHAMS}
\end{algorithm}

\section{Over-relaxation}
The several operations discussed so far, auto-regression, negation, and gradient correction,
are all about updating the momentum variable. In this section, we introduce over-relaxation for updating the state variable $s$,
motivated by the fact that Gaussian over-relaxation is one of the key steps in deriving the HAMS sampler for continuous distributions.

As a brief review, the Gaussian over-relaxation technique \citep{Adler1981overrelaxation} works as follows.
Given the current variable $x_0$ with stationary distribution $\mathcal{N}(\mu,\Sigma)$
(implicitly conditioned on other variables as in Gibbs sampling), the standard ``random-walk'' approach is to draw the next variable $x_1$ from
$\mathcal{N}(\mu,\Sigma)$ independently of $x_0$. By relaxation, $x_1$ is generated as
\begin{equation}
    x_1 = \mu + \alpha(x_0 -\mu) + \sqrt{1-\alpha^{2}}\Sigma^{1/2}Z,
    \label{eq:Gauss-overrelax}
\end{equation}
where $Z \sim \mathcal{N}(0,I)$ and $\alpha \in (-1,1)$ is a tuning parameter.
Taking $\alpha=0$ recovers the standard ``random-walk'' approach, whereas
$\alpha<0$ (or $>0$), corresponding over-relaxation (or under-relaxation), induces positive or negative correlations between $x_1$ and $x_0$.
The over-relaxation is known to be capable of suppressing random-walk behaviour in Gibbs sampling \citep{Neal1998overrelax}.

We aim to design an over-relaxation scheme for discrete distributions, while achieving properties analogous to those of Gaussian over-relaxation.
In addition to handling discrete distributions, another difference of our work from the conventional use of over-relaxation in Gibbs sampling
is that over-relaxation is exploited as a scheme to generate a proposal which can be accepted or rejected
by generalized Metropolis--Hastings sampling.
In such a scheme, $x_0$ may not be distributed as $\mathcal{N}(\mu, \Sigma)$ in stationarity, and $\mathcal{N}(\mu, \Sigma)$ serves as a reference distribution to generate proposal $x_1$. An example related to this use of over-relaxation can be found in the construction of the HAMS sampler in \cite{Song2023hams}. See Supplement Section \ref{sec:HAMS-A}.

In spite of these complications,
a simplifying feature in our previous development is that
the proposal distribution \eqref{eq:negation-prod} in Vanilla Discrete-HAMS for the state variable takes a product form, i.e., a joint distribution with independent components.
Therefore, our primary task involves designing an over-relaxation scheme for univariate discrete distributions.
For multivariate distributions in a product form, over-relaxation can be applied component by component.

\subsection{Over-relaxation for discrete distributions} \label{sec:overrelax-general}
Given a univariate discrete variable $x_0$ and a discrete reference distribution $p(x)$, we aim to construct an over-relaxation scheme,
which generates $x_1$ from a conditional probability denoted as $p(x_1|x_0)$, such that the following properties hold:
\begin{itemize}
\item[(i)] For all possible values of $(x_0,x_1)$, still denoted as $(x_0,x_1)$ by some abuse of notation,
\begin{align}
    p(x_0) p (x_1|x_0) = p(x_1) p(x_0|x_1).\label{eq:eq:overrelaxation_dbc}
\end{align}
In other words, as a transition kernel, $p(x_1|x_0)$ satisfies detailed balance (or reversibility) with respect to $p(x)$.
The property \eqref{eq:eq:overrelaxation_dbc} implies (hence is stronger than) that
if $x_0 \sim p(x)$, then $x_1 \sim p(x)$ marginally.

\item[(ii)] $x_1$ can depend on $x_0$ in a range of degrees controlled by some tuning parameter.
\end{itemize}
Equivalently, detailed balance \eqref{eq:eq:overrelaxation_dbc} can be described as a symmetry: $ p(x_0,x_1) = p(x_1,x_0)$, where $ p(x_0,x_1) = p(x_0) p(x_1|x_0)$ is
the joint probability of $(x_0,x_1)$ if $x_0\sim p(x)$.

We distinguish two settings for applying discrete over-relaxation. In the first setting, over-relaxation is used in a conventional sense, similarly as in \cite{Neal1998overrelax}. Namely, $x_0$ is the current variable with stationary distribution $p(x)$ (implicitly conditioned on other variables, for example, the remaining components of a vector-valued state in Gibbs sampling), and $x_1$ can be either drawn independently from $p(x)$ (i.e., without over-relaxation) or
drawn using the proposed scheme (i.e., with relaxation) such that $x_1$ depends on $x_0$ but admits the same stationary distribution $p(x)$.
In this setting, $x_1$ is always accepted as the next variable. Second, more generally, in the setting where $x_0$ may not be distributed as $p(x)$ in stationarity,
an over-relaxation scheme serves to generate a proposal $x_1$, which can be either accepted or rejected subsequently.

First, for discrete variables $x_j \sim p(x)$ in stationarity, we use the following representation of $x_j$ through a latent, continuous variable $w_j$ for $j=0,1$:
\begin{itemize}\label{itm:represent}
\item[(i)] $w_j \sim \Unif( [0,1) ) \label{eq:represent1}$ marginally,
\item[(ii)] $ x_j = \xi$ if and only if $ F(\xi^-) \leq w_j < F(\xi)$ for $j=0,1$ and any $\xi \in\mathbb{R} \label{eq:represent2}$,
where $F(\cdot)$ is the cumulative distribution function of $p(\cdot)$,
and $F(\xi^- ) = \lim_{x \to \xi^-} F(x)$.
\end{itemize}
Then the conditional distribution of $x_j$ given $w_j$ is degenerate, with probability 1 at $\xi$ such that $ F(\xi^-) \le w_j < F(\xi)$.
Conversely, the conditional distribution of $w_j$ given $x_j$ is $\Unif ( [ F(x_j^-), F(x_j)))$.
Such a representation was introduced in copula models for discrete variables \citep{Genest_Neslehova_2007Dcopula, Smith2012Dcopula}.

Next, we provide a sampling scheme which given $w_0 \sim \Unif ( [0,1))$, generates $w_1$ such that the transition from $w_0$ to $w_1$ is reversible with respect to $\Unif ( [0,1))$, hence
$w_1\sim \Unif ( [0,1))$ marginally, and $w_1$ exhibits various degrees of dependency on $w_0$.

\begin{figure}[tbp]
\includegraphics[width=0.4\linewidth]{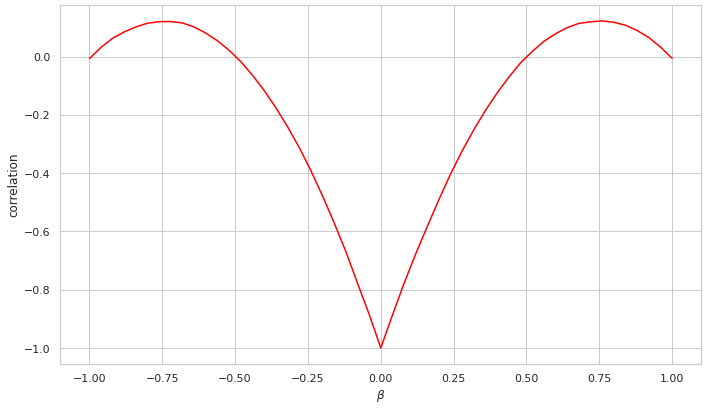}
\centering
\caption{Plot of correlations between $w_0$ and $w_1$}
\label{pic:corrplot}
\end{figure}

\begin{proposition}\label{prop:disc_embed}
Draw $w_0\sim \Unif([0,1))$ and $\tilde w \sim \Unif([0,1))$, independently of $w_0$. Set
    \begin{align}
        w_{1} = ( - w_{0} +\beta \tilde{w} )\%1 ,
        \label{eq:eq:overrelaxation-w}
    \end{align}
where $\beta\in [-1, 1]$ and $ c \% 1 = c - \lfloor c \rfloor$ for $c \in \mathbb{R}$.
Then  $w_{1}$ is also $\Unif([0,1))$. Moreover detailed balance is satisfied as follows:
\begin{align}
 p(w_0) p(w_1 | w_0) = p (w_1) p(w_0 | w_1),  \label{eq:w-symmetry}
\end{align}
where $p(w_0)$ and $p(w_1 | w_0)$ denote, respectively, the marginal density of $w_0$ and conditional density of $w_1$ given $w_0$.
Detailed balance \eqref{eq:w-symmetry} is equivalent to the symmetry $p(w_0,w_1) = p(w_1,w_0)$, where
$ p(w_0,w_1) =  p(w_0) p(w_1 | w_0) $ is the joint density of $(w_0,w_1)$.
\end{proposition}
We provide several remarks on the construction in~\eqref{eq:eq:overrelaxation-w}. See Figure~\ref{pic:corrplot} for a plot of correlations of $w_0$ and $w_1$.
First, when $\beta = \pm 1$, the variable $w_{1}$ is independent of $w_{0}$ (Supplement Section \ref{eq:random-walk}). Second, when $\beta=0$, the correlation of $w_0$ and $w_1$ attains a minimum of $-1$. Finally, as related to \eqref{eq:eq:overrelaxation-w},
the scheme defined by setting $w_{1} = ( w_{0} +\beta \tilde{w} )\%1$ also results in $w_1 \sim \Unif([0,1))$,
but the transition from $w_0$ to $w_1$ does not satisfy detailed balance \eqref{eq:w-symmetry}
and, instead, the transition from $w_0$ to $1-w_1$ satisfies detailed balance with respect to $\Unif ( [0,1))$.

We obtain an over-relaxation scheme (Algorithm \ref{algo:over-relaxation}) by combining the discrete variable representation discussed earlier and uniform variable generation in \eqref{eq:eq:overrelaxation-w}. As desired, the following properties are satisfied.
First, detailed balance \eqref{eq:eq:overrelaxation_dbc} is satisfied, so that if $x_0 \sim p(x)$, then $x_1 \sim p(x)$.
See Supplement Sections \ref{sec:uniformity} and \ref{sec:overrelaxation-symmetricity} for proofs. Second, $x_1$ can depend on $x_0$ in various ways controlled by parameter $\beta$.
Particularly, when $x_0\sim p(x)$ from a Bernoulli distribution, the correlation of $-1$ is obtained
between $x_1$ and $x_0$ by taking $\beta=0$ in \eqref{eq:eq:overrelaxation-w} (Supplement Section \ref{sec:couplings}).

\begin{algorithm}[tbp]
 \begin{itemize}
       \item Given a discrete variable $x_0$ and a discrete reference distribution $p(\cdot)$ with CDF $F(\cdot)$.
       \item Sample  $w_{0} \sim Unif([F(x_{0}^{-}), F(x_{0})))$ and $\tilde{w} \sim Unif([0,1))$ independently.
       \item Compute $w_{1} = ( -w_{0}+ \beta \tilde{w} )\%1$.
       \item Output $x_1$ such that $F(x_1^- ) \leq w_1 < F(x_1)$.
\end{itemize}
\caption{Discrete Over-relaxation}
\label{algo:over-relaxation}
\end{algorithm}

We stress that Algorithm \ref{algo:over-relaxation} can be applied to obtain $x_1$ as a ``proposal'' from $x_0$, whether or not $x_0 \sim p(x)$ in stationarity.
If $x_0 \not\sim p(x)$, then in general $x_1 \not\sim p(x)$.
In such a situation, as in the later application of Algorithm \ref{algo:over-relaxation} to Vanilla Discrete-HAMS,
$x_1$ as a proposal from $x_0$ can be accepted or rejected according to the Metropolis--Hastings probability or a generalization,
which requires evaluating the conditional distribution $p(x_1 | x_0)$ under Algorithm \ref{algo:over-relaxation}.
Fortunately, such conditional distributions implied from Algorithm \ref{algo:over-relaxation} can be directly calculated for any $F(\cdot)$ and any $\beta$,
facilitating the application of which as a proposal scheme.
See Supplement Section \ref{sec:overrelax_prob} for the explicit formulas of $p(x_1 | x_0)$.

\subsection{Over-relaxed Discrete-HAMS (O-DHAMS)}\label{sec:overrelaxed-HAMS}
We return to the Vanilla Discrete-HAMS (Algorithm \ref{algo:V-DHAMS}) and incorporate over-relaxation for updating the state variable. Specifically,
instead of the independent sampling $s^* \sim Q (s | z_t= s_t - \delta u_{t+1/2}; s_t)$,
we apply the over-relaxation scheme (Algorithm \ref{algo:over-relaxation}) to draw a proposal $s^*$ component by component
with the reference distribution $Q (s | z_t= s_t - \delta u_{t+1/2}; s_t)$,
i.e., apply Algorithm \ref{algo:over-relaxation} with
\begin{align}
 \begin{split}
  x_0 & = \text{$i$th component of $s_t$},  \\
  p(x) &= \text{$i$th component in $Q (s | z_t= s_t - \delta u_{t+1/2}; s_t)$}, \\
  x_1 &= \text{$i$th component of $s^*$}
\end{split}
\label{eq:over-HAMS0}
\end{align}
for each dimension $i$ of state $s_{t}$. This component-wise application is feasible because
$Q (s | z_t= s_t - \delta u_{t+1/2}; s_t)$ is of a product form \eqref{eq:negation-prod} , with independent components.
The resulting transition probability of $s^*$ from $s_t$, denoted as $\tilde Q(s^* | z_t= s_t - \delta u_{t+1/2}; s_t)$,
is also a product of those individual components of $s^*$ from $s_t$, where $i$th component's distribution is determined as the conditional distribution $ p(x_1|x_0)$ in Section \ref{sec:overrelax-general}.

The remaining steps of Over-relaxed Discrete-HAMS remains similar as in Algorithm \ref{algo:V-DHAMS}. We still generate the new momentum $u^*$ by \eqref{eq:mod1}
with gradient correction depending on $\phi$.
Then we either accept $(s_{t+1},u_{u+1}) = (s^*,u^*)$ with the following probability
\begin{equation}
 \min\{1, \frac{\pi(s^{*}, -u^{*}) \tilde Q_{\phi} (s_{t}, -u_{t+1/2}|s^{*}, -u^{*})}{\pi(s_{t}, u_{t+1/2}) \tilde Q_{\phi} (s^{*}, u^{*}|s_{t}, u_{t+1/2})}\},
    \label{eq:over-HAMS-acc}
\end{equation}
or reject the proposal and set $(s_{t+1},u_{t+1}) = (s_t, -u_{t+1/2})$,
where the forward transition probability $\tilde Q_{\phi}(s^{*}, u^{*}|s_{t}, u_{t+1/2})$
can be evaluated as $\tilde Q( s^* | z_t=s_t -\delta u_{t+1/2}; s_t) $,
and backward transition probability $\tilde Q_{\phi}(s_{t}, -u_{t+1/2}|s^{*}, -u^{*})$ can be evaluated as $\tilde Q (s_t | z_t = s^* + \delta u^*; s^*)$. The resulting transition can be shown to also satisfy generalized detailed balance (Supplement Section~\ref{gdbc}).

\begin{proposition}\label{prop:2-new}
Let $\tilde K_\phi (s_{t+1}, u_{t+1}|s_{t}, u_{t+1/2})$ be the over-relaxed transition kernel from $(s_t,  u_{t+1/2})$ to $(s_{t+1},u_{t+1})$ defined by
the proposal scheme \eqref{eq:over-HAMS0} and \eqref{eq:mod1} and acceptance probability \eqref{eq:over-HAMS-acc}.
Then the generalized detailed balance holds:
\begin{equation}
    \pi(s_{t}, u_{t+1/2}) \tilde K_\phi (s_{t+1}, u_{t+1}|s_{t}, u_{t+1/2}) = \pi(s_{t+1}, -u_{t+1}) \tilde K_\phi (s_{t}, -u_{t+1/2}|s_{t+1}, -u_{t+1}).
    \label{eq:overhams-dbc}
\end{equation}
Moreover, the augmented target distribution $\pi(s, u)$ is a stationary distribution of the resulting Markov chain.
\end{proposition}

We summarize, in Algorithm \ref{algo:O-DHAMS}, the Over-relaxed Discrete-HAMS method by incorporating over-relaxation as well as the operations from Sections \ref{sec:auxiliary_scheme}--\ref{sec:modification}. We also make two remarks on this new sampler. First, when $\beta = \pm1$ in \eqref{eq:eq:overrelaxation-w}, $s^*$ is drawn independently from the reference distribution, and Over-relaxed Discrete-HAMS (Algorithm \ref{algo:O-DHAMS}) reduces to the Vanilla Discrete-HAMS (Algorithm \ref{algo:V-DHAMS}). As remarked at the end of Section \ref{sec:modification}, the Vanilla Discrete-HAMS algorithm is rejection-free for
a target distribution in the product form $\pi(s) \propto \exp(a^{\T}s)$.
The Over-relaxed Discrete-HAMS can be shown to preserve this property, provided that the generalized acceptance probability \eqref{eq:over-HAMS-acc} is used.
The rejection-free property also depends on the fact that our over-relaxation scheme (Algorithm~\ref{algo:over-relaxation})
satisfies the detailed balance \eqref{eq:eq:overrelaxation_dbc}. See Supplement Section \ref{sec:ohams_rejection-free} for details.

\begin{algorithm}[tbp]
With current state and momentum $(s_{t}, u_{t})$
\begin{itemize}
\item Generate $u_{t+1/2} = -\epsilon u_{t} +\sqrt{1-\epsilon^{2}}Z$, where $Z \sim \mathcal{N}(0,I)$.
\item Compute $z_{t} = s_{t} -\delta u_{t+1/2}$.
\item  Calculate the original proposal distribution $Q(s|z_t =s_{t}-\delta u_{t+1/2}; s_t)$ from \eqref{eq:negation-s} or \eqref{eq:negation-prod} as reference distribution, and \\
Propose $s^*$  from the over-relaxed proposal probability $\tilde{Q}(s|z_t =s_{t}-\delta u_{t+1/2}; s_t)$ via component-wise application of Algorithm \ref{algo:over-relaxation}.
\item Compute $u^{*}$ by \eqref{eq:mod1}.
\item Accept $(s_{t+1}, u_{t+1}) = (s^{*}, u^{*})$ with probability \eqref{eq:over-HAMS-acc}, or otherwise set $(s_{t+1}, u_{t+1}) =(s_{t}, -u_{t+1/2})$.
\end{itemize}
\caption{Over-relaxed Discrete-HAMS}
\label{algo:O-DHAMS}
\end{algorithm}

\vspace{-.1in}
\section{Numerical Studies}

We present simulation studies comparing various samplers, including Metropolis, GWG, NCG, AVG, V-DHAMS, O-DHAMS.
Throughout, V-DHAMS refers to Vanilla Discrete-HAMS (Algorithm \ref{algo:V-DHAMS}) and O-DHAMS refers to Over-relaxed Discrete-HAMS (Algorithm \ref{algo:O-DHAMS}).
NCG and AVG are described in Sections \ref{sec:inform_proposal} and \ref{sec:AVG}.
The Metropolis and GWG samplers used for ordinal distributions as in later Section \ref{sec:exp_gauss} and \ref{sec:poly_mix} are detailed in Supplement Section~\ref{sec:prototype_variant}.

To evaluate the performance of samplers, we use the following metrics: total variation distance (TV-distance) and effective sample size (ESS).
The TV-distance is an important metric for evaluating convergence of Markov chains \citep{liu2001montetv}.
For each chain, we compute the TV-distance between the empirical distribution $\hat{\pi}$ and the target distribution $\pi$,
which for discrete distributions with support $\mathcal{S}$ is defined as
\begin{align}
    \TV(\pi, \hat{\pi}) = \frac{1}{2} \sum\limits_{s \in \mathcal{S}} |\pi(s) - \hat{\pi}(s)|. \nonumber
\end{align}
Similarly as in \cite{ma2019tv} and \cite{Jiang2020tv}, we report the mean and standard deviation of the TV-distances across multiple chains from repeated independent runs. The ESS is another important metric for evaluating the mixing efficiency of Markov chains.
Instead of relying on estimation of auto-correlations (which can be sensitive), we use the multiple chain approach, based on \cite{GelmanRubin1992ESS},
by incorporating both within-chain and between-chain variances from multiple independent chains. Suppose we are interested in the mean of a variable $x$ and have $M$ chains, each of length $T$. Let $x_{m,t}$ be the variable of interest from the $t$-th draw of the $m$-th chain. The ESS is computed by:
\begin{equation}
    \ESS (x)= T\frac{W}{B}, \; W = \frac{1}{M(T-1)}\sum\limits_{m,t} (x_{m,t}-\bar{x}_{m,\cdot})^{2},\; B = \frac{T}{M-1}\sum\limits_{m}(\bar{x}_{m,\cdot}-\bar{x})^{2},
    \label{eq:ess}
\end{equation}
where $\bar{x}_{m,\cdot} = \frac{1}{T}\sum\limits_{t}x_{m,t}$ is the mean of the $m$-th chain and $\bar{x} = \frac{1}{M}\sum\limits_{m} \bar{x}_{m,\cdot}$ is the overall mean across all chains. This ESS estimator \eqref{eq:ess} is also known as the \textit{multivariate batch mean ESS estimator} \citep{vats2018Ess}.

\subsection{Discrete Gaussian}
\label{sec:exp_gauss}
\begin{figure} [tbp]
    \centering
    \includegraphics[width=0.55\linewidth]{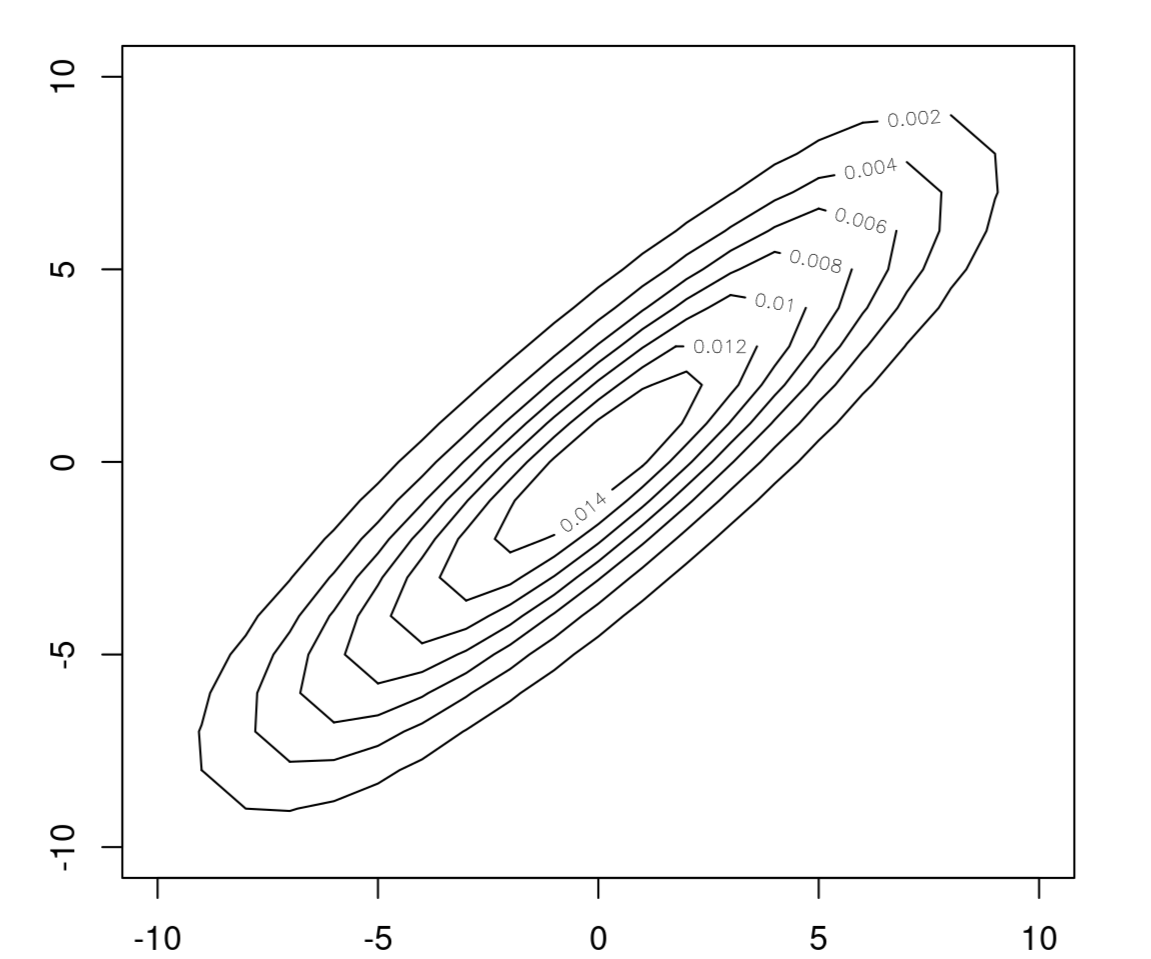} 
    \caption{Contour plot of discrete Gaussian distribution}
    \label{fig:heatmap} 
\end{figure}

The discrete Gaussian distribution mimics Gaussian: it is restricted to a $d$-dimensional lattice $\mathcal{S} \in \mathbb{R}^{d}$, $\mathcal{S} = \{-k, -(k-1), \cdots, -1,0, 1, \cdots, (k-1), k\}^{d}$, and the negative potential function $f(s)$ is of the quadratic form
\begin{equation}
   f(s) = -\frac{1}{2}s^{\T}\Sigma^{-1}s, \nonumber
\end{equation}
where $\Sigma$ is a positive definite matrix.
This distribution is also known as the lattice Gaussian distribution, which has important applications to cryptography \citep{Kschischang1993Dgaussian, Micciancio2007Dgaussian}. In our experiment, $\Sigma$ is specified in an equi-correlation form as
\begin{equation}
    \Sigma = \sigma^{2}[\rho\textbf{1}\textbf{1}^{\T}+(1-\rho)I],
    \label{eq:disgau_Sigma}
\end{equation}
where $\sigma^2$ mimics the variance of each coordinate, and $\rho$ mimics correlation. In our example, $d=8$, $k=10$, $\sigma=5$ and $\rho = 0.9$. The marginal distribution of the first two coordinates is visualized using a contour plot in Figure \ref{fig:heatmap}. Given the equi-correlation structure of \eqref{eq:disgau_Sigma}, the marginal distribution of any two coordinates is also visualized by the same contour.
For our discrete Gaussian distribution, there are a total of $21^8$ possible states. By direct enumeration, the exact distributions of up to any four-dimensional marginals, $\pi(s_{i_{1}}, s_{i_{2}}, s_{i_{3}}, s_{i_{4}})$, can be computed at some manageable computational cost.

For parameter tuning with each sampler, we search over 50 different parameters. For each parameter, we run 50 independent chains, each of length 5,500
after 1,000 burn-in draws.
As in \cite{sun2023discs}, we select the best parameter by the highest ESS of the negative potential function $f(s)$. Details of the tuning process are provided in Supplement Section \ref{sec:ordinal_tuning}. After completing parameter tuning, we conduct 100 independent chains for each sampler using the optimal parameter setting. For each chain, the initial 1,000 draws were discarded as burn-in, and the subsequent 15,000 draws were retained for analysis.

\begin{figure} [tbp]
    \centering
    \begin{subfigure}{0.45\textwidth}
        \centering
        \includegraphics[width=\linewidth]{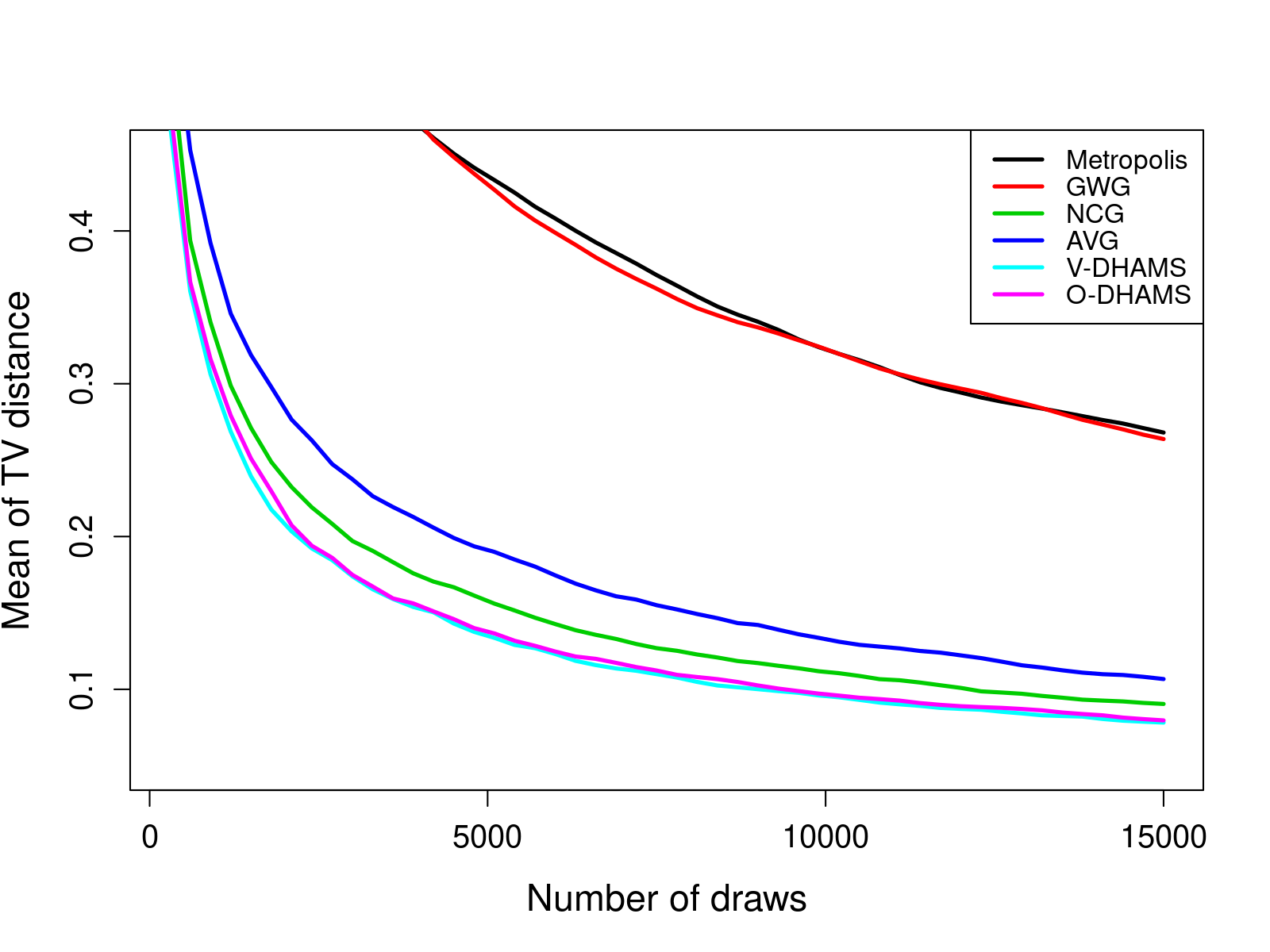}
        \caption{Average mean of TV-distances for all two-dimensional marginal distributions}
        \label{fig:tv2}
    \end{subfigure}
     \begin{subfigure}{0.45\textwidth}
        \centering
        \includegraphics[width=\linewidth]{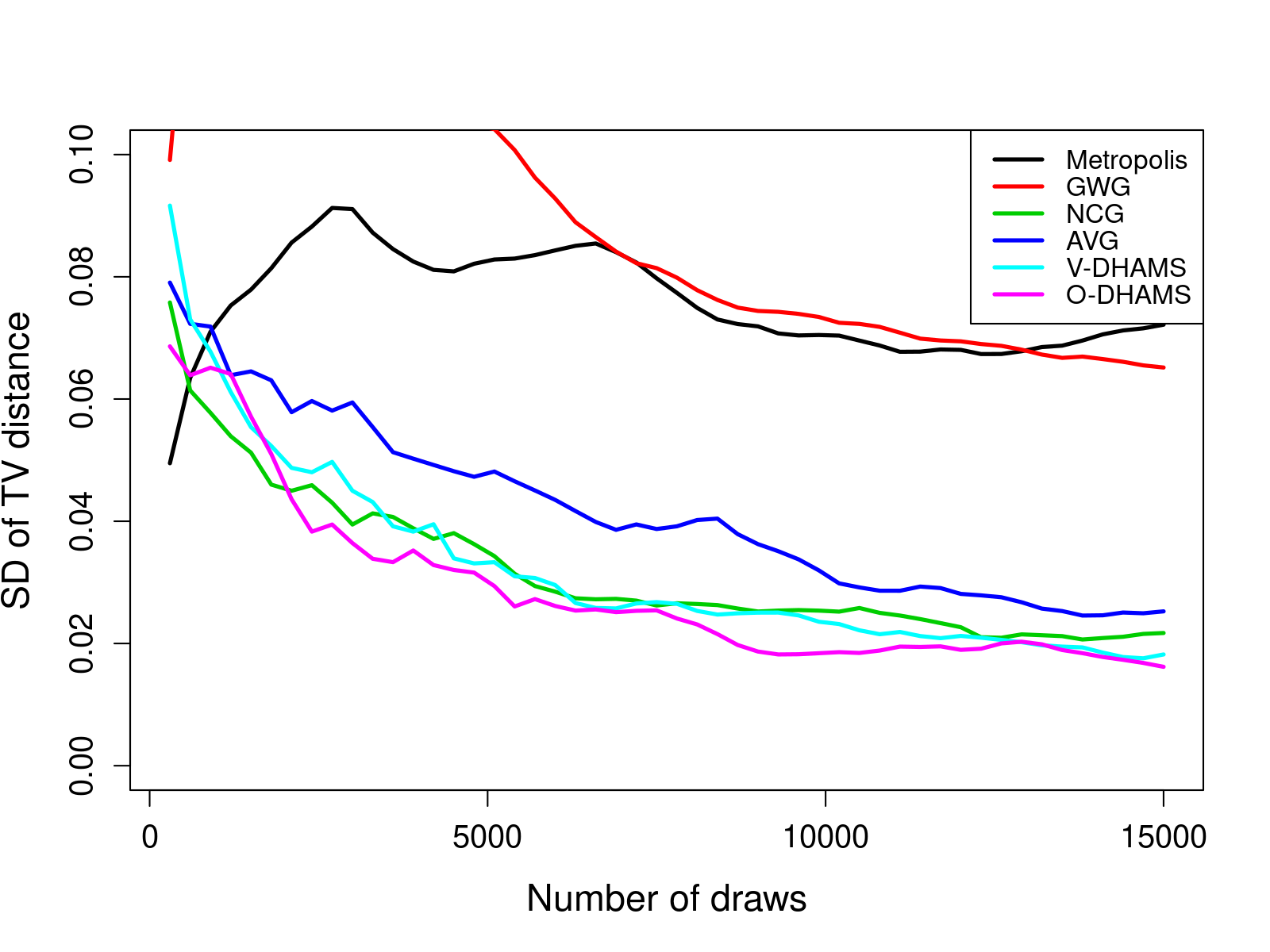}
        \caption{Average standard deviation of TV-distances for all two-dimensional marginal distributions}
        \label{fig:tv2sd}
    \end{subfigure}
    \begin{subfigure}{0.45\textwidth}
        \centering
        \includegraphics[width=\linewidth]{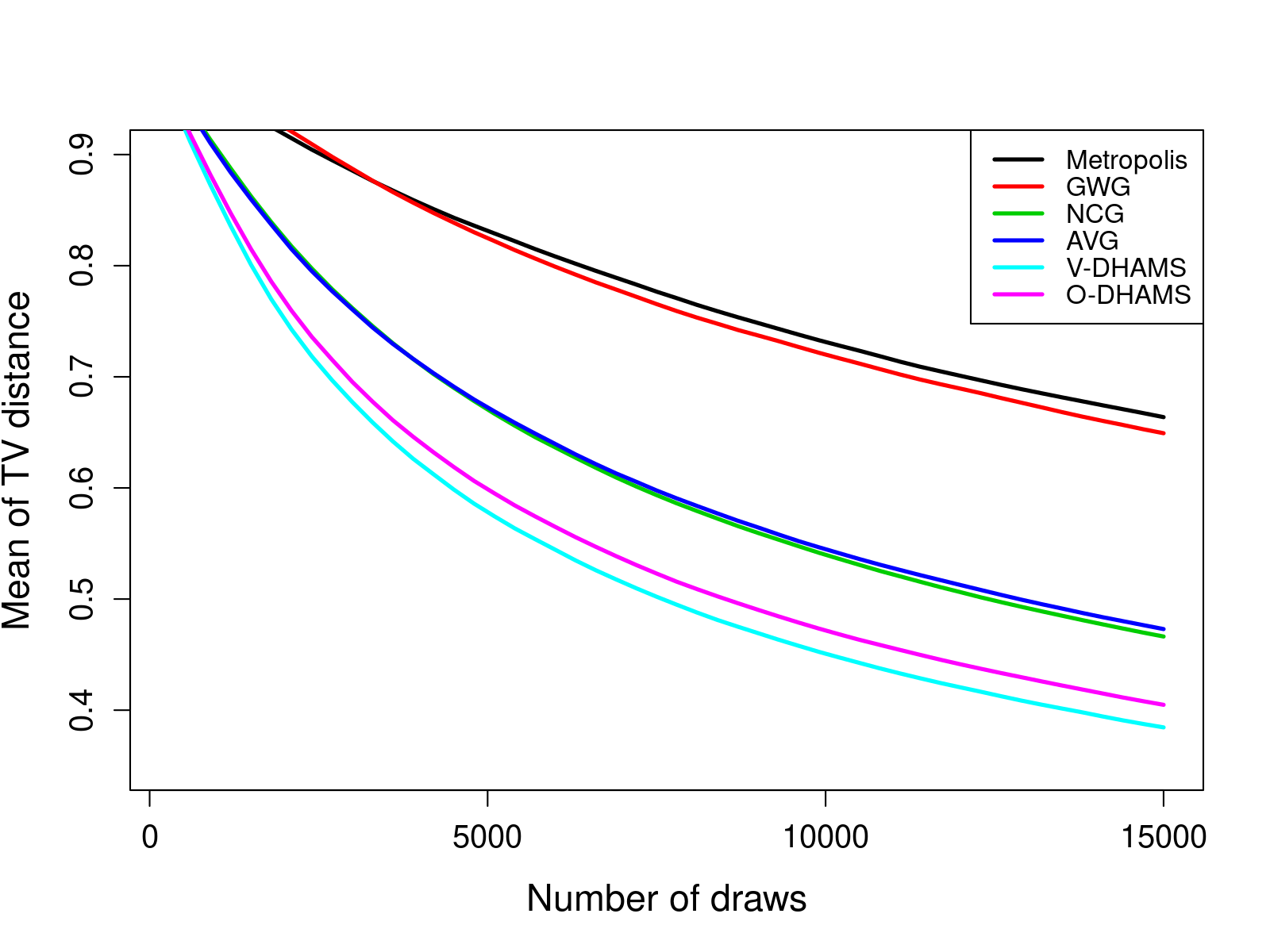}
        \caption{Average mean of TV-distances for all four-dimensional marginal distributions}
        \label{fig:tv4}
    \end{subfigure}
     \begin{subfigure}{0.45\textwidth}
        \centering
        \includegraphics[width=\linewidth]{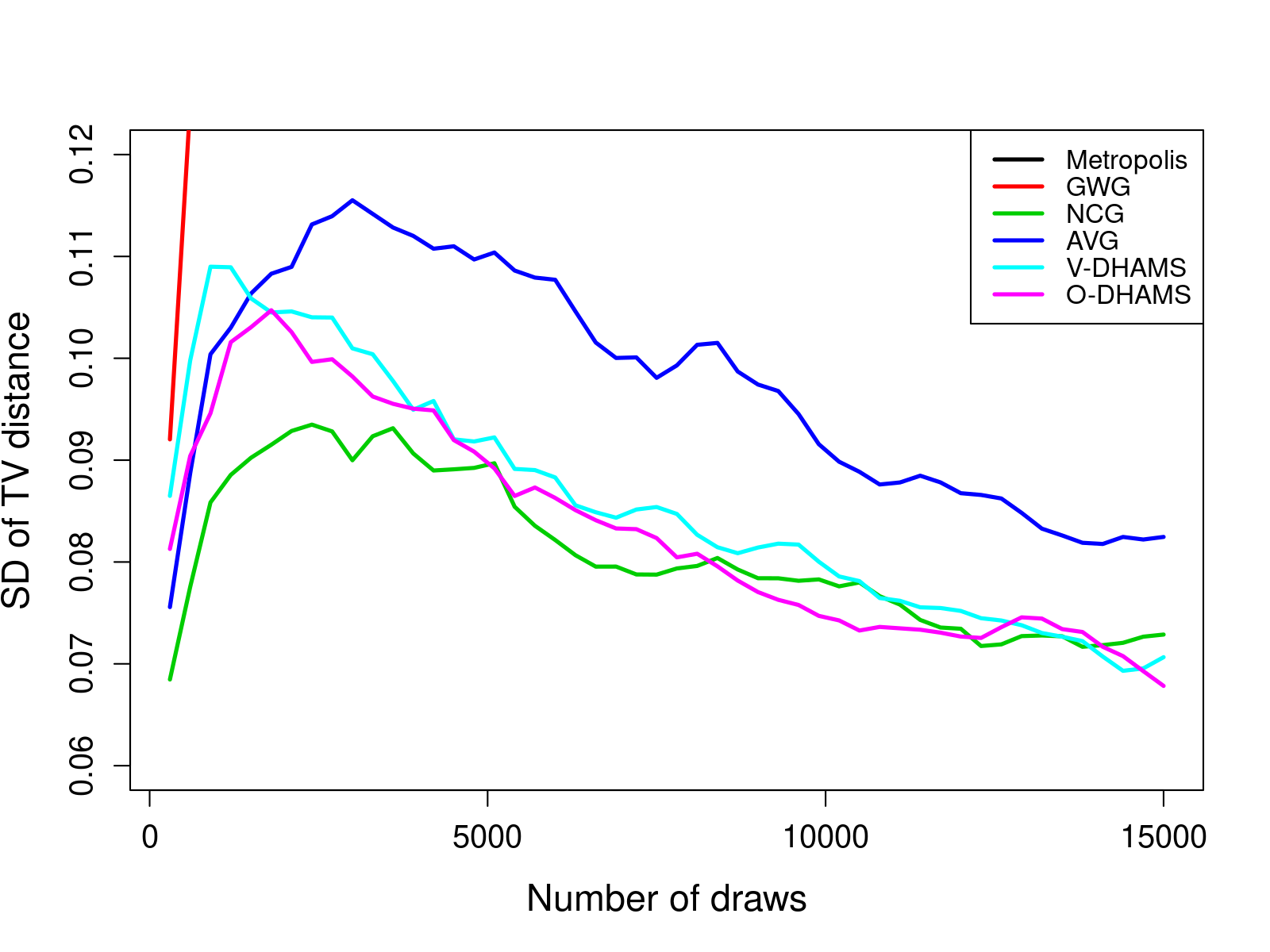}
        \caption{Average standard deviation of TV-distances for all four-dimensional marginal distributions}
        \label{fig:tv4sd}
    \end{subfigure}\caption{TV-distance results for discrete Gaussian distribution}
    \label{fig:tvs}
\end{figure}

In our analysis, we focus on the following marginal distributions $\pi(s_{i_1}, s_{i_2})$ and  $\pi(s_{i_{1}}, s_{i_{2}}, s_{i_{3}}, s_{i_{4}})$. Due to the equi-correlation structure in \eqref{eq:disgau_Sigma}, the joint distribution is invariant under permutations of the indices. As a result, we further average the means and standard deviations of the TV-distances (each over 100 chains) across all index pairs $(s_{i_1}, s_{i_2})$ for the bivariate marginals, and across all index combinations  $(s_{i_{1}}, s_{i_{2}}, s_{i_{3}}, s_{i_{4}})$ for the four-dimensional marginals. The aggregated results are presented in Figure~\ref{fig:tvs}, plotted against the number of draws.

\begin{figure}[tbp]
    \begin{subfigure}{0.45\textwidth}
        \centering
        \includegraphics[width=\linewidth]{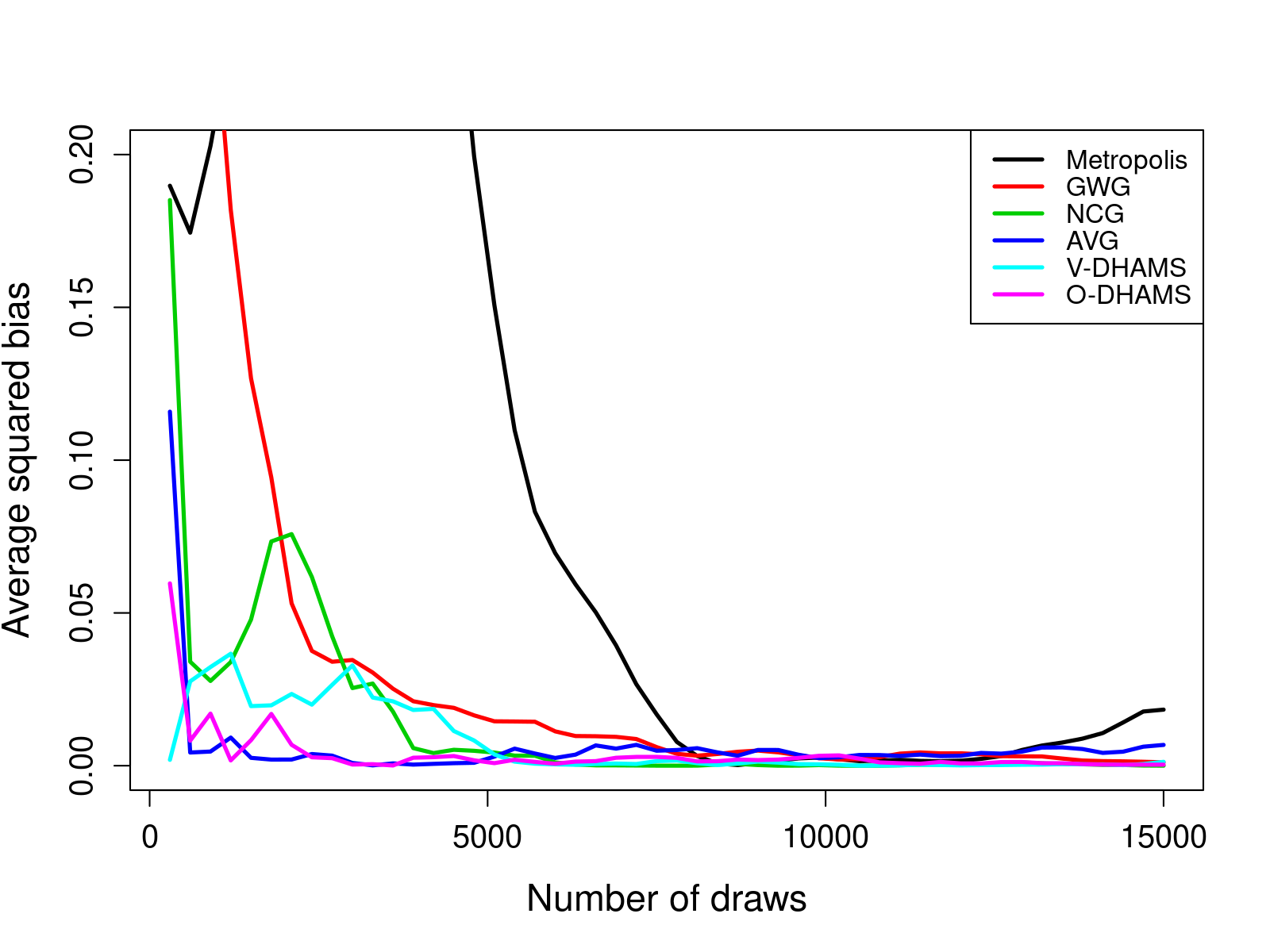}
        \caption{Average squared bias for $E[s_{i}]$}
        \label{fig:biass1}
    \end{subfigure}
    \hfill
    \begin{subfigure}{0.45\textwidth}
        \centering
        \includegraphics[width=\linewidth]{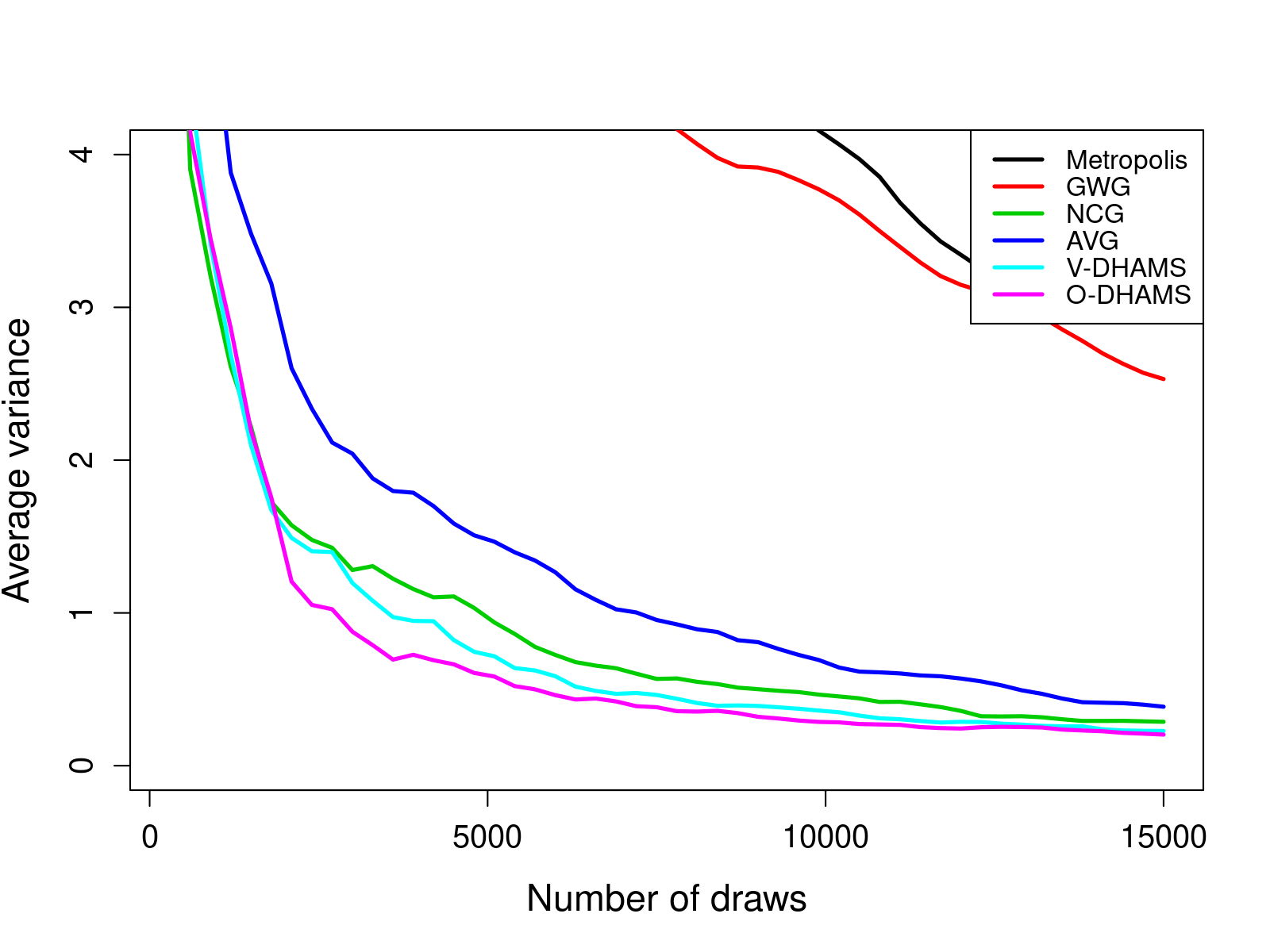}
        \caption{Average variance for $E[s_{i}]$}
        \label{fig:vars1}
    \end{subfigure}

    \par\medskip

    \begin{subfigure}{0.45\textwidth}
        \centering
        \includegraphics[width=\linewidth]{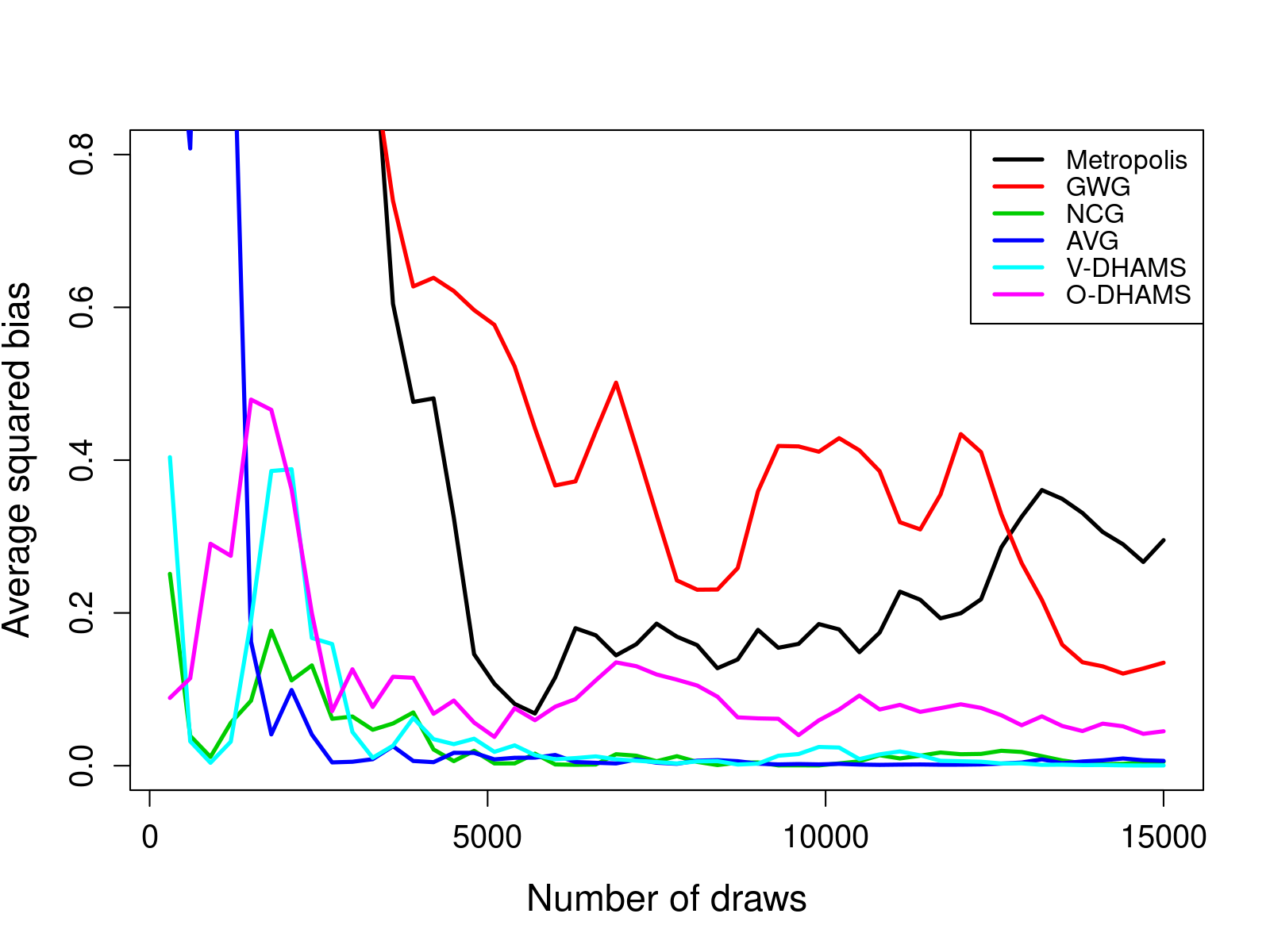}
        \caption{Average squared bias for $E[s_{i}^{2}]$}
        \label{fig:biass12}
    \end{subfigure}
    \hfill
    \begin{subfigure}{0.45\textwidth}
        \centering
        \includegraphics[width=\linewidth]{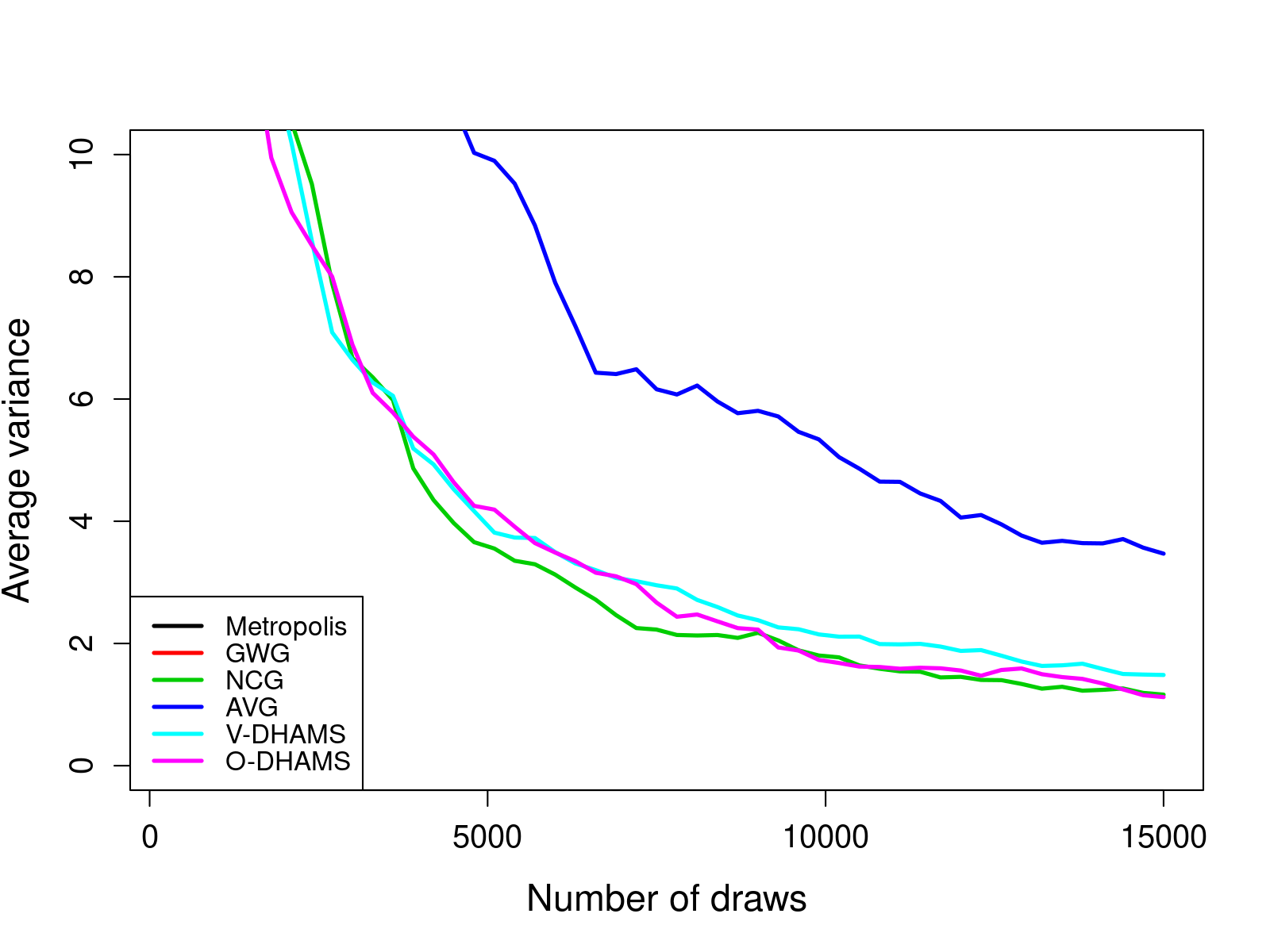}
        \caption{Average variance for $E[s_{i}^{2}]$}
        \label{fig:vars12}
    \end{subfigure}

    \par\medskip

    \begin{subfigure}{0.45\textwidth}
        \centering
        \includegraphics[width=\linewidth]{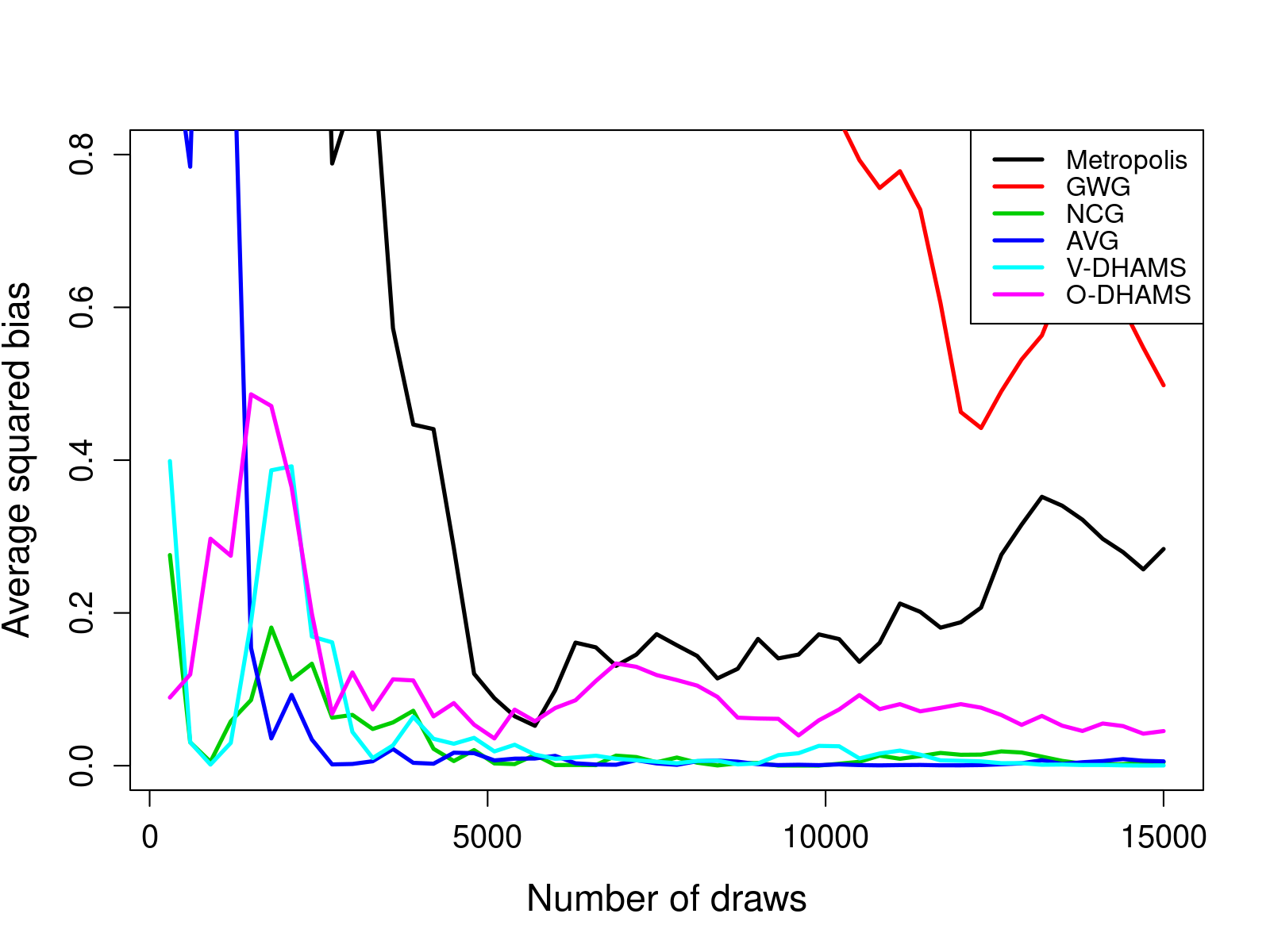}
        \caption{Average squared bias for $E[s_{i_1}s_{i_2}]$}
        \label{fig:biass1s2}
    \end{subfigure}
    \hfill
    \begin{subfigure}{0.45\textwidth}
        \centering
        \includegraphics[width=\linewidth]{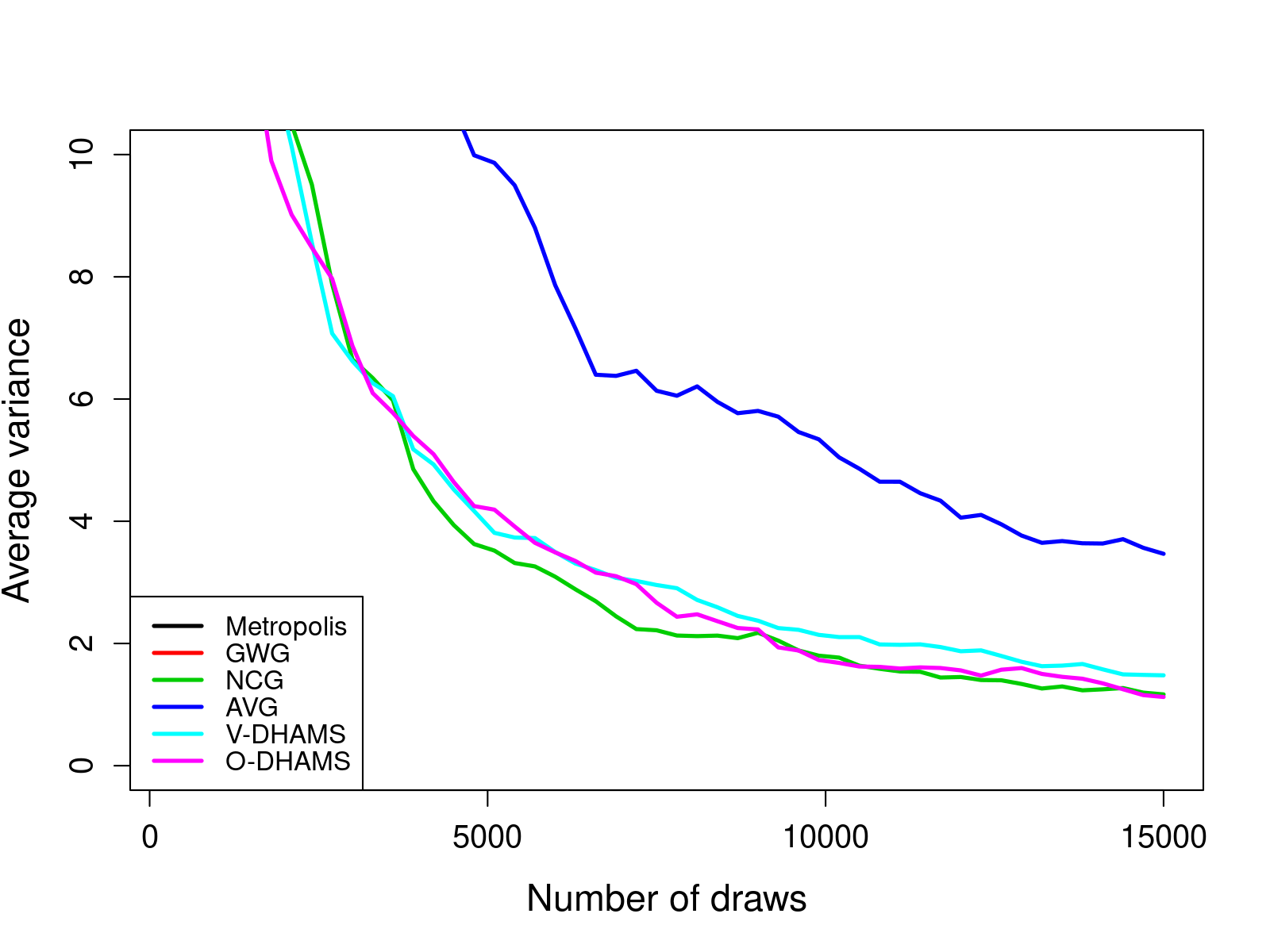}
        \caption{Average variance for $E[s_{i_1}s_{i_2}]$}
        \label{fig:vars1s2}
    \end{subfigure}

    \caption{Estimation results for discrete Gaussian distribution}
    \label{fig:estimations_disgau}
\end{figure}

We are also interested in the estimation of the following three quantities, $\E[s_i]$, $\E[s_i^2]$ and $\E[s_{i_1}s_{i_2}]$. For each chain, we compute estimates of these quantities, and then evaluate the squared bias and variance, using the 100 estimates obtained from the 100 repeated chains for each sampler. The squared bias and variance for $\E[s_i]$, $\E[s_i^2]$ are further averaged over all dimensions, while those for $\E[s_{i_1}s_{i_2}]$ are averaged over all index pairs. The results are plotted against the number of draws in Figure \ref{fig:estimations_disgau}. We also report the minimum, median, and maximum of ESS across all coordinates and the ESS for the negative potential function $f(s)$, and present them in Table \ref{tab:ess_gaussian}.

Both V-DHAMS and O-DHAMS outperform all other samplers in terms of convergence speed, consistently achieving lower TV-distances and small standard deviations, particularly for four-dimensional marginal distributions. In the estimation tasks, O-DHAMS exhibits lower variances than V-DHAMS. From Table~\ref{tab:ess_gaussian}, V-DHAMS and O-DHAMS also achieve superior performance in terms of ESS across all coordinates.
The overall performance in simulating the discrete Gaussian distribution follows the order:
\[
\text{O-DHAMS} > \text{V-DHAMS} > \text{NCG} > \text{AVG} > \text{Metropolis} > \text{GWG}.
\]
Additional results (including selected parameter settings, average acceptance rates, and trace, frequency and auto-correlation plots from individual runs) are presented in Supplement Section~\ref{sec:gaussian_results}.
\begin{table}[tbp]
\centering
\begin{tabular}{|c|c|c|c|c|}
\hline
    Sampler & ESS Minimum & ESS Median & ESS Maximum & ESS Energy\\
    \hline
    Metropolis & 4.67 & 4.72 &  4.81 & 180.50\\
    GWG & 5.70 & 6.01 &  6.26 & 10.12\\

    NCG & 58.50 & 58.97 & 59.55 & 3388.48\\
       AVG & 43.02 & 43.67 & 43.94 & 2254.74\\
   V-DHAMS & 73.87 & 75.09 & 76.14 & 3841.09\\
    O-DHAMS & 82.25 & 82.73 & 83.78 & 3167.07\\ \hline
\end{tabular}
\caption{ESS table for discrete Gaussian}
\label{tab:ess_gaussian}
\end{table}

\subsection{Quadratic Mixture} \label{sec:poly_mix}
\begin{figure}[tbp]
    \centering
    \includegraphics[width=0.55\linewidth]{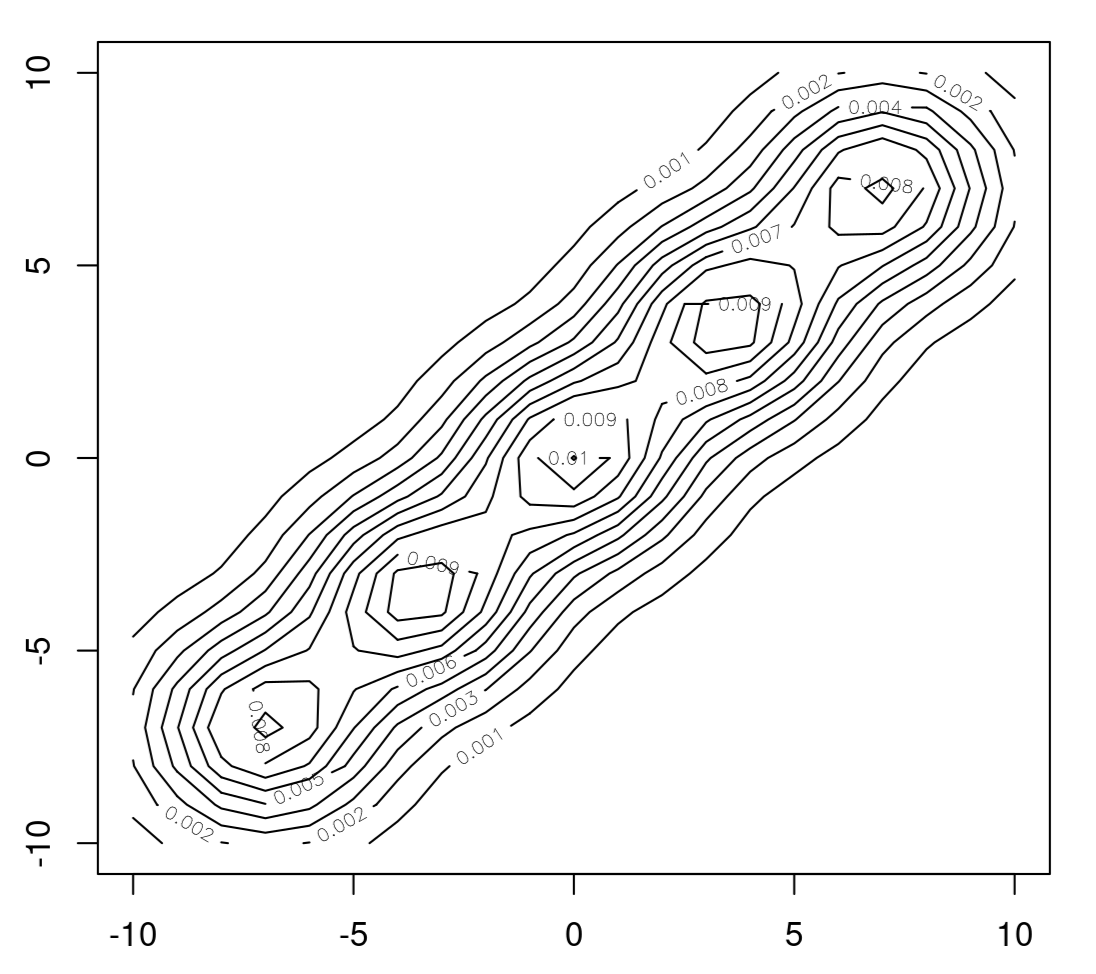} 
    \caption{Contour plot of quadratic mixture distribution}
    \label{fig:poly_heatmap} 
\end{figure}

We consider a quadratic mixture distribution similarly as in \cite{Rhodes2022GradientMC}, but with more separated mixture components.
The negative potential function $f(s)$ is defined as
\begin{align}
    f(s) = \log\left(\sum\limits_{m=1}^{M} \exp\left(-\frac{1}{2}(s - \mu_m)^{\T}\Sigma_m^{-1}(s - \mu_m)\right)\right).
    \label{eq:poly_mix_f}
\end{align}
The distribution is defined over a $d$-dimensional lattice $\mathcal{S} \subset \mathbb{R}^d$, given by $\mathcal{S} = \{-k, -(k-1), \ldots, -1, 0, 1, \ldots, k-1, k\}^d$. In our experiment, we set $d = 8$, $k = 10$, and $M = 5$. Each component has mean $\mu_m = (-10.5 + 3.5m)\mathbf{1}$ and covariance $\Sigma_m = \frac{25}{49}I$, ensuring that all components share the same variance and the means are equally spaced. The resulting distribution is symmetric along each coordinate axis and invariant under permutations of coordinates. For visualization, we present the contour plot over any two selected dimensions in Figure~\ref{fig:poly_heatmap}. Compared to discrete Gaussian distribution, the quadratic mixture exhibits considerable multi-modality.

For each sampler, 50 different parameters are searched for tuning as in the discrete Gaussian experiment. We run 50 independent chains, each of length 7,500 after 1,000 burn-in draws for each parameter. The best parameter is then selected by the highest ESS of negative potential function $f(s)$.
After tuning, we conduct 100 independent chains for each sampler using the optimal parameter setting. For each chain, the initial 1,000 draws were discarded as burn-in, and the subsequent 24,000 draws were retained for analysis.

\begin{figure}[tbp]
    \centering
    \begin{subfigure}{0.45\textwidth}
        \centering
        \includegraphics[width=\linewidth]{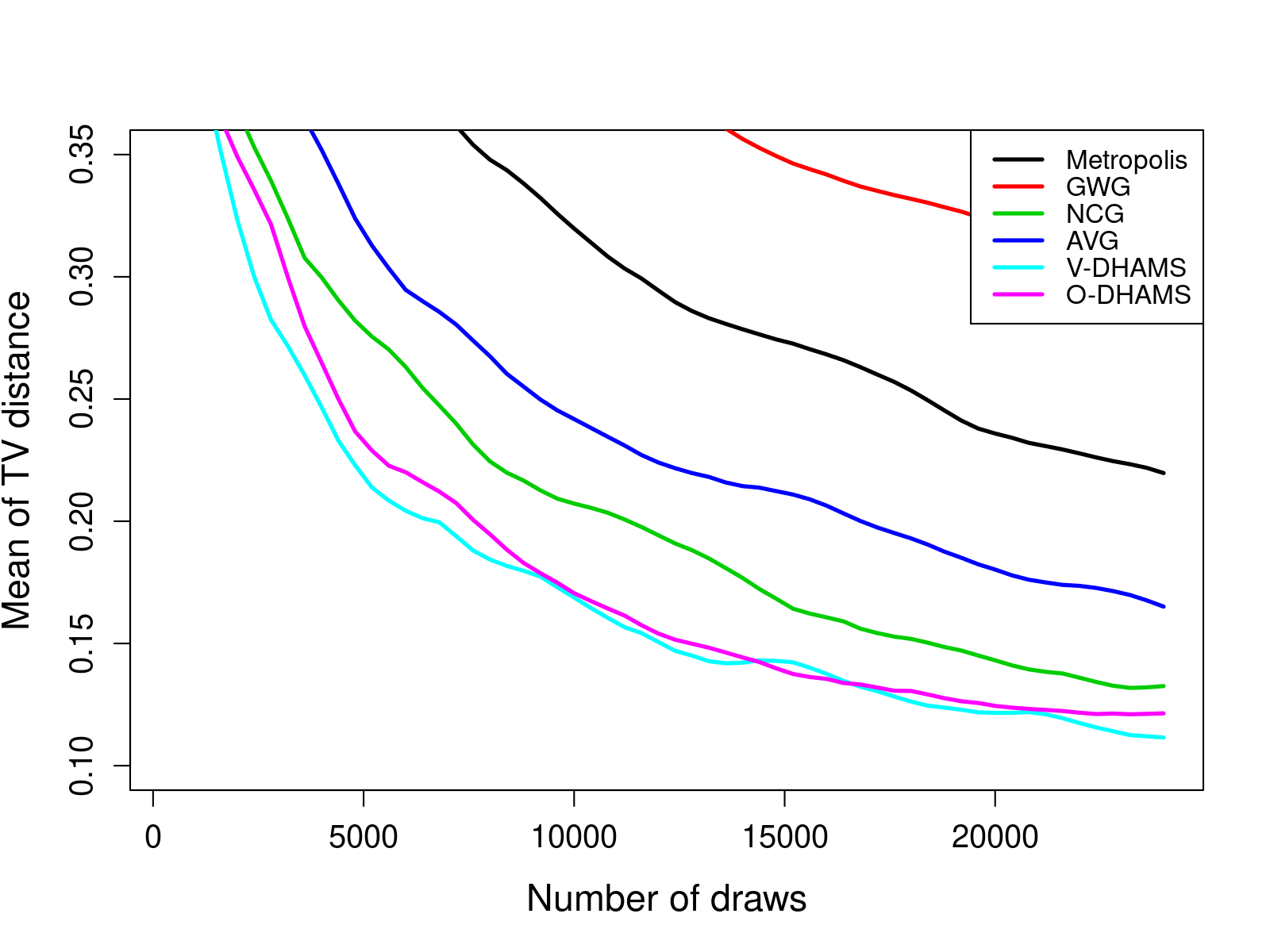}
        \caption{Average mean of TV-distance for all univariate marginal distributions}
        \label{fig:poly_tv2}
    \end{subfigure}
     \begin{subfigure}{0.45\textwidth}
        \centering
        \includegraphics[width=\linewidth]{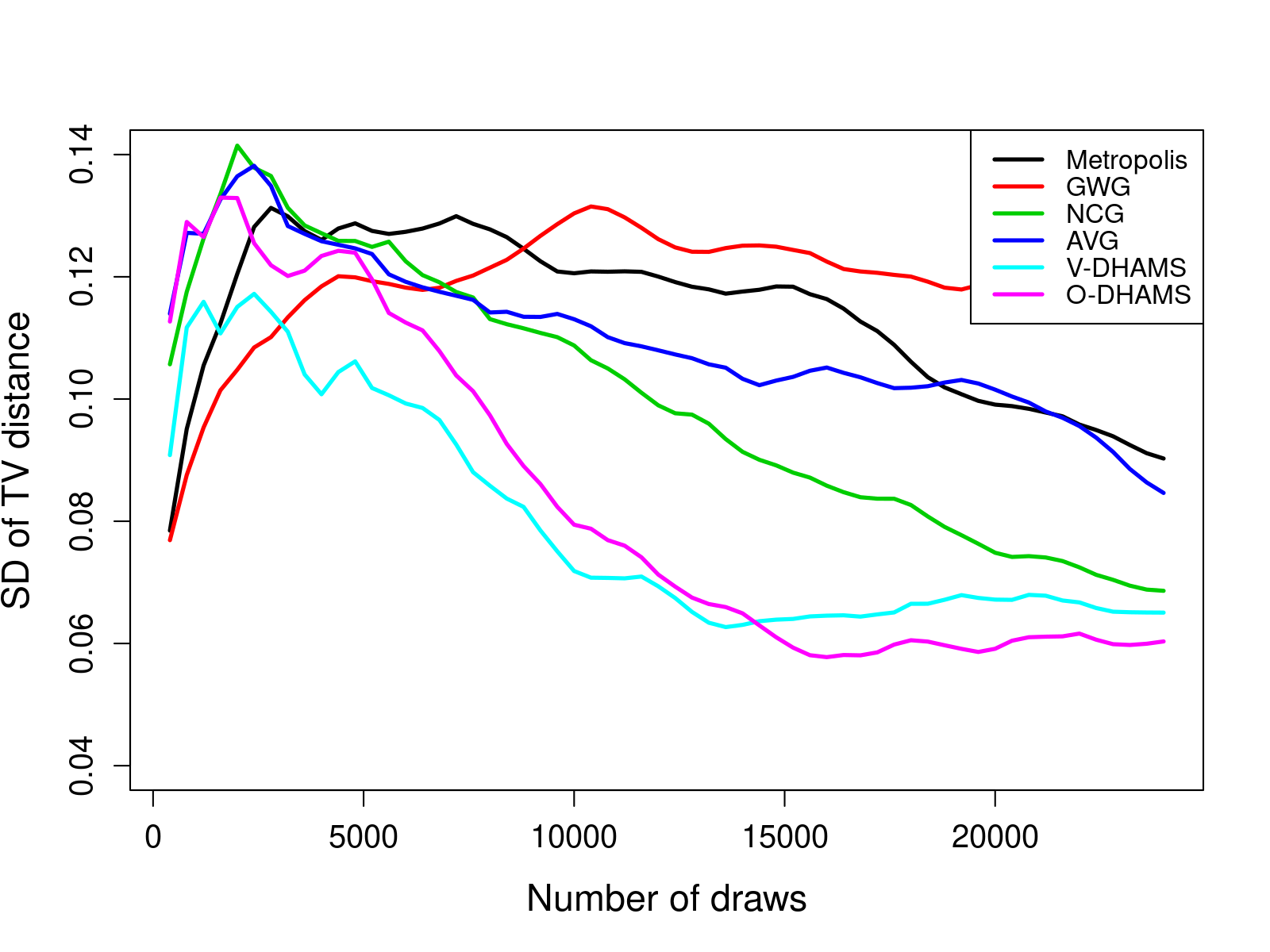}
        \caption{Average standard deviation of TV-distance for all univariate marginal distributions}
        \label{fig:poly_tv2sd}
    \end{subfigure}
    \begin{subfigure}{0.45\textwidth}
        \centering
        \includegraphics[width=\linewidth]{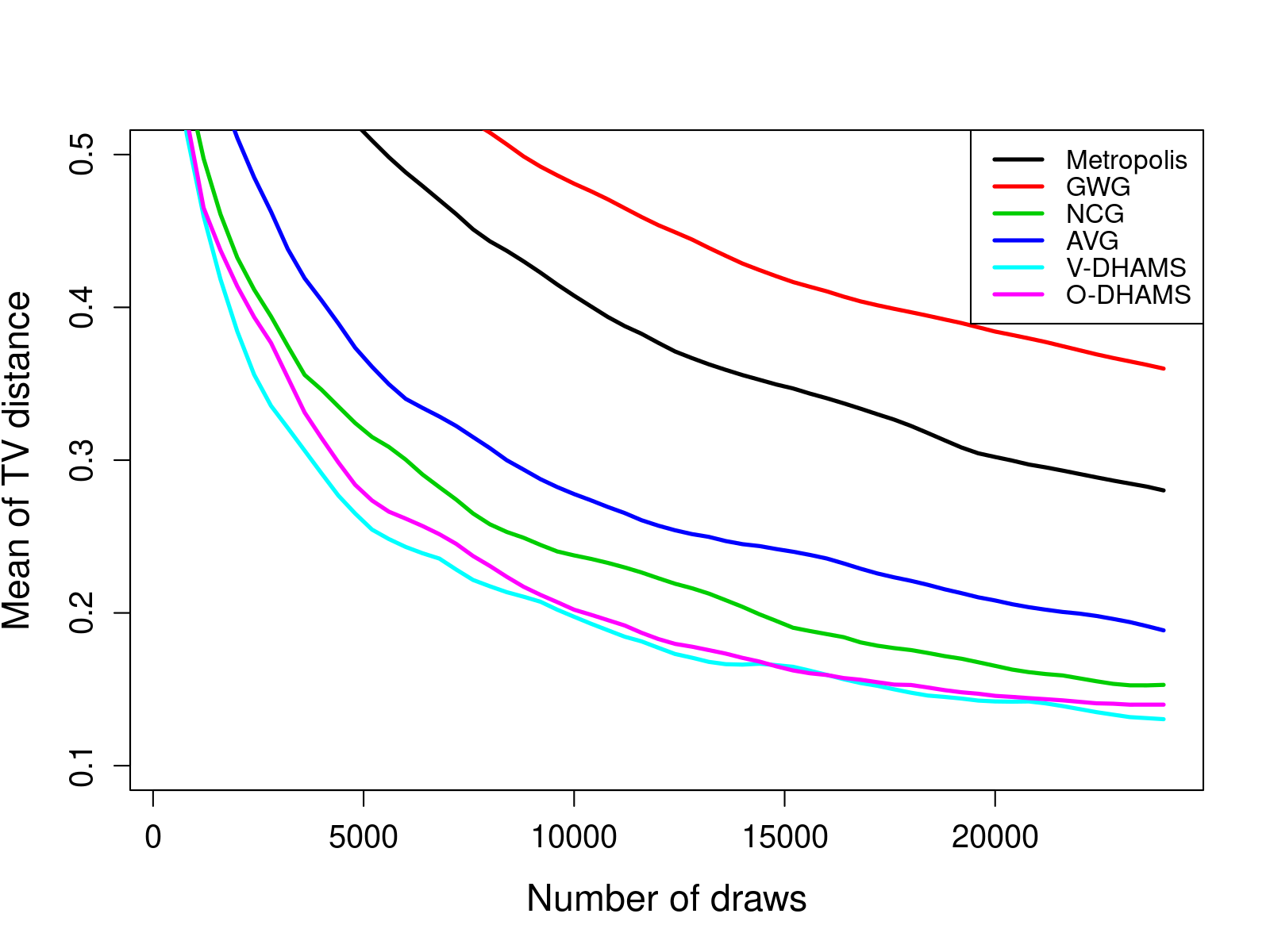}
        \caption{Average mean of TV-distance for all bivariate marginal distributions}
        \label{fig:poly_tv4}
    \end{subfigure}
     \begin{subfigure}{0.45\textwidth}
        \centering
        \includegraphics[width=\linewidth]{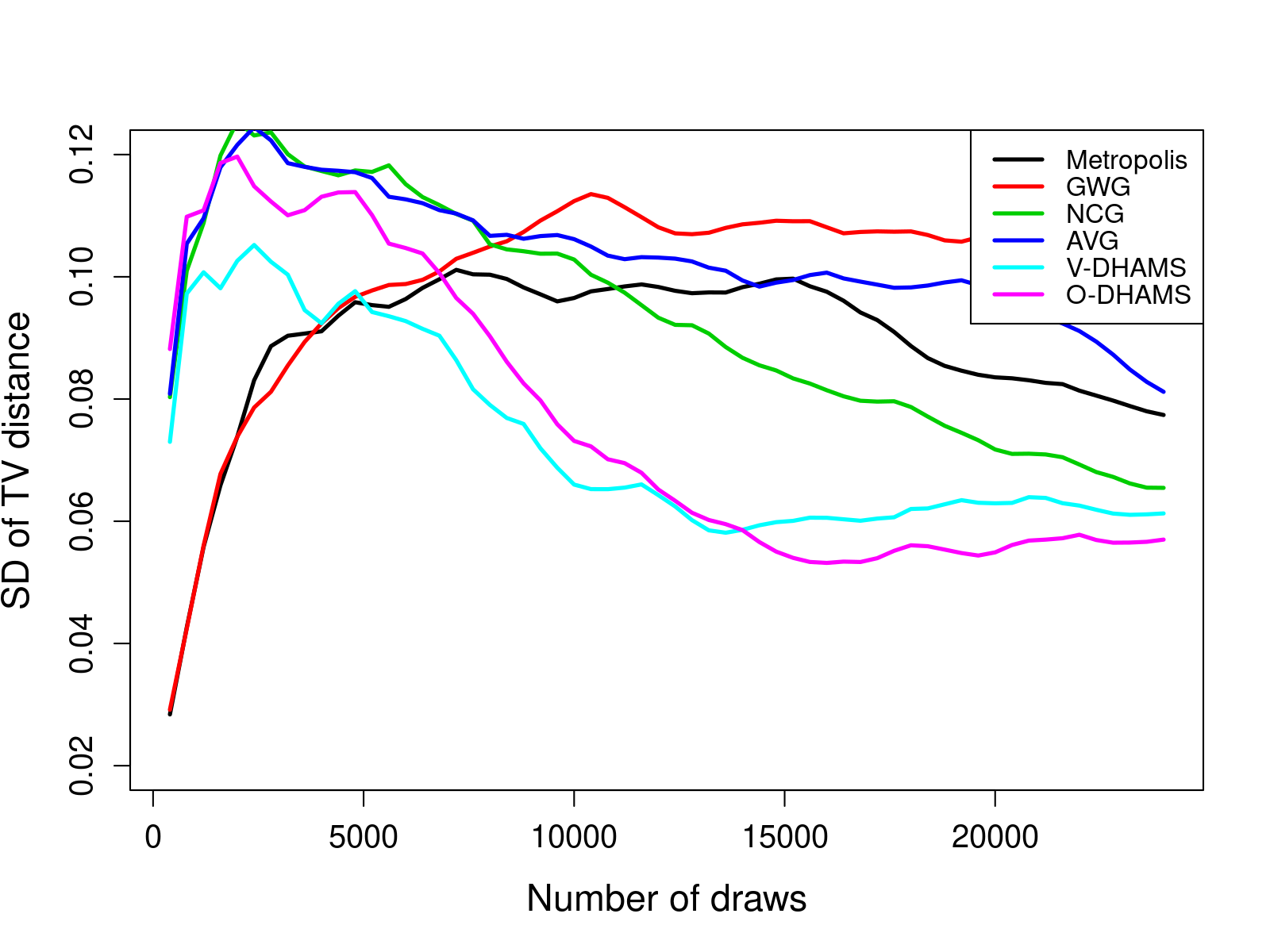}
        \caption{Average standard deviation of TV-distance for all bivariate marginal distributions}
        \label{fig:poly_tv4sd}
    \end{subfigure}\caption{TV-distances results for quadratic mixture distribution}
    \label{fig:poly_tvs}
\end{figure}

In the quadratic mixture experiment, we focus on the analysis of TV-distances for univariate and bivariate marginal distributions $\pi(s_i)$ and $\pi(s_{i_1}, s_{i_2})$. The higher-dimensional marginals are not calculated due to high computational and memory cost. The joint distribution \eqref{eq:poly_mix_f} is also invariant under permutations of the indices as the discrete Gaussian distribution in Section~\ref{sec:exp_gauss}. We then average the means and standard deviations of the TV-distances over 100 chains across all dimensions $s_i$ for the univariate marginals, and across all pairs  $(s_{i_1}, s_{i_2})$ for bivariate marginals. The aggregated results are presented in Figure~\ref{fig:poly_tvs}, plotted against the number of draws.

\begin{figure}[tbp]
    \begin{subfigure}{0.45\textwidth}
        \centering
        \includegraphics[width=\linewidth]{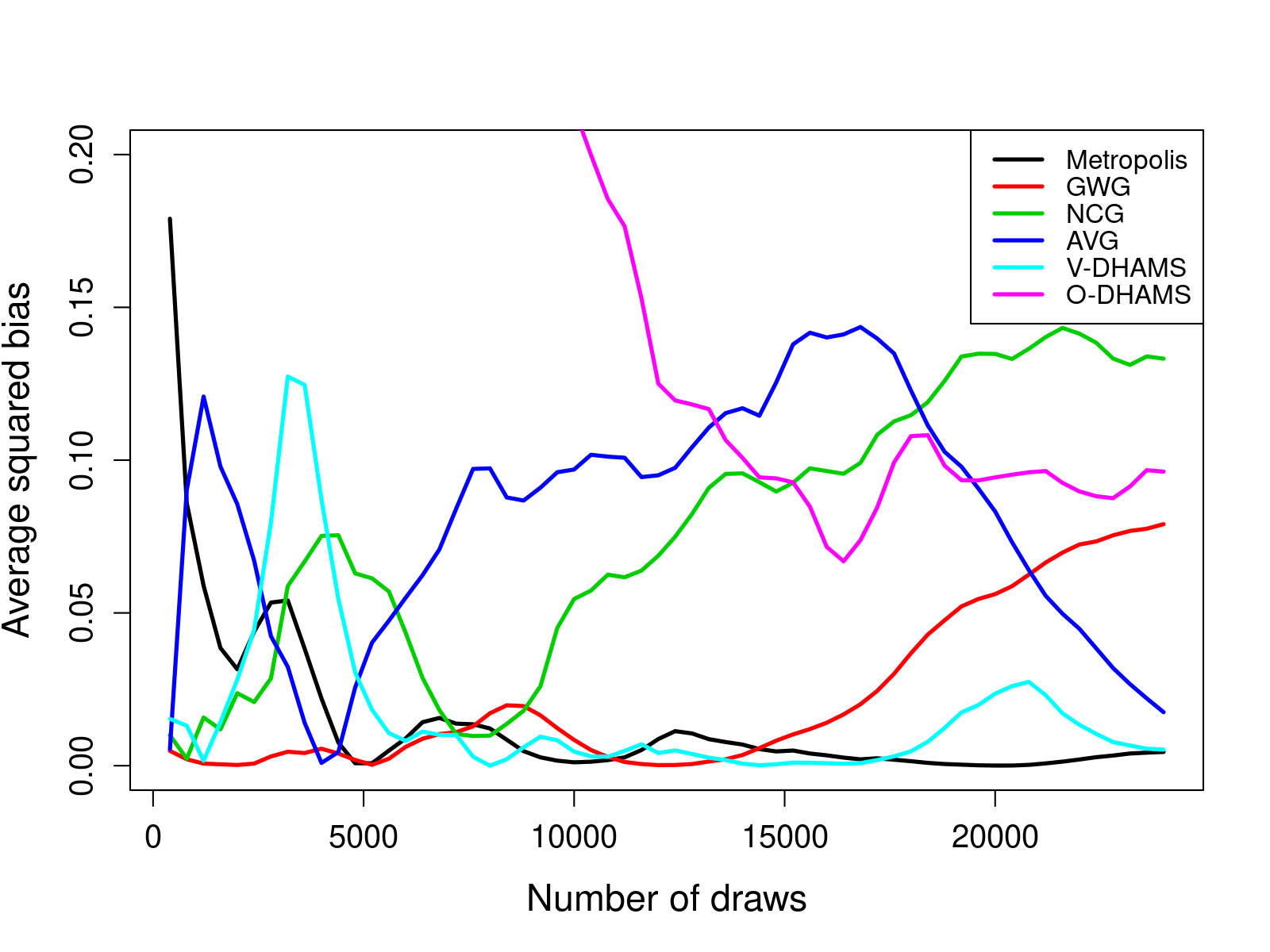}
        \caption{Average squared bias of $E[s_{i}]$}
        \label{fig:poly_biass1}
    \end{subfigure}
    \hfill
    \begin{subfigure}{0.45\textwidth}
        \centering
        \includegraphics[width=\linewidth]{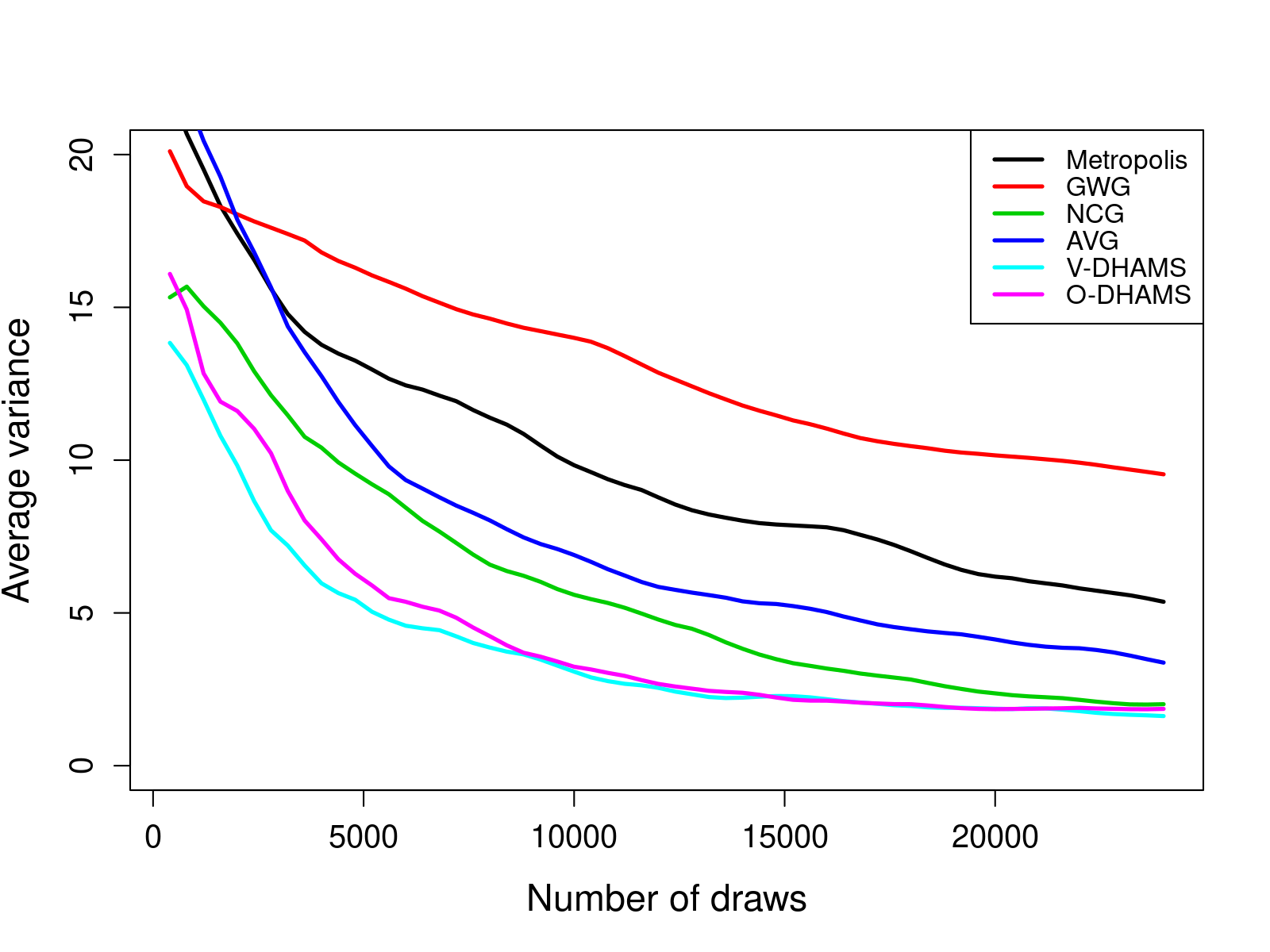}
        \caption{Average variance of $E[s_{i}]$}
        \label{fig:poly_vars1}
    \end{subfigure}

    \par\medskip

    \begin{subfigure}{0.45\textwidth}
        \centering
        \includegraphics[width=\linewidth]{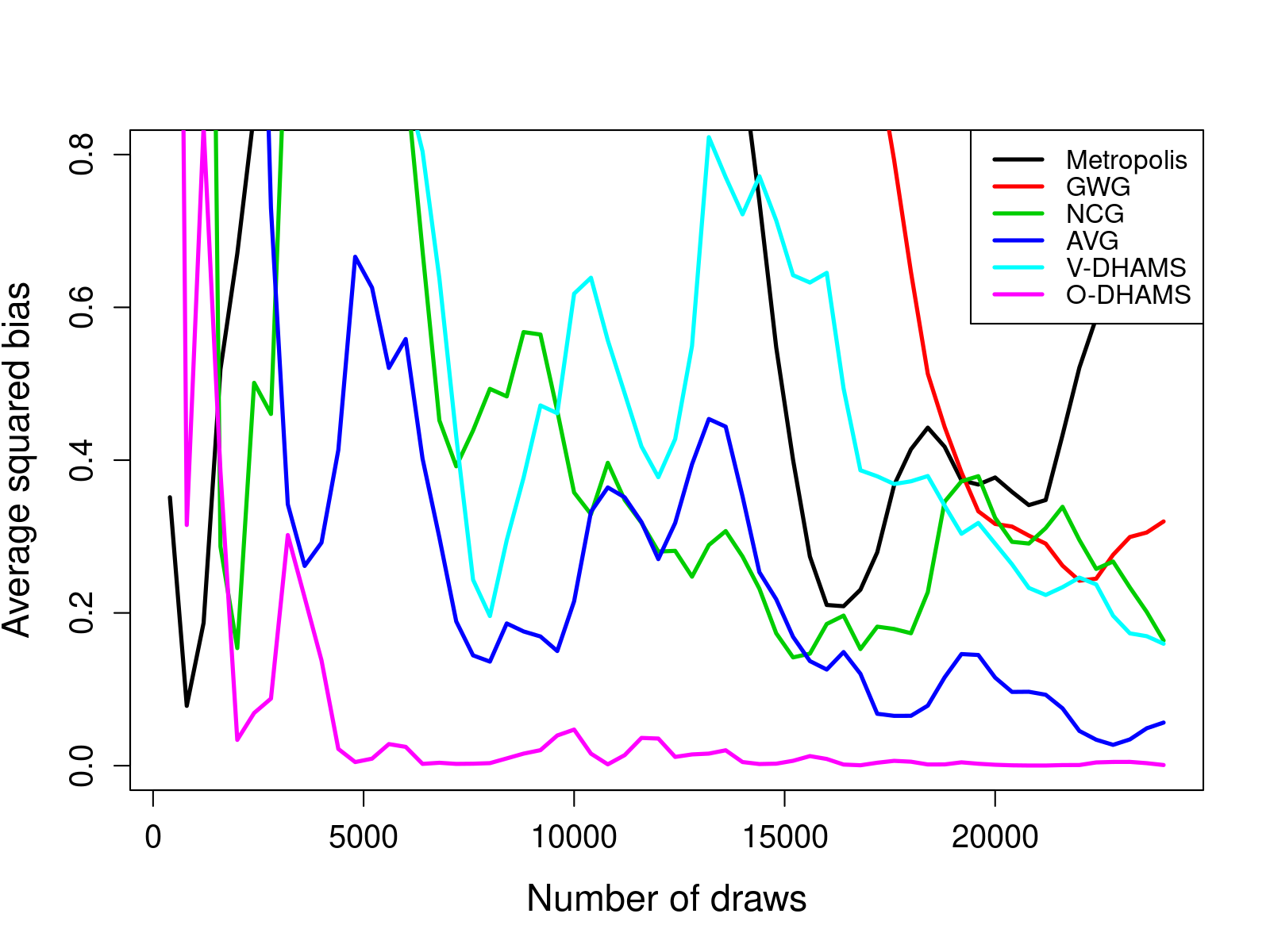}
        \caption{Average squared bias of $E[s_{i}^{2}]$}
        \label{fig:poly_biass12}
    \end{subfigure}
    \hfill
    \begin{subfigure}{0.45\textwidth}
        \centering
        \includegraphics[width=\linewidth]{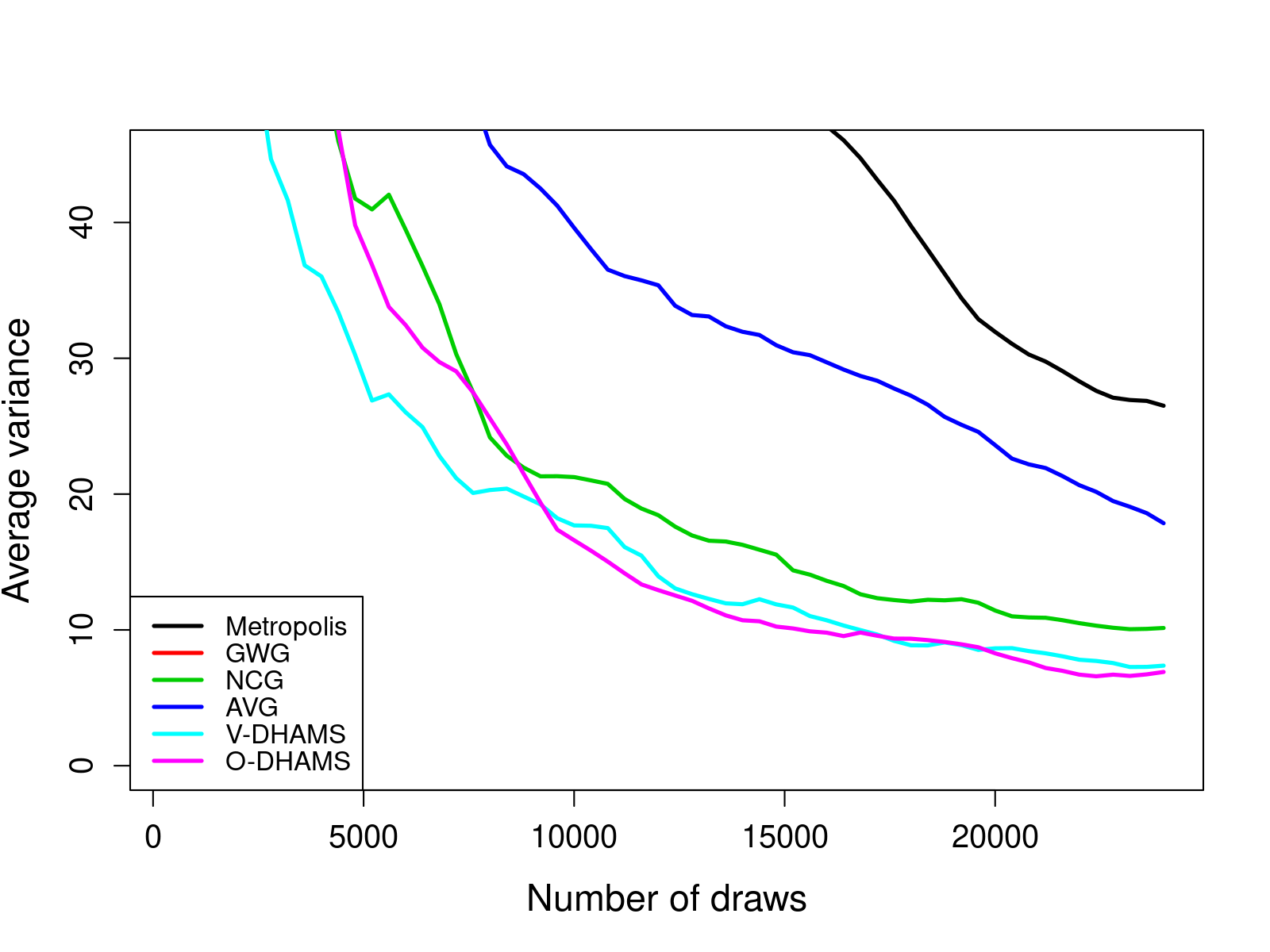}
        \caption{Average variance of $E[s_{i}^{2}]$}
        \label{fig:poly_vars12}
    \end{subfigure}

    \par\medskip

    \begin{subfigure}{0.45\textwidth}
        \centering
        \includegraphics[width=\linewidth]{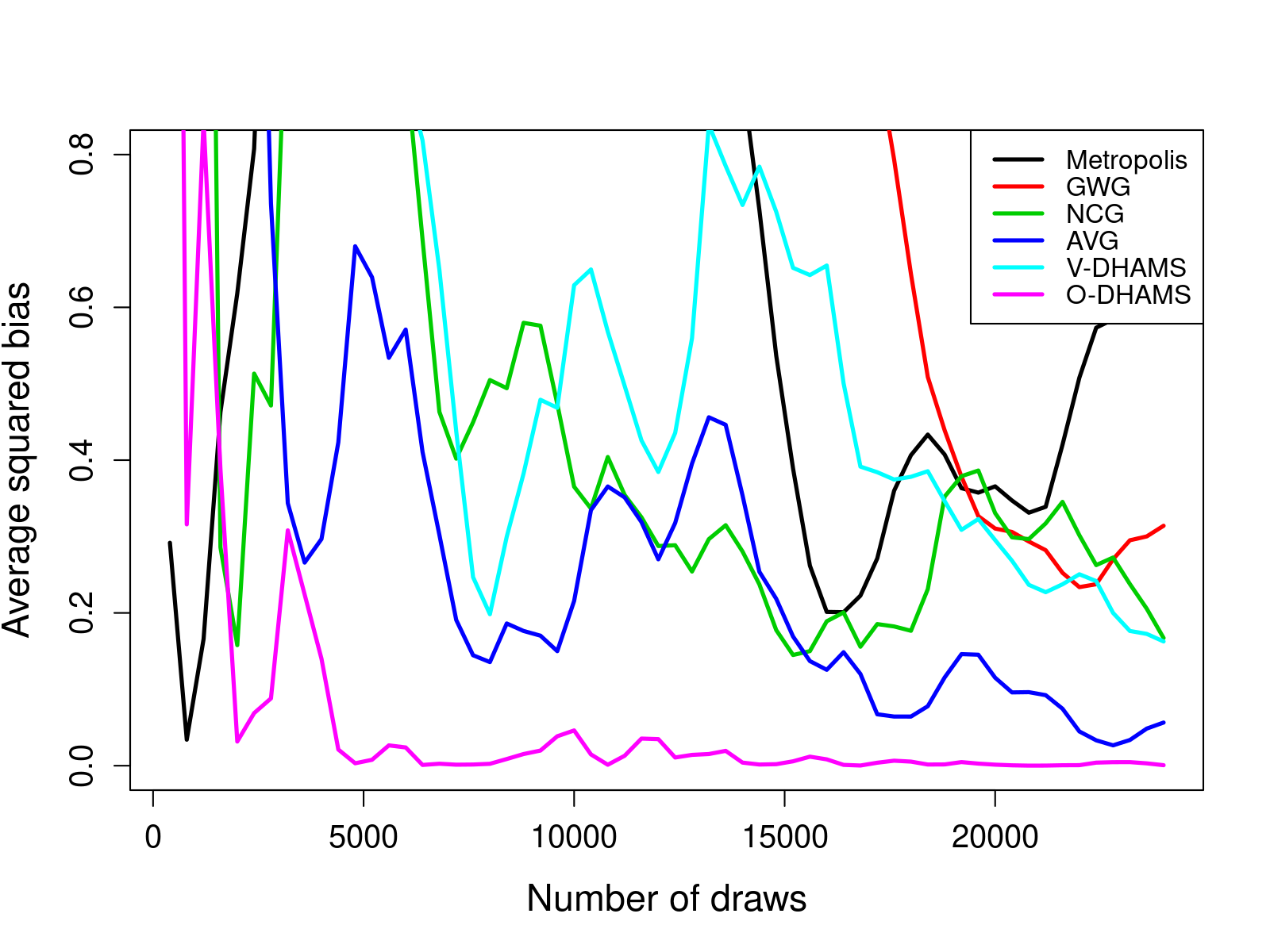}
        \caption{Average squared bias of $E[s_{i_1}s_{i_2}]$}
        \label{fig:poly_biass1s2}
    \end{subfigure}
    \hfill
    \begin{subfigure}{0.45\textwidth}
        \centering
        \includegraphics[width=\linewidth]{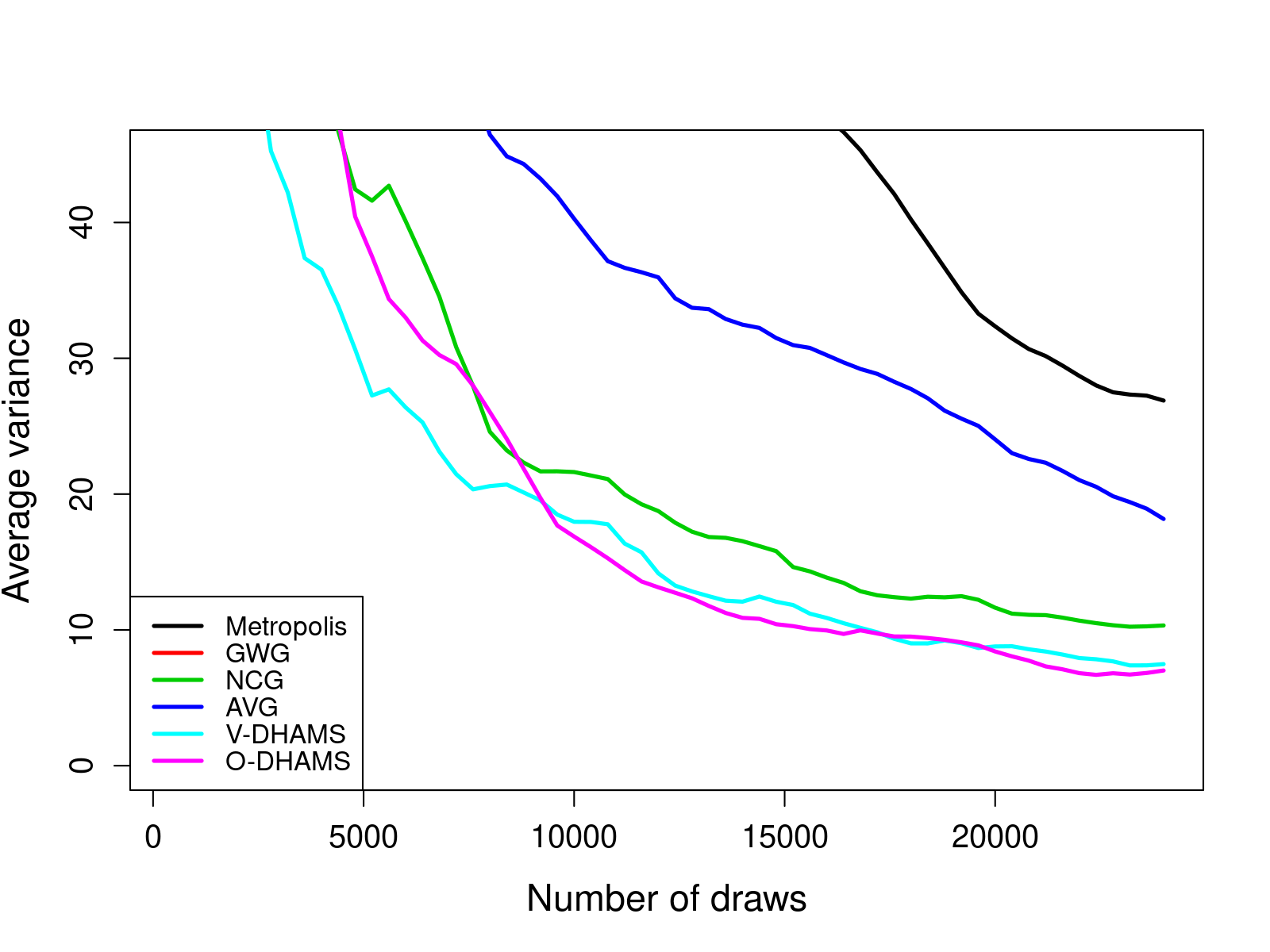}
        \caption{Average variance of $E[s_{i_1}s_{i_2}]$}
        \label{fig:poly_vars1s2}
    \end{subfigure}

    \caption{Estimation results for quadratic mixture distribution}
    \label{fig:estimations_poly}
\end{figure}

We also report the estimation results of $\E[s_i]$, $\E[s_i^2]$ and $\E[s_{i_1}s_{i_2}]$. The squared bias and variance for $\E[s_i]$, $\E[s_i^2]$ are averaged over all dimensions, and those for $\E[s_{i_1}s_{i_2}]$ are averaged over all index pairs. The averaged results are plotted against the number of draws in Figure \ref{fig:estimations_poly}.
The minimum, median, and maximum of ESS across all coordinates, as well as the ESS for the negative potential function $f(s)$ are reported in Table \ref{tab:ess_poly}.

Both Discrete-HAMS samplers demonstrate significantly less mean and standard deviation of TV-distances and lower variance in estimation problems. Moreover, O-DHAMS shows lower biases and higher marginal ESS. The overall performance ranking of all samplers is summarized below:
\[
\text{O-DHAMS} > \text{V-DHAMS} > \text{NCG} > \text{AVG} > \text{Metropolis} > \text{GWG}.
\]
Additional results (including parameter settings, average acceptance rates, and trace, frequency and auto-correlation plots from individual runs) are presented in Supplement Section \ref{sec:mixture_results}.
\begin{table}[tbp]
\centering
\begin{tabular}{|c|c|c|c|c|}
\hline
    Sampler & ESS Minimum & ESS Median & ESS Maximum & ESS Energy\\
    \hline
    Metropolis & 3.50 & 3.54 &  3.57 & 351.73\\
    GWG & 1.57 & 1.58 &  1.60 & 245.82\\
    NCG & 11.58 & 11.66 & 11.71 & 1863.44\\
       AVG & 5.74 & 5.79 & 5.86 & 1328.76\\
   V-DHAMS & 14.21 & 14.26 & 14.36 & 2828.46\\
    O-DHAMS & 12.21 & 12.28 & 12.29 & 2046.33\\ \hline
\end{tabular}
\caption{ESS table for quadratic mixture distribution}
\label{tab:ess_poly}
\end{table}

\subsection{Bayesian Sparse Regression}\label{sec:bayesian_linear}
\subsubsection{Model Selection Framework}

We consider the following model selection problem in Bayesian linear regression, which is studied in \cite{Titsias2017Hamming} and \cite{Rhodes2022GradientMC}.
Let $y \in \mathbb{R}^{n}$ be a response vector and $X \in \mathbb{R}^{n\times d}$ be a covariate matrix with $d$ covariates.
The linear regression model states
\begin{align}
    y = Xw +\sigma \epsilon, \quad \epsilon \sim \mathcal{N}(0,I_{n}). \nonumber
\end{align}
For variable selection, we further introduce a binary mask vector $s$ such that $w_{i} = 0$ if $s_{i} =0$ and $w_{i} \neq 0$ if $s_{i} = 1$. Given a binary mask $s$, we denote $X_{s}\in \mathbb{R}^{n\times d_s} $ as the columns of $X$ corresponding to elements of $s$ being 1, with $d_{s} = \sum\limits_{i=1}^{d}s_{i} $,
and denote $w_{s}$ as the non-zero elements of $w$ indicated by $s$. The reduced model based on $s$ is then
\begin{align}
    y = X_{s} w_{s} +\sigma \epsilon, \quad \epsilon \sim \mathcal{N}(0,I_{n}). \nonumber
\end{align}
The conditional distribution of the response given the selected variables is
\begin{equation}
    \pi(y|X, s, w_{s}, \sigma^{2}) \sim \mathcal{N}(y|X_{s}w_{s}, \sigma^{2}I_{n}).
    \label{eq:y_prior}
\end{equation}
An inverse-gamma prior is typically imposed on $\sigma^{2}$,
and given $\sigma^{2}$ and $s$, a ridge-type g-prior \eqref{eq:generalized-gprior} is introduced for $w_s$, as follows:
    \begin{align}
    &\pi(\sigma^{2}) \sim \text{InvGamma} (\sigma^{2} ; \alpha_{\sigma}, \beta_{\sigma}), \label{eq:sigma2-prior} \\
    &\pi(w_{s}| X, s, \sigma^{2} ) = \mathcal{N}(w_{s};0, g\sigma^{2}(\kappa X_{s}^{\T}X_{s}+\lambda I_{d_{s}})^{-1}), \label{eq:generalized-gprior}
    \end{align}
where $(\alpha_\sigma, \beta_\sigma)$ and $(g,\kappa,\lambda)$ are hyper-parameters.
The ridge-type g-prior \eqref{eq:generalized-gprior} is a perturbed version of Zellner's g-prior \citep{Zellner1986gprior} where $\pi(w_{s}|\sigma^{2}, s) = \mathcal{N}(w_{s};0, g\sigma^{2}( X_{s}^{\T}X_{s})^{-1}) $. The reason for adding the perturbation is that when $X_{s}$ is of rank lower than $d_s$ due to high dimensionality or multi-collinearity, the g-prior would fail due to singular matrix $X_{s}^{\T}X_{s}$. This ridge-type g-prior is also discussed in \cite{Gupta2009ridgeinfoprior} and \cite{Baragatti2012ridgegprior}.
We also introduce a Beta-Bernoulli prior for the binary mask vector $s$ as in \cite{Titsias2017Hamming}:
\begin{align}
    \pi(\psi) \sim \text{Beta} (\psi; \alpha_{\psi}, \beta_{\psi})
    \label{eq:sprior1},\\
    \pi(s|\psi) \sim \prod\limits_{i=1}^{d} \text{Bernoulli} (s_{i}; \psi), \label{eq:sprior2}
\end{align}
depending on hyper-parameters $(\alpha_{\psi}, \beta_{\psi})$. With the likelihood and prior specification in \eqref{eq:y_prior}--\eqref{eq:generalized-gprior}, our task is to compute the posterior distribution $\pi(s|y,X)$, which can be calculated in a closed form as follows by marginalizing over the parameter pair $(w, \sigma^2)$:
    \begin{align}
      \pi(s|y,X) & =  \int \pi(s, w_s, \sigma^2|y, X) dw_s d\sigma^{2} \nonumber\\
        & \propto \int \pi(s) \pi(\sigma^{2}) \pi(w_{s}|X, s,\sigma^2) \pi(y| X, s, w_{s}, \sigma^{2})  dw_s d\sigma^{2} \nonumber\\
        & \propto \pi(s)\frac{\sqrt{|X_{s}^{\T}X_{s}+\lambda I_{d_{s}}|}}{\sqrt{|(g+\kappa)X_{s}^{\T}X_{s}+\lambda I_{d}|}}(2\beta_{\sigma}+y^{\T}y-gy^{\T}X_{s}[(g+\kappa)X_{s}^{\T}X_{s}+\lambda I_{d_{s}}]^{-1}X_{s}^{\T}y)^{-\frac{2\alpha_{\sigma}+n}{2}}  .   \label{eq:full_posterior}
    \end{align}
See Supplement Section \ref{sec:ridge_prior} for details, along with the corresponding gradients.

For Bayesian variable selection, an important statistic is the Posterior Inclusion Probability (PIP), as introduced in \cite{george1993PIP}.
The PIP for $i$-th covariate is the posterior probability of $s_i=1$ (i.e., the $i$-th covariate is selected),
and hence provides a measure of its relative importance in the Bayesian regression model. For a simulated chain of length $T$ from the posterior distribution, the PIP for $i$-th covariate (or for $s_i$) is estimated by
\begin{equation}
    \text{PIP}(s_{i}) = \pi(s_{i} = 1|y, X)\approx \frac{1}{T}\sum\limits_{t=1}^{T} \mathds{1} \{s_{t,i} = 1\}, \nonumber
\end{equation}
where $s_{t,i}$ denotes the $t$-th draw for $s_i$.

In our numerical experiment, we set $\alpha_{\psi} = 0.1$ and $\beta_{\psi} = 10$ to encourage a sparse prior on the binary mask, and use $\alpha_{\sigma} = \beta_{\sigma} = 0.1$ to impose a vague prior on $\sigma^2$. The choice of the triplet $(g, \kappa, \lambda)$ is detailed in Supplement Section~\ref{sec:ridge_prior_calib}.

\subsection{Experiment with eQTL Data}
In our experiment, we extract a subset from the eQTL data in \cite{Myers2007snp} consisting of 100 observations and 1200 covariates to construct the covariate matrix $X$ as in \cite{Titsias2017Hamming}. Each covariate corresponds to the genotype expression of a distinct single nucleotide polymorphism (SNP). The covariate matrix $X$ contains only three distinct values, $0$, $1$, and $2$, with approximately equal proportions (about one-third each). The response vector $y$ is synthetically constructed to depend solely on the first covariate, defined as
\begin{equation}
    y_j = x_{j,1} + \eta_j, \quad \eta_j \sim \mathcal{N}(0, 0.01), \quad j = 1, \dots, 100. \label{eq:sparse_design}
\end{equation}
To introduce redundancy, the 601-st covariate is manually reset to be identical to the first, i.e., $x_{j,1} = x_{j,601}$ for all $j$, similarly as in \cite{Titsias2017Hamming}. Consequently, the PIP for $s_1$ and $s_{601}$ are approximately each $1/2$, and that for $s_i$ is close to 0 for $i\not\in\{1,601\}$. [Think about if need change other places.
Although being designed to depend solely on these two covariates, the response also exhibits correlations greater than 0.4 with 14 other covariates.

The Metropolis algorithm requires evaluating \eqref{eq:full_posterior} for all neighboring states when generating proposals, which becomes computationally prohibitive in high-dimensional settings. Although the GWG algorithm relies only on a single gradient evaluation of \eqref{eq:full_posterior}, it still requires normalization over all neighboring states, rendering it similarly infeasible with high dimensional $s$. Therefore, we exclude both the Metropolis and GWG algorithms from our Bayesian linear regression experiments due to their computational inefficiency.

For the Bayesian linear regression experiment, the parameter tuning strategy is different from previous experiments.
First, running multiple independent chains for each candidate parameter becomes computationally infeasible. Instead, we run a single chain with 15,000 draws after 3,000 burn-in draws for each parameter. Second, the optimal parameter is selected using acceptance rates and average flips; see Supplement Section \ref{sec:binary_tuning} for details.
Then we simulate 50 independent chains using the selected parameters, for each sampler. Each chain is initialized independently from an i.i.d.\ Bernoulli distribution with equal probability. The first 8,000 draws are discarded as burn-in, and the remaining 65,000 draws are retained for analysis.
\begin{figure}[tbp]
    \centering
    \begin{subfigure}{0.45\textwidth}
        \centering
        \includegraphics[width=\linewidth]{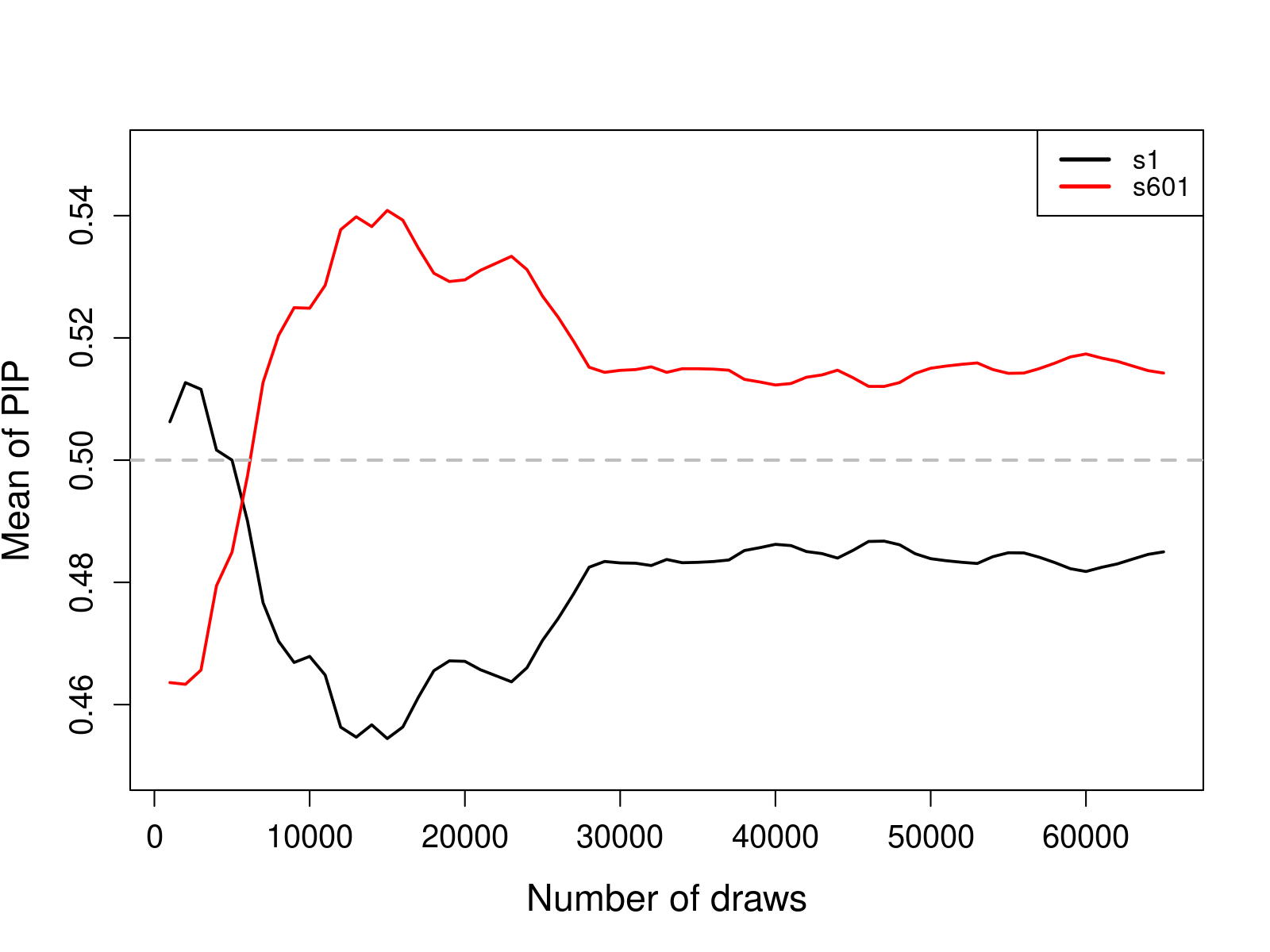}
        \caption{PIP from NCG}
        \label{fig:sparse_high_pip_ncg}
    \end{subfigure}
    \begin{subfigure}{0.45\textwidth}
        \centering
        \includegraphics[width=\linewidth]{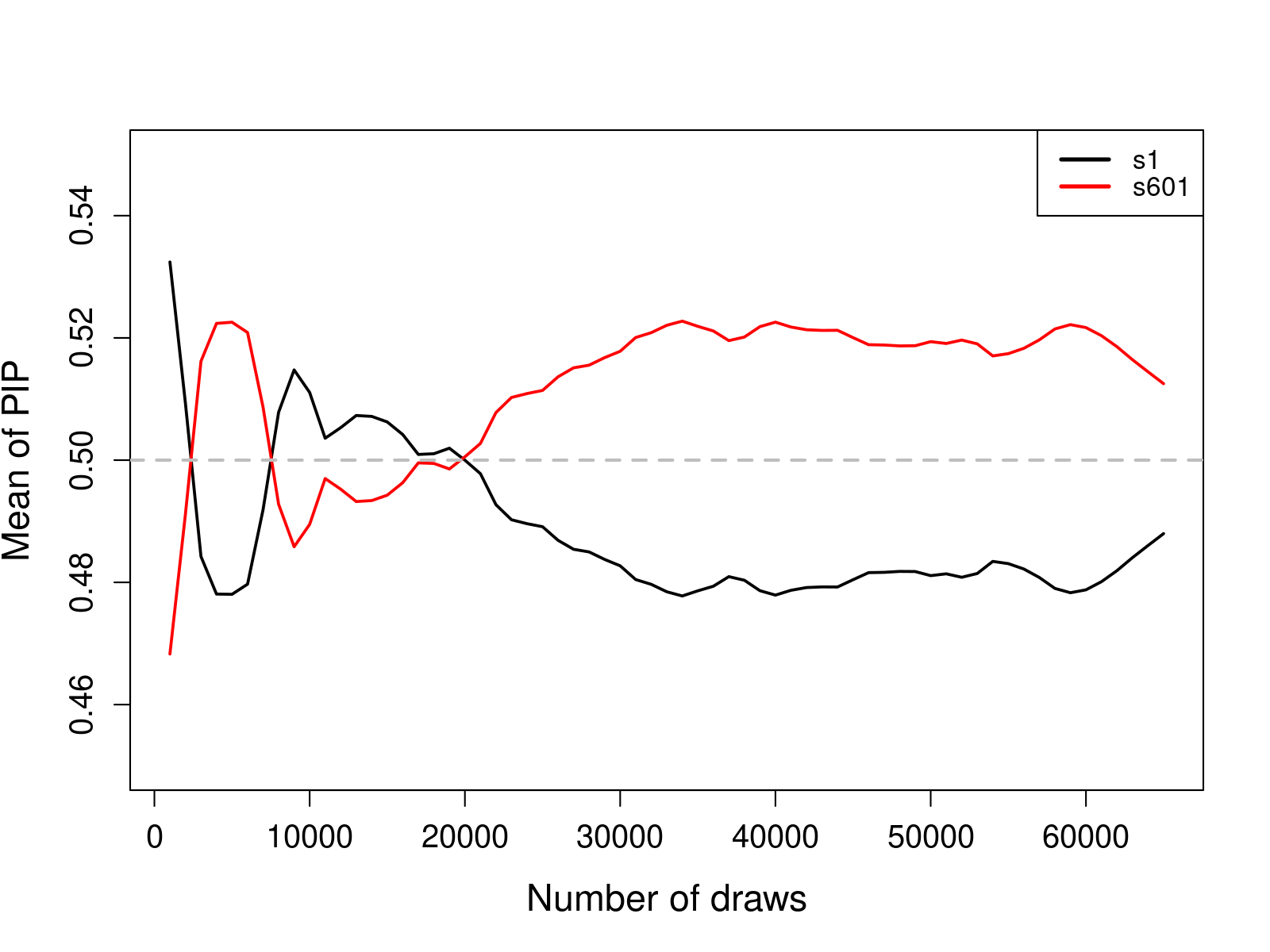}
        \caption{PIP from AVG}
        \label{fig:sparse_high_pip_avg}
    \end{subfigure}
    \begin{subfigure}{0.45\textwidth}
        \centering
        \includegraphics[width=\linewidth]{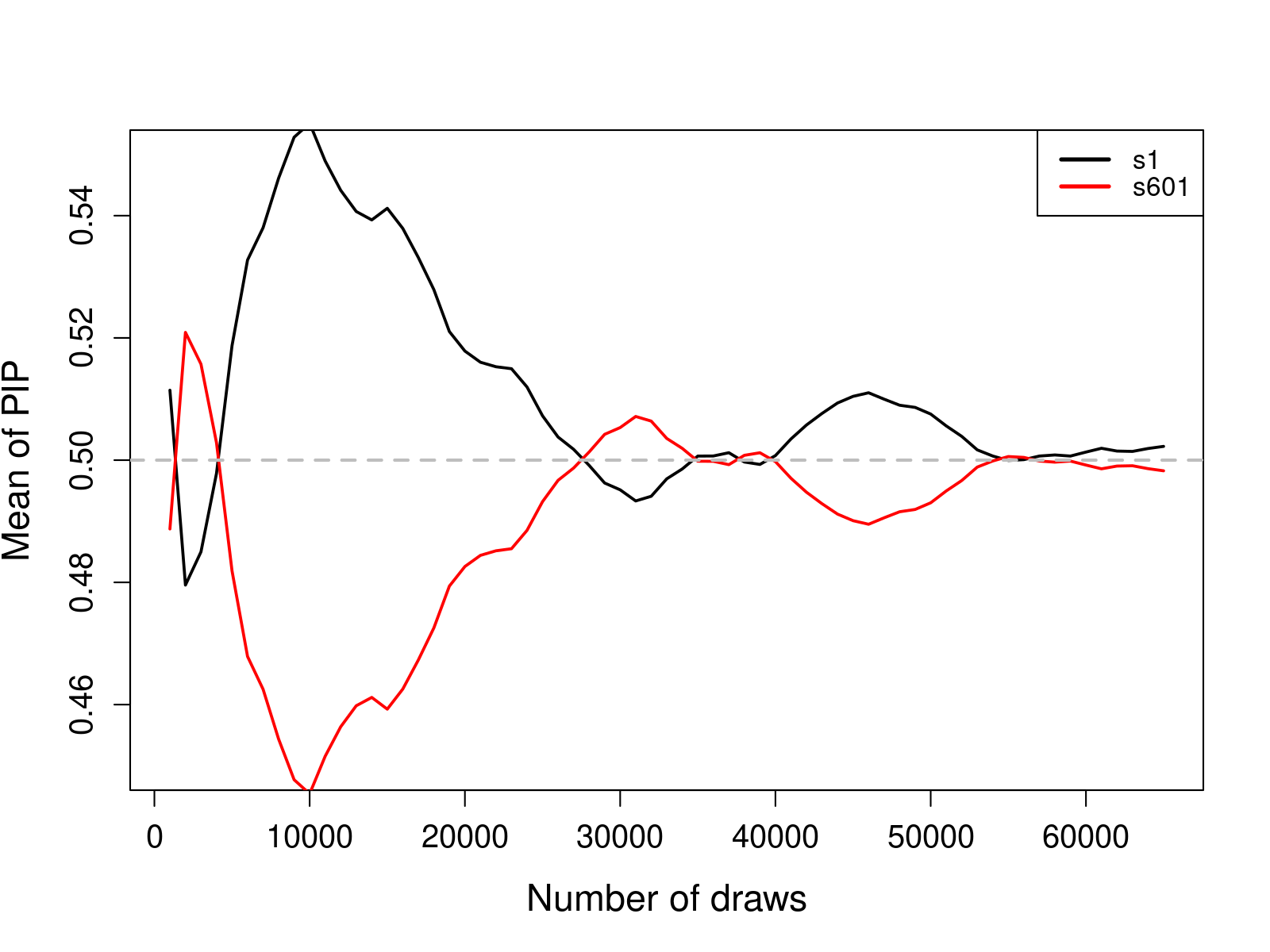}
        \caption{PIP from V-DHAMS}
        \label{fig:sparse_high_pip_hams}
    \end{subfigure}
    \begin{subfigure}{0.45\textwidth}
        \centering
        \includegraphics[width=\linewidth]{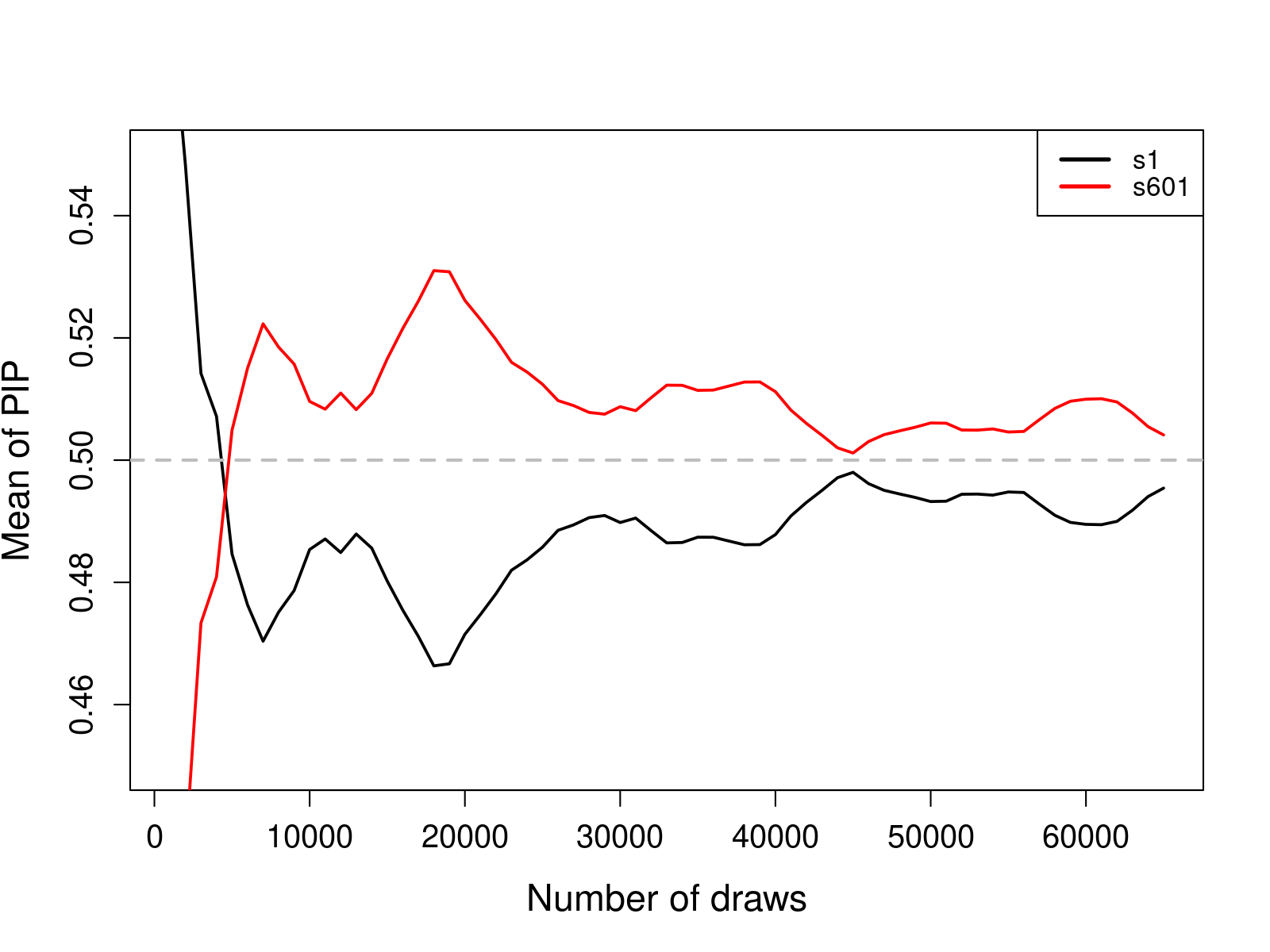}
        \caption{PIP from O-DHAMS}
        \label{fig:sparse_high_pip_overhams}
    \end{subfigure}
    \caption{Mean of PIP for pair $s_{1}$ and $s_{601}$ for Bayesian linear regression}
    \label{fig:sparse_high_pip}
\end{figure}
\begin{figure}[tbp]
    \centering
    \begin{subfigure}{0.45\textwidth}
        \centering
        \includegraphics[width=\linewidth]{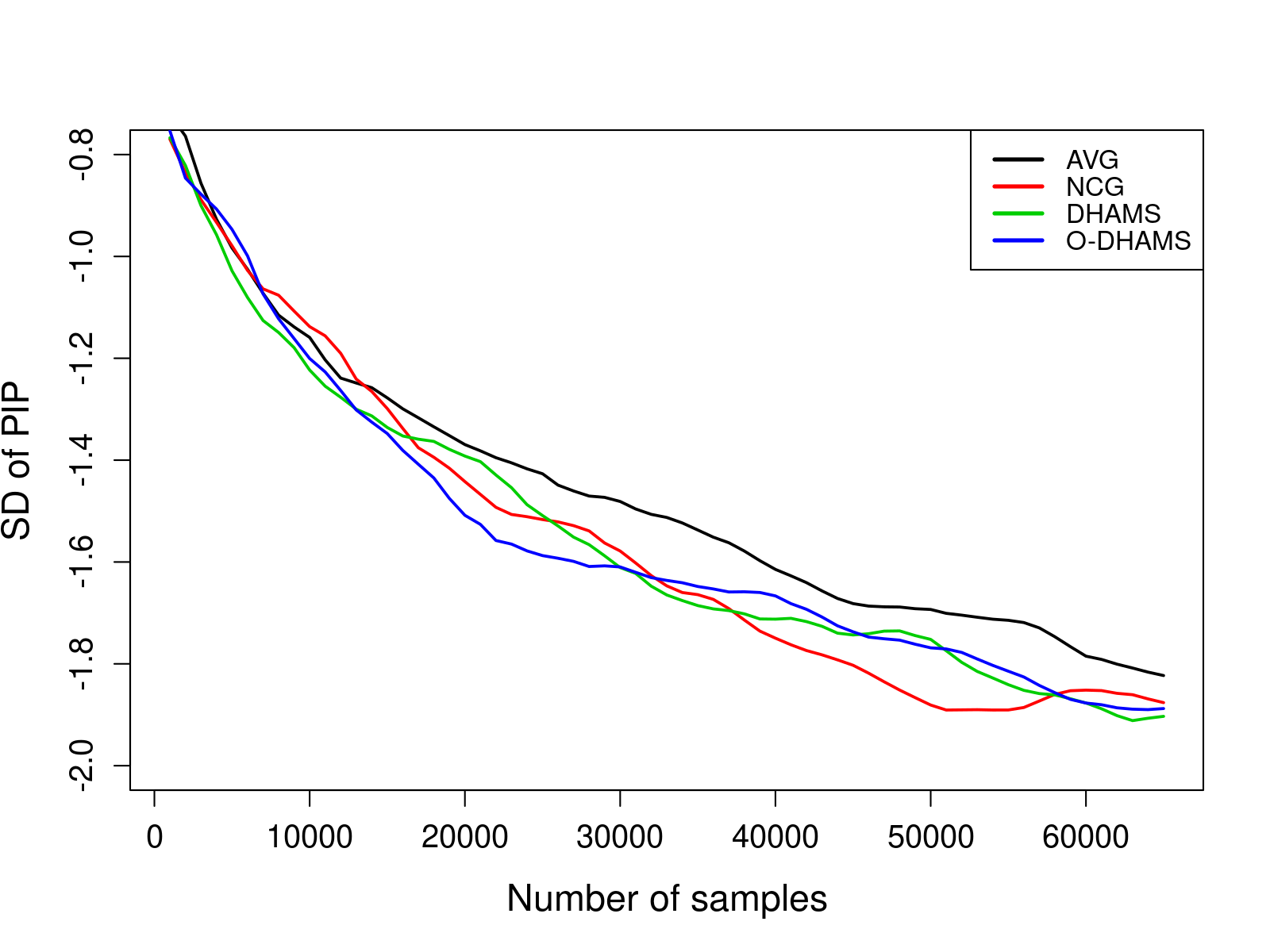}
        \caption{PIP standard deviation for $s_{1}$}
        \label{fig:sparse_high_pip_sd_s1}
    \end{subfigure}
    \begin{subfigure}{0.45\textwidth}
        \centering
        \includegraphics[width=\linewidth]{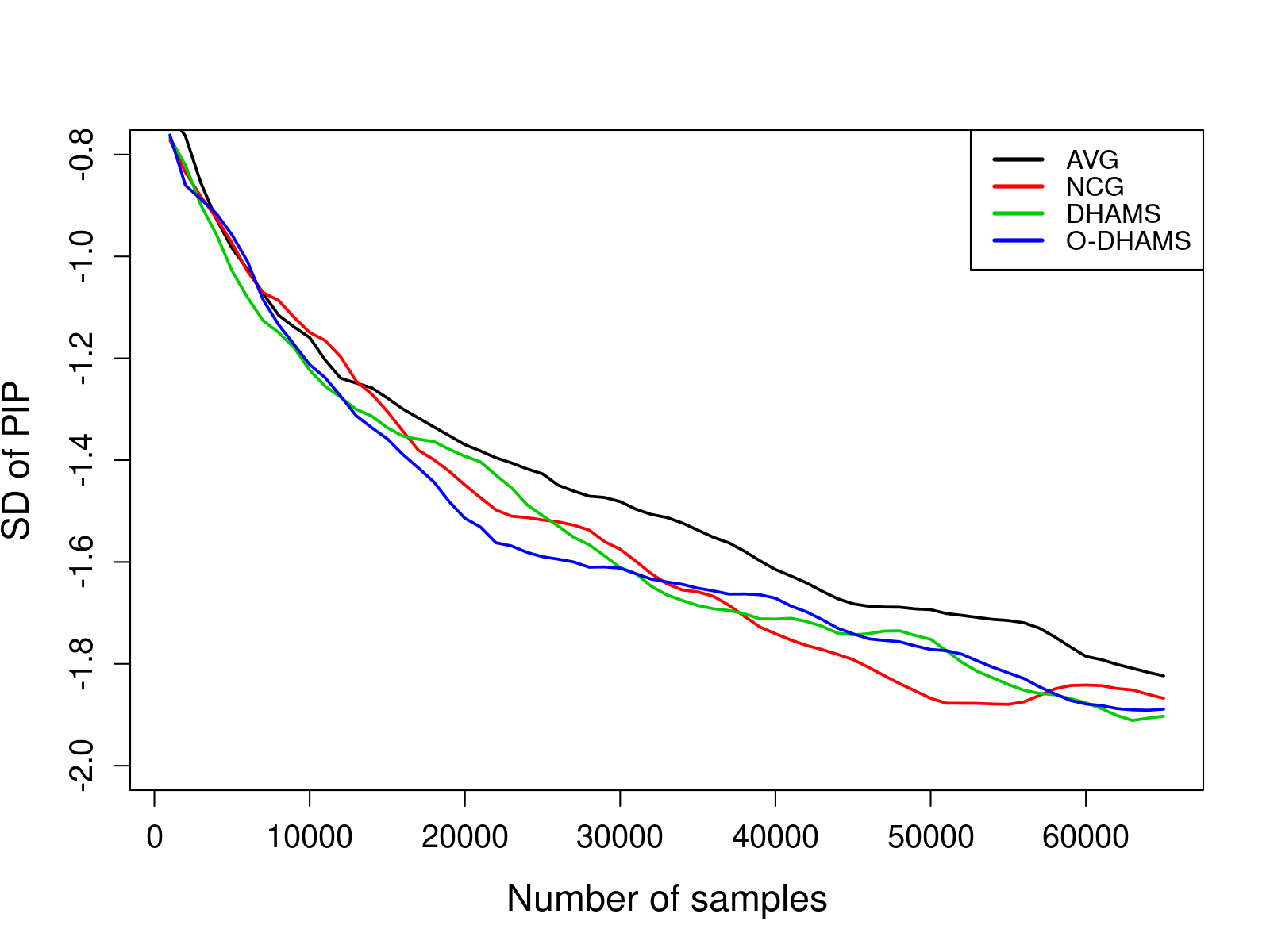}
        \caption{PIP standard deviation for $s_{601}$}
        \label{fig:sparse_high_pip_sd_s601}
    \end{subfigure}
    \caption{Standard deviation of PIP for pair $s_{1}$ and $s_{601}$ for Bayesian linear regression}
    \label{fig:sparse_high_pip_sd}
\end{figure}

Given high dimensionality $d$, computing the exact posterior probabilities is intractable. This makes comparison via TV-distances impractical.
Nevertheless, it is feasible to compare the PIP for relevant covariates of interest, along with their corresponding ESS.
The plots illustrating the mean of PIP from multiple chains for the pair $s_{1}$ and $s_{601}$
are shown in Figure~\ref{fig:sparse_high_pip}, while the plots depicting the standard deviation are presented in Figure~\ref{fig:sparse_high_pip_sd}.
Due to the sparse design \eqref{eq:sparse_design}, any mask other than the pair $s_1$ and $s_{601}$ have a very low posterior probability of being 1, resulting in highly unstable ESS estimates. The ESS of the negative potential function is also excluded as $f(s)$ takes only a few distinct values throughout most simulated chains. Therefore we only report ESS for the identical pair in Table~\ref{tab:highdim_sparse}.
\begin{table}[tbp]
\centering
\begin{tabular}{|c|c|c|c|}
\hline
    Sampler & ESS of $s_{1}$ & ESS of $s_{601}$  \\
    \hline
        NCG & 9.66 & 9.48 \\
    AVG  & 8.59 & 8.60\\
    V-DHAMS  & 10.26 & 10.26 \\
    O-DHAMS  & 9.92 & 9.95\\ \hline
\end{tabular}
\caption{ESS table for Bayesian linear regression}
\label{tab:highdim_sparse}
\end{table}

Figure ~\ref{fig:sparse_high_pip} indicates that both Discrete-HAMS sampler exhibits significantly smaller gaps between PIP of the identical pair $s_{1}$ and $s_{601}$,
while Figure~\ref{fig:sparse_high_pip_sd} suggests that the standard deviations of PIP estimates are similar across the four methods. In summary, the results of PIP suggests that Discrete-HAMS samplers achieve better mixing, compared to NCG and AVG. Moreover, V-DHAMS also achieves a higher ESS. The overall performance follows the order:
\[
\text{V-DHAMS} \approx \text{O-DHAMS} > \text{NCG} > \text{AVG}.
\]
Additional results including acceptance rates, parameter settings and auto-correlation plots are presented in Supplement Section~\ref{sec:linear_results}.

\section{Conclusion}

We introduce the first discrete sampler which not only exploits gradient information but also
samples an augmented distribution (or a Hamiltonian) with a Gaussian moment variable, such that generalized detailed balance is achieved.
Our numerical experiments demonstrate superior performance of DHAMS compared with existing algorithms.
It is interesting to further study theoretical properties of DHAMS and pursue potential improvements and extensions.
For example, a promising topic we currently investigate is
to incorporate a second-order quadratic term into the approximation of the potential function, as seen from \cite{Rhodes2022GradientMC} and \cite{sun2023anyscale}.
We expect that the rejection-free property of DHAMS can be extended to any target distribution with a quadratic potential (not just a linear potential).

\bibliographystyle{apalike}
\bibliography{hams}

\clearpage
\appendix

\begin{center}
{\Large Supplementary Material for}\\
{\Large ``Discrete Hamiltonian-Assisted Metropolis Sampling''}\\
\author{{Yuze Zhou} \text{and} {Zhiqiang Tan}}
\end{center}

\section{Generalized Metropolis-Hastings Sampling}\label{sec:GMH}
\cite{Song2023hams} presents the following generalized framework for Metropolis--Hastings samplers. Suppose that a target distribution $\pi(y)$ is invariant under some orthogonal transformation $J$ such that $\pi(J^{-1}y) = \pi(y)$, or equivalently
\begin{equation}
    \int_{C} d \pi(y) = \int_{J(C)}d \pi(y), \quad \forall C \subset \mathbb{R}^{d}.
    \label{eq:GMH_assumption}
\end{equation}
The Generalized Metropolis--Hastings sampler is as follows:
    \begin{itemize}
        \item[(i)] Given current state $y_{t}$, propose $y^{*}$ from some proposal distribution $Q(\cdot|y_{t})$.
        \item[(ii)] Accept $y_{t+1}=y^{*}$ with following probability,
        \begin{align}
               \min\{1, \frac{\pi(J^{-}y{*})Q(Jy_{t}|J^{-}y{*})}{\pi(y_{t})Q(y^{*}|y_{t})}\}. \nonumber
        \end{align}
         or reject $y^*$ and set $y_{t+1} = Jy_{t}$.
    \end{itemize}
This generalization can be shown to leave the target distribution $\pi(y)$ invariant.
With the corresponding transition kernel denoted as $K(\cdot|\cdot)$, the sampler also satisfies the following generalized detailed balance condition,
\begin{equation}
    \pi(y_{t})K(y_{t+1}|y_{t}) = \pi(J^{-1}y_{t+1})K(Jy_{t}|J^{-1}y_{t+1}). \nonumber
\label{eq:GMH}
\end{equation}

The framework does not require the target distribution of $\pi(y)$ to be continuous. For an augmented distribution with Gaussian momentum such that $\pi(y) \propto \exp( f(s) -\frac{1}{2}\|u\|_{2}^{2})$ with $y = (s,u) $, a natural choice of
\begin{align}
J = \begin{pmatrix}
    I & 0 \\
    0 & -I
\end{pmatrix} \label{eq:J-choice}
\end{align}
satisfies the assumption \eqref{eq:GMH_assumption}.
The associated Generalized Metropolis--Hastings sampling corresponds to the momentum negation technique as described in Section \ref{sec:negation}.  Discrete-HAMS (Algorithms \ref{algo:V-DHAMS} and \ref{algo:O-DHAMS}), HAMS (Algorithm \ref{algo:continuousHAMS}) and G2MS (Algorithm \ref{algo:G2MS}) are all examples of such samplers.

\section{Relevant Continuous Samplers}
In this section, we present two continuous samplers relevant to our study: MALA and HAMS. Then we present an alternative approach that derives the HAMS-A proposal using auxiliary variable and Gaussian over-relaxation techniques, analogous to those used to develop O-DHAMS (Algorithm \ref{algo:O-DHAMS}). For completeness, we also discuss two distinct ways of incorporating an auto-regression step into continuous HAMS.
Throughout this section, the target density $\pi(s) \propto \exp(f(s))$ is a continuous probability density with support $\mathds{R}^{d}$.

\subsection{Metropolis-adjusted Langevin Algorithm (MALA)} \label{sec:MALA}
The Metropolis-adjusted Langevin algorithm (MALA) \citep{Besag1994mala,Roberts1996mala} for sampling from $\pi(s)$ proceeds as follows:
    \begin{itemize}
        \item[(i)] Propose
        \begin{align} s^{*}  = s_{t} +\frac{\epsilon^{2}}{2} \nabla f(s_{t}) +\epsilon Z, \quad  Z \sim \mathcal{N}(0, I). \label{eq:MALA-proposal}
        \end{align}
        \item[(ii)] Accept $s_{t+1} = s^{*}$ with Metropolis--Hastings probability \eqref{eq:metropolis_prob}, or reject $s^*$  and set $s_{t+1} = s_{t}$.
    \end{itemize}

An interesting variant of the MALA sampler, called the modified-MALA \citep{Song2023hams}, modifies the proposal \eqref{eq:MALA-proposal} to
\begin{align} s^{*} & = s_{t} +\frac{\epsilon^{2}}{1+\sqrt{1-\epsilon^2}} \nabla f(s_{t}) +\epsilon Z, \quad  Z \sim \mathcal{N}(0, I). \label{eq:mMALA-proposal}\end{align}
The proposal is then accepted or rejected using standard Metropolis--Hastings probability.

\subsection{Hamiltonian-assisted Metropolis Sampling (HAMS)}\label{sec:HAMS}

In HAMS \citep{Song2023hams, Song2022reversible}, we introduce a momentum variable $u \in \mathds{R}^{d}$ from a $d$-dimensional standard Gaussian distribution and we sample from the augmented density $\pi(s,u) \sim \exp(f(s)-\frac{1}{2}\|u\|_{2}^{2})$. The sampler is summarized in Algorithm \ref{algo:continuousHAMS}.
\begin{algorithm}[htb]
With current state and momentum $(s_{t}, u_{t})$:
\begin{itemize}
    \item Propose
\begin{align}
\begin{pmatrix}
s^{*} \\
u^{*}
\end{pmatrix}
&=
\begin{pmatrix}
s_{t} \\
- u_{t}
\end{pmatrix}
- A
\begin{pmatrix}
-\nabla f(s_{t}) \\
- u_{t}
\end{pmatrix}
+
\begin{pmatrix}
Z_{1} \\
Z_{2}
\end{pmatrix}, \label{eq:hams_proposal}
\end{align}
   where $(Z_{1}^\T, Z_{2}^\T)^{\T} \sim \mathcal{N}(0, 2A-A^{2})$ and
\begin{align}
A =
\begin{pmatrix}
a_{1}I & a_{2}I \\
a_{2}I & a_{3}I
\end{pmatrix}. \nonumber
\end{align}
    With gradient correction on momentum, redefine
   \begin{equation}
       u^{*} = -u_{t} +a_{2} \nabla f(s_{t}) + a_{3}u_{t} +Z_{2} + \phi(s^{*}-s_{t}-\nabla f(s_{t}) + \nabla f(s^{*})). \nonumber
   \end{equation}
   \item Accept $(s_{t+1}, u_{t+1}) = (s^{*}, u^{*})$ with the following generalized Metropolis--Hastings probability
   \begin{equation}
         \min\{1, \frac{\pi(s^{*}, -u^{*})Q_{\phi}(s_{t}, -u_{t}|s^{*}, -u^{*})}{\pi(s_{t}, u_{t})Q_{\phi}(s^{*}, u^{*}|s_{t}, u_{t})}\}, \nonumber
   \end{equation}
 or otherwise set $(s_{t+1}, u_{t+1}) = (s_{t}, -u_{t})$.
\end{itemize}
\caption{Hamiltonian-assisted Metropolis Sampling (HAMS)}
\label{algo:continuousHAMS}
\end{algorithm}

A special class of the HAMS algorithm, known as the HAMS-A \citep{Song2023hams}, is of particular interest, which can be directly derived using auxiliary variable and Gaussian over-relaxation techniques \citep{Adler1981overrelaxation}. In HAMS-A, the matrix $A$ is set to be singular such that
\begin{align}
    A =
    \begin{pmatrix}
        aI & \sqrt{ab}I\\
        \sqrt{ab}I &  bI
    \end{pmatrix}. \nonumber
\end{align}
From \cite{Song2023hams, Song2022reversible}, for the choice $a = \frac{\epsilon^2}{1+\sqrt{1-\epsilon^2}}$, $b=0$ and $\phi=0$ in HAMS-A, the proposal $s^*$ becomes
\begin{align}
    s^{*} &= s_{t}+a\nabla f(s_{t})-\sqrt{ab}u_{t}+Z_1, \nonumber\\
    &= s_{t}+ \frac{\epsilon^2}{1+\sqrt{1-\epsilon^2}} \nabla f(s_{t})+Z_1, \nonumber \\
    \var(Z_1) &= a(2-a-b)I = \frac{\epsilon^2}{1+\sqrt{1-\epsilon^2}}(2-\frac{\epsilon^2}{1+\sqrt{1-\epsilon^2}}) I = \epsilon^2I,
\end{align}
which is the same as the proposal in modified-MALA \eqref{eq:mMALA-proposal}.

Another sampler of interest related to HAMS is the Generalized Gradient-Guided Metropolis Sampling (G2MS), as described in \cite{Song2023hams}. The key difference between this sampler and HAMS lies in the proposal step, where the symmetric matrix $A$ in HAMS (Algorithm \ref{algo:continuousHAMS}) is replaced by an asymmetric matrix $B$. This sampler is presented in its general version in Algorithm~\ref{algo:G2MS}, provided that invariance \eqref{eq:GMH_assumption} is satisfied. For sampling from the augmented distribution $\pi(y) \propto \exp( f(s) -\frac{1}{2}\|u\|_{2}^{2})$  with $y = (s,u) $,
we take $f(y) = f(s) -\frac{1}{2}\|u\|_{2}^{2}$ by some abuse of notation, and $J$ to be the diagonal matrix \eqref{eq:J-choice}.

\begin{algorithm}[htb]
With current state variable $y_{t}$:

\begin{enumerate}
    \item Propose
     \begin{equation}
           y^{*} = y_{t}+B \nabla f(s_{t}) +Z, \nonumber
       \end{equation}
       where $Z \sim N(0, 2A-A^{2})$, $B= I-(I-A)J$ and $ 2A-A^{2} = B+B^{\T}-BB^{\T}$.
       \item  Compute backward proposal $Z^{*}$ from
       \begin{equation}
           Jy_{t} = J^{-1}y^{*}+B\nabla f(J^{-1}y^{*}) +Z^{*} \nonumber
       \end{equation}
       by replacing $(y_{t}, y^{*})$ with $(J^{-1}y^{*}, Jy_{t})$ in Step 1 above.

       \item Accept $y_{t+1} = y^{*}$ with following generalized Metropolis--Hastings probability
       \begin{equation}
           \min\{1, \frac{\pi(y^{*})\mathcal{N}(Z^{*}|0, 2A-A^{2})}{\pi(y_{t})\mathcal{N}(Z|0, 2A-A^{2})}\},\nonumber
       \end{equation}
       or otherwise set $y_{t+1} = Jy_{t}$.
\end{enumerate}
\caption{Generalized Gradient-Guided Metropolis Sampling (G2MS)}
\label{algo:G2MS}

\end{algorithm}

\subsection{Alternative Derivation for HAMS-A proposal} \label{sec:HAMS-A}

To compare with the ideas in the derivation of O-HAMS,
we present an alternative approach to derive the HAMS-A proposal (which is a degenerate case in the class of HAMS proposals)
using auxiliary variable and over-relaxation techniques. See \cite{Song2023hams}, Supplement Section I, for a derivation of HAMS proposals
with marginalization over auxiliary variables.

With current state and momentum $(s_{t}, u_{t})$, the density for state $s$ is approximated as $\tilde{\pi}(s;s_{t}) \propto \mathcal{N}(s;s_{t}+\nabla f(s_{t}), I)$.
First, we construct the auxiliary variable $z = s+\delta u$ as in Section~\ref{sec:auxiliary_scheme}. Then the joint distribution of $(s,u,z)$ is approximated as
\begin{align}
\pi(s, u, z; s_{t}) \sim \mathcal{N} \left(
\begin{pmatrix}
s_{t} + \nabla f(s_{t}) \\
0 \\
s_{t} + \nabla f(s_{t})
\end{pmatrix},
\begin{pmatrix}
I & 0 & I \\
0 & I & \delta I \\
I & \delta I & (1 + \delta^{2}) I
\end{pmatrix}
\right).
\label{eq:HAMS1}
\end{align}
From \eqref{eq:HAMS1}, with $z_{t} = s_t +\delta u_t$, the conditional distribution of $(s,u)$ given $z_t$ is
\begin{align}
\begin{pmatrix}
s \\
u
\end{pmatrix}
\Big| z_{t} \sim \mathcal{N} \left(
\begin{pmatrix}
\frac{1}{1 + \delta^{2}} z_{t} + \frac{\delta^{2}}{1 + \delta^{2}} (s_{t} + \nabla f(s_{t})) \\
\frac{\delta}{1 + \delta^{2}} z_{t} - \frac{\delta}{1 + \delta^{2}} (s_{t} + \nabla f(s_{t}))
\end{pmatrix},
\begin{pmatrix}
\frac{\delta^{2}}{1 + \delta^{2}} I & -\frac{\delta}{1 + \delta^{2}} I \\
-\frac{\delta}{1 + \delta^{2}} I & \frac{1}{1 + \delta^{2}} I
\end{pmatrix}
\right).
\label{eq:HAMS2}
\end{align}
The state and momentum $(s^{*}, u^{*})$ is proposed from the conditional distribution \eqref{eq:HAMS2}.
By substituting $z_{t} = s_{t}+\delta u_{t}$ into \eqref{eq:HAMS2}
and introducing Gaussian noise $(Z_{1}^\T, Z_{2}^\T)^{\T} \sim \mathcal{N}(0, A)$, the proposal $Q(s^{*}, u^{*}|s_{t}, u_{t})$ can be expressed as
\begin{align}
\begin{pmatrix}
s^{*} \\
u^{*}
\end{pmatrix}
=
\begin{pmatrix}
s_{t} \\
u_{t}
\end{pmatrix}
- A
\begin{pmatrix}
- \nabla f(s_{t}) \\
u_{t}
\end{pmatrix}
+
\begin{pmatrix}
Z_{1} \\
Z_{2}
\end{pmatrix},
\label{eq:HAMS3}
\end{align}
where
\begin{align}
A = A^{2} =
\begin{pmatrix}
\frac{\delta^{2}}{1 + \delta^{2}} I & -\frac{\delta}{1 + \delta^{2}} I \\
-\frac{\delta}{1 + \delta^{2}} I & \frac{1}{1 + \delta^{2}} I
\end{pmatrix}.
\label{eq:A-matrix}
\end{align}
which is easily verified to be singular. Note that $2A-A^2 = A $ by \eqref{eq:A-matrix}.

Next, by treating $(s_{t}, u_{t})$ as $x_0$ and the proposal distribution \eqref{eq:HAMS3} as the Gaussian reference distribution $\mathcal{N}(\mu, \Sigma)$,
we apply Gaussian over-relaxation in \eqref{eq:Gauss-overrelax} \citep{Adler1981overrelaxation} and obtain
\begin{align}
\begin{pmatrix}
s^{*} \\
u^{*}
\end{pmatrix}
=
\begin{pmatrix}
s_{t} \\
u_{t}
\end{pmatrix}
- (1 - \alpha) A
\begin{pmatrix}
\nabla U(x_{t}) \\
u_{t}
\end{pmatrix}
+ \sqrt{1 - \alpha^{2}}
\begin{pmatrix}
Z_{1} \\
Z_{2}
\end{pmatrix}, \nonumber
\end{align}
where the following holds, with $\tilde{A} = (1-\alpha)A$:
\begin{align}
    \var \begin{pmatrix}
        \sqrt{1-\alpha^2}Z_1 \\
         \sqrt{1-\alpha^2}Z_1
    \end{pmatrix} = (1-\alpha^2)A = 2\tilde{A}-\tilde{A}^2. \nonumber
\end{align}
To introduce generalized reversibility, we apply momentum negation to $u_{t}$, yielding
\begin{align}
\begin{pmatrix}
s^{*} \\
u^{*}
\end{pmatrix}
=
\begin{pmatrix}
s_{t} \\
- u_{t}
\end{pmatrix}
- (1 - \alpha) A
\begin{pmatrix}
- \nabla f(s_{t}) \\
- u_{t}
\end{pmatrix}
+ \sqrt{1 - \alpha^{2}}
\begin{pmatrix}
Z_{1} \\
Z_{2}
\end{pmatrix}.
\label{eq:HAMS5}
\end{align}

Finally, by the singularity of the variance matrix $A$ for $(Z_{1}^\T, Z_{2}^\T)^\T$ in \eqref{eq:A-matrix},
we rewrite $Z_{1} = \sqrt{\frac{\delta^{2}}{1+\delta^{2}}}Z$ and $Z_{2} = -  \sqrt{\frac{1}{1+\delta^{2}}}Z$ for some common Gaussian variable $Z \sim \mathcal{N}(0, I)$.
Then expanding all the terms in \eqref{eq:HAMS5}, we have
\begin{align}
     s^{*} &= s_{t}+(1-\alpha)\frac{\delta^{2}}{1+\delta^{2}}\nabla f(s_{t})-(1-\alpha)\frac{\delta}{1+\delta^{2}}u_{t}+\sqrt{1-\alpha^{2}}\sqrt{\frac{\delta^{2}}{1+\delta^{2}}}Z, \nonumber \\
    u^{*} &= -u_{t}-(1-\alpha)\frac{\delta}{1+\delta^{2}}\nabla f(s_{t}) +(1-\alpha)\frac{1}{1+\delta^{2}}u_{t} -\sqrt{1-\alpha^{2}}\sqrt{\frac{1}{1+\delta^{2}}}Z.\label{eq:HAMS6}
\end{align}
The final form of the proposal in \eqref{eq:HAMS6} is exactly the HAMS-A proposal without the gradient correction term on momentum:
\begin{align}
     s^{*} &= s_{t}+a\nabla f(s_{t})-\sqrt{ab}u_{t}+\sqrt{a(2-a-b)}Z, \nonumber \\
    u^{*} &= -u_{t}-\sqrt{ab}\nabla f(s_{t}) +bu_{t} - \sqrt{b(2-a-b)}Z.\nonumber
\end{align}
by setting $a= \frac{1-\alpha}{1+\delta^2}$ and $b= \frac{(1-\alpha)\delta^2}{1+\delta^2}$, or equivalently $\delta = -\sqrt{\frac{a}{b}}$ and $\alpha = 1-a-b$. This mapping between parameters $(a,b)$ and $(\delta, \alpha)$ is one-to-one.

\subsection{Auto-regression Step for HAMS-A}\label{sec:auto_regressiveHAMS}

In this and next sections, we study two distinct ways of incorporating an auto-regression step into HAMS.
Here we incorporate auto-regression inside the derivation of HAMS-A in Supplement Section \ref{sec:HAMS-A}.
Specifically, we construct the auxiliary variable $z = s + \delta(\epsilon u + \sqrt{1 - \epsilon^{2}} Z)$, where $Z \sim \mathcal{N}(0, I)$ is a Gaussian noise independent of $s$ and $u$. Combined with the approximation $\tilde{\pi}(s; s_{t}) = \mathcal{N}(s; s_{t} + \nabla f(s_{t}), I)$, the joint distribution of $(s,u,z)$ is approximated by
\begin{align}
\pi(s, u, z; s_{t}) \sim \mathcal{N} \left(
\begin{pmatrix}
s_{t} + \nabla f(s_{t}) \\
0 \\
s_{t} + \nabla f(s_{t})
\end{pmatrix},
\begin{pmatrix}
I & 0 & I \\
0 & I & \delta \epsilon I \\
I & \delta \epsilon I & (1 + \delta^{2}) I
\end{pmatrix}
\right).
\label{eq:auto_regressiveHAMS1}
\end{align}
With the current state $s_{t}$ and momentum $ u_{t}$, we have $z_{t} = s_{t}+\delta u_{t+1/2} = s_{t} +\delta(\epsilon u_{t} +\sqrt{1-\epsilon^{2}}Z)$, where $u_{t+1/2}$ is the intermediate momentum as in \eqref{eq:auto-regressive}. From \eqref{eq:auto_regressiveHAMS1}, the conditional distribution of $(s,u)$ given $z_t$ is
\begin{align}
\begin{pmatrix}
s \\
u
\end{pmatrix}
\Big| z_{t} \sim \mathcal{N} \left(
\begin{pmatrix}
\frac{1}{1 + \delta^{2}} z_{t} + \frac{\delta^{2}}{1 + \delta^{2}} (s_{t} + \nabla f(s_{t})) \\
\frac{\delta \epsilon}{1 + \delta^{2}} z_{t} - \frac{\delta \epsilon}{1 + \delta^{2}} (s_{t} + \nabla f(s_{t}))
\end{pmatrix},
\begin{pmatrix}
\frac{\delta^{2}}{1 + \delta^{2}} I & -\epsilon \frac{\delta}{1 + \delta^{2}} I \\
-\epsilon \frac{\delta}{1 + \delta^{2}} I & \left(1 - \frac{\epsilon^{2} \delta^{2}}{1 + \delta^{2}} \right) I
\end{pmatrix}
\right).
\label{eq:auto_regressiveHAMS2}
\end{align}
The proposal $(s^{*}, u^{*})$ is then obtained from the conditional distribution \eqref{eq:auto_regressiveHAMS2}. By introducing the Gaussian noise $(Z_{1}^\T, Z_{2}^\T)^{\T} \sim \mathcal{N}(0, A)$,  the proposal can be expressed as
\begin{align}
   & s^{*} = \frac{1}{1+\delta^{2}}z_{t}+\frac{\delta^{2}}{1+\delta^{2}}(s_{t}+\nabla f(s_{t}))+Z_{1}, \nonumber \\
& u^{*} = \frac{\delta\epsilon}{1+\delta^{2}}z_{t}-\frac{\delta\epsilon}{1+\delta^{2}}(s_{t}+\nabla f(s_{t}))+Z_{2}, \label{eq:auto_regressiveHAMS3}
\end{align}
where
\begin{align}
A =
\begin{pmatrix}
\frac{\delta^{2}}{1 + \delta^{2}} I & -\epsilon \frac{\delta}{1 + \delta^{2}} I \\
-\epsilon \frac{\delta}{1 + \delta^{2}} I & \left(1 - \frac{\epsilon^{2} \delta^{2}}{1 + \delta^{2}} \right) I
\end{pmatrix}.
\label{eq:auto_regressiveHAMS4}
\end{align}

Substituting $z_{t} = s_{t}+\delta u_{t+1/2} = s_{t} +\delta(\epsilon u_{t} +\sqrt{1-\epsilon^{2}}Z)$ into \eqref{eq:auto_regressiveHAMS3}:
\begin{align}
    & s^{*} = s_{t} + \frac{\delta^{2}}{1+\delta^{2}} \nabla f(s_{t}) +\epsilon \frac{\delta}{1+\delta^{2}} u_{t}+\sqrt{1-\epsilon^{2}}\frac{\delta}{1+\delta^{2}}Z +Z_{1}, \nonumber \\
    & u^{*} = -\epsilon \frac{\delta}{1+\delta^{2}} \nabla f(s_{t}) +\epsilon^{2}\frac{\delta^{2}}{1+\delta^{2}}u_{t} +\epsilon\sqrt{1-\epsilon^{2}} \frac{\delta^{2}}{1+\delta^{2}} Z +Z_{2}. \label{eq:auto_regressiveHAMS5}
\end{align}
Define the new noise pair $\tilde{Z_{1}} =\sqrt{1-\epsilon^{2}}\frac{\delta}{1+\delta^{2}}Z +Z_{1}$ and $\tilde{Z_{2}} = \epsilon\sqrt{1-\epsilon^{2}} \frac{\delta^{2}}{1+\delta^{2}} Z +Z_{2}$.
The proposal \eqref{eq:auto_regressiveHAMS5} can be rewritten as
\begin{align}
\begin{pmatrix}
s^{*} \\
u^{*}
\end{pmatrix}
=
\begin{pmatrix}
s_{t} \\
u_{t}
\end{pmatrix}
-
A
\begin{pmatrix}
- \nabla f(s_{t}) \\
u_{t}
\end{pmatrix}
+
\begin{pmatrix}
\tilde{Z}_{1} \\
\tilde{Z}_{2}
\end{pmatrix}.
\label{eq:auto_regressiveHAMS6}
\end{align}
The new noise pair $(\tilde Z_{1}^\T, \tilde Z_{2}^\T)^{\T}$ is Gaussian with zero mean and variance directly calculated as
\begin{align}
\var
\begin{pmatrix}
\tilde{Z}_{1} \\
\tilde{Z}_{2}
\end{pmatrix}
&=
\var
\begin{pmatrix}
\sqrt{1 - \epsilon^{2}} \, \frac{\delta}{1 + \delta^{2}} Z \\
\epsilon \sqrt{1 - \epsilon^{2}} \, \frac{\delta^{2}}{1 + \delta^{2}} Z
\end{pmatrix}
+
\var
\begin{pmatrix}
Z_{1} \\
Z_{2}
\end{pmatrix} \nonumber \\
&=
\begin{pmatrix}
(1 - \epsilon^{2}) \frac{\delta^{2}}{1 + \delta^{2}} + \frac{\delta^{2}}{1 + \delta^{2}} &
\quad -\frac{\epsilon \delta (1 + \epsilon^{2} \delta^{2})}{(1 + \delta^{2})^{2}} \\
-\frac{\epsilon \delta (1 + \epsilon^{2} \delta^{2})}{(1 + \delta^{2})^{2}} &
\quad 1 - \frac{\epsilon^{2} \delta^{2} (1 + \epsilon^{2} \delta^{2})}{(1 + \delta^{2})^{2}}
\end{pmatrix}
\otimes I \nonumber \\
&= 2A - A^{2}. \nonumber
\end{align}

Applying momentum negation in \eqref{eq:auto_regressiveHAMS6} leads to exactly the proposal \eqref{eq:hams_proposal} (before gradient correction on $u^*$) in the HAMS sampler with the choice of $A$ in \eqref{eq:auto_regressiveHAMS4}.
This choice of $A$ is in general non-singular and hence the resulting HAMS sampler is no longer HAMS-A.

The operation of combining the terms of $(Z_{1}^\T, Z_{2}^\T)^{\T}$ and $Z$ into new noise pair $(\tilde Z_{1}^\T, \tilde Z_{2}^\T)^{\T}$
effectively marginalizes out the auxiliary variable $z_t$ and yields a marginalized proposal  $(s^*, u^*)$ from $(s_t, u_t)$. Such a marginalization depends on additivity of jointly Gaussian variables, and does not apply to the proposal \eqref{eq:auto-regressive-s}--\eqref{eq:auto-regressive-u} in Discrete-HAMS.

\subsection{Auto-regression Step for HAMS}\label{sec:insymmetricHAMS}
In this section, we demonstrate that the auto-regression step can also be combined with any HAMS proposal (outside the auxiliary variable derivation), and a G2MS proposal is obtained after marginalizing out the intermediate momentum $u_{t+1/2}$.
First, recall that a HAMS proposal without negation and gradient correction on momentum in Section \ref{sec:HAMS} is
\begin{align}
\begin{pmatrix}
s^{*} \\
u^{*}
\end{pmatrix}
&=
\begin{pmatrix}
s_{t} \\
u_{t}
\end{pmatrix}
- A
\begin{pmatrix}
-\nabla f(s_{t}) \\
u_{t}
\end{pmatrix}
+ \begin{pmatrix}
Z_{1} \\
Z_{2}
\end{pmatrix},
\label{eq:asymmetric1}
\end{align}
\\
where
\begin{align*}
    A &=
    \begin{pmatrix}
    a_{1}I & a_{2}I \\
    a_{2}I & a_{3}I
    \end{pmatrix},
    \quad
    \begin{pmatrix}
    Z_{1} \\
    Z_{2}
    \end{pmatrix}
    \sim \mathcal{N}(0,\, 2A - A^{2}).
\end{align*}
In the proposal \eqref{eq:asymmetric1}, we replace the initial momentum $u_{t}$ with the intermediate momentum
 $u_{t+1/2} = \epsilon u_{t}+\sqrt{1-\epsilon^{2}}Z$, where $Z \sim \mathcal{N}(0,I)$ is drawn independently.
The new proposal is
\begin{align*}
\begin{pmatrix}
s^{*} \\
u^{*}
\end{pmatrix}
&=
\begin{pmatrix}
s_{t} \\
u_{t+1/2}
\end{pmatrix}
- A
\begin{pmatrix}
-\nabla f(s_{t}) \\
u_{t+1/2}
\end{pmatrix}
+ \begin{pmatrix}
Z_{1} \\
Z_{2}
\end{pmatrix},
\end{align*}
and can be rewritten as
\begin{align}
\begin{pmatrix}
s^{*} \\
u^{*}
\end{pmatrix}
&=
\begin{pmatrix}
s_{t} \\
u_{t}
\end{pmatrix}
- B
\begin{pmatrix}
-\nabla f(s_{t}) \\
u_{t}
\end{pmatrix}
+
\begin{pmatrix}
\tilde{Z}_{1} \\
\tilde{Z}_{2}
\end{pmatrix}, \nonumber
\end{align}
where $\tilde{Z}_{1} = -a_{2}\sqrt{1-\epsilon^{2}}Z+Z_{1}$, $\tilde{Z}_{2} = (1-a_{3})\sqrt{1-\epsilon^{2}}Z+Z_{2}$, and the matrix $B$ is
\begin{align}
B &= A \begin{pmatrix}
I & 0 \\
0 & \epsilon I
\end{pmatrix}
+ \begin{pmatrix}
0 & 0 \\
0 & (1 - \epsilon)I
\end{pmatrix} \nonumber \\
&= \begin{pmatrix}
a_1 & a_2 \epsilon \\
a_2 & a_3 \epsilon + (1 - \epsilon)
\end{pmatrix} \otimes I. \nonumber
\end{align}
The new noise pair $(\tilde{Z}_{1}^\T, \tilde{Z}_{2}^\T)^{\T}$ is Gaussian with zero mean and variance calculated as
\begin{align}
    \operatorname{Var} \left(
    \begin{pmatrix}
         \tilde{Z}_{1} \\
         \tilde{Z}_{2}
    \end{pmatrix}
    \right)
    &=
    \operatorname{Var}\left(
    \begin{pmatrix}
        -a_{2} \sqrt{1-\epsilon^{2}} Z \\
        (1 - a_{3}) \sqrt{1-\epsilon^{2}} Z
    \end{pmatrix}
    \right)
    +
    \operatorname{Var} \left(
    \begin{pmatrix}
        Z_{1} \\
        Z_{2}
    \end{pmatrix}
    \right) \nonumber \\
    &=
    \begin{pmatrix}
        a_{2}^{2}(1 - \epsilon^{2}) & -a_{2}(1 - a_{3})(1 - \epsilon^{2}) \\
        -a_{2}(1 - a_{3})(1 - \epsilon^{2}) & (1 - a_{3})^{2}(1 - \epsilon^{2})
    \end{pmatrix} \otimes I + 2A - A^{2} \nonumber \\
    &= B + B^\T - BB^\T.
    \label{eq:asymmetric2}
\end{align}

By combining the terms of $(Z_{1}^\T, Z_{2}^\T)^{\T}$ and $Z$ into new noise pair $(\tilde{Z}_{1}^\T, \tilde{Z}_{2}^\T)^{\T}$, we effectively marginalize out $u_{t+1/2}$ and obtain the proposal $(s^*, u^*)$ directly from $(s_t, u_t)$. The relation between $B$ and the variance of the new Gaussian noise pair $(\tilde{Z}_{1}, \tilde{Z}_{2})^{\T}$ in \eqref{eq:asymmetric2} indicates that the proposed state and momentum $(s^{*}, u^{*})$ represents a G2MS proposal (Algorithm \ref{algo:G2MS}).

\section{GWG and Metropolis}\label{sec:prototype_variant}
For multivariate ordinal distributions on a large lattice (as encoded by $k$ in Sections \ref{sec:exp_gauss} and \ref{sec:poly_mix}), we specify the neighborhood structure in both the Gibbs-with-Gradient and Metropolis--Hastings algorithms to guide the proposal mechanism.
The corresponding algorithms are described as follows.

\subsection{Ordinal-GWG}
We adopt the neighbor structure for Ordinal-GWG as in \cite{Rhodes2022GradientMC}. The original GWG algorithm \citep{Grathwohl2021gwg} is designed for multivariate binary distributions, where the proposal $s^*$ from $s_t$ automatically satisfies
$\|s_t - s^*\|_\infty \leq 1$. However, in a multivariate ordinal distribution,
the proposal $s^*$ may differ from $s_t$ by multiple levels along each coordinate, even when restricted by the Hamming ball.
To control the neighborhood size, the proposal is restricted by $\|s-s_{t}\|_{\infty} \leq r$
for a window size parameter $r$, in addition to being restricted by the Hamming ball.
The corresponding Ordinal-GWG proposal is then
\begin{equation}
   s^{*} \sim Q(s|s_{t}) \propto \exp(\frac{1}{2}\nabla f(s_{t})^{\T}(s-s_{t})) \mathds{1}
\{d_{H}(s,s_{t})\leq \delta , \|s-s_{t}\|_{\infty} \leq r \}. \nonumber
\label{eq:ordinal_GWG}
\end{equation}
\subsection{Metropolis with $L_{\infty}$ Neighborhood}
We implement the Metropolis--Hastings sampler for multivariate ordinal distributions with
the neighbors of $s_{t}$ specified by
$\{s : \|s - s_t\|_{\infty} \leq r\}$, for a window size parameter $r$. The corresponding proposal distribution is defined as
\begin{equation}
 s^{*} \sim  Q(s|s_{t}) \propto \mathds{1} \{\|s-s_{t}\|_{\infty} \leq r\}. \nonumber
\end{equation}

\begin{figure}[tbp]
    \centering
    \begin{subfigure}{0.3\textwidth}
        \centering
        \includegraphics[width=\linewidth]{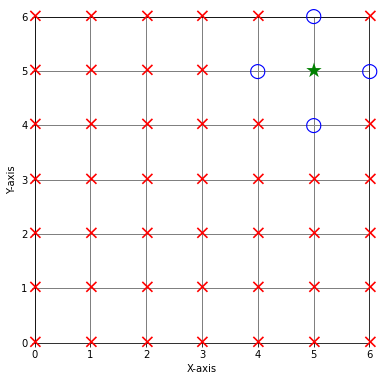}
        \caption{GWG}
        \label{fig:gwg_neighbors}
    \end{subfigure}
    \begin{subfigure}{0.3\textwidth}
        \centering
        \includegraphics[width=\linewidth]{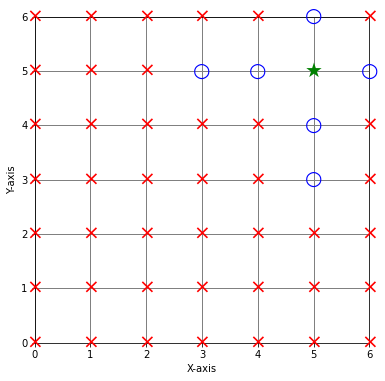}
        \caption{ordinal-GWG }
        \label{fig:gwg_ordinal_neighbors}
    \end{subfigure}
    \begin{subfigure}{0.3\textwidth}
        \centering
        \includegraphics[width=\linewidth]{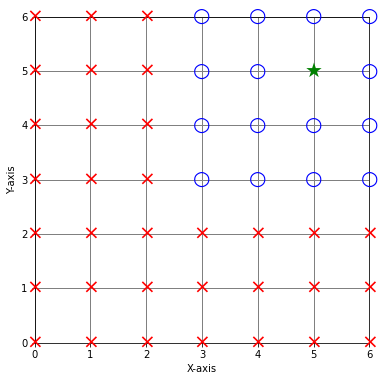}
        \caption{Metropolis}
        \label{fig:metropolis_neighbors}
    \end{subfigure}
    \caption{Illustration of proposal candidates}
    \label{fig:neighbor_states}
\end{figure}

Figure~\ref{fig:neighbor_states} demonstrates the possible candidates for the next proposal for GWG, ordinal-GWG, and Metropolis with $L_\infty$ neighborhood, assuming the current state is $(5,5)$ on a $6 \times 6$ lattice. GWG has a Hamming ball radius $\delta=1$; ordinal-GWG has a Hamming ball radius $\delta=1$ and a window size $r=2$; Metropolis has a window size $r=2$. The figure features:
\begin{itemize}
    \item A star (★) marking the \textit{current state} at $(5,5)$.
    \item A circle ($\circ$) marking the \textit{valid positions} for the next proposal.
    \item A cross (×) marking the \textit{invalid positions} for the next proposal.
\end{itemize}

\section{Continuous Analogues}
\subsection{Continuous Analogue for NCG} \label{NCG_cont}
Assume the state $s$ is continuous with support $\mathbb{R}^{d}$, and $s^{*}$ is proposed from the following distribution with the same form as in \eqref{eq:NCG1-new},
\begin{equation}
  s^* \sim  Q(s|s_{t}) \propto \exp(\frac{1}{2}\nabla f(s_{t})^{\T}(s-s_{t}))\exp(-\frac{1}{2 \delta}\|s-s_{t}\|_{2}^{2}), \nonumber
\end{equation}
which can be further calculated:
\begin{align}
        Q(s|s_{t}) & \propto \exp(\frac{1}{2}\nabla f(s_{t})^{\T}(s-s_{t}))\exp(-\frac{1}{2\delta}\|s-s_{t}\|_{2}^{2}) \nonumber \\
        & \propto \exp(\frac{1}{2}\nabla f(s_{t})^{\T}s - \frac{1}{2\delta}\|s\|_{2}^{2} +\frac{1}{\delta}s_{t+1}^{\T}s) \nonumber \\
        &\propto \exp(-\frac{1}{2\delta} \|s-\frac{\delta}{2}\nabla f(s_{t}) -s_{t}\|_{2}^{2}) \nonumber \\
        &\propto \mathcal{N}(s;s_{t}+\frac{\delta}{2}\nabla f(s_{t}), \delta I). \nonumber
\end{align}
The new proposal is the MALA proposal \eqref{eq:MALA-proposal}, equivalently written as
\begin{equation}
    s^{*} = s_{t}+\frac{\delta}{2}\nabla f(s_{t}) +\sqrt{\delta}Z , \quad Z \sim \mathcal{N}(0, I). \nonumber
\end{equation}
\subsection{Continuous Analogue for AVG}\label{AVG_cont}
Recall the auxiliary sampler \eqref{eq:AVG-2z} and \eqref{eq:AVG-2} corresponding to AVG where
\begin{align}
    z_{t} & \sim \mathcal{N}(z;s_{t}, \delta^{2}I), \nonumber\\
   s^{*} & \sim Q(s|z_{t}, s_{t})  \propto \exp( \nabla f(s_{t})^{\T}(s-s_{t}) - \frac{1}{2\delta^{2}}\|z_{t}-s\|_{2}^{2}). \nonumber
\end{align}
When $s$ is a continuous variable, the auxiliary variable $z_{t}$ can be integrated out similarly as in a marginalized sampler \citep{Titsias2018auxavg}.
The marginalized proposal is
\begin{align}
        s^{*} \sim Q(s|s_{t}) &= \int Q(s|z_{t},s_{t})\pi(z_{t}|s_{t}) dz_{t} \nonumber\\
        & \propto \int  \exp(\nabla f(s_{t})^{\T}(s-s_{t})-\frac{1}{2\delta^{2}}\|z_{t}-s\|_{2}^{2}) \mathcal{N}(z_{t};s_{t}, \delta^{2}I)dz_{t} \nonumber \\
        & \propto \int \exp(-\frac{1}{2\delta^{2}} \|s -(z_{t}+\delta^{2} \nabla f(s_{t}))\|_{2}^{2})\exp(-\frac{1}{2\delta^{2}}\|z_{t}-s_{t}\|_{2}^{2}) dz_{t} \nonumber \\
        & \propto \exp(-\frac{1}{4\delta^{2}}\|s-s_{t}-\delta^{2} \nabla f(s_{t})\|_{2}^{2}). \nonumber
\end{align}
The new proposal is the MALA proposal \eqref{eq:MALA-proposal}, equivalently written as
\begin{equation}
    s^{*} = s_{t}+ \delta^{2} \nabla f(s_{t}) + \sqrt{2}\delta Z \quad Z \sim \mathcal{N}(0, I). \nonumber
\end{equation}

\section{Properties for AVG}
\subsection{Exact Formula for AVG Acceptance Probability} \label{exact_accAVG}
For the acceptance probability in \eqref{eq:AVG_acc}, we calculate the denominator and the numerator as
\begin{align}
   \mathcal{N}(z_{t}|s^{*},\delta^{2}I)Q(s_{t}|z_{t},s^{*}) &= \exp(-\frac{1}{2\delta^{2}} \|z_{t}-s^{*}\|_{2}^{2}) Q(s_{t}|z_{t},s^{*}) \nonumber \\
   &= \exp(-\frac{1}{2\delta^{2}} \|z_{t}-s^{*}\|_{2}^{2}) \frac{\exp(\nabla f(s^{*})^{\T}(s_{t}-s^{*}) - \frac{1}{2\delta^{2}}\|s_{t}-z_{t}\|_{2}^{2})}{\sum\limits_{s \in \mathcal{S}}\exp(\nabla f(s^{*})^{\T}(s-s^{*})-\frac{1}{2\delta^{2}}\|s-z_{t}\|_{2}^{2})},
   \label{eq:exact_acc1AVG}
\end{align}
\begin{align}
   \mathcal{N}(z_{t}| s_{t},\delta^{2}I)Q(s^{*}|z_{t},s_{t}) &= \exp(-\frac{1}{2\delta^{2}} \|z_{t}-s_{t}\|_{2}^{2})Q(s^{*}|z_{t},s_{t}) \nonumber \\
   &= \exp(-\frac{1}{2\delta^{2}} \|z_{t}-s_{t}\|_{2}^{2}) \frac{\exp(\nabla f(s_{t})^{\T}(s^{*}-s_{t}) - \frac{1}{2\delta^{2}}\|s^{*}-z_{t}\|_{2}^{2})}{\sum\limits_{s \in \mathcal{S}}\exp(\nabla f(s_{t})^{\T}(s-s_{t})-\frac{1}{2\delta^{2}}\|s-z_{t}\|_{2}^{2})}.
   \label{eq:exact_acc2AVG}
\end{align}
Substituting \eqref{eq:exact_acc1AVG} and \eqref{eq:exact_acc2AVG} into \eqref{eq:AVG_acc}, we obtain the exact formula for acceptance probability as
\begin{align}
 & \min\{1, \exp(f(s^{*})-f(s_{t}))\frac{\mathcal{N}(z_{t}| s^{*},\delta^{2}I)Q(s_{t}|z_{t},s^{*})}{\mathcal{N}(z_{t} | s_{t}, \delta^{2}I)Q(s^{*}|z_{t},s_{t})}\} \nonumber \\
 &= \min \{1, \frac{\sum\limits_{s \in \mathcal{S}}\exp(\nabla f(s_{t})^{\T}(s-s_{t})-\frac{1}{2\delta^{2}}\|s-z_{t}\|_{2}^{2})}{\sum\limits_{s \in \mathcal{S}}\exp(\nabla f(s^{*})^{\T}(s-s^{*})-\frac{1}{2\delta^{2}}\|s-z_{t}\|_{2}^{2})} \nonumber \\
&\cdot \exp(f(s^{*})-f(s_{t}) + (\nabla f(s^{*}) + \nabla f(s_{t}))^{\T} (s_{t}-s^{*}) )\}. \label{eq:exact_acc3AVG}
\end{align}
When $\mathcal{S}$ is a product space, for example, a $d$-dimensional lattice $\mathcal{S} = \mathcal{A}^{d}$, the two sums over $\mathcal{S}$ in \eqref{eq:exact_acc3AVG} can be simplified as follows:
\begin{align*}
    \sum\limits_{s \in \mathcal{S}}\exp(\nabla f(s_{t})^{\T}(s-s_{t})-\frac{1}{2\delta^{2}}\|s-z_{t}\|_{2}^{2}) &= \prod\limits_{i=1}^{d}[\sum\limits_{s_i \in \mathcal{A}}\exp(\nabla f(s_{t})_i(s_i-s_{t,i})-\frac{1}{2\delta^{2}}(s_i-z_{t,i})^{2} ) ], \\
    \sum\limits_{s \in \mathcal{S}}\exp(\nabla f(s^{*})^{\T}(s-s^{*})-\frac{1}{2\delta^{2}}\|s-z_{t}\|_{2}^{2}) &= \prod\limits_{i=1}^{d}[\sum\limits_{s_i \in \mathcal{A}}\exp(\nabla f(s^*)_i(s_i-s^{*}_{i})-\frac{1}{2\delta^{2}}(s_i-z_{t,i})^{2} ) ] ,
\end{align*}
where $s_i$ is the $i$th component of $s$.

\subsection{Rejection-free Property of AVG} \label{sec:AVGrejectfree}
A target distribution with a linear potential function can be written as
\begin{align}
    \pi(s) & \propto \exp(a^{\T}s) \nonumber \\
    & \propto \prod_{i=1}^{d}\exp(a_{i}s_{i}). \label{eq:f_product}
\end{align}
Under this distribution, $\nabla f(s) \equiv a$ is a constant vector.
Substituting this into the exact formula of acceptance probability \eqref{eq:exact_acc3AVG}, we have
\begin{align}
 &  \exp(f(s^{*})-f(s_{t}))\frac{\mathcal{N}(z_{t}| s^{*},\delta^{2}I)Q(s_{t}|z_{t},s^{*})}{\mathcal{N}(z_{t} | s_{t}, \delta^{2}I)Q(s^{*}|z_{t},s_{t})} \nonumber \\
 &=  \frac{\sum\limits_{s \in \mathcal{S}}\exp(\nabla f(s_{t})^{\T}(s-s_{t})-\frac{1}{2\delta^{2}}\|s-z_{t}\|_{2}^{2})}{\sum\limits_{s \in \mathcal{S}}\exp(\nabla f(s^{*})^{\T}(s-s^{*})-\frac{1}{2\delta^{2}}\|s-z_{t}\|_{2}^{2})}
         \cdot \exp(f(s^{*})-f(s_{t}) + (\nabla f(s^{*}) + \nabla f(s_{t}))^{\T} (s_{t}-s^{*}) )  \nonumber \\
&= \frac{\sum\limits_{s \in \mathcal{S}}\exp(a^{\T}(s-s_{t})-\frac{1}{2\delta^{2}}\|s-z_{t}\|_{2}^{2})}{\sum\limits_{s \in \mathcal{S}}\exp(a^{\T}(s-s^{*})-\frac{1}{2\delta^{2}}\|s-z_{t}\|_{2}^{2})} \cdot \exp(a^{\T}(s^{*}-s_{t}) +2 a^{\T}(s_{t}-s^{*})) \nonumber \\
&= \frac{\exp(-a^{\T}s_{t})\sum\limits_{s \in \mathcal{S}}\exp(a^{\T}s-\frac{1}{2\delta^{2}}\|s-z_{t}\|_{2}^{2})}{\exp(-a^{\T}s^{*})\sum\limits_{s \in \mathcal{S}}\exp(a^{\T}s-\frac{1}{2\delta^{2}}\|s-z_{t}\|_{2}^{2})} \cdot \exp(f(s^{*})-f(s_{t}) + (\nabla f(s^{*}) + \nabla f(s_{t}))^{\T} (s_{t}-s^{*}) ) \nonumber \\
&=1. \label{eq:reject_free}
\end{align}
Equation \eqref{eq:reject_free} indicates that the acceptance probability is always 1, confirming the rejection-free property of AVG.

\section{Properties for V-DHAMS}
\subsection{Negation Equivalency} \label{eq:negation_equi}
We show that the transition from $(s_t,u_t)$ to $(s_t,u_{t+1/2})$ and then to $(s_{t+1},u_{t+1})$ in Sections \ref{sec:auxiliary_scheme} and \ref{sec:negation} can be equivalently presented as a transition which applies an auto-regression step \eqref{eq:auto-regressive} with $\epsilon\in (-1,0]$,
uses the proposal \eqref{eq:V-DHAMS-s}--\eqref{eq:V-DHAMS-u} with $u_t$ replaced by $u_{t+1/2}$ and performs acceptance-rejection with the standard Metropolis--Hastings probability \eqref{eq:auto-regressive-acc}.

The update on $u_{t+1/2}$ with $\epsilon\in (-1,0]$ can be equivalently written as
\begin{align}
        u_{t+1/2} &= -v_{t+1/2}, \nonumber \\
        v_{t+1/2} &= -\epsilon u_{t} + \sqrt{1-\epsilon^{2}} (-Z), \quad (-Z) \sim \mathcal{N}(0,I).
            \label{eq:negation_equi1}
\end{align}
Then $v_{t+1/2}$ follows the same update of $u_{t+1/2}$ as in \eqref{eq:auto-regressive} with a positive parameter $-\epsilon$. Substituting $u_{t+1/2}$ for $u_t$ in the proposal  \eqref{eq:V-DHAMS-s}--\eqref{eq:V-DHAMS-u} leads to
\begin{align}
 s^* & \sim Q( s | z_{t}=s_t+\delta u_{t+1/2}; s_t), \nonumber \\
 & \quad \propto
        \exp(\nabla f(s_{t})^{\T}(s-s_{t})-\frac{1}{2}\|\frac{ s_t+\delta u_{t+1/2} -s}{\delta}\|_{2}^{2}) \nonumber \\
 & \quad =
        \exp(\nabla f(s_{t})^{\T}(s-s_{t})-\frac{1}{2}\|\frac{ s_t-\delta v_{t+1/2} -s}{\delta}\|_{2}^{2}) \label{eq:negation-equis}, \\
        u^{*} &=  u_{t+1/2} + \frac{s_{t} -s^{*}}{\delta} + \phi (\nabla f(s^*) -\nabla f(s_t)) \nonumber \\
        &= -v_{t+1/2} + \frac{s_{t} -s^{*}}{\delta} + \phi (\nabla f(s^*) -\nabla f(s_t)). \label{eq:negation-equiu}
\end{align}
The proposal following \eqref{eq:negation-equis}--\eqref{eq:negation-equiu} is exactly the proposal with negation in \eqref{eq:negation-s}--\eqref{eq:negation-u}, after matching $v_{t+1/2}$ and $u_{t+1/2}$ in the respective expressions. As for the backward proposal, we have
\begin{align}
     s_t & \sim Q(s|z_t = s^* +\delta u^*; s^*) \nonumber \\
     & \quad \propto
        \exp(\nabla f(s^*)^{\T}(s-s_{t})-\frac{1}{2}\|\frac{ s^*+\delta u^* -s}{\delta}\|_{2}^{2}), \nonumber\\
    -v_{t+1/2} = u_{t+1/2} & = u^{*} + \frac{s^*-s_t}{\delta} + \phi (\nabla f(s^*) -\nabla f(s_t)).\nonumber
\end{align}
The backward proposal is also the same as used in \eqref{eq:negation-acc}, after matching $v_{t+1/2}$ and $u_{t+1/2}$ in the respective expressions. We accept $(s^*, u^*)$ from \eqref{eq:negation-equis}--\eqref{eq:negation-equiu} with the standard Metropolis--Hastings acceptance probability,
which is the same as acceptance probability \eqref{eq:negation-acc}. Upon rejection, we  set $s_{t+1}=s_t$ and $u_{t+1} = -v_{t+1/2}$.
In summary, we obtain the same transition as in Sections \ref{sec:auxiliary_scheme} and \ref{sec:negation}.

\subsection{Exact Formula for V-DHAMS Acceptance Probability} \label{exact_accHAMS}

Denote the normalizing constant over the augmented target distribution as $C = \int \pi(s,u)dsdu = \sum\limits_{s \in \mathcal{S}}\pi(s) \int \exp(-\frac{1}{2}\|u\|_{2}^{2})du$. The denominator of the ratio in acceptance probability \eqref{eq:mod-acc} is
    \begin{align}
        &\pi(s_{t}, u_{t})Q_{\phi}(s^{*}, u^{*}|s_{t}, u_{t}) \nonumber \\
        &= \frac{\exp(f(s_{t})-\frac{1}{2}\|u_{t}\|_{2}^{2})}{C} \frac{\exp(\nabla f(s_{t})^{\T}s^{*} - \frac{1}{2\delta^{2}}\|s_{t}-\delta u_{t+1/2}-s^{*}\|_{2}^{2})}{\sum\limits_{s \in \mathcal{S}}\exp(\nabla f(s_{t})^{\T}s - \frac{1}{2\delta^{2}} \|s_{t}-\delta u_{t+1/2} -s\|_{2}^{2})}.  \nonumber
    \end{align}
Similarly, the numerator of the ratio is
    \begin{align}
        &\pi(s^{*}, -u^{*})Q_{\phi}(s_{t}, -u_{t+1/2}|s^{*}, -u^{*}) \nonumber\\
        &= \frac{\exp(f(s^{*})-\frac{1}{2}\|u^{*}\|_{2}^{2})}{C} \frac{\exp(\nabla f(s^{*})^{\T}s_{t} - \frac{1}{2\delta^{2}}\|s^{*}+\delta u^{*}-s_{t}\|_{2}^{2})}{\sum\limits_{s \in \mathcal{S}}\exp(\nabla f(s^{*})^{\T}s - \frac{1}{2\delta^{2}} \|s^{*}+\delta u^{*} -s\|_{2}^{2})}.  \nonumber
    \end{align}
From these expressions, the ratio inside the acceptance probability \eqref{eq:mod-acc} is
    \begin{align}
        &  \frac{\pi(s^{*},-u^{*})Q_{\phi}(s_{t}, -u_{t+1/2}|s^{*} -u^{*})}{\pi(s_{t}, u_{t+1/2})Q_{\phi}(s^{*}, u^{*}|s_{t}, u_{t+1/2})} \nonumber \\
        &= \frac{\exp(f(s^{*})-\frac{1}{2}\|u^{*}\|_{2}^{2}+\nabla f(s^{*})^{\T}s_{t} - \frac{1}{2\delta^{2}}\|s^{*}+\delta u^{*}-s_{t}\|_{2}^{2})}{\exp(f(s_{t})-\frac{1}{2}\|u_{t+1/2}\|_{2}^{2}+\nabla f(s_{t})^{\T}s^{*} - \frac{1}{2\delta^{2}}\|s_{t}-\delta u_{t+1/2}-s^{*}\|_{2}^{2})} \nonumber\\
        &\cdot \frac{\sum\limits_{s \in \mathcal{S}}\exp(\nabla f(s_{t})^{\T}s - \frac{1}{2\delta^{2}} \|s_{t}-\delta u_{t+1/2} -s\|_{2}^{2})}{\sum\limits_{s \in \mathcal{S}}\exp(\nabla f(s^{*})^{\T}s - \frac{1}{2\delta^{2}} \|s^{*}+\delta u^{*} -s\|_{2}^{2})} \nonumber \\
        &= \exp(f(s^{*})-f(s_{t})+\nabla f(s^{*})^{\T}s_{t} - \nabla f(s_{t})^{\T}s^{*} - \|u^{*}\|_{2}^{2} +\|u_{t+1/2}\|_{2}^{2} \nonumber \\
        &- \frac{1}{\delta}(s^{*}-s_{t})^{\T}(u^{*}+u_{t+1/2})) \nonumber \\
        &\cdot \frac{\sum\limits_{s \in \mathcal{S}}\exp(\nabla f(s_{t})^{\T}s - \frac{1}{2\delta^{2}} \|s_{t}-\delta u_{t+1/2} -s\|_{2}^{2})}{\sum\limits_{s \in \mathcal{S}}\exp(\nabla f(s^{*})^{\T}s - \frac{1}{2\delta^{2}} \|s^{*}+\delta u^{*} -s\|_{2}^{2})}.     \label{eq:exact_acc_hams}
\end{align}
From \eqref{eq:exact_acc_hams}, the $\phi$ term is only implicitly incorporated into $(s^{*}, u^{*})$.

When $\mathcal{S}$ is a product space, for example, a $d$-dimensional lattice, $\mathcal{S} = \mathcal{A}^d$, the two sums over $\mathcal{S}$ in \eqref{eq:exact_acc_hams} can be simplified as follows:
\begin{align*}
    \sum\limits_{s \in \mathcal{S}}\exp(\nabla f(s_{t})^{\T}s - \frac{1}{2\delta^{2}} \|s_{t}-\delta u_{t+1/2} -s\|_{2}^{2}) &= \prod\limits_{i=1}^{d}[\sum\limits_{s_i \in \mathcal{A}}\exp(\nabla f(s_{t})_is_i-\frac{1}{2\delta^{2}}(s_{t,i}-\delta u_{t+1/2,i}-s_i)^{2} ) ] , \\
    \sum\limits_{s \in \mathcal{S}}\exp(\nabla f(s^{*})^{\T}s - \frac{1}{2\delta^{2}} \|s^{*}+\delta u^{*} -s\|_{2}^{2}) &= \prod\limits_{i=1}^{d}[\sum\limits_{s_i \in \mathcal{A}}\exp(\nabla f(s^*)_is_i-\frac{1}{2\delta^{2}}(s^{*}_i+\delta u^{*}_{i}-s_i)^{2} ) ] ,
\end{align*}
where $s_i$ is the $i$th component of $s$.

\subsection{V-DHAMS and AVG}\label{limitations}
We show that when $\epsilon = 0$ and $\phi=0$, V-DHAMS reduces to AVG. If $\epsilon =0$, the intermediate momentum becomes $u_{t+1/2}=Z$, which is an independent draw from $\mathcal{N}(0, I)$
and hence the momentum update can be ignored.
Consequently, $z_{t}$ is constructed in the same manner as in AVG, $z_{t} = s_{t}+\delta Z$. Furthermore if $\phi=0$, the gradient correction term vanishes, and the proposal becomes
\begin{align}
    & s^{*}  \sim Q(s|z_{t};s_t)\propto\exp(\nabla f(s_{t})^{\T}(s-s_{t})-\frac{1}{2}\|\frac{z_t-s}{\delta}\|_{2}^{2}), \nonumber \\
   &u^{*}= \frac{z_{t}-s^{*}}{\delta} = \frac{s_{t}-s^{*}}{\delta}+Z.
   \label{eq:dhams_limit1}
\end{align}
The proposal $s^*$ is the same as \eqref{eq:AVG-2}. The Metropolis--Hastings ratio  in \eqref{eq:exact_acc_hams} becomes
\begin{align}
    & \frac{\pi(s^{*},-u^{*})Q_{\phi}(s_{t}, -u_{t+1/2}|s^{*},-u^{*})}{\pi(s_{t}, u_{t+1/2})Q_{\phi}(s^{*}, u^{*}|s_{t}, u_{t+1/2})} \nonumber \\
    &  = \frac{\sum\limits_{s \in \mathcal{S}}\exp(\nabla f(s_{t})^{\T}(s-s_{t})-\frac{1}{2\delta^{2}}\|s-z_{t}\|_{2}^{2})}{\sum\limits_{s \in \mathcal{S}}\exp(\nabla f(s^{*})^{\T}(s-s^{*})-\frac{1}{2\delta^{2}}\|s-z_{t}\|_{2}^{2})} \nonumber \\
    & \cdot \frac{\exp(f(s^{*})-\frac{1}{2}\|u^{*}\|_{2}^{2})}{\exp(f(s_{t})-\frac{1}{2}\|u_{t+1/2}\|_{2}^{2})}\cdot\frac{\exp(-\frac{1}{2}\|s_{t}-s^{*}-\nabla f(s^{*})\|_{2}^{2}-\frac{1}{2\delta^{2}}\|z_{t}-s_{t}\|_{2}^{2}}{\exp(-\frac{1}{2}\|s^{*}-s_{t}-\nabla f(s_{t})\|_{2}^{2}-\frac{1}{2\delta^{2}}\|z_{t}-s^{*}\|_{2}^{2})}{} \nonumber \\
    &  = \frac{\sum\limits_{s \in \mathcal{S}}\exp(-\frac{1}{2}\|s-s_{t}-\nabla f(s_{t})\|_{2}^{2}-\frac{1}{2\delta^{2}}\|s-z_{t}\|_{2}^{2})}{\sum\limits_{s \in \mathcal{S}}\exp(-\frac{1}{2}\|s-s^{*}-\nabla f(s^{*})\|_{2}^{2}-\frac{1}{2\delta^{2}}\|s-z_{t}\|_{2}^{2})} \nonumber \\
    & \cdot \frac{\exp(f(s^{*})-\frac{1}{2}\|u^{*}\|_{2}^{2})}{\exp(f(s_{t})-\frac{1}{2}\|u_{t+1/2}\|_{2}^{2})}\cdot\frac{\exp(-\frac{1}{2}\|s_{t}-s^{*}-\nabla f(s^{*})\|_{2}^{2}-\frac{1}{2\delta^{2}}\|\delta u_{t+1/2}\|_{2}^{2}}{\exp(-\frac{1}{2}\|s^{*}-s_{t}-\nabla f(s_{t})\|_{2}^{2}-\frac{1}{2\delta^{2}}\|\delta u^{*}\|_{2}^{2})}{} \nonumber \\
     & = \frac{\sum\limits_{s \in \mathcal{S}}\exp(\nabla f(s_{t})^{\T}(s-s_{t})-\frac{1}{2\delta^{2}}\|s-z_{t}\|_{2}^{2})}{\sum\limits_{s \in \mathcal{S}}\exp(\nabla f(s^{*})^{\T}(s-s^{*})-\frac{1}{2\delta^{2}}\|s-z_{t}\|_{2}^{2})} \nonumber \\
    & \cdot \exp(f(s^{*})-f(s_{t}) + (\nabla f(s^{*}) + \nabla f(s_{t}))^{\T} (s_{t}-s^{*})), \label{eq:dhams_limit2}
\end{align}
which is also exactly the Metropolis--Hastings ratio in acceptance probability \eqref{eq:exact_acc3AVG} for AVG.
Summarizing \eqref{eq:dhams_limit1} and \eqref{eq:dhams_limit2}, we see that V-DHAMS reduces to AVG when $\epsilon=0$ and $\phi=0$.

\subsection{Rejection-free Property of V-DHAMS} \label{rejection_free}

For completeness, we show that the auto-regression step \eqref{eq:auto-regressive} is reversible to the augmented target distribution $\pi(s,u)$
and hence always accepted by the standard Metropolis--Hastings probability. Recall the auto-regressive step,
\begin{align}
    u_{t+1/2} = \epsilon u_t +\sqrt{1-\epsilon^2} Z, \quad Z \sim \mathcal{N}(0,I), \nonumber
\end{align}
or equivalently,
\begin{align}
    Q(u_{t+1/2}|u_t) = \mathcal{N}(u_{t+1/2}; \epsilon u_t,(1-\epsilon^2)I). \nonumber
\end{align}
The following can be directly verified:
\begin{align}
    \pi(s_t, u_t)Q(s_t, u_{t+1/2}|s_t, u_t) &=  \pi(s_t, u_t) Q(u_{t+1/2}|u_t) \nonumber, \\
    & \propto \exp(f(s_t) - \frac{1}{2(1-\epsilon^2)}(\|u_t\|_2^2+\|u_{t+1/2}\|_2^2)-\frac{\epsilon}{1-\epsilon^2}u_{t}^{\T}u_{t+1/2}),  \label{eq:auto-regressive-reversible-f}
\end{align}
\begin{align}
    \pi(s_t, u_{t+1/2})Q(s_t, u_t|s_t, u_{t+1/2}) &=  \pi(s_t, u_{t+1/2}) Q(u_t|u_{t+1/2}) \nonumber, \\
    & \propto \exp(f(s_t) - \frac{1}{2(1-\epsilon^2)}(\|u_t\|_2^2+\|u_{t+1/2}\|_2^2)-\frac{\epsilon}{1-\epsilon^2}u_{t}^{\T}u_{t+1/2}).  \label{eq:auto-regressive-reversible-b}
\end{align}
The two expressions in \eqref{eq:auto-regressive-reversible-f} and \eqref{eq:auto-regressive-reversible-b} are equivalent, thus confirming the reversibility of auto-regression step.

Next, we show the rejection-free property for V-DHAMS.
A distribution of the product form (or equivalently with a linear potential) is presented in \eqref{eq:f_product} in Supplement Section \ref{sec:AVGrejectfree}. Then the gradient $\nabla f(s) \equiv a$ is a constant vector, so that the gradient correction term in \eqref{eq:mod1} vanishes,
and hence  $z_{t} = s_{t}-\delta u_{t+1/2} = s^{*}+\delta u^{*}$.
From the exact formula \eqref{eq:exact_acc_hams}, the acceptance probability can be calculated as follows:
\begin{align}
         & \min\{1, \frac{\pi(s^{*}, -u^{*})Q_{\phi}(s_{t}, -u_{t+1/2}| s^{*}, -u^{*})}{\pi(s_{t}, u_{t+1/2})Q_{\phi}(s^{*}, u^{*}|s_{t}, -u_{t+1/2})} \} \nonumber\\
         & = \min \{1, \frac{\exp(f(s^{*})+\nabla f(s^{*})^{\T}s_{t})}{\exp(f(s_{t}) +\nabla f(s_{t})^{\T}s^{*})} \frac{\sum\limits_{s \in  \mathcal{S}}\exp(\nabla f(s_{t})^{\T}s-\frac{1}{2\delta^{2}}\|s_{t}-\delta u_{t+1/2}-s\|_{2}^{2})}{\sum\limits_{s \in  \mathcal{S}}\exp(\nabla f(s^{*})^{\T}s-\frac{1}{2\delta^{2}}\|s^{*}+\delta u^{*}-s\|_{2}^{2})}\} \nonumber \\
        & = \min \{1, \frac{\exp(a^{\T}s^{*}+a^{\T}s_{t})}{\exp(a^{\T}s_{t} +a^{\T}s^{*})} \frac{\sum\limits_{s \in  \mathcal{S}}\exp(a^{\T}s-\frac{1}{2\delta^{2}}\|z_{t}-s\|_{2}^{2})}{\sum\limits_{s \in  \mathcal{S}}\exp(a^{\T}s-\frac{1}{2\delta^{2}}\|z_{t}-s\|_{2}^{2})}\}  \nonumber \\
        &= 1, \nonumber
    \end{align}
which confirms the rejection-free property for V-DHAMS.

\section{Over-relaxation for Discrete Distributions}
\subsection{Proof of Uniformity}\label{sec:uniformity}

For completeness, we give a direct proof for the result that
if $w_{0} \sim \Unif([0,1))$ and $\tilde w \sim \Unif([0,1))$ independently, then $w_{1} = (- w_{0}+ \beta \tilde{w} )\%1$ is also $\Unif([0,1))$ for any $\beta \in [-1, 1]$.
This result can also be deduced from the symmetry (or detailed balance) in Section \ref{sec:overrelaxation-symmetricity}.

By construction, $w_{1}$ is a continuous variable with support on $[0,1)$.
If $\beta \in [0,1]$, the corresponding density of $w_1$, $p(w_{1})$, can be calculated as
    \begin{align}
     p(w_{1})  &= \int_{0}^{1} p(w_{1}|w_{0}) p(w_{0}) dw_{0}   \nonumber \\
     & = \int_{0}^{1} \frac{1}{\beta} \mathds{1} \{w_{1} \in (-w_{0}, -w_{0}+\beta)\%1 \} d w_{0} \nonumber \\
     & = \int_{0}^{1} \frac{1}{\beta} \mathds{1} \{ -w_{0} \in (w_{1}-\beta, w_{1})\%1 \} d w_{0} \nonumber \\
     & = \frac{1}{\beta} P( -w_{0} \in (w_{1}-\beta, w_{1})\%1 ) \nonumber \\
     & =1.                  \label{eq:uniformity1}
    \end{align}
If $\beta \in [-1,0)$,
    \begin{align}
         w_{1} & = (-w_{0}+ \beta \tilde{w} )\%1 \nonumber \\
    &= (-w_{0} - \beta (1-\tilde{w}) +\beta )\%1 \nonumber \\
    &= ((-w_{0} - \beta (1-\tilde{w}))\%1 + \beta)\%1. \label{eq:uniformity2}
    \end{align}
The conclusion in \eqref{eq:uniformity1} implies that $(-w_{0} - \beta (1-\tilde{w}))\%1$ is $\Unif([0,1))$, because $(1-\tilde{w})$ is $\Unif([0,1))$,
independent of $w_0$.
Note that if $v \sim \Unif([0,1))$ then $(v +\beta)\%1 \sim \Unif([0,1))$ for any $\beta \in \mathbb{R}$.
Therefore, from \eqref{eq:uniformity2}, $w_{1}  = (w_{0}+ \beta \tilde{w} )\%1$ is $\Unif([0,1)$ when $\beta <0$.

\subsection{Proof of Symmetry (or Detailed Balance)}\label{sec:overrelaxation-symmetricity}

We show that the joint distribution of $(w_{0}, w_{1})$ is symmetric when $w_{1} = (-w_{0}+ \beta \tilde{w} )\%1$.
The joint density function of $(w_{0}, w_{1})$ can be calculated as follows.

If $\beta \in [0,1]$:
\begin{equation}
    p(w_0, w_1) = \frac{1}{\beta} \quad \text{if} \quad\left\{
    \begin{aligned}
        &0 \leq w_1 < \beta-w_0, 1-w_0 \leq w_1 < 1 \quad and \quad0 \leq w_0 < \beta, \\
        & 1-w_0 \leq w_1 < 1-w_0+\beta \quad and \quad \beta \leq w_0 <1,
    \end{aligned}
    \right.
\label{eq:symmetric_pdf1}
\end{equation}
or $p(w_0, w_1) = 0$ otherwise.

If $\beta \in [-1,0)$:
\begin{equation}
    p(w_0, w_1) = \frac{1}{\beta} \quad \text{if} \quad\left\{
    \begin{aligned}
        &1-w_0+\beta < w_1 \leq 1-w_0 \quad and \quad0 \leq w_0 < 1+\beta, \\
        & 0 \leq w_1 < 1-w_0, 2-w_0+\beta < w_1 < 1 \quad and \quad 1+\beta \leq w_0 <1,
    \end{aligned}
    \right.
\label{eq:symmetric_pdf2}
\end{equation}
or $p(w_0, w_1) = 0$ otherwise.

\begin{figure}[htb]
    \centering
    \includegraphics[width=\linewidth]{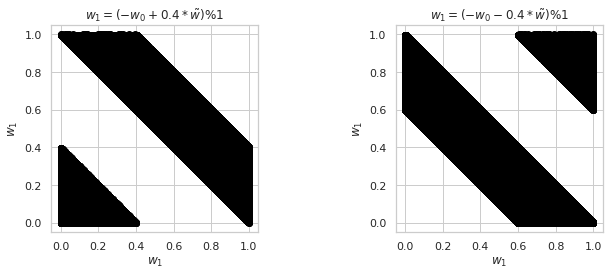}
    \caption{Probability density function plots of $(w_0, w_1)$. In both plots, the black parts indicate a density of $1/|\beta|$ and the white parts indicate a density of $0$ with $\beta=\pm 0.4$.}
    \label{fig:wpdf}
\end{figure}

Both density functions \eqref{eq:symmetric_pdf1} and \eqref{eq:symmetric_pdf2} are symmetric, confirming the symmetry of the joint distribution for $(w_0, w_1)$.
Figure~\ref{fig:wpdf} illustrates the density functions with $\beta =\pm 0.4$.

The symmetry in the joint distribution of  $(w_0,w_1)$ also implies that if $x_0 \sim p(x)$ from the reference distribution and $x_1$ is obtained using Algorithm \ref{algo:over-relaxation}, then the joint distribution $p(x_0, x_1)$ is also symmetric such that $p(x_0, x_1) = p(x_1, x_0)$.
The symmetric joint distribution of $(x_0,x_1)$ is essential for showing the rejection-free property of O-DHAMS.

\subsection{Recovery of ``Random Walk'' Proposal} \label{eq:random-walk}
We show that $w_{1}$ is independent of $w_{0}$ when $\beta = \pm 1$ in \eqref{eq:eq:overrelaxation-w}.
In fact, the conditional distribution of $w_{1}$ given $w_0$ can be calculated as
\begin{align}
    p(w_{1} \mid w_{0})    &= \left\{
    \begin{array}{cl}
    \frac{1}{|\beta|} \mathds{1} \left\{ w_{1} \in [ -w_{0}, -w_{0} + 1) \% 1 \right\} & \text{if } \beta = 1 \\
      \frac{1}{|\beta|} \mathds{1} \left\{ w_{1} \in (-w_{0}-1, -w_{0}] \% 1 \right\} & \text{if } \beta = -1 \
     \end{array}
     \right. \nonumber \\
    &= \mathds{1} \left\{ w_{1} \in [0,1) \right\} , \nonumber
\end{align}
where the last step follows because $[ -w_0, -w_0 + 1) \% 1 = [0,1)$ and $ (-w_{0}-1, -w_{0}]\%1 = [0,1)$  for any $w_0 \in [0,1)$.
This shows that when $\beta =\pm1$, the conditional distribution $p(w_{1}|w_{0})$ is $\Unif([0,1))$, and hence $w_{1}$ is independent of $w_{0}$.

\subsection{Recovering Correlation of $-1$ for Bernoulli Distribution}\label{sec:couplings}

In Algorithm \ref{algo:over-relaxation},
suppose that $x_0 \sim p(x)$ from a Bernoulli distribution with $p(0)= \alpha$ for some $0 \leq \alpha \leq 1$.
Then $x_1 \sim p(x)$, marginally from the same Bernoulli distribution.
We show that the lowest correlation of $-1$ between $x_0$ and $x_1$ is obtained by taking $\beta=0$.
In fact, the lowest correlation of two identically distributed Bernoulli variables is achieved when $p(x_0=0, x_1=0)$ reaches minimum \citepSupp{thorisson2000coupling}.
Note that $\max\{0 , 2\alpha-1\} \leq p(x_0=0, x_1=0)$.
When $\beta=0$, we have $w_1 = (-w_0)\%1$, and
\begin{align}
   p(x_0=0, x_1 =0) &= P(w_0 < \alpha, w_1 < \alpha) \nonumber \\
    &= P(w_0 < \alpha, 1-w_0 < \alpha) \nonumber \\
    &= P(1-\alpha < w_0 < \alpha) \nonumber \\
    &= \max\{0, 2\alpha-1\}, \label{eq:bernoulli_lwr} \nonumber
\end{align}
which is the minimum possible value for $p(x_0=0, x_1=0)$. Therefore, the corresponding lowest correlation between $x_0$ and $x_1$ (which is $-1$) is achieved.

\subsection{Explicit Formulas for Over-relaxation Proposal Probability} \label{sec:overrelax_prob}
We present the formulas for the over-relaxed proposal probability from Algorithm \ref{algo:over-relaxation}.
These formulas can be evaluated in $\mathcal{O}(k)$ time for a univariate reference distribution with $k$ possible values. Such over-relaxed proposal probabilities for univariate distributions are required to calculate the acceptance probability in O-DHAMS (Algorithm \ref{algo:O-DHAMS}).

We use the following notation for the reference distribution $p(x)$. Assume that $p(x)$ takes values in $\{0,1,\cdots, k\}$ with probabilities $ p(i) = p_{i}$ such that $\sum\limits_{i=0}^{k}p_{i} =1$. Let $F(i)$ be the CDF evaluated at $x=i$, and $R_i$ be the interval corresponding to $x=i$ such that
\begin{align}
    F(i) &= P(x \leq i) = \sum\limits_{i=0}^{k}p_{i}, \nonumber\\
    F(-1) &=0. \nonumber
\end{align}
Then we compute $p(x_1|x_0)$ in the two cases of $\beta$ for $w_{1} = (- w_{0}+\beta \tilde{w})\%1$.

\textbf{Case 1:} $w_{1} = (-w_{0}+\beta \tilde{w})\%1$, $0 \le \beta \le 1$
\begin{flalign}
    p(x_{1} = j |x_{0} =i) = \frac{f_{4}(1-F(i),1-F(i-1), F(j-1), F(j), \beta)}{\beta(F(i)-F(i-1))}. \nonumber
\end{flalign}

\textbf{Case 2:} $w_{1} = (-w_{0}+\beta \tilde{w})\%1$, $-1 \le \beta  < 0$
\begin{flalign}
    & \quad p(x_{1} = j |x_{0} =i)= \nonumber \\
    & \left\{
    \begin{aligned}
        & \frac{f_4(2-F(i)+\beta, 2-F(i-1)+\beta, F(j-1), F(j), -\beta}{-\beta(F(i)-F(i-1))} \quad \text{if} \quad 1-F(i-1)+\beta \leq 0, \\
        & \frac{f_4(1-F(i)+\beta, 1-F(i-1)+\beta, F(j-1), F(j), -\beta)}{-\beta(F(i)-F(i-1))} \quad \text{if} \quad 1-F(i)+\beta \geq 0, \\
        & \frac{f_4(2-F(i)+\beta, 1, F(j-1), F(j), -\beta) + f_4(0, 1-F(i-1)+\beta, F(j-1), F(j), -\beta)}{-\beta(F(i)-F(i-1))}  \quad \text{otherwise}.
    \end{aligned}
    \right.\nonumber
\end{flalign}

The formulas for intermediate functions $f_1$, $f_2$, $f_3$ and $f_4$ are as follows.
\begin{flushleft}
       $ f_1(p_{il}, p_{ir}, p_{jl}, p_{jr}, \beta) = $
\end{flushleft}
\begin{equation*}
\begin{aligned}
\begin{cases}
0 & \text{if } p_{il} + \beta < p_{jl},\; p_{ir} + \beta < p_{jl}, \\[1.2ex]
\frac{(p_{ir} + \beta - p_{jl})^2}{2}
& \text{if } p_{il} + \beta < p_{jl} \leq p_{ir} + \beta < p_{jr}, \\[1.2ex]
\frac{(p_{jr} - p_{jl})^2}{2} + (p_{jr} - p_{jl})(p_{ir} + \beta - p_{jr})
& \text{if } p_{il} + \beta < p_{jl},\; p_{jr} \leq p_{ir} + \beta < p_{jl}, \\[1.2ex]
\frac{(p_{jr} - p_{jl})^2}{2} + (p_{jr} - p_{jl})(p_{ir} + \beta - p_{jr}) + \frac{(p_{ir} + \beta - p_{jl})^2}{2}
& \text{if } p_{il} + \beta < p_{jl} \leq p_{ir} + \beta < p_{jr}, \\[1.2ex]
\frac{(p_{jr} - p_{jl})^2}{2} + (p_{jr} - p_{jl})(2p_{ir} + 2\beta - 2p_{jr} - 1)
& \text{if } p_{il} + \beta < p_{jl},\; p_{jr} \leq p_{ir} + \beta, \\[1.2ex]
\frac{(p_{ir} + \beta - p_{jl})^2}{2} - \frac{(p_{il} + \beta - p_{jl})^2}{2}
& \text{if } p_{jl} \leq p_{il} + \beta < p_{jr},\; p_{jl} \leq p_{ir} + \beta < p_{jr}, \\[1.2ex]
\frac{(p_{jr} - p_{jl})^2}{2} - \frac{(p_{il} + \beta - p_{jl})^2}{2} + (p_{jr} - p_{jl})(p_{ir} + \beta - p_{jr})
& \text{if } p_{jl} \leq p_{il} + \beta < p_{jr},\; p_{jr} \leq p_{ir} + \beta < p_{jl}, \\[1.2ex]
\frac{(p_{jr} - p_{jl})^2}{2} - \frac{(p_{il} + \beta - p_{jl})^2}{2} + (p_{jr} - p_{jl})(p_{ir} + \beta - p_{jr}) + \frac{(p_{ir} + \beta - p_{jl})^2}{2}
& \text{if } p_{jl} \leq p_{il} + \beta < p_{jr},\; p_{jl} \leq p_{ir} + \beta < p_{jr}, \\[1.2ex]
\frac{(p_{jr} - p_{jl})^2}{2} - \frac{(p_{il} + \beta - p_{jl})^2}{2} + (p_{jr} - p_{jl})(2p_{ir} + 2\beta - 2p_{jr} - 1)
& \text{if } p_{jl} \leq p_{il} + \beta < p_{jr},\; p_{jr} \leq p_{ir} + \beta, \\[1.2ex]
(p_{jr} - p_{jl})(p_{ir} - p_{il})
& \text{if } p_{jr} \leq p_{il} + \beta < p_{jl},\; p_{jr} \leq p_{ir} + \beta < p_{jl}, \\[1.2ex]
(p_{jr} - p_{jl})(p_{ir} - p_{il}) + \frac{(p_{ir} + \beta - p_{jl})^2}{2}
& \text{if } p_{jr} \leq p_{il} + \beta < p_{jl},\; p_{jl} \leq p_{ir} + \beta < p_{jr}, \\[1.2ex]
(p_{jr} - p_{jl})(p_{ir} - p_{il}) + \frac{(p_{jr} - p_{jl})^2}{2} + (p_{jr} - p_{jl})(p_{ir} + \beta - p_{jr})
& \text{if } p_{jr} \leq p_{il} + \beta < p_{jl},\; p_{jr} \leq p_{ir} + \beta, \\[1.2ex]
\frac{(p_{il} + p_{ir} - 4p_{jl} + 2p_{jr} - 2 + 2\beta)(p_{ir} - p_{il})}{2}
& \text{if } p_{jl} \leq p_{il} + \beta < p_{jr},\; p_{jl} \leq p_{ir} + \beta < p_{jr}, \\[1.2ex]
\frac{(3p_{jr} + p_{ir} - 4p_{jl} - 1 + \beta)(p_{jr} + 1 - p_{il} - \beta)}{2} + 2(p_{jr} - p_{jl})(p_{ir} + \beta - p_{jr})
& \text{if } p_{jl} \leq p_{il} + \beta < p_{jr},\; p_{jr} \leq p_{ir} + \beta, \\[1.2ex]
2(p_{jr} - p_{jl})(p_{ir} - p_{il})
& \text{if } p_{jr} \leq p_{il} + \beta.
\end{cases}
\end{aligned}
\end{equation*}


\begin{flushleft}
    $f_2(p_{il}, p_{ir}, p_{jl}, p_{jr}, \beta) = $
\end{flushleft}
\begin{equation*}
\begin{aligned}
\begin{cases}
\beta(p_{ir} - p_{il})
& \text{if } p_{ir} + \beta \leq p_{jr}, \\[1.2ex]

\beta(p_{jr} - \beta - p_{il}) + \dfrac{(\beta + p_{jr} - p_{ir})(p_{ir} - p_{jr} + \beta)}{2}
& \text{if } p_{jr} < p_{ir} + \beta \leq 1 + p_{jl},\; p_{il} + \beta \leq p_{jr}, \\[1.2ex]

\dfrac{(p_{ir} - p_{il})(2p_{jr} - p_{il} - p_{ir})}{2}
& \text{if } p_{jr} < p_{ir} + \beta \leq 1 + p_{jl},\; p_{jr} < p_{il} + \beta, \\[1.2ex]

\beta(p_{jr} - \beta - p_{il}) + \dfrac{(1 - p_{jr} + p_{jl})(2\beta + p_{jr} - p_{jl} - 1)}{2} & \\+ (p_{jr} - p_{jl} + \beta - 1)(-1 + p_{ir} + \beta - p_{jl})
& \text{if } p_{ir} + \beta > p_{jl} + 1,\; p_{il} + \beta \leq p_{jr}, \\[1.2ex]

\dfrac{(2p_{jr} - p_{jl} - p_{il} + \beta - 1)(p_{jl} - \beta - p_{il} + 1)}{2} & \\+ (p_{jr} - p_{jl} + \beta - 1)(-1 + p_{ir} - p_{jl} + \beta)
& \text{if } p_{ir} + \beta > p_{jl} + 1,\; -1 + p_{il} + \beta \leq p_{jl} < p_{il} + \beta, \\[1.2ex]

(p_{jr} - p_{jl} + \beta - 1)(p_{ir} - p_{il})
& \text{if } p_{ir} + \beta > p_{jl} + 1,\; p_{jl} \leq -1 + p_{il} + \beta.
\end{cases}
\end{aligned}
\end{equation*}


\begin{flushleft}
    $f_3(p_{il}, p_{ir}, p_{jl}, p_{jr}, \beta) = $
\end{flushleft}
\begin{equation*}
\begin{aligned}
\begin{cases}
0
& \text{if } -1 + p_{ir} + \beta \leq p_{jl}, \\[1.2ex]

\dfrac{(-1 + p_{ir} + \beta - p_{jl})^2}{2}
& \text{if } p_{jl} \leq -1 + p_{ir} + \beta < p_{jr},\; -1 + p_{il} + \beta \leq p_{jl}, \\[1.2ex]

\dfrac{(-1 + p_{ir} + \beta - p_{jl})^2}{2} - \dfrac{(-1 + p_{il} + \beta - p_{jl})^2}{2}
& \text{if } p_{jl} \leq -1 + p_{ir} + \beta < p_{jr},\; p_{jl} < -1 + p_{il} + \beta, \\[1.2ex]

\dfrac{(p_{jr} - p_{jl})^2}{2} + (p_{jr} - p_{jl})(-1 + p_{ir} + \beta - p_{jr})
& \text{if } p_{jl} \leq -1 + p_{ir} + \beta,\; -1 + p_{il} + \beta \leq p_{jl}, \\[1.2ex]

\dfrac{(p_{jr} - p_{jl})^2}{2} - \dfrac{(-1 + p_{il}+ \beta - p_{jl})^2}{2} & \\+(p_{jr} - p_{jl})(-1 + p_{ir} + \beta - p_{jr})
& \text{if } p_{jl} \leq -1 + p_{ir} + \beta,\; p_{jl} \leq -1 + p_{il} + \beta < p_{jr}, \\[1.2ex]

(p_{jr} - p_{jl})(p_{ir} - p_{il})
& \text{if } p_{jl} \leq -1 + p_{ir} + \beta,\; p_{jr} \leq -1 + p_{il} + \beta.
\end{cases}
\end{aligned}
\end{equation*}

\begin{flushleft}
    $f_4(p_{il}, p_{ir}, p_{jl}, p_{jr}, \beta) =
$
\end{flushleft}
\begin{equation*}
\begin{aligned}
\begin{cases}
f_1(p_{il}, p_{ir}, p_{jl}, p_{jr}, \beta)
& \text{if } p_{il} \leq p_{jl},\; p_{ir} \leq p_{jl}, \\[1.2ex]

f_1(p_{il}, p_{ir}, p_{jl}, p_{jr}, \beta) + f_2(p_{jl}, p_{ir}, p_{jl}, p_{jr}, \beta)
& \text{if } p_{il} \leq p_{jl} < p_{ir} \leq p_{jr}, \\[1.2ex]

f_1(p_{il}, p_{ir}, p_{jl}, p_{jr}, \beta) + f_2(p_{jl}, p_{jr}, p_{jl}, p_{jr}, \beta) + f_3(p_{jr}, p_{ir}, p_{jl}, p_{jr}, \beta)
& \text{if } p_{il} \leq p_{jl},\; p_{jr} < p_{ir}, \\[1.2ex]

f_2(p_{il}, p_{ir}, p_{jl}, p_{jr}, \beta)
& \text{if } p_{jl} < p_{il} \leq p_{jr},\; p_{jl} < p_{ir} \leq p_{jr}, \\[1.2ex]

f_2(p_{il}, p_{jr}, p_{jl}, p_{jr}, \beta) + f_3(p_{jr}, p_{ir}, p_{jl}, p_{jr}, \beta)
& \text{if } p_{jl} < p_{il} \leq p_{jr},\; p_{jr} < p_{ir}, \\[1.2ex]

f_3(p_{il}, p_{ir}, p_{jl}, p_{jr}, \beta)
& \text{if } p_{jr} < p_{il}.
\end{cases}
\end{aligned}
\end{equation*}

\section{Properties of O-DHAMS}\label{sec:overrelaxeV-DHAMS}

\subsection{Generalized Detailed Balance} \label{gdbc}
As mentioned in Supplement Section \ref{sec:GMH}, O-DHAMS is an example of generalized Metropolis--Hastings sampling.
Hence the generalized detailed balance for O-DHAMS follows from that for generalized Metropolis--Hastings sampling in \cite{Song2023hams}, Proposition 3.
For discrete distributions, density functions can be defined with respect to counting measures.

For completeness, we give a direct proof.
We compute both the left-hand side and the right-hand side to verify the generalized detailed balance \eqref{eq:overhams-dbc} as stated in Proposition \ref{prop:2-new}.
For $s_{t} \neq s_{t+1}$, the left-hand side of \eqref{eq:overhams-dbc} is
\begin{align}
    & \pi(s_{t}, u_{t+1/2})\tilde{K}_{\phi}(s_{t+1}, u_{t+1}|s_{t}, u_{t+1/2}) \nonumber \\
    &= \pi(s_{t}, u_{t+1/2})\tilde{Q}_{\phi}(s_{t+1}, u_{t+1}|s_{t}, u_{t+1/2}) \min\left\{1, \frac{\pi(s_{t+1}, -u_{t+1})\tilde{Q}_{\phi}(s_{t}, -u_{t+1/2}|s_{t+1}, -u_{t+1})}{\pi(s_{t}, u_{t+1/2})\tilde{Q}_{\phi}(s^{*}, u^{*}|s_{t}, u_{t+1/2})}\right\} \nonumber \\
    &= \min\left\{ \pi(s_{t}, u_{t+1/2})\tilde{Q}_{\phi}(s_{t+1}, u_{t+1}|s_{t}, u_{t+1/2}), \pi(s_{t+1}, -u_{t+1})\tilde{Q}_{\phi}(s_{t}, -u_{t+1/2}|s_{t+1}, -u_{t+1}) \right\}. \label{eq:gdbc_neq_l}
\end{align}
The corresponding right-hand side is
    \begin{align}
       & \pi(s_{t+1}, -u_{t+1})\tilde{K}_{\phi}(s_{t}, -u_{t+1/2}|s_{t+1}, -u_{t+1}) \nonumber \\
      &= \pi(s_{t+1}, -u_{t+1})\tilde{Q}_{\phi}(s_{t}, -u_{t+1/2}|s_{t+1}, -u_{t+1})\min\{1,
    \frac{\pi(s_{t}, u_{t+1/2})\tilde{Q}_{\phi}(s_{t+1}, u_{t+1}|s_{t}, u_{t+1/2})}{\pi(s_{t+1}, -u_{t+1})\tilde{Q}_{\phi}(s_{t}, -u_{t+1/2}|s_{t+1}, -u_{t+1})}\} \nonumber \\
    &= \min\{ \pi(s_{t}, u_{t+1/2})\tilde{Q}_{\phi}(s_{t+1}, u_{t+1}|s_{t}, u_{t+1/2}), \pi(s_{t+1}, -u_{t+1})\tilde{Q}_{\phi}(s_{t}, -u_{t+1/2}|s_{t+1}, -u_{t+1})\}.\label{eq:gdbc_neq_r}
    \end{align}
For $s_{t} = s_{t+1}$, the left hand side and the right hand side of \eqref{eq:overhams-dbc} become respectively
\begin{align}
    & \pi(s_{t}, u_{t+1/2})\tilde{K}_{\phi}(s_{t+1}, u_{t+1}|s_{t}, u_{t+1/2}) = \pi(s_{t}, u_{t+1/2})\tilde{K}_{\phi}(s_{t}, -u_{t+1/2}|s_{t}, u_{t+1/2}), \label{eq:gdbc_eq_l}\\
    & \pi(s_{t+1}, -u_{t+1})\tilde{K}_{\phi}(s_{t}, -u_{t+1/2}|s_{t+1}, -u_{t+1}) = \pi(s_{t}, u_{t+1/2})\tilde{K}_{\phi}(s_{t}, -u_{t+1/2}|s_{t}, u_{t+1/2}). \label{eq:gdbc_eq_r}
\end{align}
Comparing \eqref{eq:gdbc_neq_l} and \eqref{eq:gdbc_neq_r}, the equality in \eqref{eq:overhams-dbc} is verified when $s_t \neq s_{t+1}$; comparing \eqref{eq:gdbc_eq_l} and \eqref{eq:gdbc_eq_r}, it is also verified in the case $s_{t} = s_{t+1}$. In summary, the detailed balance \eqref{eq:overhams-dbc} is confirmed for O-DHAMS.
The detailed balance of V-DHAMS (Proposition \ref{prop:2}) is also confirmed because it is a special case of O-DHAMS with over-relaxation parameter $\beta = \pm1$, where the proposal is of ``random-walk" type.

Next, we show that the transition kernel $\tilde{K}_{\phi}$ admits $\pi(s,u)$ as a stationary distribution,
similarly as in \citeSupp{Tierney1994MC} and \cite{Song2023hams}.
By some abuse of notation, denote $\pi(s,u) = \pi(s)\pi(u)$, where $\pi(s) \propto \exp(f(s))$ and $\pi(u) \propto \exp(-\frac{1}{2}\|u\|_{2}^{2})$.
It suffices to show that for any subset $\bar{S} \subset \mathcal{S}$ and any $C \subset \mathbb{R}^{d}$,
\begin{align*}
    & \quad \sum\limits_{s_{t+1} \in \bar{S}} \pi(s_{t+1}) \int\limits_{C} \pi(u_{t+1})du_{t+1}\\
    & =  \sum\limits_{s_{t+1} \in \bar{S}}\int\limits_{C}du_{t+1} ( \sum\limits_{s_t \in \mathcal{S}}\int \pi(s_t) \pi (u_{t+1/2})\tilde{K}_{\phi}(s_{t+1}, u_{t+1}|s_{t}, u_{t+1/2}) du_{t+1/2}).
\end{align*}
The right hand side can be directly calculated as
\begin{align}
     & \quad \sum\limits_{s_{t+1} \in \bar{S}}\int\limits_{C}du_{t+1} ( \sum\limits_{s_t \in \mathcal{S}}\int \pi(s_t) \pi (u_{t+1/2})\tilde{K}_{\phi}(s_{t+1}, u_{t+1}|s_{t}, u_{t+1/2}) du_{t+1/2}) \nonumber\\
&= \sum\limits_{s_{t+1} \in \bar{S}}\int\limits_{C}du_{t+1} ( \sum\limits_{s_t \in \mathcal{S}}\int\pi(s_{t+1}, -u_{t+1})\tilde{K}_{\phi}(s_{t}, -u_{t+1/2}|s_{t+1}, -u_{t+1}) du_{t+1/2}) \nonumber\\
&= \sum\limits_{s_{t+1} \in \bar{S}}\int\limits_{C}du_{t+1} ( \pi(s_{t+1}, -u_{t+1})) \nonumber \\
&=  \sum\limits_{s_{t+1} \in \bar{S}} \pi(s_{t+1}) \int\limits_{C} \pi(-u_{t+1})du_{t+1} \nonumber \\
&=  \sum\limits_{s_{t+1} \in \bar{S}} \pi(s_{t+1}) \int\limits_{C} \pi(u_{t+1})du_{t+1}, \nonumber
\end{align}
which gives the left-hand side. This confirms that O-DHAMS (and V-DHAMS as a special case) admits $\pi(s,u) = \pi(s)\pi(u)$ as a stationary distribution.

\subsection{Rejection-free Property of O-DHAMS}\label{sec:ohams_rejection-free}

As in Supplement Section \ref{rejection_free}, for a distribution with a linear potential function \eqref{eq:f_product}, the gradient correction term vanishes, and hence $z_{t} = s_{t}-\delta u_{t+1/2} = s^{*}+\delta u^{*}$, which indicates that
\begin{align}
    \pi(s^{*}, -u^{*}) & \propto \exp(a^{\T}s^{*} -\frac{1}{2\delta^2}\|z_{t}-s^{*}\|_{2}^{2}), \nonumber \\
    \pi(s_{t}, u_{t+1/2}) & \propto \exp(a^{\T}s_{t} -\frac{1}{2\delta^2}\|z_{t}-s_{t}\|_{2}^{2}). \nonumber
\end{align}
For O-DHAMS in Algorithm \ref{algo:O-DHAMS}, the forward proposal is $\tilde{Q}(s^*|z_t = s_t-\delta u_{t+1/2};s_t)$ and the backward proposal is $\tilde{Q}(s_t|z_t = s^* +\delta u^*;s_t)$. Their corresponding reference distribution for over-relaxation are the components of $Q(s|z_t = s_t- \delta u_{t+1/2})$ and $Q(s|s^*+\delta u^*; s^*)$ in \eqref{eq:negation-s}, i.e.,
\begin{align}
 & Q( s | z_{t}=s_t- \delta u_{t+1/2}; s_t) \nonumber \\
 & \propto
        \exp(\nabla f(s_{t})^{\T}(s-s_{t})-\frac{1}{2\delta^2}\|\frac{ s_t-\delta u_{t+1/2} -s}{\delta}\|_{2}^{2}) \nonumber \\
   & \propto \exp(a^{\T}s-\frac{1}{2\delta^2}\|\frac{z_t-s}{\delta}\|_2^2), \label{eq:forward_ref}\\
 & Q( s | z_{t}=s^* + \delta u^*; s^*) \nonumber \\
 & \propto
        \exp(\nabla f(s^*)^{\T}(s-s^*)-\frac{1}{2}\|\frac{ s^*+\delta u^* -s}{\delta}\|_{2}^{2}) \nonumber \\
   & \propto \exp(a^{\T}s-\frac{1}{2}\|\frac{z_t-s}{\delta}\|_2^2).\label{eq:backward_ref}
\end{align}
As shown by \eqref{eq:forward_ref} and \eqref{eq:backward_ref}, both forward and backward proposals share the same reference distribution, which we denote as $p(s)$. We also observe that $\pi(s^{*}, -u^{*}) \propto p(s^{*})$ and  $\pi(s_{t}, u_{t+1/2}) \propto p(s_{t})$. Using the notation for over-relaxation in Section \ref{sec:overrelax-general} with reference $p(s)$, we have
    \begin{align}
    \tilde{Q}(s^{*}| s_{t}, u_{t+1/2})  &= p(s^{*}|s_{t}), \nonumber\\
        \tilde{Q}(s_{t}| s^{*}, -u^{*}) &= p(s_{t}|s^{*}).
    \label{eq:eq:reference_overrelax1}
    \end{align}
By substituting \eqref{eq:eq:reference_overrelax1} into \eqref{eq:over-HAMS-acc}, the acceptance probability of O-DHAMS can be written as
\begin{align}
& \min\{1, \frac{\pi(s^{*}, -u^{*})\tilde{Q}_{\phi}(s_{t}, -u_{t+1/2}| s^{*}, -u^{*})}{\pi(s_{t}, u_{t+1/2})\tilde{Q}_{\phi}(s^{*}, u^{*}|s_{t}, u_{t+1/2})} \} \nonumber \\
& = \min\{1, \frac{p(s^{*})p(s_{t}| s^{*})}{p(s_{t})p(s^{*}| s_{t})}\} \nonumber \\
& = \min\{1, \frac{p(s^{*}, s_{t})}{p(s_{t}, s^{*})}\},
\label{eq:eq:reference_overrelax2}
\end{align}
where the notation of $p(\cdot, \cdot)$ represents the joint distribution of two variables in the over-relaxation Algorithm \ref{algo:over-relaxation}.
From Supplement Section \ref{sec:overrelaxation-symmetricity}, the joint distribution $p(\cdot, \cdot)$ is symmetric, so that $p(s^{*}, s_{t})=p(s_{t}, s^{*})$ in \eqref{eq:eq:reference_overrelax2}. Therefore, the acceptance probability is always $1$ for O-DHAMS, confirming the rejection-free property.

\section{Tuning Procedures in Numerical Studies}

The NCG and AVG samplers require tuning of only a single parameter, stepsize $\delta$, which can be optimized via a single grid search. In contrast, the V-DHAMS and O-DHAMS samplers require tuning multiple parameters: the auto-regression parameter $\epsilon$, the stepsize $\delta$, the gradient correction parameter $\phi$, and the over-relaxation parameter $\beta$ (for O-DHAMS). From our numerical experience, when the support of each coordinate of a multivariate discrete distribution consists of integer values, a reasonable value for $\delta$ typically falls within the range $(0, 2)$, and $\phi$ has a reasonable value between $0$ and $1$. Guided by these observations, we implement the tuning procedure for Discrete-HAMS samplers as shown in Algorithm \ref{algo:tuning}.

\begin{algorithm}[tbp]
\begin{enumerate}
    \item Fix the auto-regressive parameter $\epsilon$ close to $1$, typically around $0.9$,
    \item Fix the over-relaxation parameter $\beta$ to achieve the desired proposal behavior: $\beta = 1$ for a random-walk proposal or $\beta \approx 0$ for an over-relaxed proposal,
    \item Perform a grid search over $\delta$ (with $\phi=0$) to identify the optimal stepsize based on a chosen criteria,
    \item With $\delta$ fixed from the previous step, select $\phi$ from a set of candidate values between $0$ and $1$ based on the same criteria.
\end{enumerate}
\caption{Tuning procedure for Discrete-HAMS}
\label{algo:tuning}
\end{algorithm}

In the NCG, AVG, and Discrete-HAMS algorithms, we identify a range for the stepsize $\delta$ to target a range of acceptance rates. We observe that the acceptance rate tends to decrease as the stepsize increases. In our experiments, when targeting a specific acceptance rate $\alpha$, we adopt the stepsize tuning procedure shown in Algorithm~\ref{algo:target_acc}, inspired by \citeSupp{Andrieu2008tuning}. For each candidate stepsize, a chain of length 1000
is run (after burn-in) to obtain the corresponding acceptance rate.
\begin{algorithm}[tbp]
\caption{Stepsize tuning to target a certain acceptance rate}
\label{algo:target_acc}
\begin{algorithmic}
    \State \textbf{Require:} Initial stepsize $\delta_0$, initial acceptance rate $\alpha_0$, target acceptance rate $\alpha$,
    \For{$m = 0$ to $M$}
        \State Obtain acceptance rate $\alpha_m$ with stepsize $\delta_m$
        \If{$\alpha_m > \alpha$}
            \State $\delta_{m+1} \gets \delta_m \exp((1 + m)^{-a})$
        \ElsIf{$\alpha_m < \alpha$}
            \State $\delta_{m+1} \gets \delta_m \exp(-(1 + m)^{-a})$
        \EndIf
    \EndFor,
    \State \Return $\delta_m$ with smallest $|\alpha_m -\alpha|$.
\end{algorithmic}

\end{algorithm}

\subsection{Tuning Procedure for Discrete Gaussian Distribution and Quadratic Mixture
Model}\label{sec:ordinal_tuning}
For both the discrete Gaussian distribution and quadratic mixture distribution, we observe that ESS of $f(s)$ initially increases and then decreases as the stepsize $\delta$ increases for all samplers, while the acceptance rate consistently decreases. We then apply Algorithm~\ref{algo:target_acc} to identify an approximate range of $\delta$ values that achieve an acceptance rate between $0.5$ and $0.9$.

For NCG and AVG, we select $\delta$ from 50 equally spaced values within that identified range using repeated chains. For Discrete-HAMS, we use the procedure outlined in Algorithm~\ref{algo:tuning} by initially selecting $\delta$ from a sequence of 40 values and subsequently conducting a grid search over 10 values for $\phi$ with the fixed $\delta$.

\subsection{Tuning Procedure for Bayesian Linear Regression}\label{sec:binary_tuning}

As mentioned in Section \ref{sec:bayesian_linear}, tuning based on ESS of $f(s)$ may not be reliable for Bayesian linear regression simulation. To address these challenges, we adopt a tuning strategy based on acceptance rate and average flips, which provide more stable and informative criteria for parameter tuning in the context of Bayesian linear regression simulation.

\textbf{Acceptance Rate} As suggested by \citeSupp{sun2022optimal}, acceptance rate is an important criterion in parameter tuning for sampling binary distributions. Empirically, the best results in terms of ESS of the relevant mask variables $s_1$ and $s_{601}$ for the Bayesian linear regression simulation are achieved when the acceptance rate stabilizes after the burn-in period within specific ranges: $0.15$--$0.35$ for AVG, $0.2$--$0.4$ for NCG, and $0.2$--$0.35$ for Discrete-HAMS. In general, we observe that the acceptance rate decreases as the stepsize $\delta$ increases. The approximate range for the stepsize $\delta$ to target these acceptance rates is obtained via Algorithm~\ref{algo:target_acc}.

\textbf{Average Flips} After determining the approximate range of $\delta$ that achieves the desired acceptance rate, we proceed with a grid search within that range. Depending on the available computational resources, the grid search can be executed with a single chain or multiple parallel chains for each parameter. We then identify the optimal parameter by selecting the one that maximizes the average $L_1$-distance, $\|s_{t+1}-s_{t}\|_{1}$, between successive states across all time steps and chains. Since $s_{t}$ are binary vectors, the $L_1$-distance corresponds to the number of coordinates that change their values (i.e., flip) between time steps. Hence, in this context, we refer to the average $L_1$-distance as the \textit{average flips} for binary distributions. Tuning using average flips is also employed in \cite{Rhodes2022GradientMC}. For AVG and NCG, we run a grid search over $\delta$ with 50 values. For Discrete-HAMS, we first run a grid search over $\delta$ with 40 values and then over $\phi$ with 10 values after fixing $\delta$ following Algorithm \ref{algo:tuning}.

\section{Additional Experiment Results}\label{sec:exp_results}
\subsection{Additional Results for Discrete Gaussian}\label{sec:gaussian_results}
\begin{figure}[tbp]
    \centering
    \begin{subfigure}[b]{0.32\textwidth}
        \centering
        \includegraphics[width=0.8\linewidth]{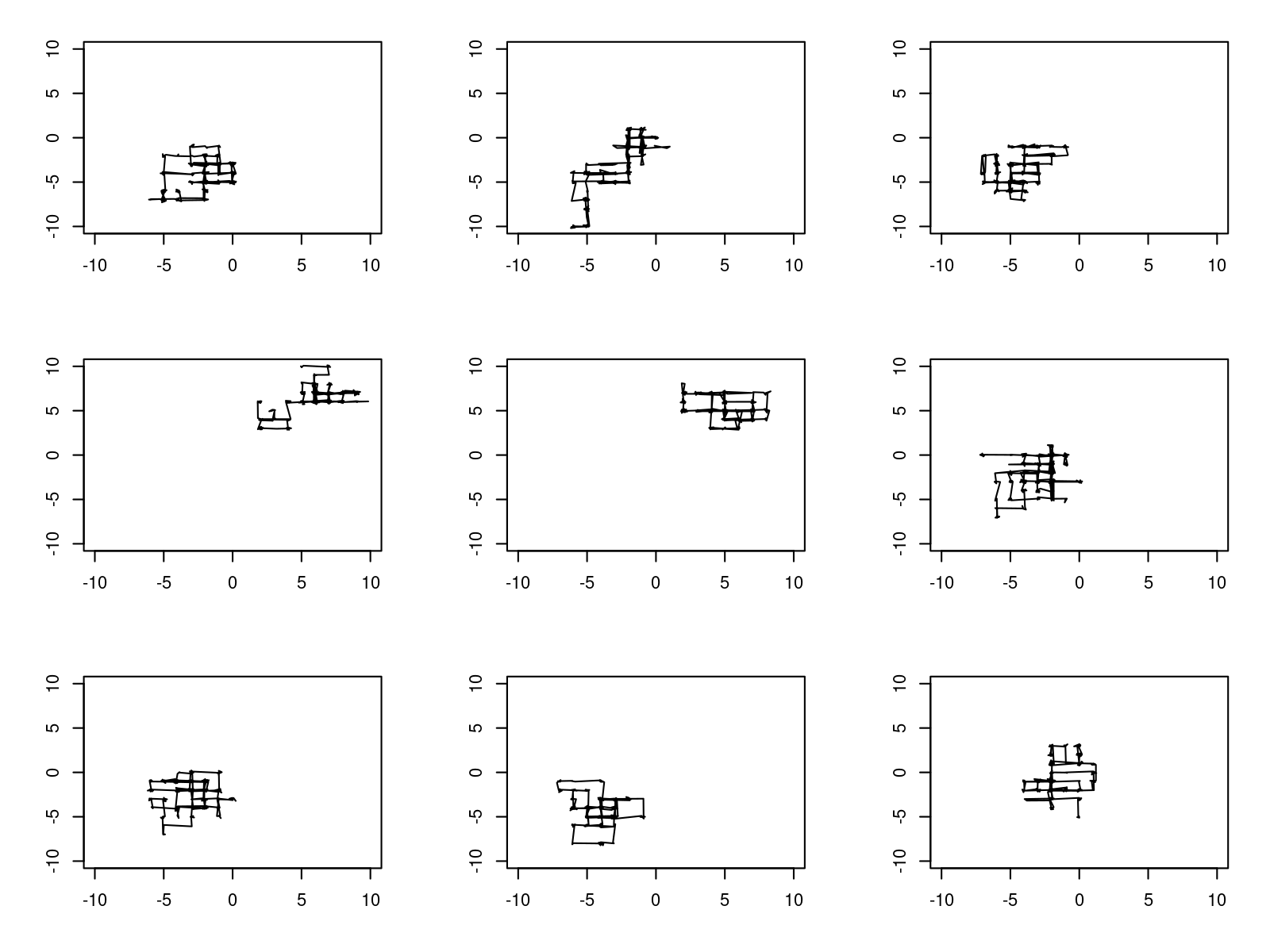}
        \caption{Trace plots from Metropolis}
        \label{fig:trace_MH}
    \end{subfigure}
    \hfill
    \begin{subfigure}[b]{0.32\textwidth}
        \centering
        \includegraphics[width=0.8\linewidth]{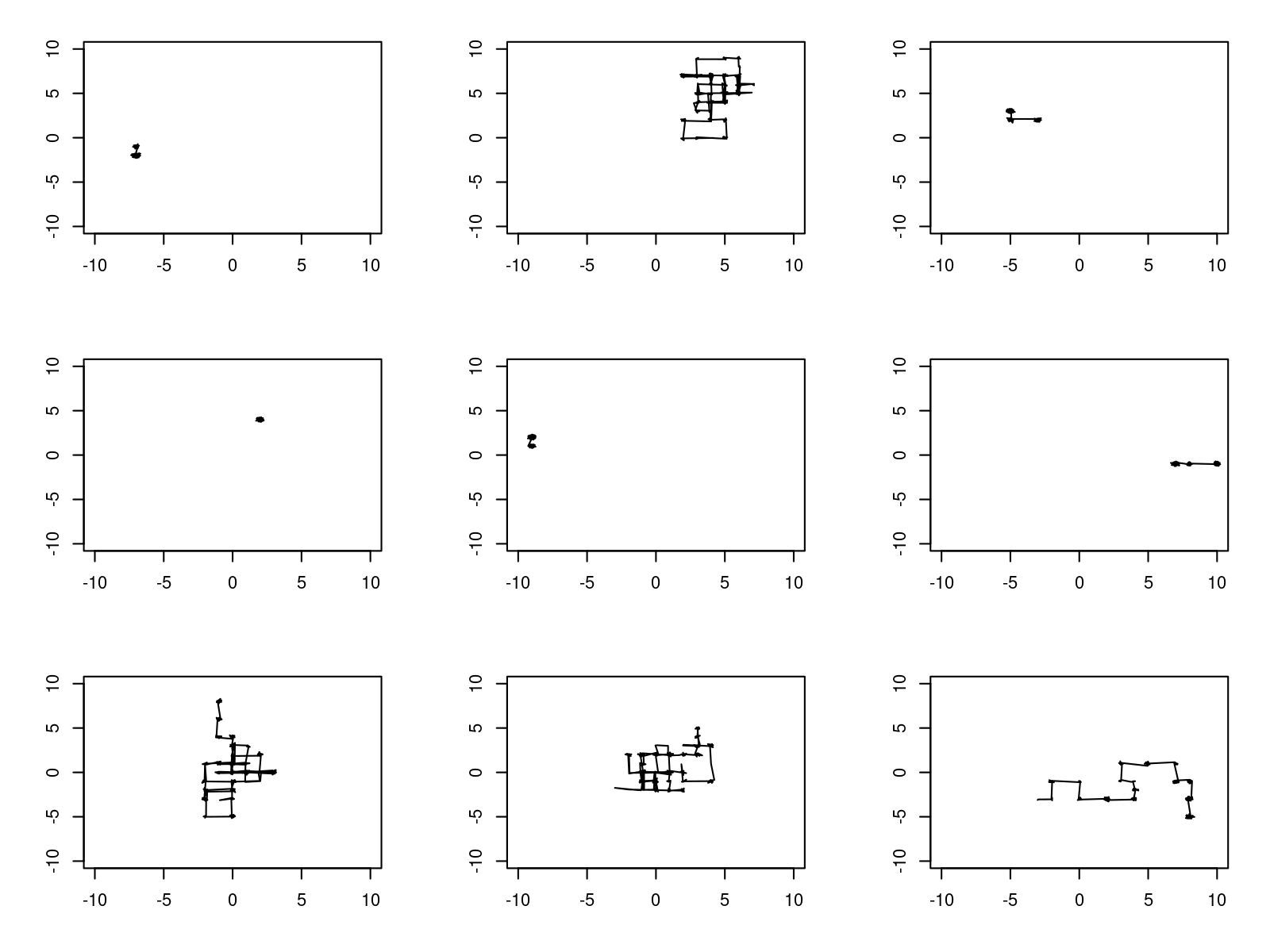}
        \caption{Trace plots from GWG}
        \label{fig:trace_GWG}
    \end{subfigure}
    \hfill
     \begin{subfigure}[b]{0.32\textwidth}
        \centering
        \includegraphics[width=0.8\linewidth]{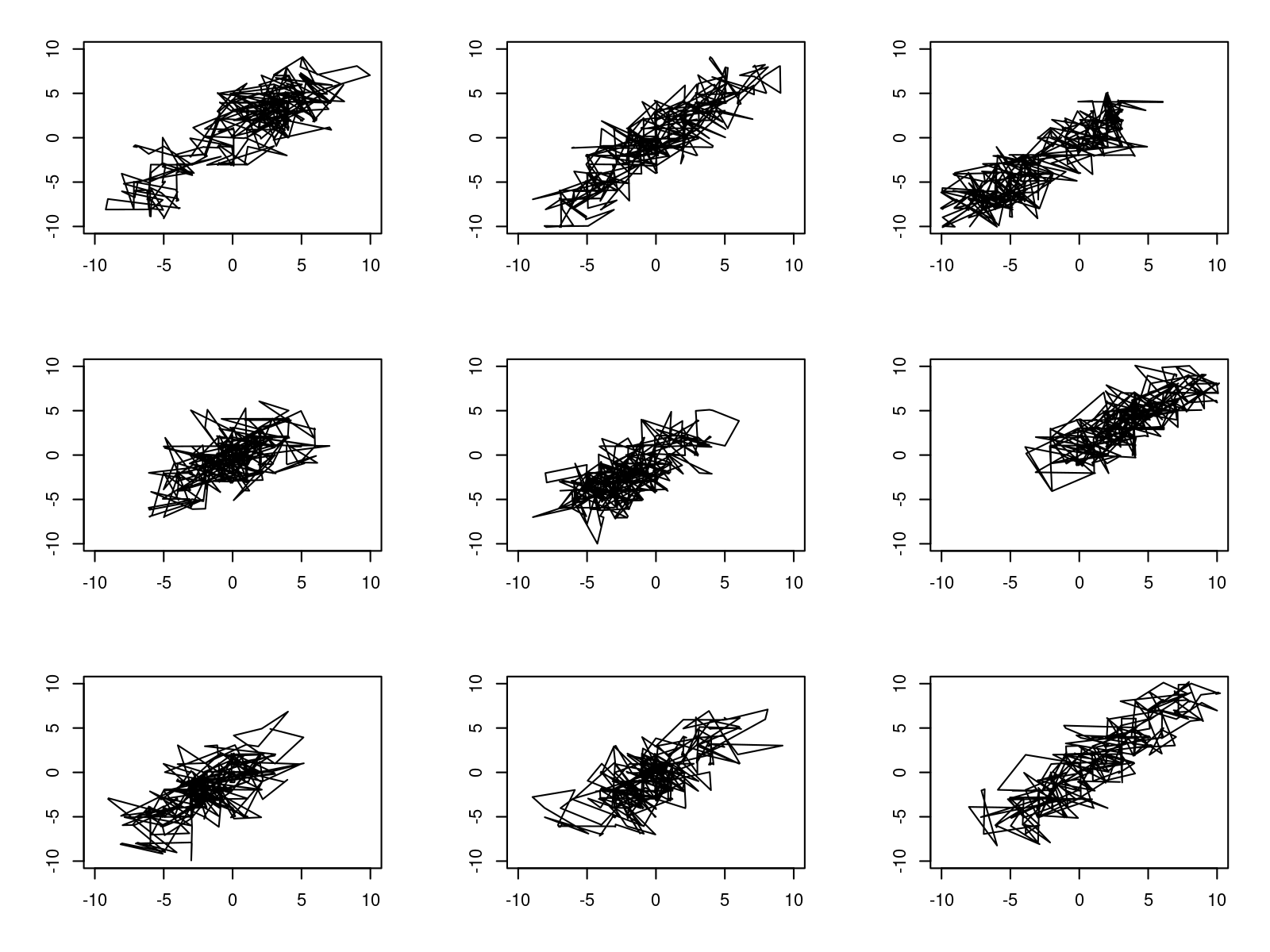}
        \caption{Trace plots from NCG}
        \label{fig:trace_NCG}
    \end{subfigure}
\hfill
     \begin{subfigure}[b]{0.32\textwidth}
        \centering
        \includegraphics[width=0.8\linewidth]{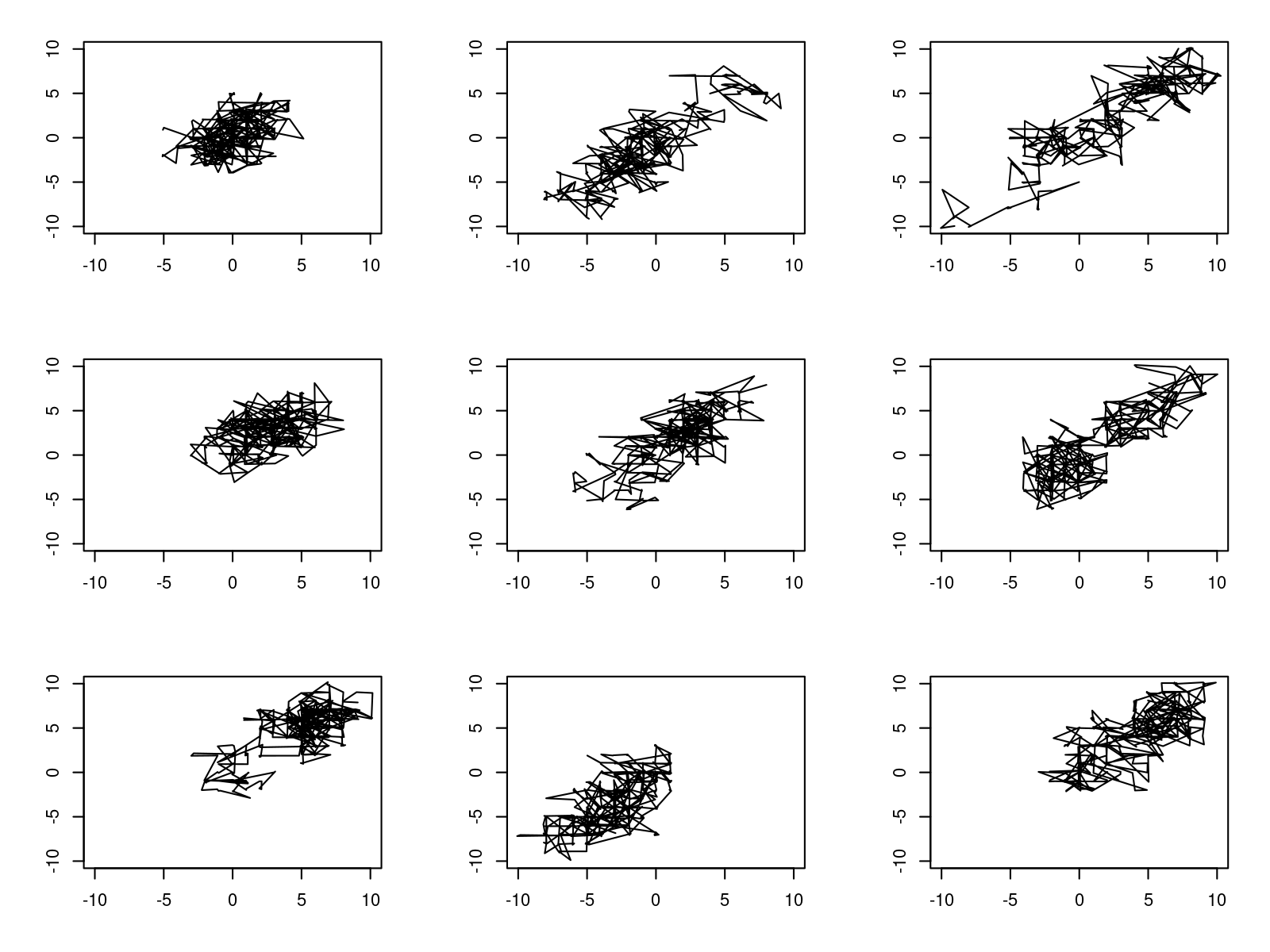}
        \caption{Trace plots from AVG}
        \label{fig:trace_AVG}
    \end{subfigure}
    \hfill
     \begin{subfigure}[b]{0.32\textwidth}
        \centering
        \includegraphics[width=0.8\linewidth]{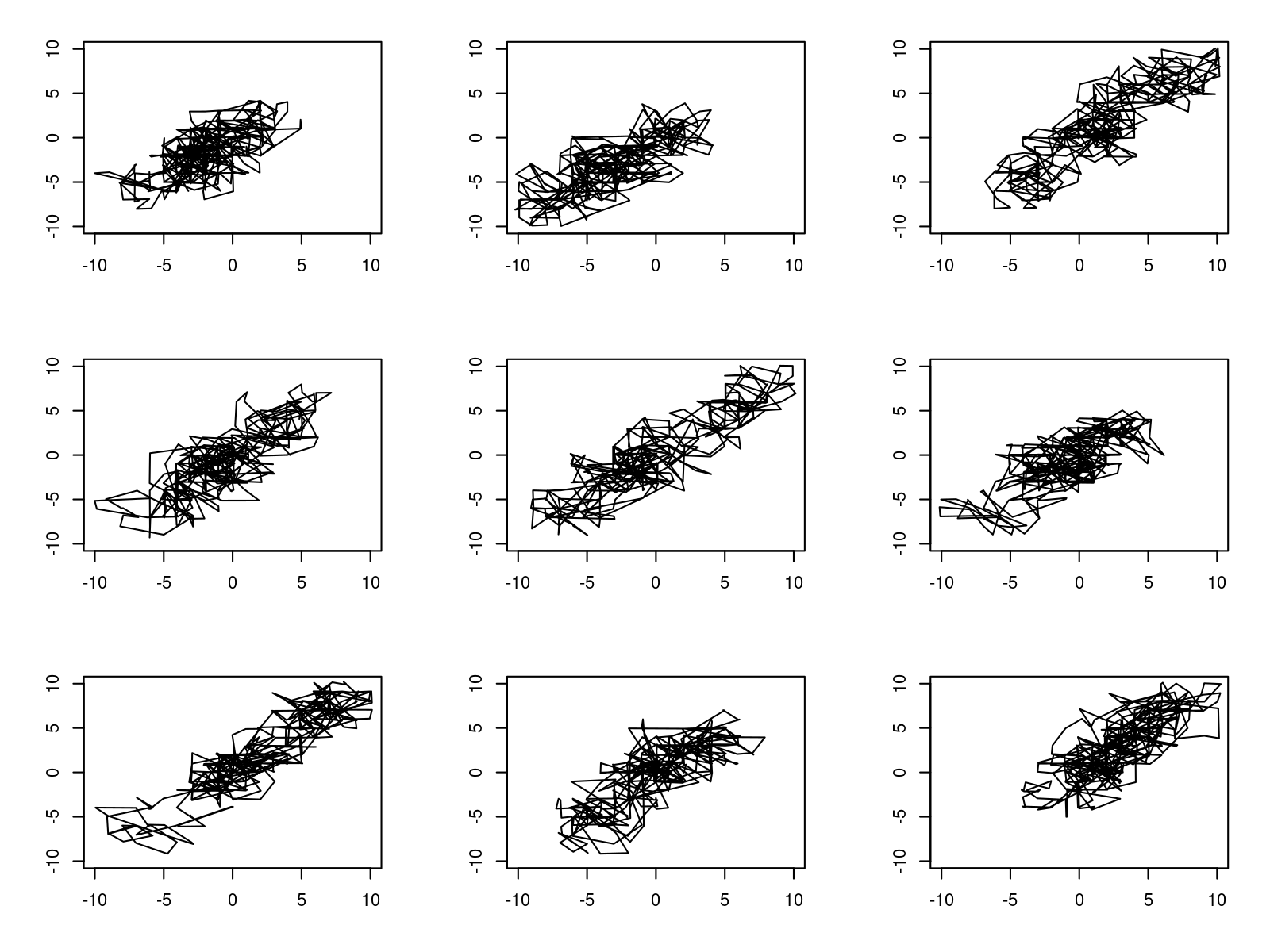}
        \caption{Trace plots from V-DHAMS}
        \label{fig:trace_Hams}
    \end{subfigure}
    \hfill
     \begin{subfigure}[b]{0.32\textwidth}
        \centering
        \includegraphics[width=0.8\linewidth]{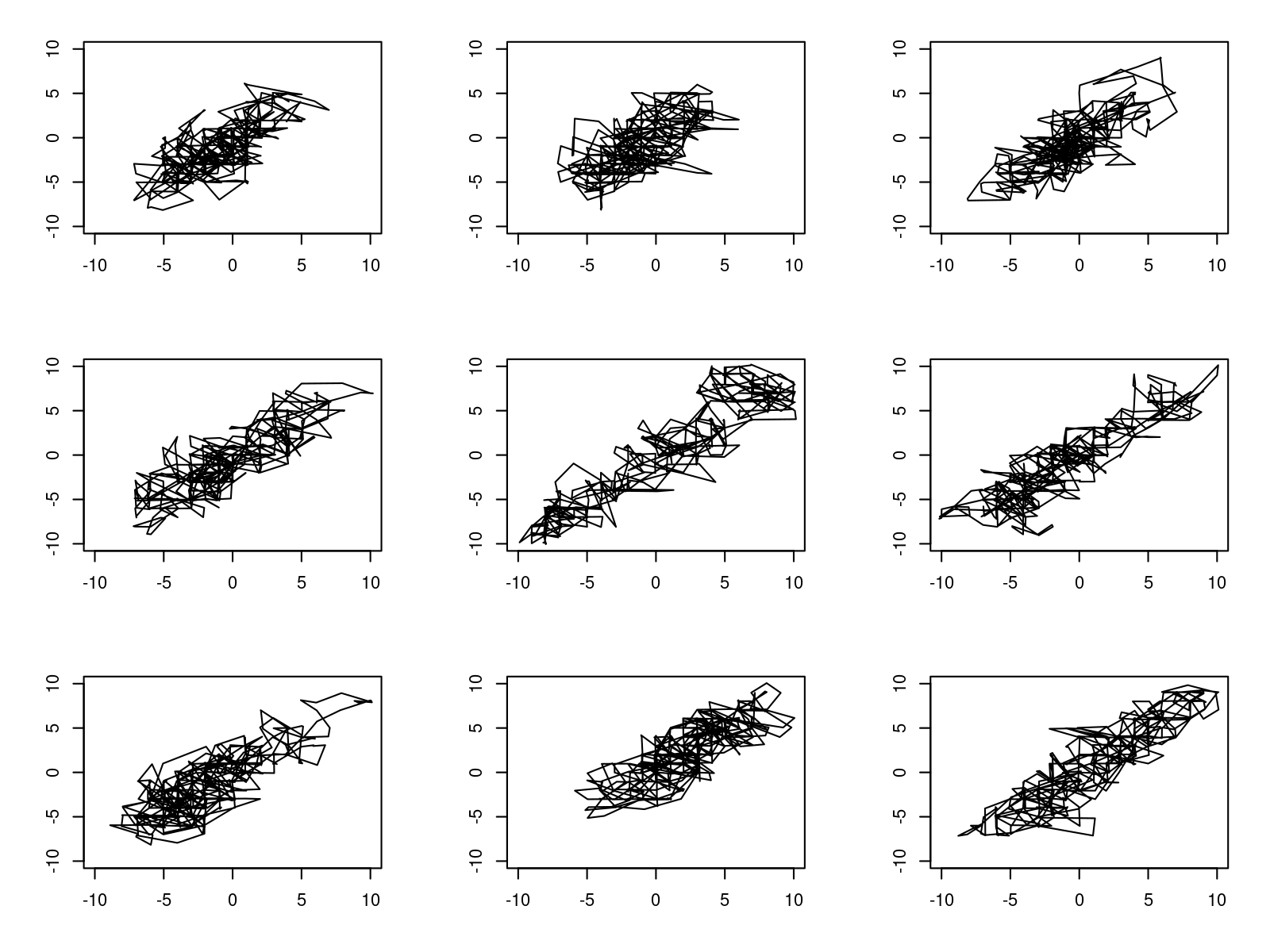}
        \caption{Trace plots from O-DHAMS}
        \label{fig:trace_Overhams}
    \end{subfigure}
\caption{Trace plots for discrete Gaussian distribution}
\label{fig:trace_plots}
\end{figure}

The optimal parameters and associated acceptance rates for each method are presented in Table~\ref{tab:parma_gaussian}.
\begin{table}[H]
\centering
\begin{tabular}{|c|c|c|c|c|}
\hline
    Sampler & Parameter & Acceptance Rate\\
    \hline
    Metropolis & $r=2$ & 0.73\\
    GWG  & $\sigma=1, r =2$ & 0.71\\
    NCG & $\delta=3.5$ & 0.61\\
        AVG & $\delta=1.88$ & 0.58\\
   V-DHAMS & $\epsilon = 0.9, \delta=0.9, \phi=0.5$ & 0.86\\
    O-DHAMS & $\epsilon = 0.9, \delta=0.75, \phi=0.5, \beta=0.7$ & 0.80\\ \hline
\end{tabular}
\caption{Parameters for discrete Gaussian distribution}
\label{tab:parma_gaussian}
\end{table}
We select 9 chains from the 100 parallel chains for each sampler and present their trace plots of the first two covariates for the first 450 draws after burn-in in Figure \ref{fig:trace_plots}. Discrete-HAMS exhibits superior exploration capability in traversing the probability landscape.

\begin{figure}[tbp]
    \centering
    \begin{subfigure}[b]{0.32\textwidth}
        \centering
        \includegraphics[width=0.8\linewidth]{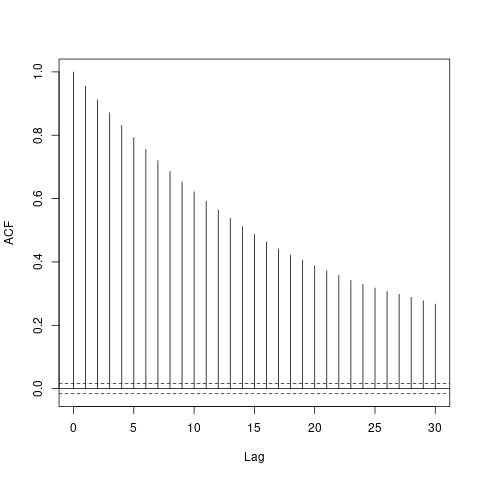}
        \caption{ACF from Metropolis}
        \label{fig:acf_MH}
    \end{subfigure}
    \hfill
    \begin{subfigure}[b]{0.32\textwidth}
        \centering
        \includegraphics[width=0.8\linewidth]{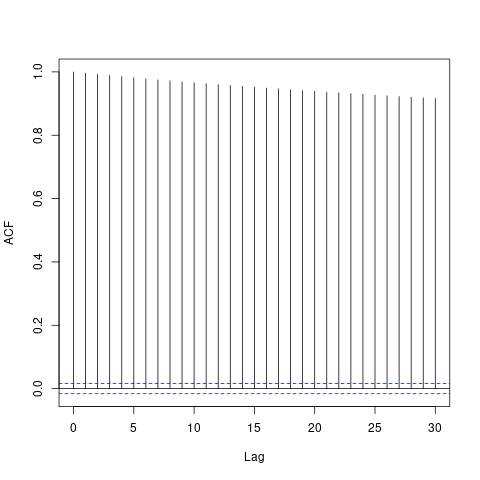}
        \caption{ACF from GWG}
        \label{fig:acf_gwg}
    \end{subfigure}
    \hfill
     \begin{subfigure}[b]{0.32\textwidth}
        \centering
        \includegraphics[width=0.8\linewidth]{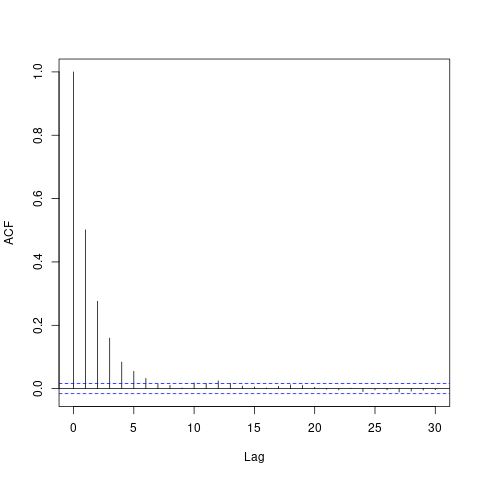}
        \caption{ACF from NCG}
        \label{fig:acf_NCG}
    \end{subfigure}
\hfill
     \begin{subfigure}[b]{0.32\textwidth}
        \centering
        \includegraphics[width=0.8\linewidth]{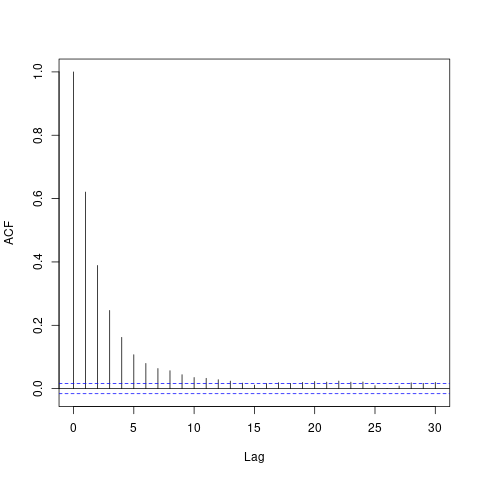}
        \caption{ACF from AVG}
        \label{fig:acf_avg}
    \end{subfigure}
    \hfill
     \begin{subfigure}[b]{0.32\textwidth}
        \centering
        \includegraphics[width=0.8\linewidth]{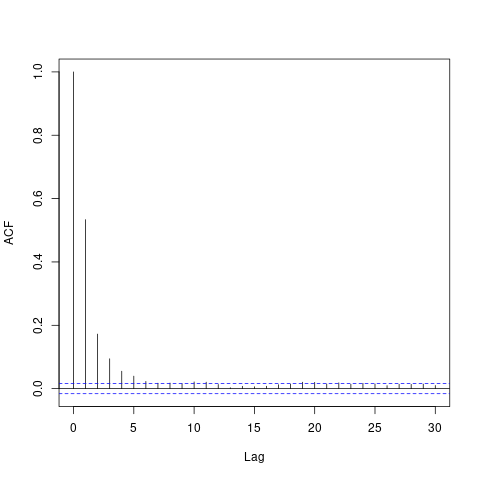}
        \caption{ACF from V-DHAMS}
        \label{fig:acf_Hams}
    \end{subfigure}
    \hfill
     \begin{subfigure}[b]{0.32\textwidth}
        \centering
        \includegraphics[width=0.8\linewidth]{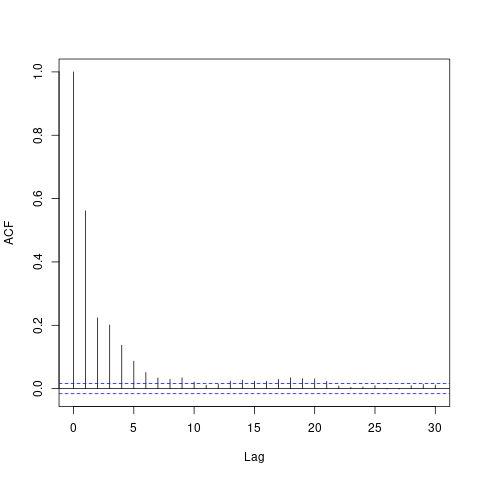}
        \caption{ACF from O-DHAMS}
        \label{fig:acf_overhams}
    \end{subfigure}
\caption{ACF plots for discrete Gaussian distribution}
\label{fig:acf_plots}
\end{figure}
\begin{figure}[tbp]
    \centering
    \includegraphics[width=0.4\linewidth]{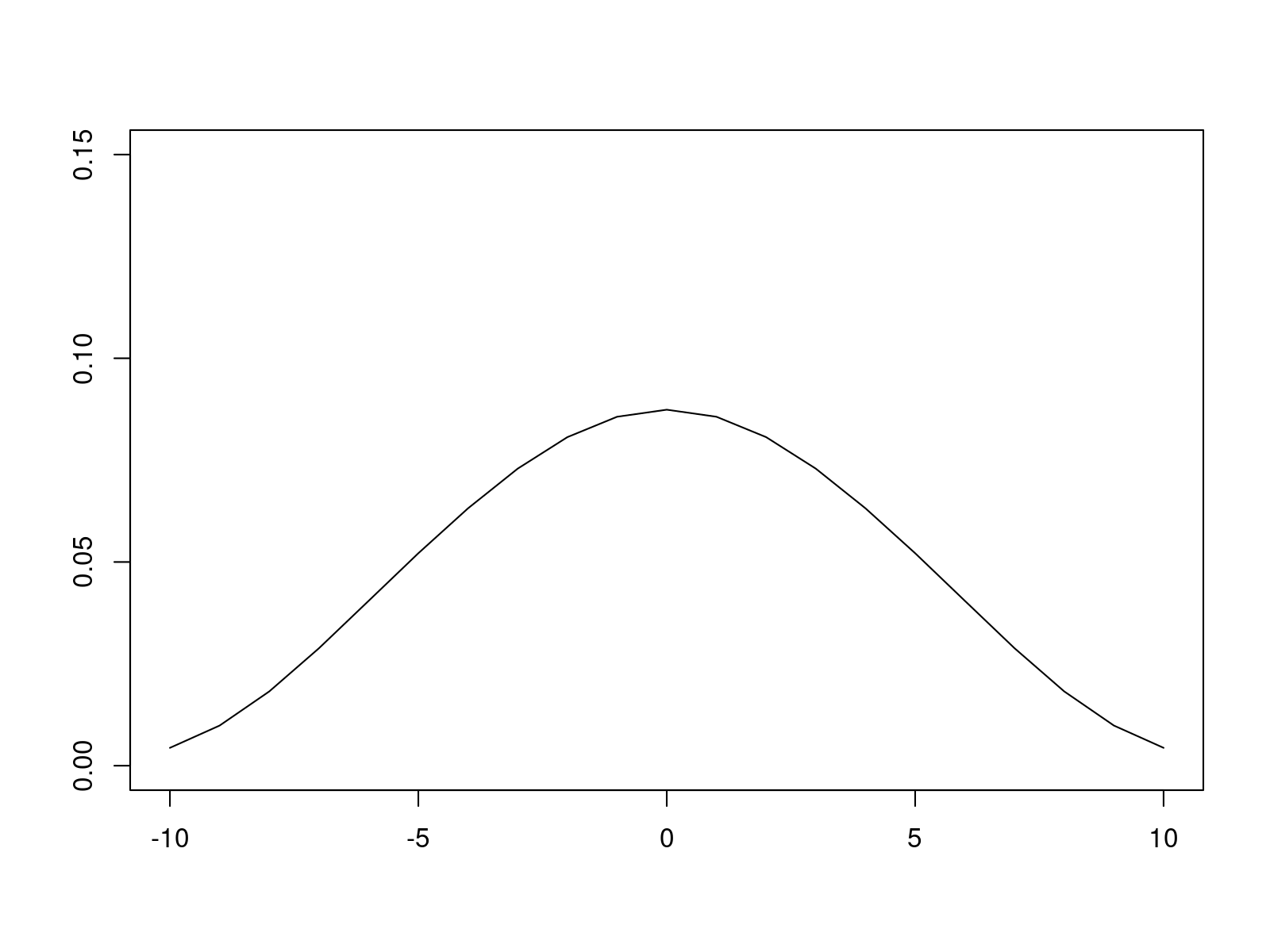}
    \caption{True marginal distribution of the first coordinate in discrete Gaussian distribution}
    \label{fig:freq_truth}
\end{figure}
The plots of the auto-correlation functions (ACF) of the negative potential function $f(s)$ from a single chain are presented in Figure~\ref{fig:acf_plots}. Discrete-Hams exhibits auto-correlations lower than NCG and significantly lower than both AVG and GWG, indicating reduced dependencies among draws and improved mixing efficiency.

\begin{figure}[tbp]
    \centering
    \begin{subfigure}[b]{0.32\textwidth}
        \centering
        \includegraphics[width=0.8\linewidth]{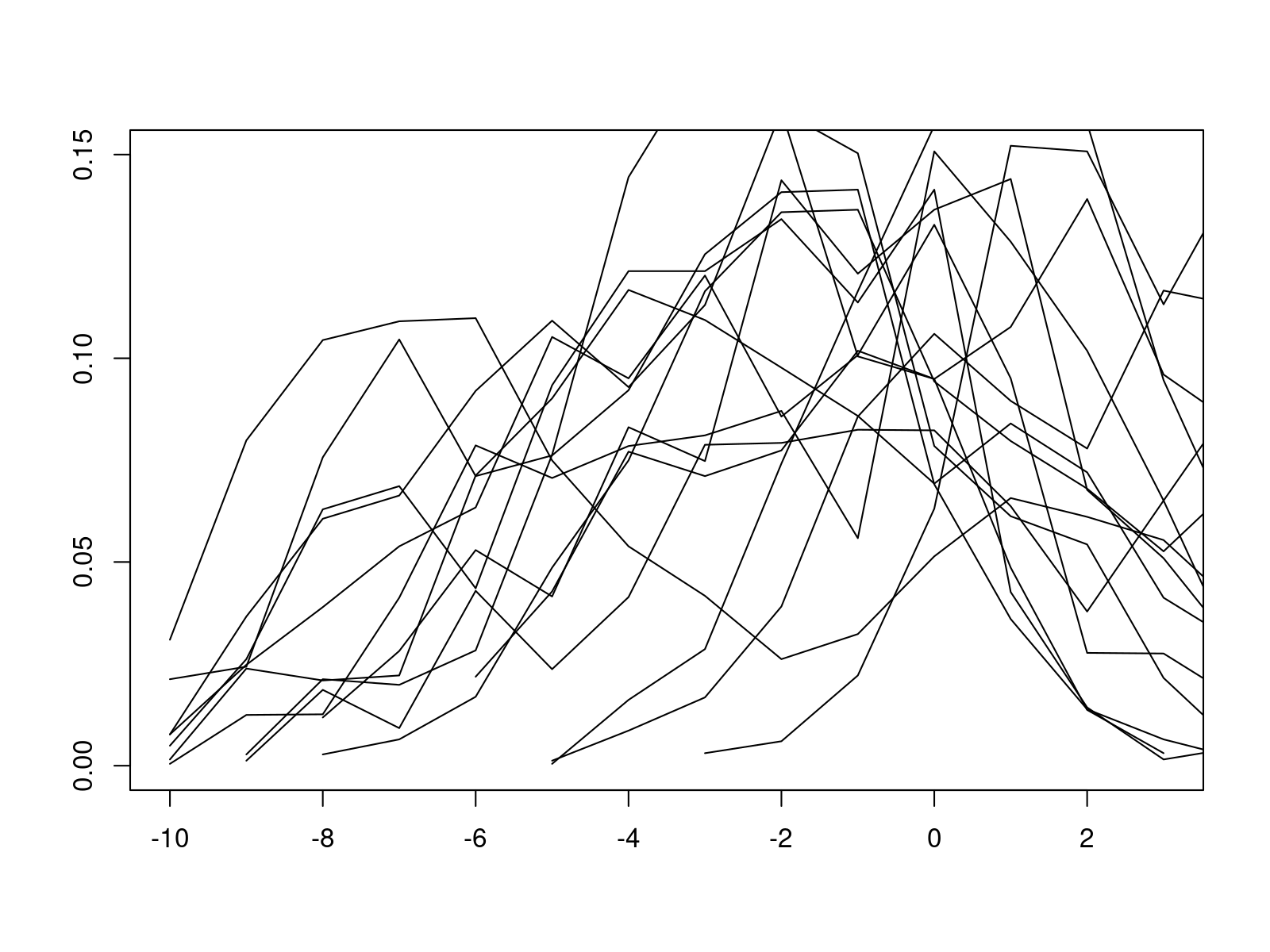}
        \caption{ Metropolis}
        \label{fig:freq_MH}
    \end{subfigure}
    \begin{subfigure}[b]{0.32\textwidth}
        \centering
        \includegraphics[width=0.8\linewidth]{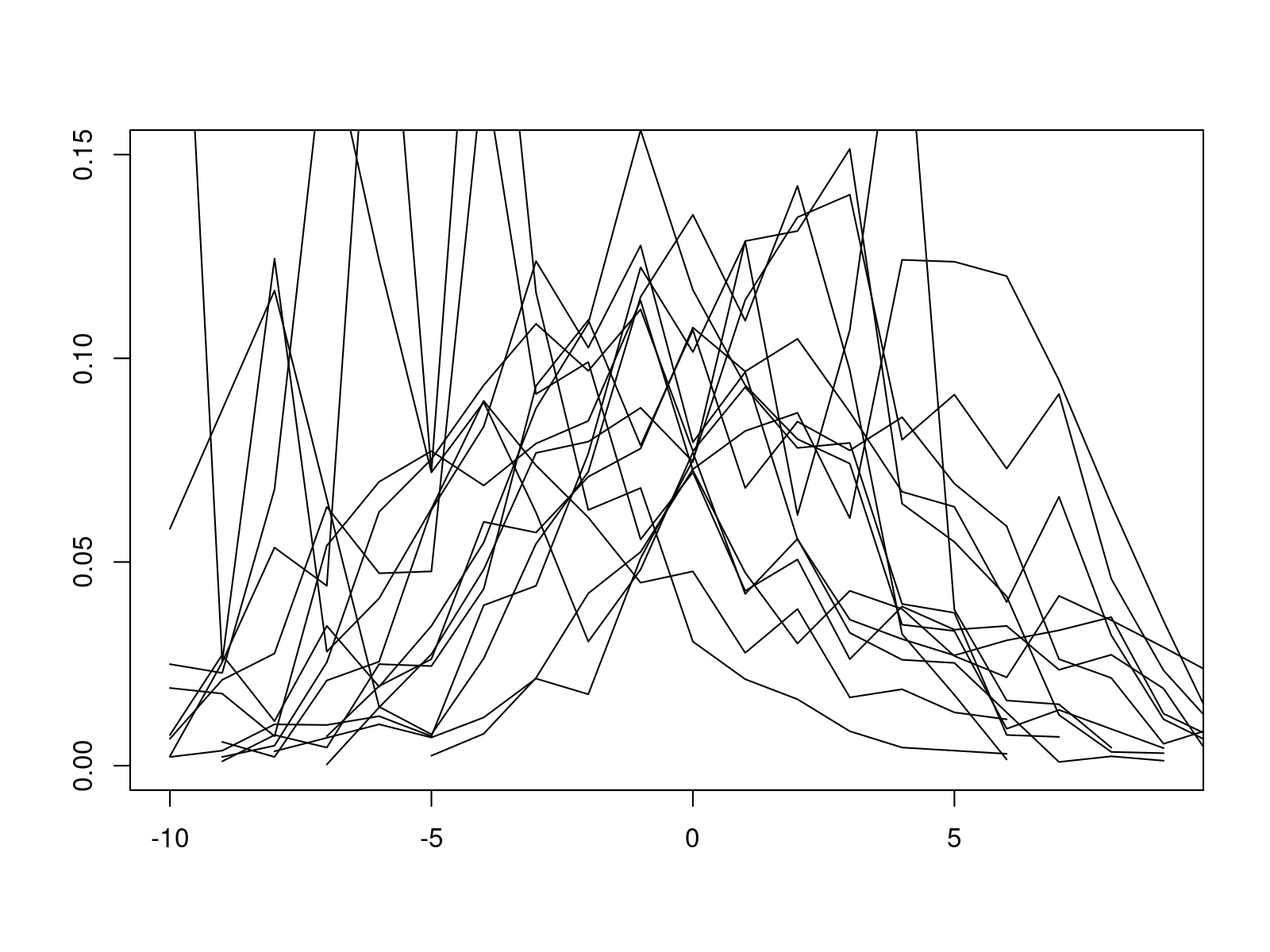}
        \caption{ GWG}
        \label{fig:freq_gwg}
    \end{subfigure}
     \begin{subfigure}[b]{0.32\textwidth}
        \centering
        \includegraphics[width=0.8\linewidth]{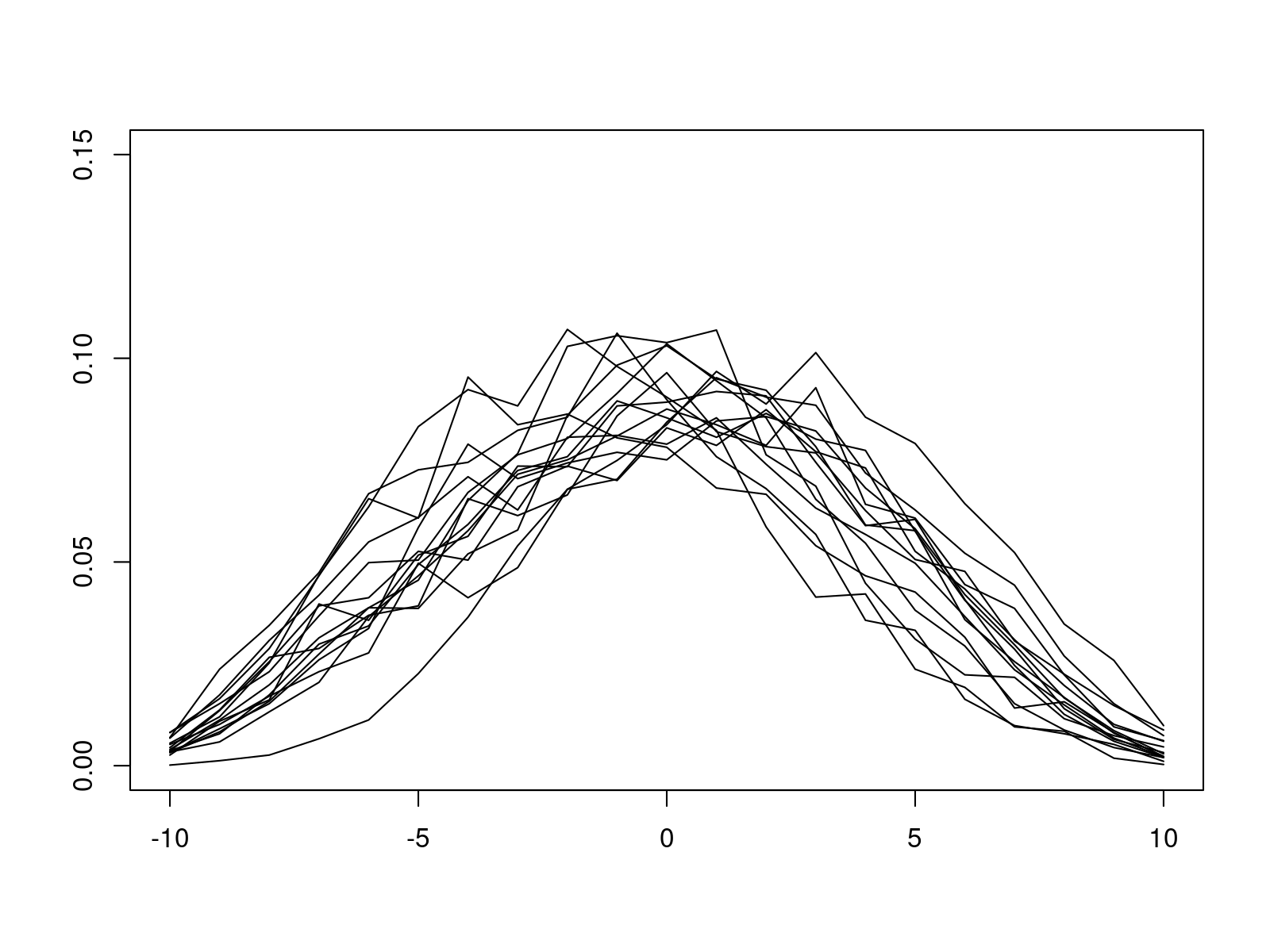}
        \caption{ NCG}
        \label{fig:freq_NCG}
    \end{subfigure}
     \begin{subfigure}[b]{0.32\textwidth}
        \centering
        \includegraphics[width=0.8\linewidth]{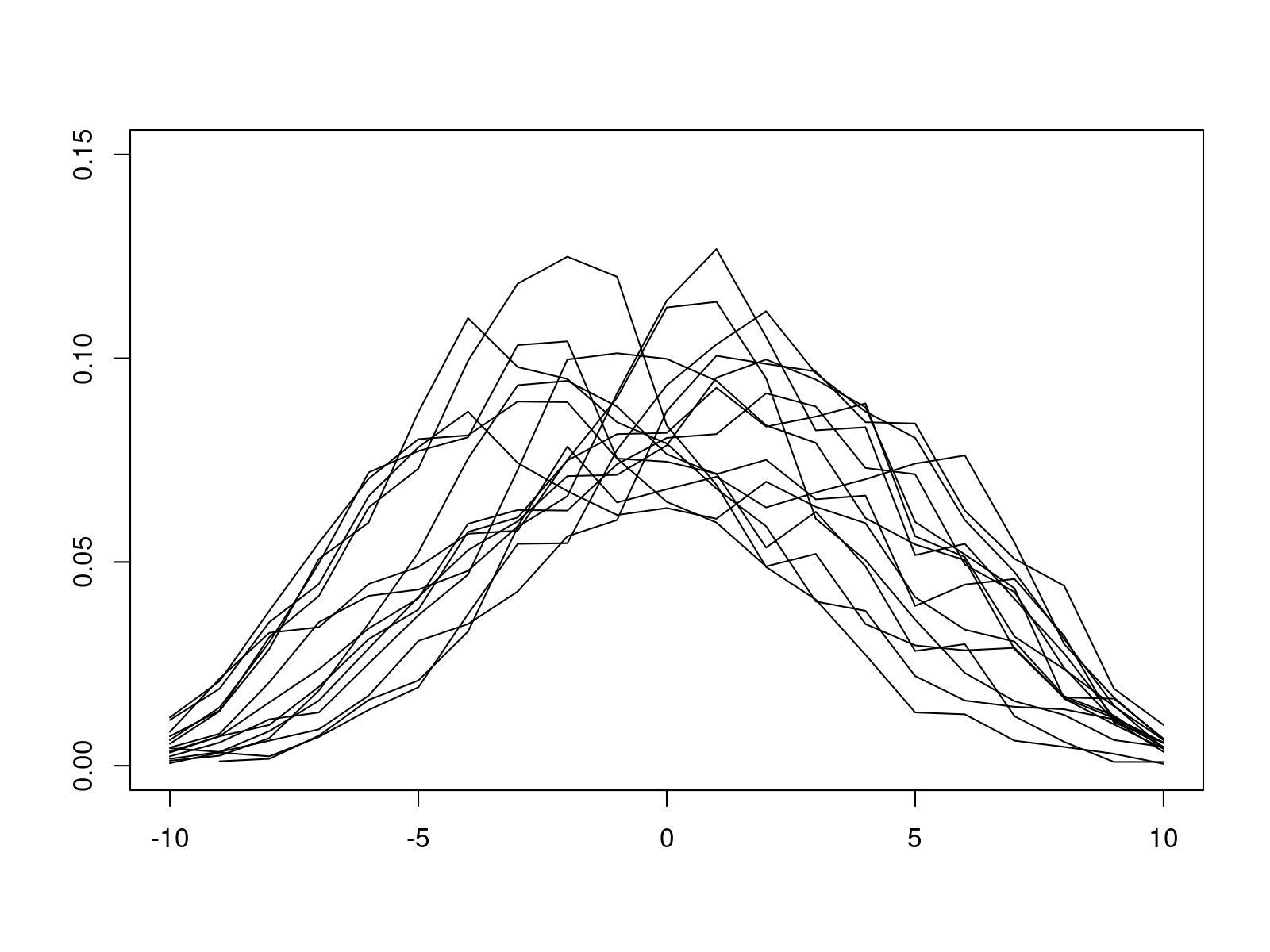}
        \caption{ AVG}
        \label{fig:freq_avg}
    \end{subfigure}
     \begin{subfigure}[b]{0.32\textwidth}
        \centering
        \includegraphics[width=0.8\linewidth]{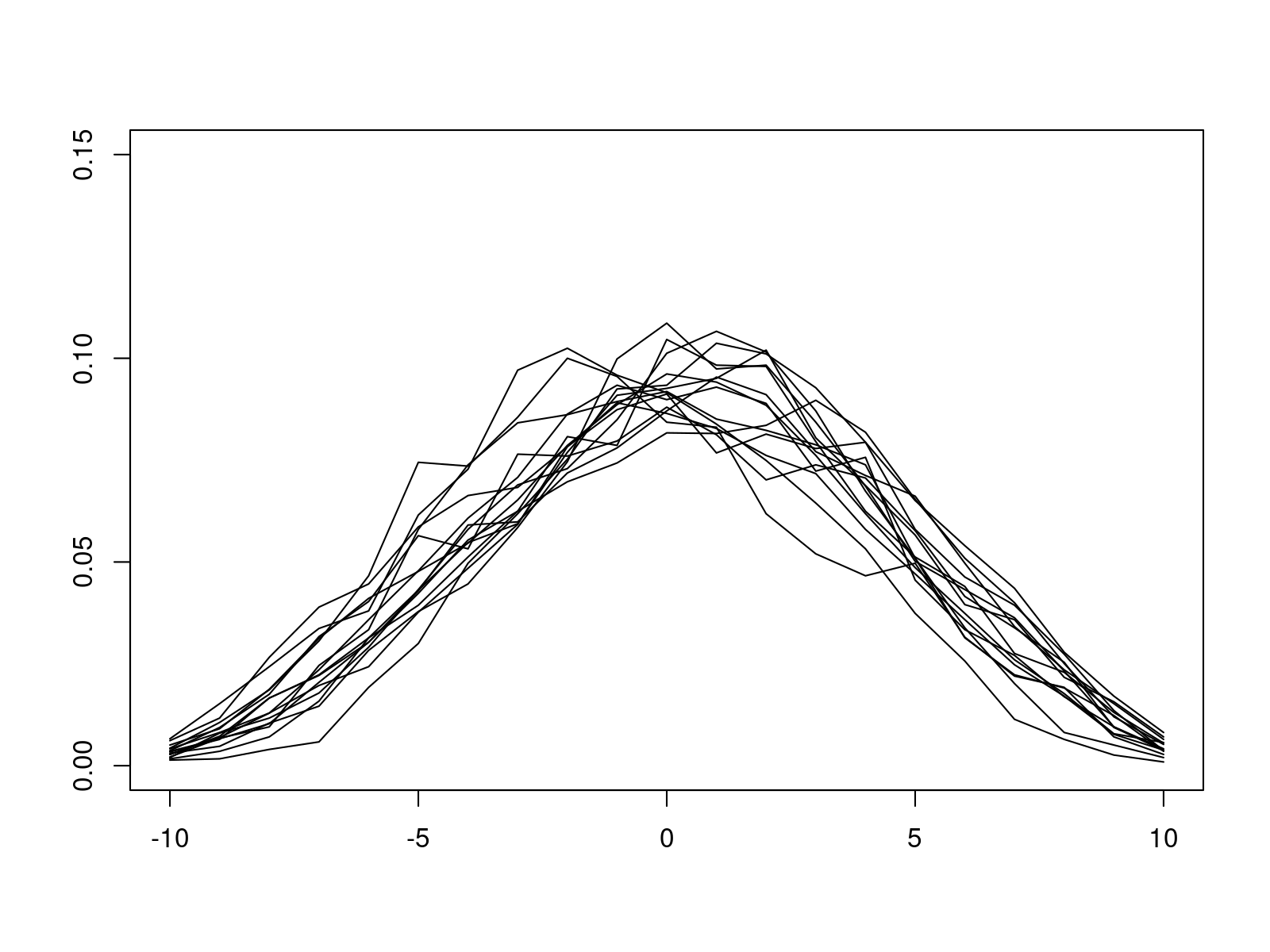}
        \caption{ V-DHAMS}
        \label{fig:freq_Hams}
    \end{subfigure}
     \begin{subfigure}[b]{0.32\textwidth}
        \centering
        \includegraphics[width=0.8\linewidth]{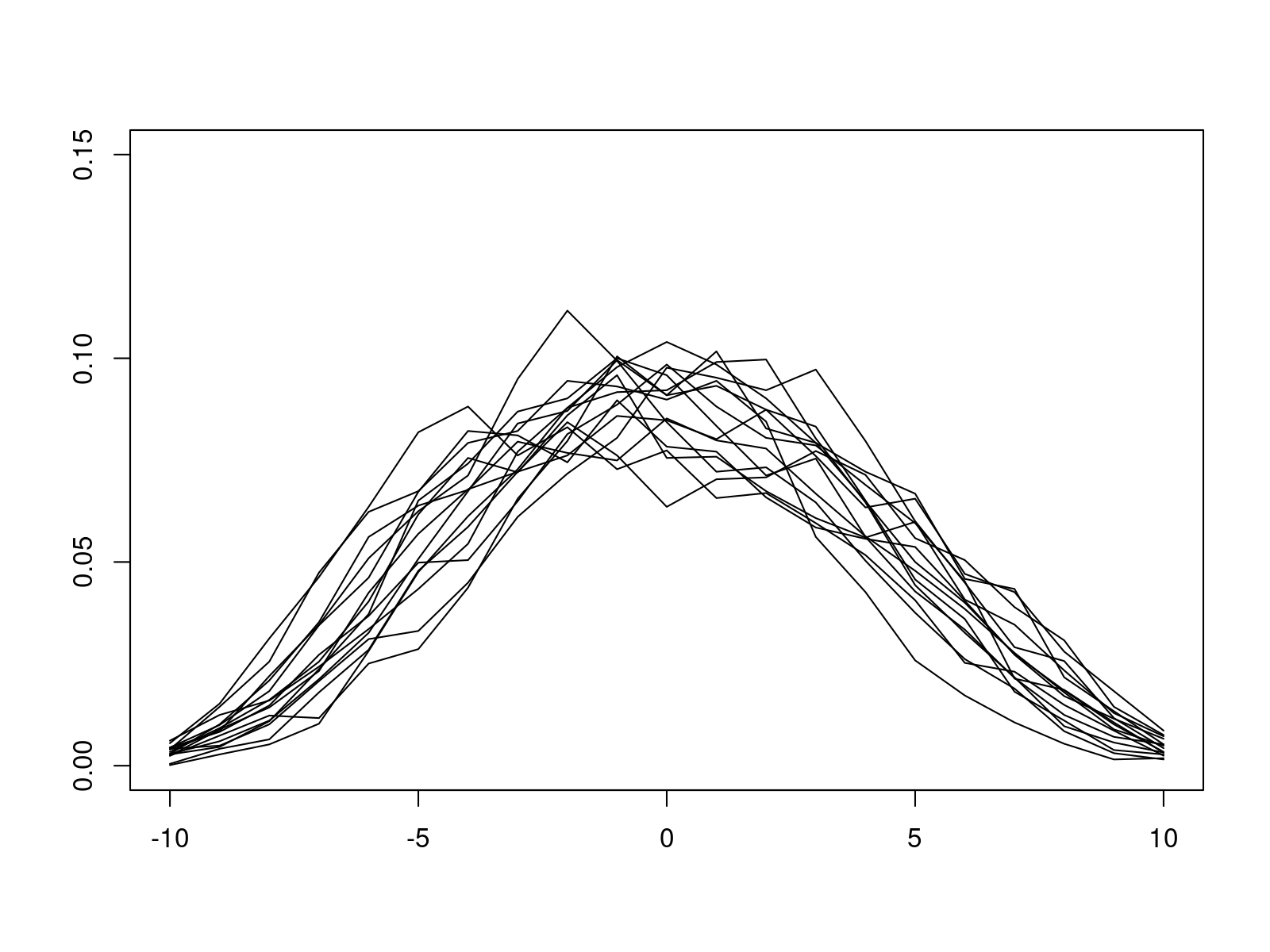}
        \caption{O-DHAMS}
        \label{fig:freq_overhams}
    \end{subfigure}
\caption{Frequency plots of the first coordinate in discrete Gaussian distribution}
\label{fig:freq_plots}
\end{figure}

For each sampler, we present frequency plots of the first coordinate across 10 independent chains, each based on 6,500 draws, as shown in Figure~\ref{fig:freq_plots}.
The corresponding true marginal distribution is displayed in Figure~\ref{fig:freq_truth}. Among all methods, the frequency plots generated by Discrete-HAMS are the closest to the ground truth, indicating more accurate sampling performance.

\subsection{Additional Results for Quadratic Mixture}\label{sec:mixture_results}
The optimal parameters and associated acceptance rates for each method are presented in Table \ref{tab:parma_poly}:

\begin{figure}[tbp]
    \centering
    \begin{subfigure}[b]{0.32\textwidth}
        \centering
        \includegraphics[width=0.8\linewidth]{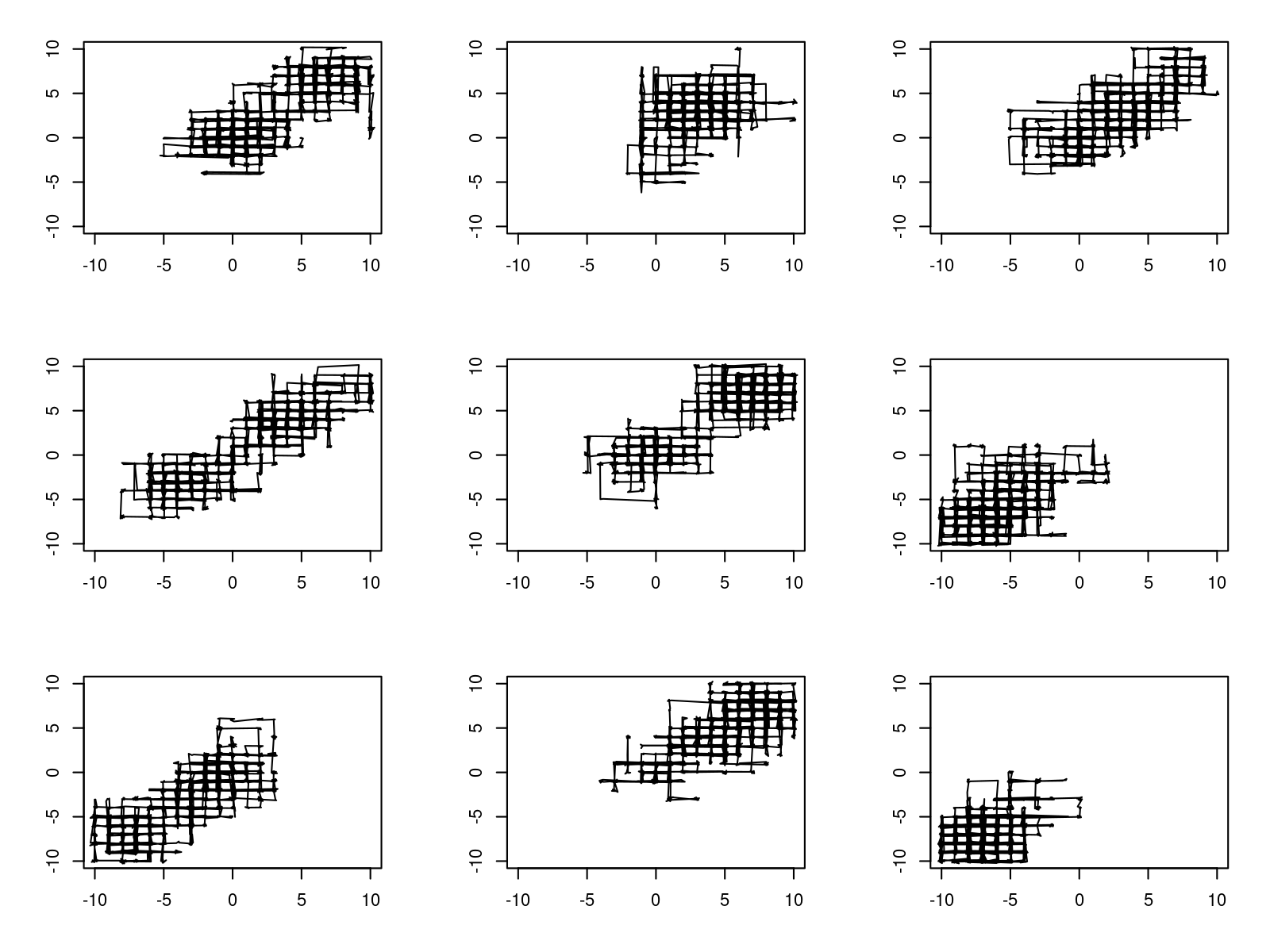}
        \caption{Trace plots from Metropolis}
        \label{fig:poly_trace_MH}
    \end{subfigure}
    \begin{subfigure}[b]{0.32\textwidth}
        \centering
        \includegraphics[width=0.8\linewidth]{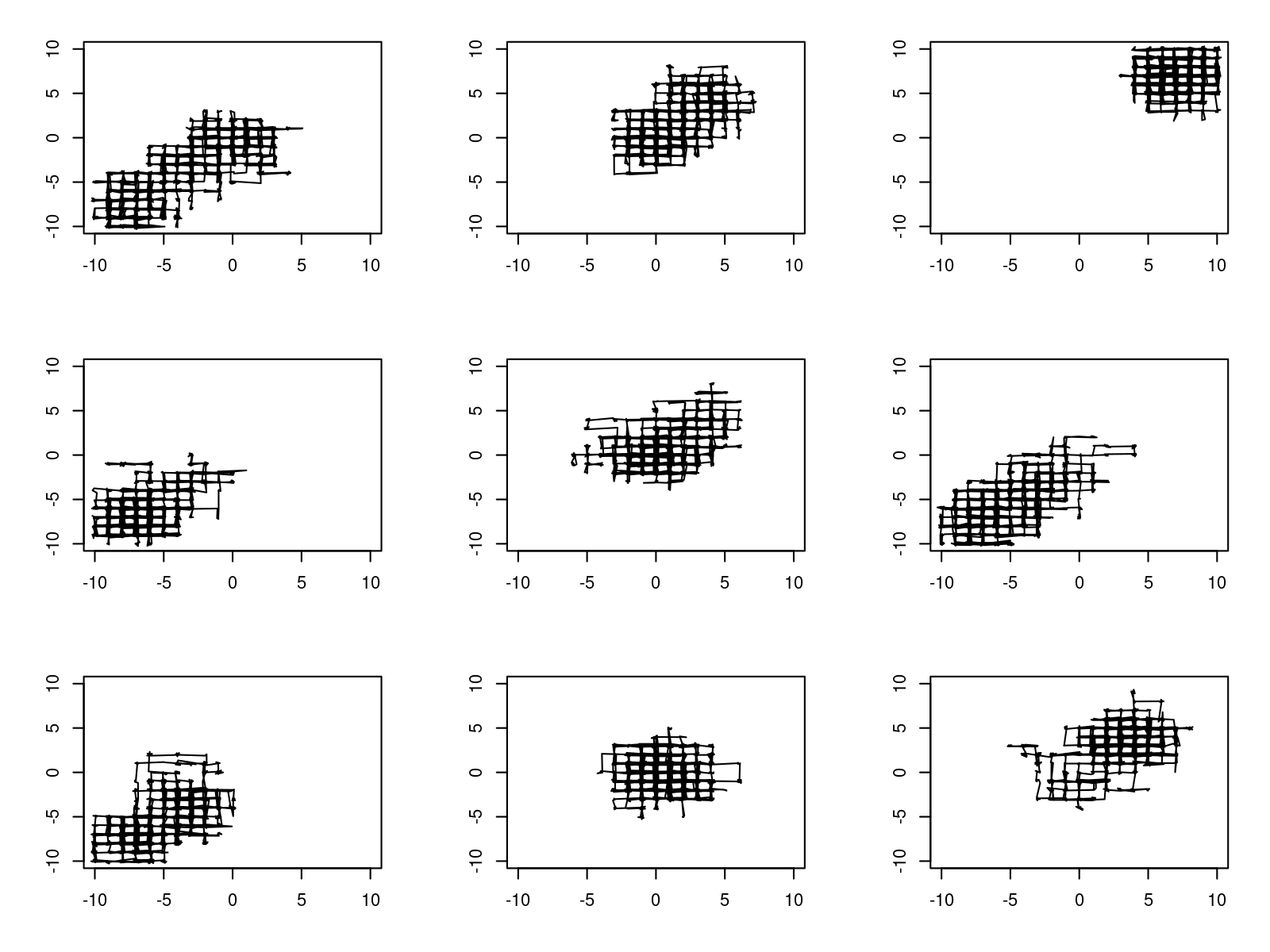}
        \caption{Trace plots from GWG}
        \label{fig:poly_trace_GWG}
    \end{subfigure}
     \begin{subfigure}[b]{0.32\textwidth}
        \centering
        \includegraphics[width=0.8\linewidth]{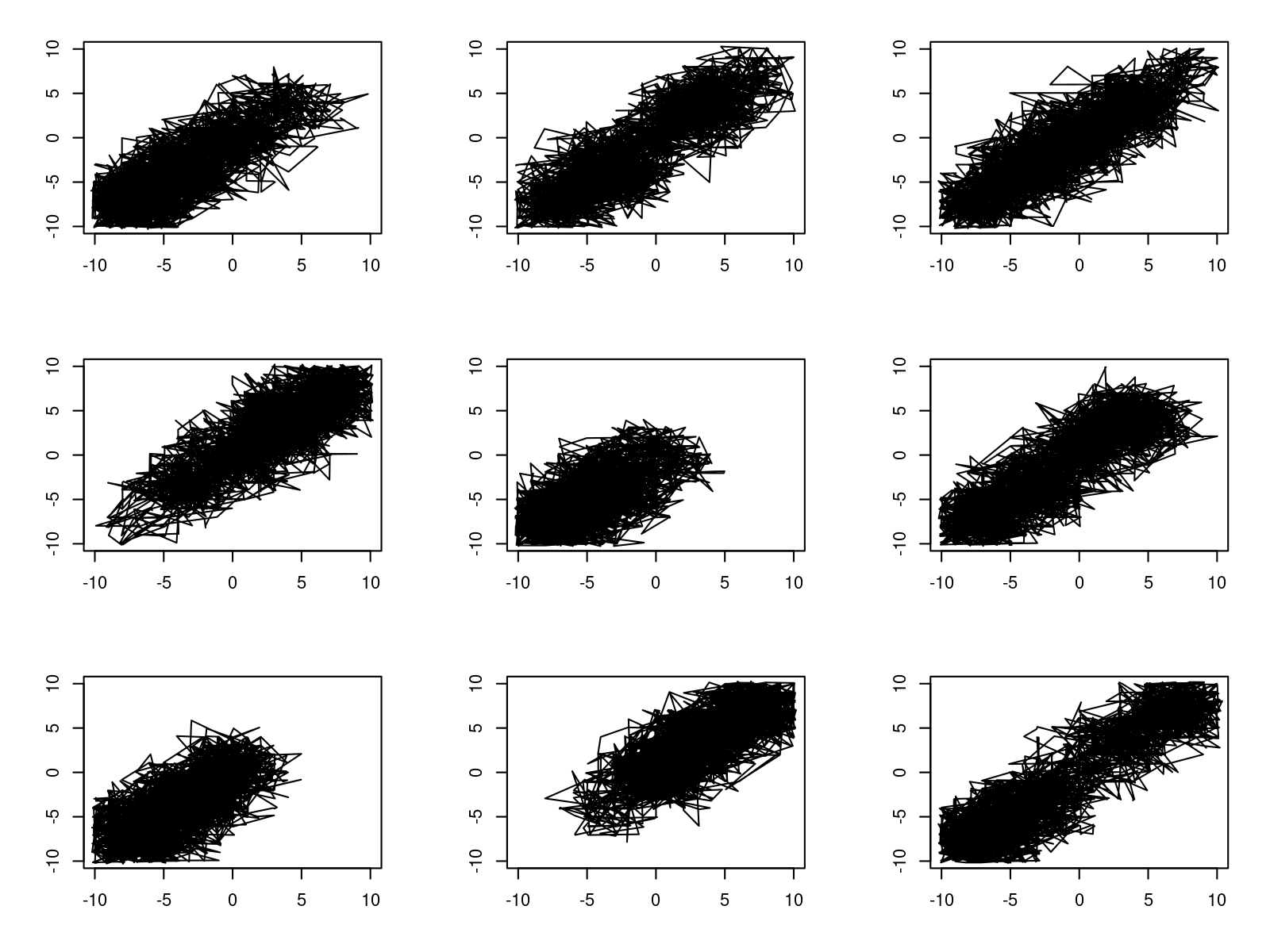}
        \caption{Trace plots from NCG}
        \label{fig:poly_trace_NCG}
    \end{subfigure}
     \begin{subfigure}[b]{0.32\textwidth}
        \centering
        \includegraphics[width=0.8\linewidth]{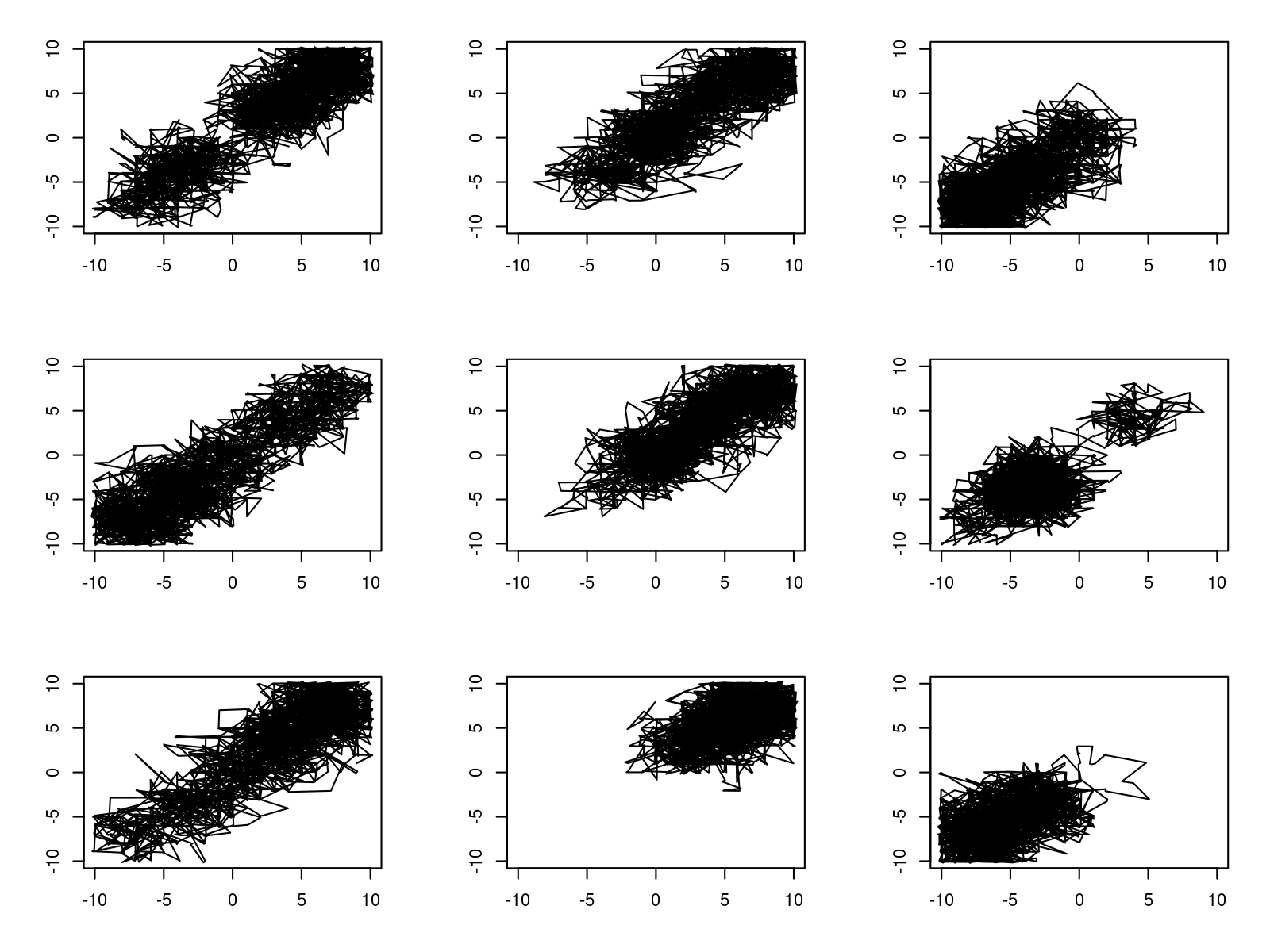}
        \caption{Trace plots from AVG}
        \label{fig:poly_trace_AVG}
    \end{subfigure}
     \begin{subfigure}[b]{0.32\textwidth}
        \centering
        \includegraphics[width=0.8\linewidth]{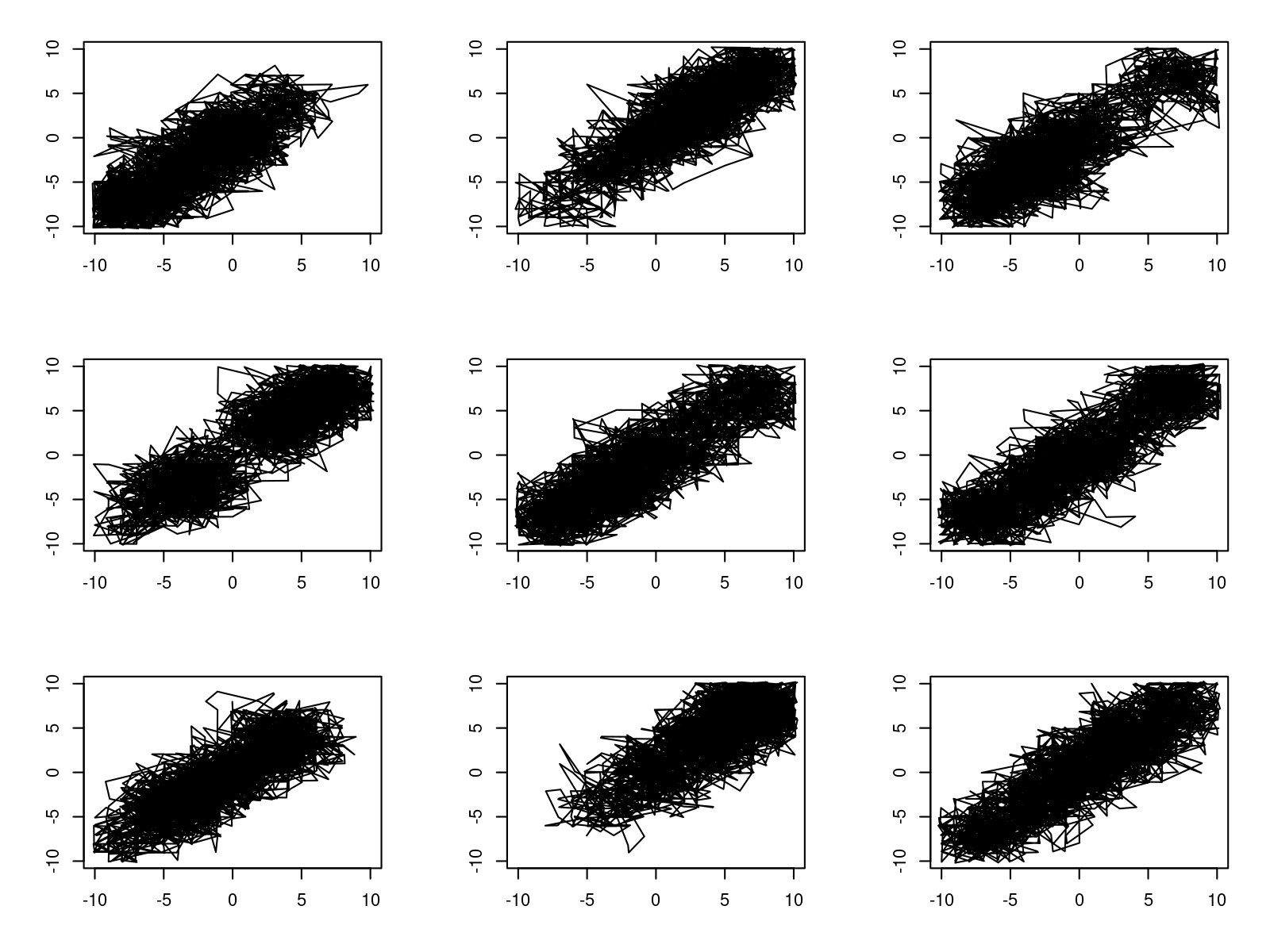}
        \caption{Trace plots from V-DHAMS}
        \label{fig:poly_trace_Hams}
    \end{subfigure}
     \begin{subfigure}[b]{0.32\textwidth}
        \centering
        \includegraphics[width=0.8\linewidth]{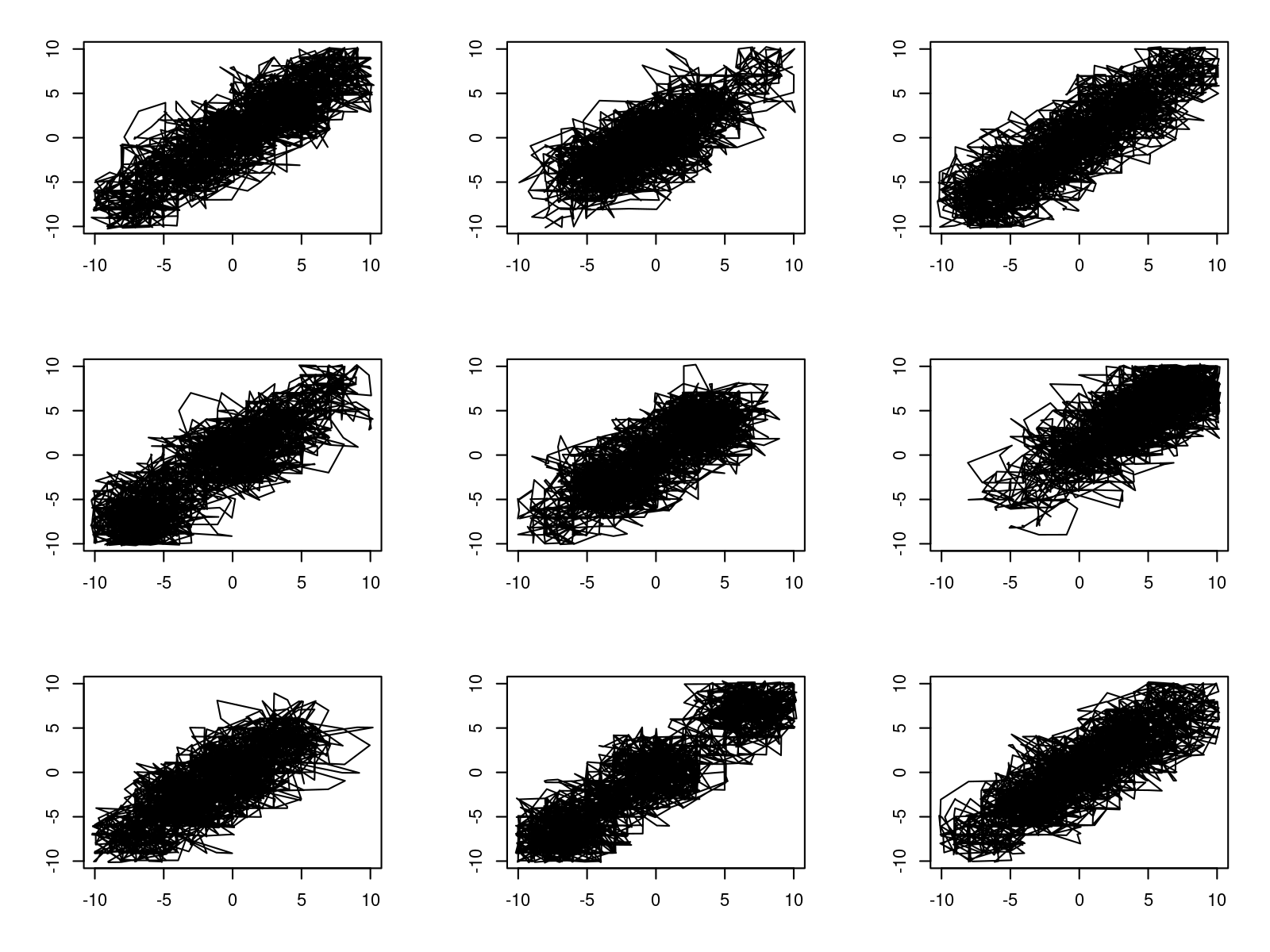}
        \caption{Trace plots from O-DHAMS}
        \label{fig:poly_trace_Overhams}
    \end{subfigure}
\caption{Trace plots for quadratic mixture distribution}
\label{fig:trace_plots_poly}
\end{figure}

\begin{table}[H]

\centering
\begin{tabular}{|c|c|c|c|c|}
\hline
    Sampler & Parameter & Acceptance Rate \\
    \hline
    Metropolis & r=4 & 0.59 \\
    GWG  & $\delta=1, r =2$ & 0.75\\
    NCG & $\delta=3.30$ & 0.74\\
        AVG & $\delta=1.86$ & 0.66\\
   V-DHAMS & $\epsilon = 0.9, \delta=1.07, \phi=0.5$& 0.84\\
    O-DHAMS & $\epsilon = 0.9, \delta=0.77, \phi=0.7, \beta=0.1$ & 0.80\\ \hline
\end{tabular}
\caption{Parameters for quadratic mixture distribution}
\label{tab:parma_poly}
\end{table}
\begin{figure}[tbp]
    \centering
    \begin{subfigure}[b]{0.32\textwidth}
        \centering
        \includegraphics[width=0.8\linewidth]{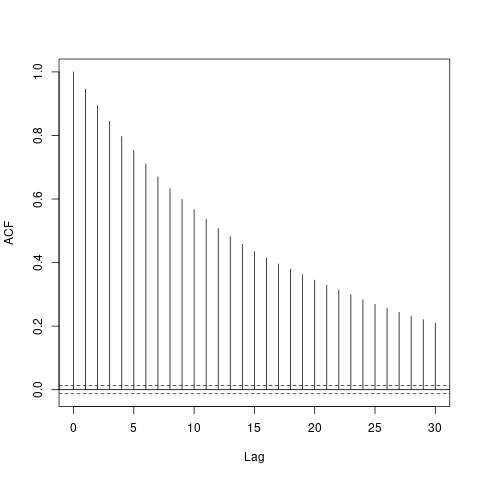}
        \caption{ACF from Metropolis}
        \label{fig:acf_MH_poly}
    \end{subfigure}
    \begin{subfigure}[b]{0.32\textwidth}
        \centering
        \includegraphics[width=0.8\linewidth]{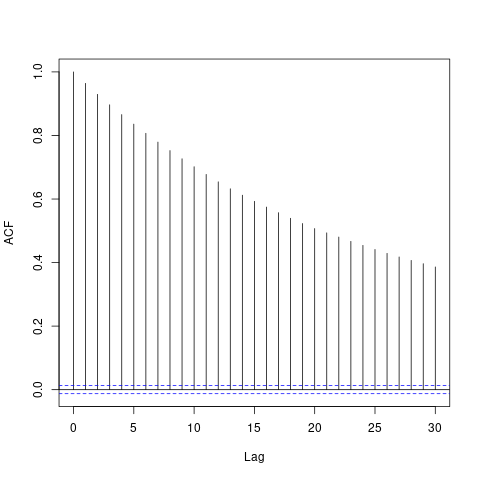}
        \caption{ACF from GWG}
        \label{fig:acf_gwg_poly}
    \end{subfigure}
     \begin{subfigure}[b]{0.32\textwidth}
        \centering
        \includegraphics[width=0.8\linewidth]{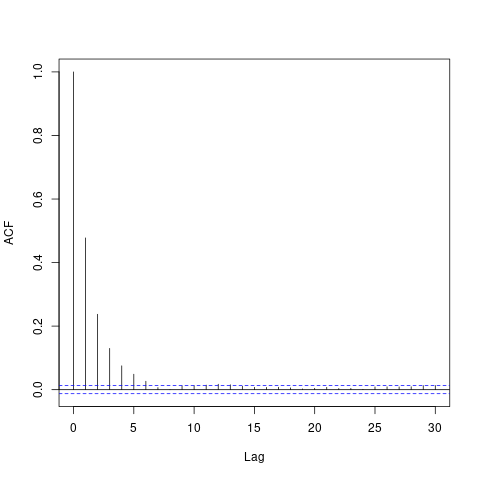}
        \caption{ACF from NCG}
        \label{fig:acf_NCG_poly}
    \end{subfigure}
     \begin{subfigure}[b]{0.32\textwidth}
        \centering
        \includegraphics[width=0.8\linewidth]{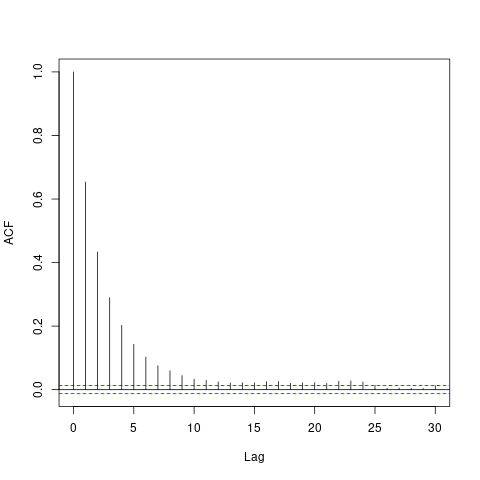}
        \caption{ACF from AVG}
        \label{fig:acf_AVG_poly}
    \end{subfigure}
     \begin{subfigure}[b]{0.32\textwidth}
        \centering
        \includegraphics[width=0.8\linewidth]{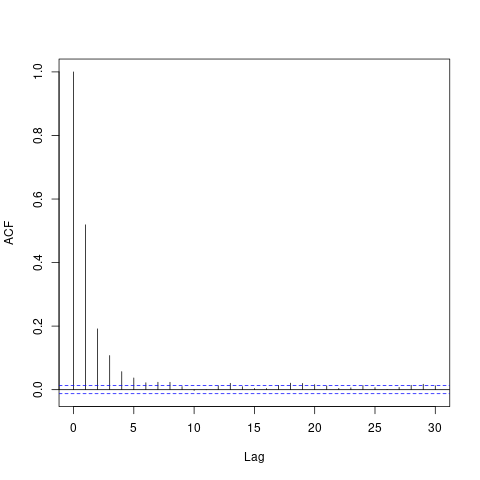}
        \caption{ACF from V-DHAMS}
        \label{fig:acf_Hams_poly}
    \end{subfigure}
     \begin{subfigure}[b]{0.32\textwidth}
        \centering
        \includegraphics[width=0.8\linewidth]{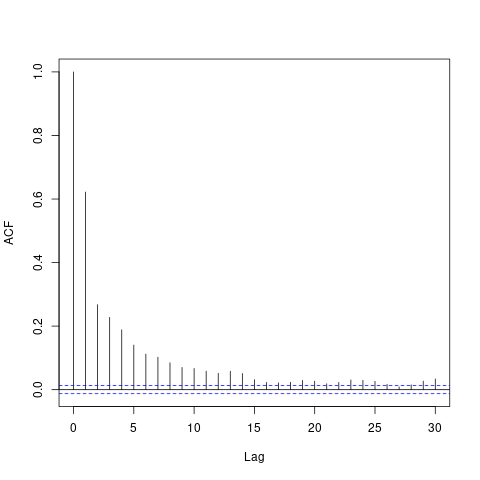}
        \caption{ACF from O-DHAMS}
        \label{fig:acf_overhams_poly}
    \end{subfigure}
\caption{ACF plots for quadratic mixture distribution}
\label{fig:acf_plots_poly}
\end{figure}
We select 9 chains from the 100 parallel chains for each sampler and present their trace plots of the first two covariates for the first 4,000 draws after burn-in in
Figure~\ref{fig:trace_plots_poly}. Both V-DHAMS and O-DHAMS exhibits significantly better exploration capability than NCG and much better than the other samplers,
in traversing the probability landscape.

\begin{figure}[tbp]
    \centering
    \includegraphics[width=0.4\linewidth]{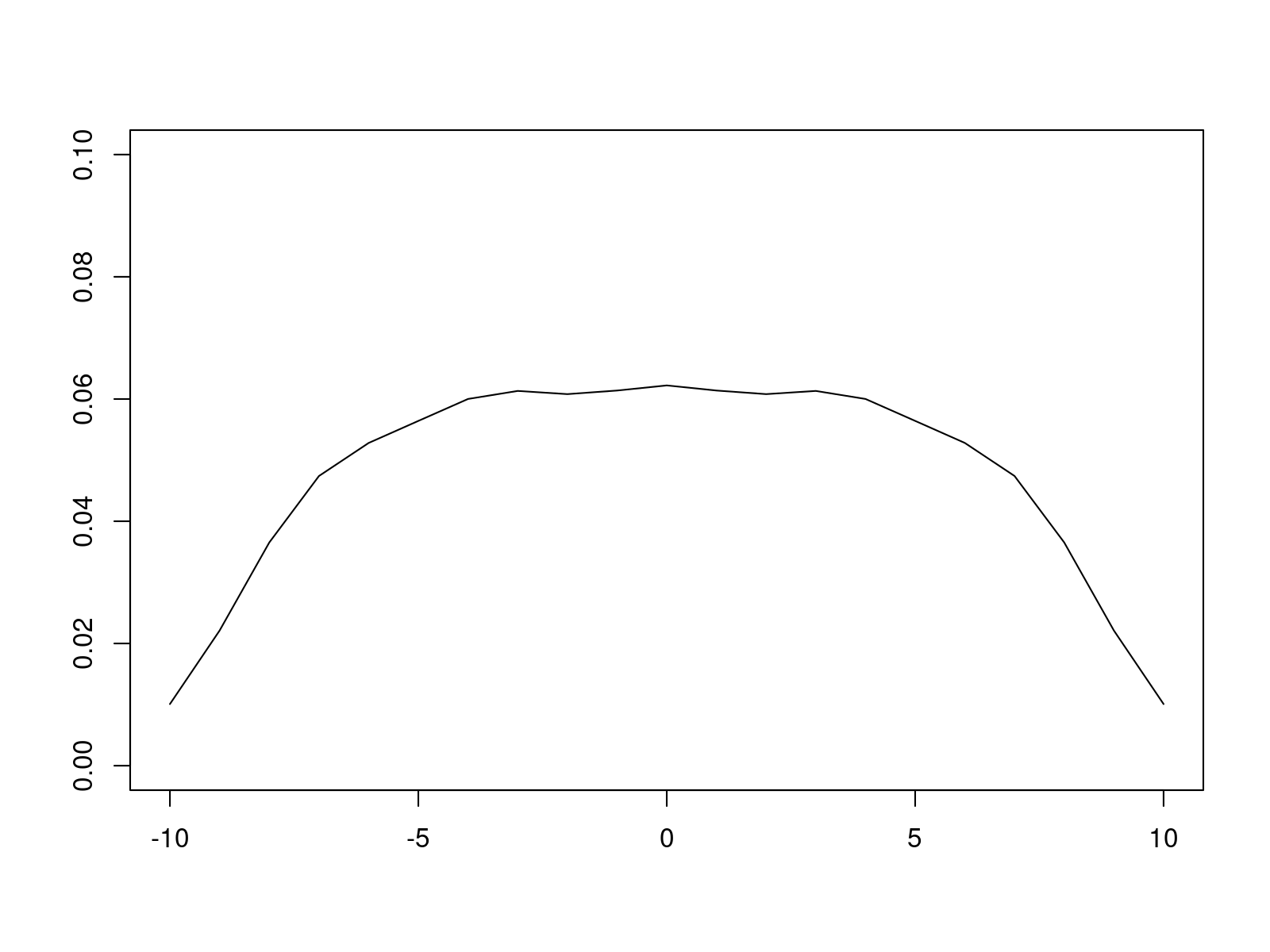}
    \caption{True marginal distribution of the first coordinate in quadratic mixture distribution}
    \label{fig:freq_truth_poly}
\end{figure}
The plots of the auto-correlation functions (ACF) of $f(s)$ from a single chain are presented in Figure~\ref{fig:acf_plots_poly}. NCG, V-DHAMS and O-DHAMS exhibit auto-correlations lower than other samplers, indicating reduced dependencies among draws.

\begin{figure}[tbp]
    \centering
    \begin{subfigure}[b]{0.32\textwidth}
        \centering
        \includegraphics[width=0.8\linewidth]{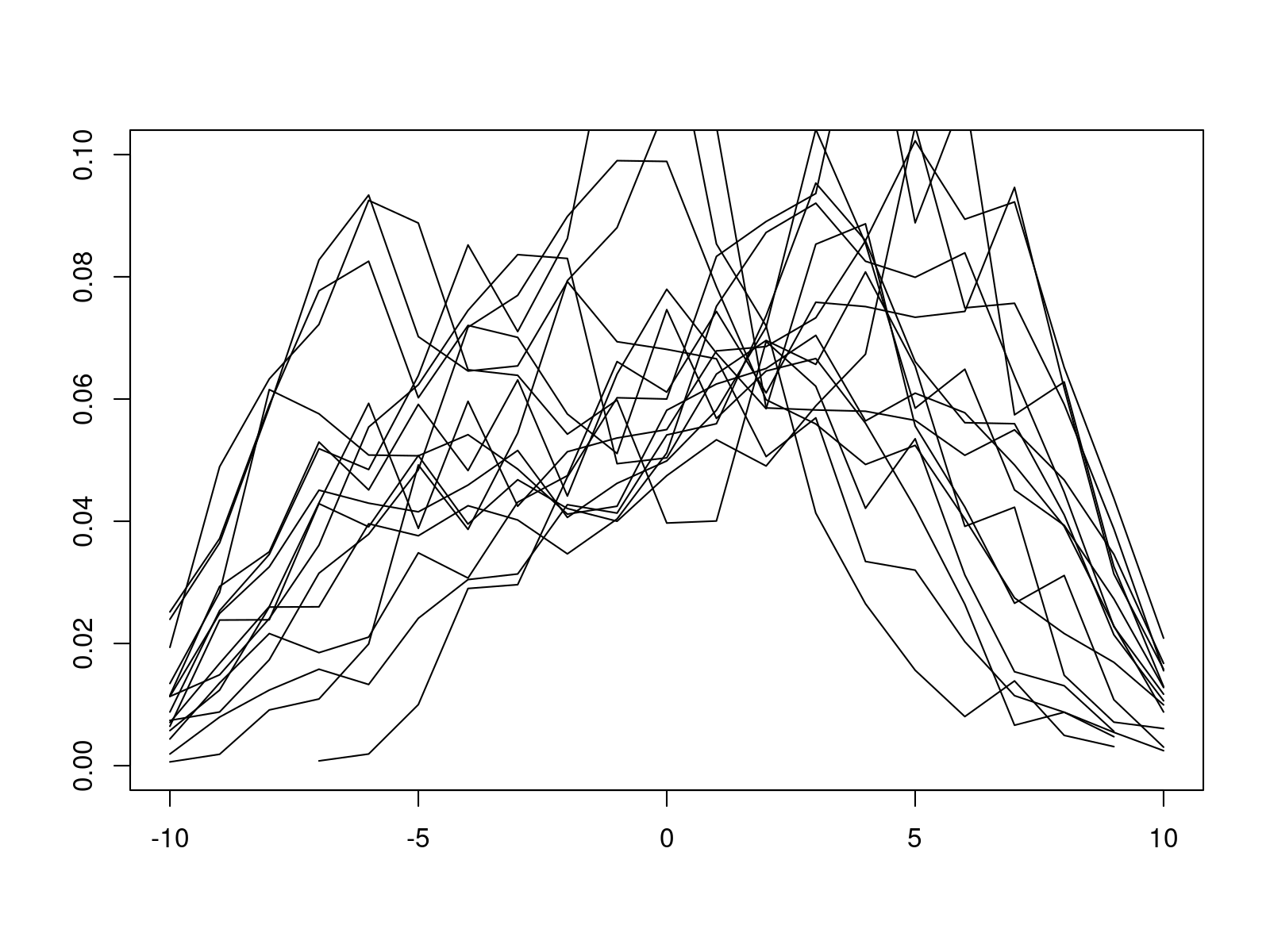}
        \caption{ Metropolis}
        \label{fig:freq_MH_poly}
    \end{subfigure}
    \begin{subfigure}[b]{0.32\textwidth}
        \centering
        \includegraphics[width=0.8\linewidth]{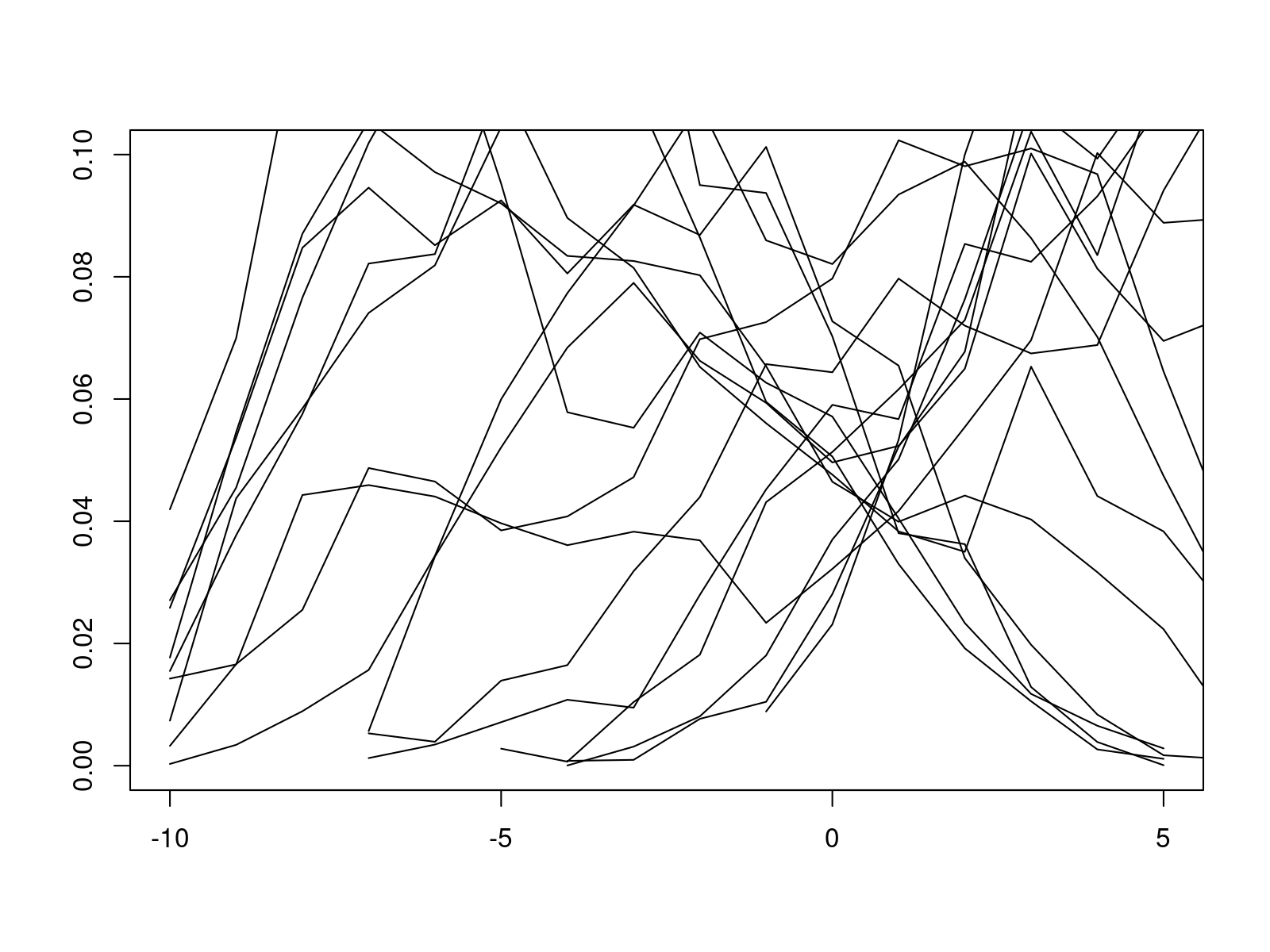}
        \caption{ GWG}
        \label{fig:freq_gwg_poly}
    \end{subfigure}
     \begin{subfigure}[b]{0.32\textwidth}
        \centering
        \includegraphics[width=0.8\linewidth]{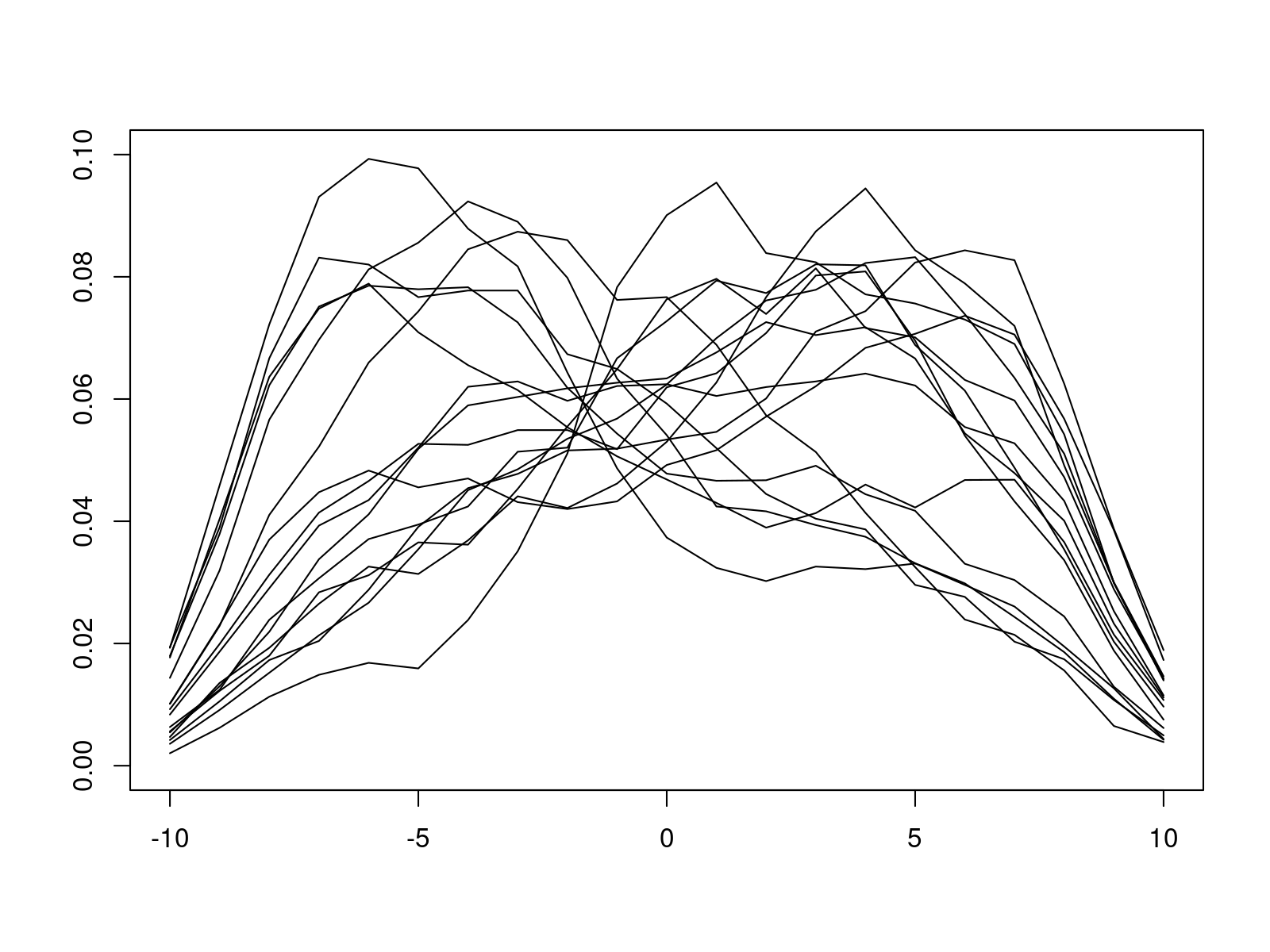}
        \caption{ NCG}
        \label{fig:freq_NCG_poly}
    \end{subfigure}
     \begin{subfigure}[b]{0.32\textwidth}
        \centering
        \includegraphics[width=0.8\linewidth]{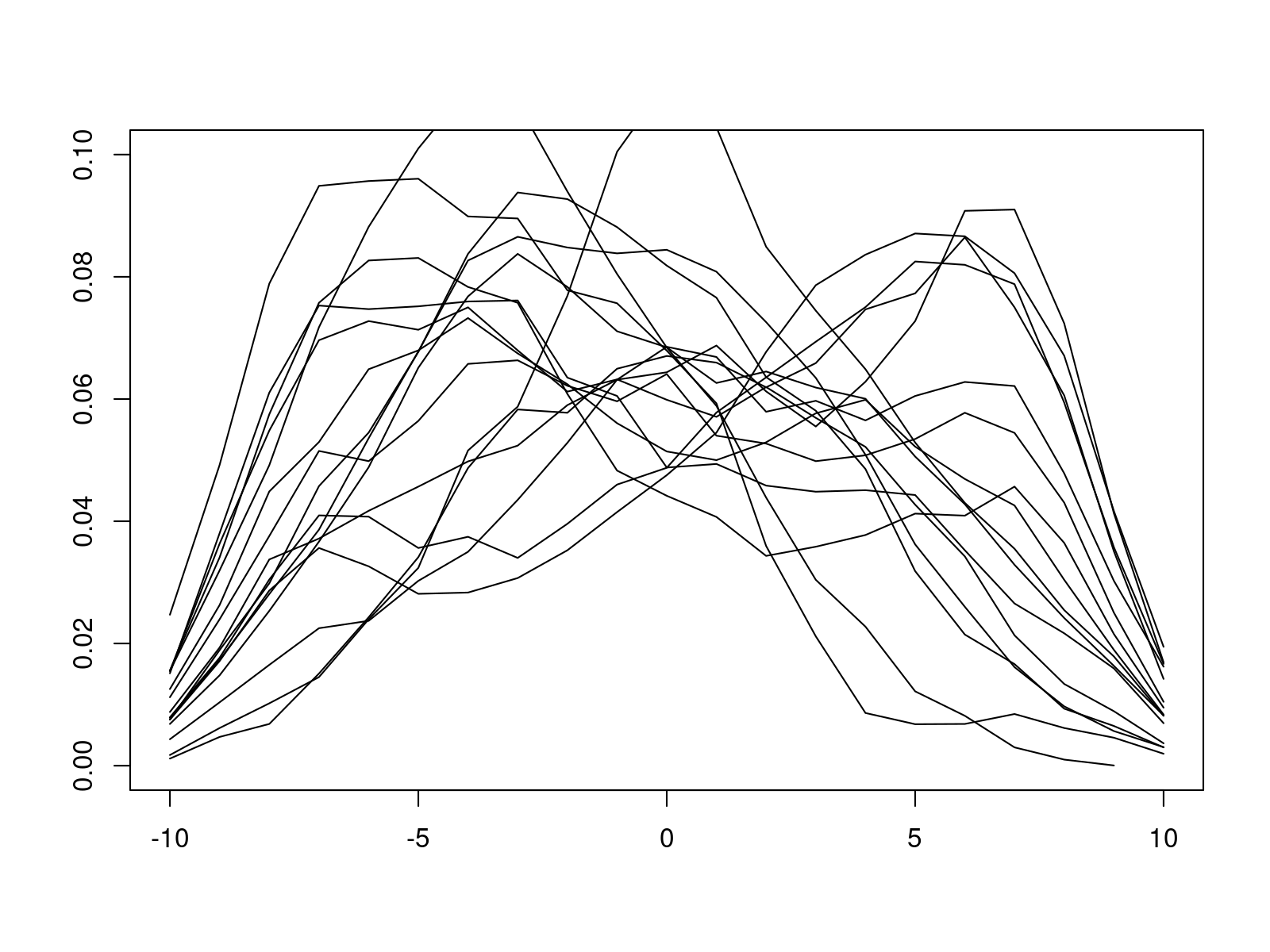}
        \caption{ AVG}
        \label{fig:freq_avg_poly}
    \end{subfigure}
     \begin{subfigure}[b]{0.32\textwidth}
        \centering
        \includegraphics[width=0.8\linewidth]{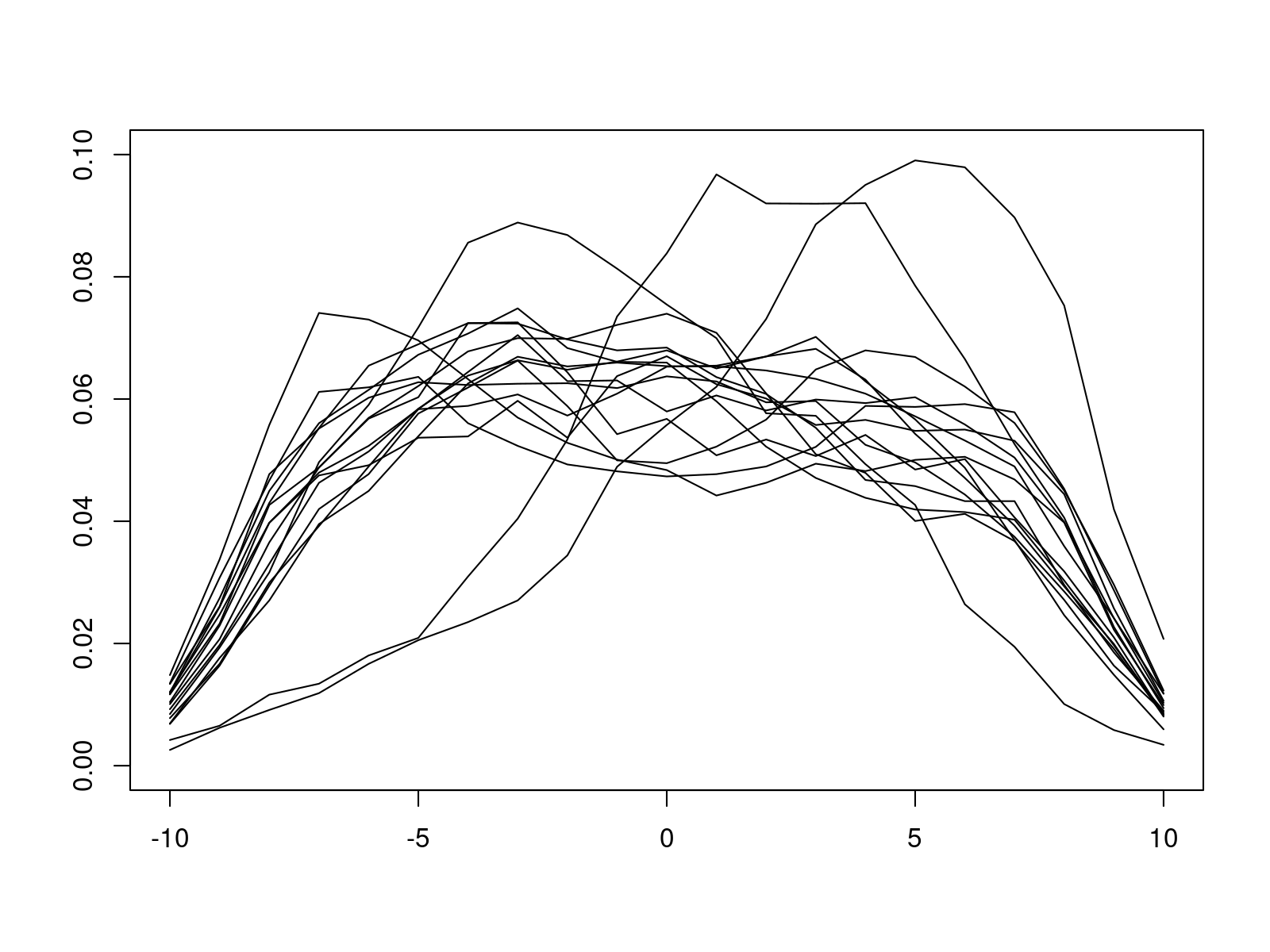}
        \caption{ V-DHAMS}
        \label{fig:freq_Hams_poly}
    \end{subfigure}
     \begin{subfigure}[b]{0.32\textwidth}
        \centering
        \includegraphics[width=0.8\linewidth]{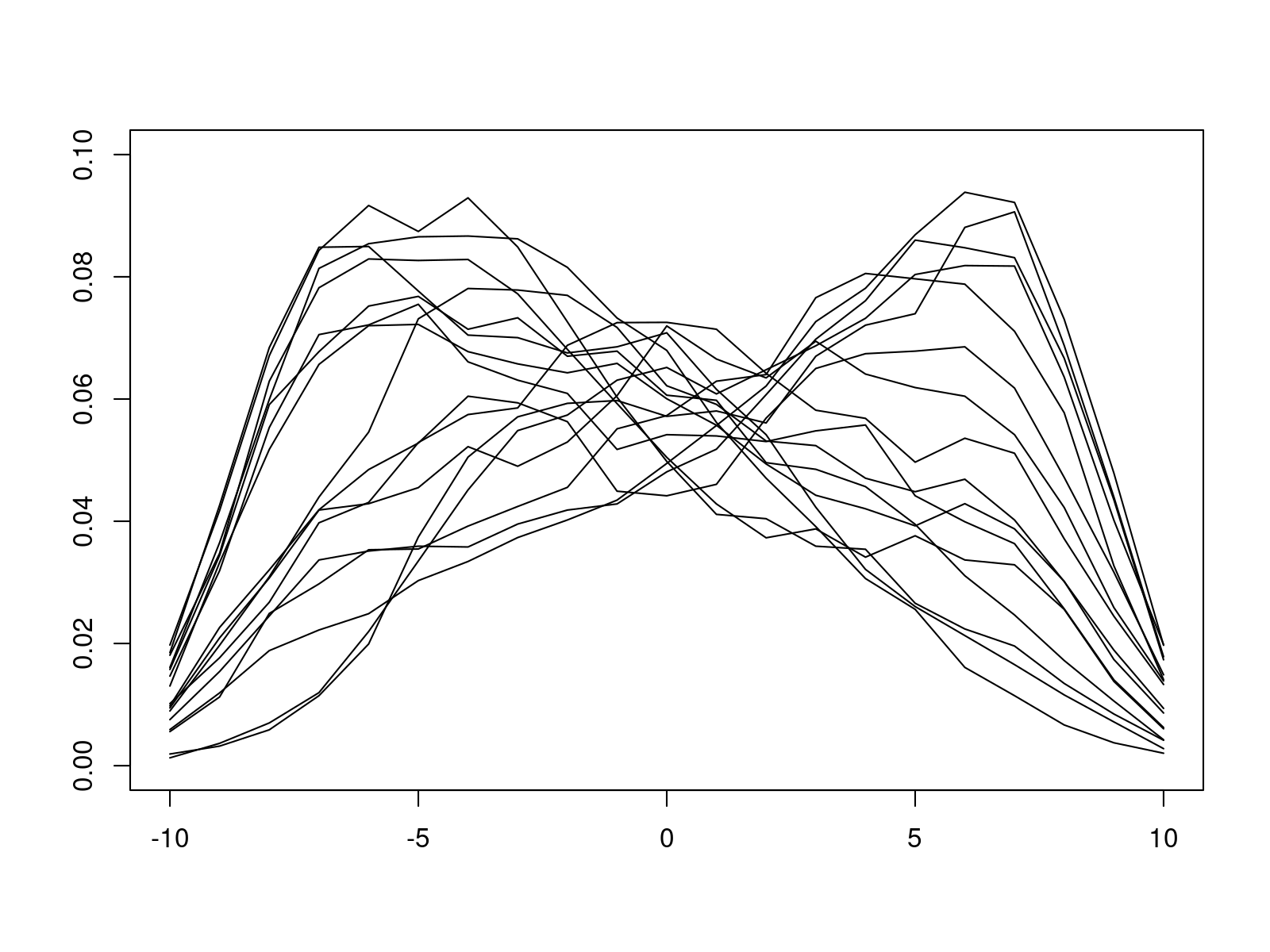}
        \caption{O-DHAMS}
        \label{fig:freq_overhams_poly}
    \end{subfigure}
\caption{Frequency plots of the first coordinate in quadratic mixture distribution}
\label{fig:freq_plots_poly}
\end{figure}
For each sampler, we present frequency plots of the first coordinate across 10 independent chains, each based on all 24,000 draws, as shown in Figure~\ref{fig:freq_plots_poly}. The corresponding true marginal distribution is depicted in Figure~\ref{fig:freq_truth_poly}. Due to the multi-modal nature of the distribution, Figure~\ref{fig:freq_truth_poly} appears relatively flat in the middle region, where the probability mass is highest. Among all methods, the frequency plots produced by Discrete-HAMS most closely resemble the ground truth, indicating superior exploration of mixture components in the presence of multi-modality.

\subsection{Additional Results for Bayesian Sparse Regression}\label{sec:linear_results}
We first present the optimal parameters and associated acceptance rates in Table \ref{tab:parma_highdim}.

\begin{figure}[tbp]
    \centering
     \begin{subfigure}[b]{0.32\textwidth}
        \centering
        \includegraphics[width=0.8\linewidth]{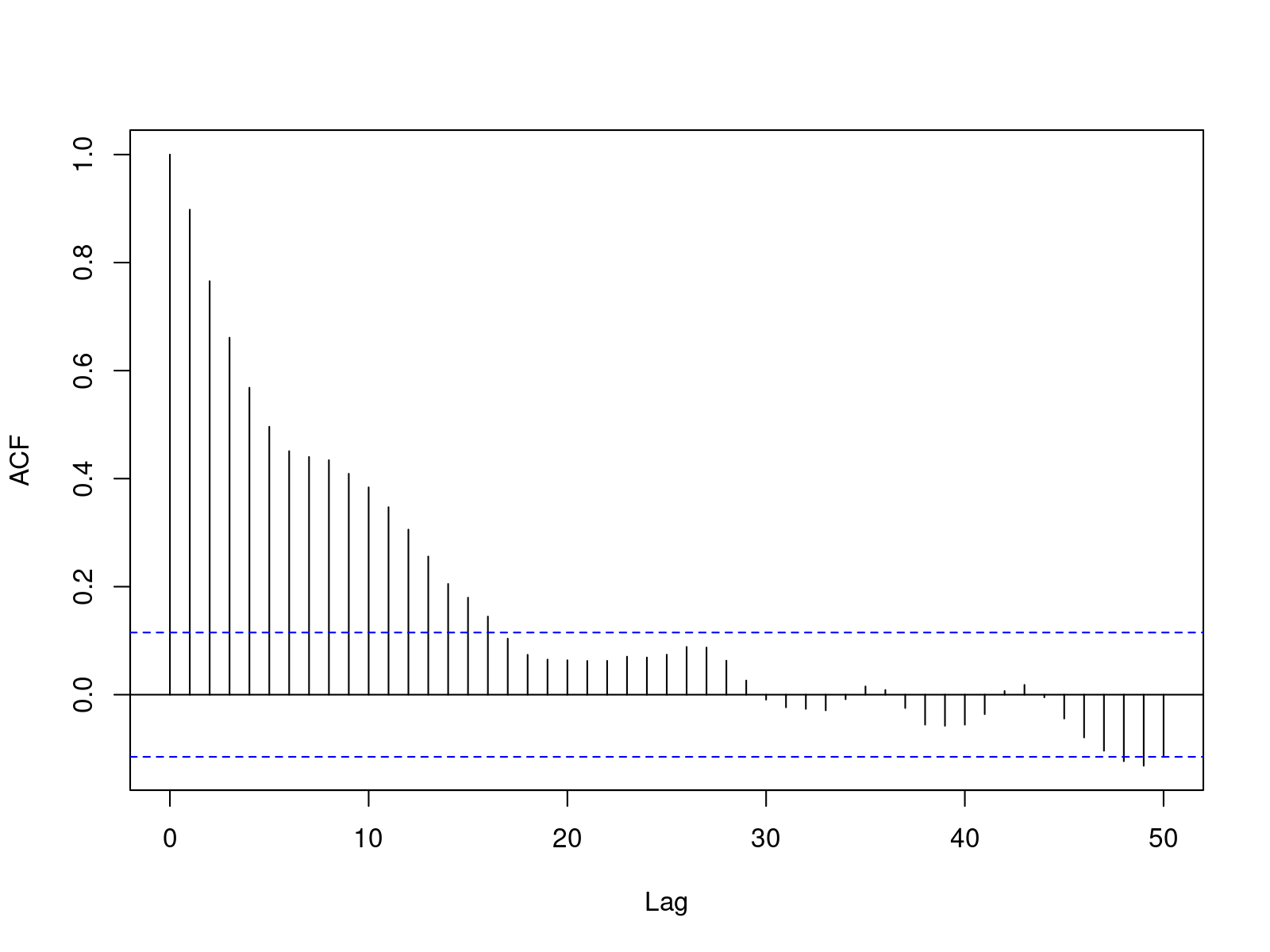}
        \caption{ACF from NCG}
        \label{fig:acf_NCG_linear}
    \end{subfigure}
     \begin{subfigure}[b]{0.32\textwidth}
        \centering
        \includegraphics[width=0.8\linewidth]{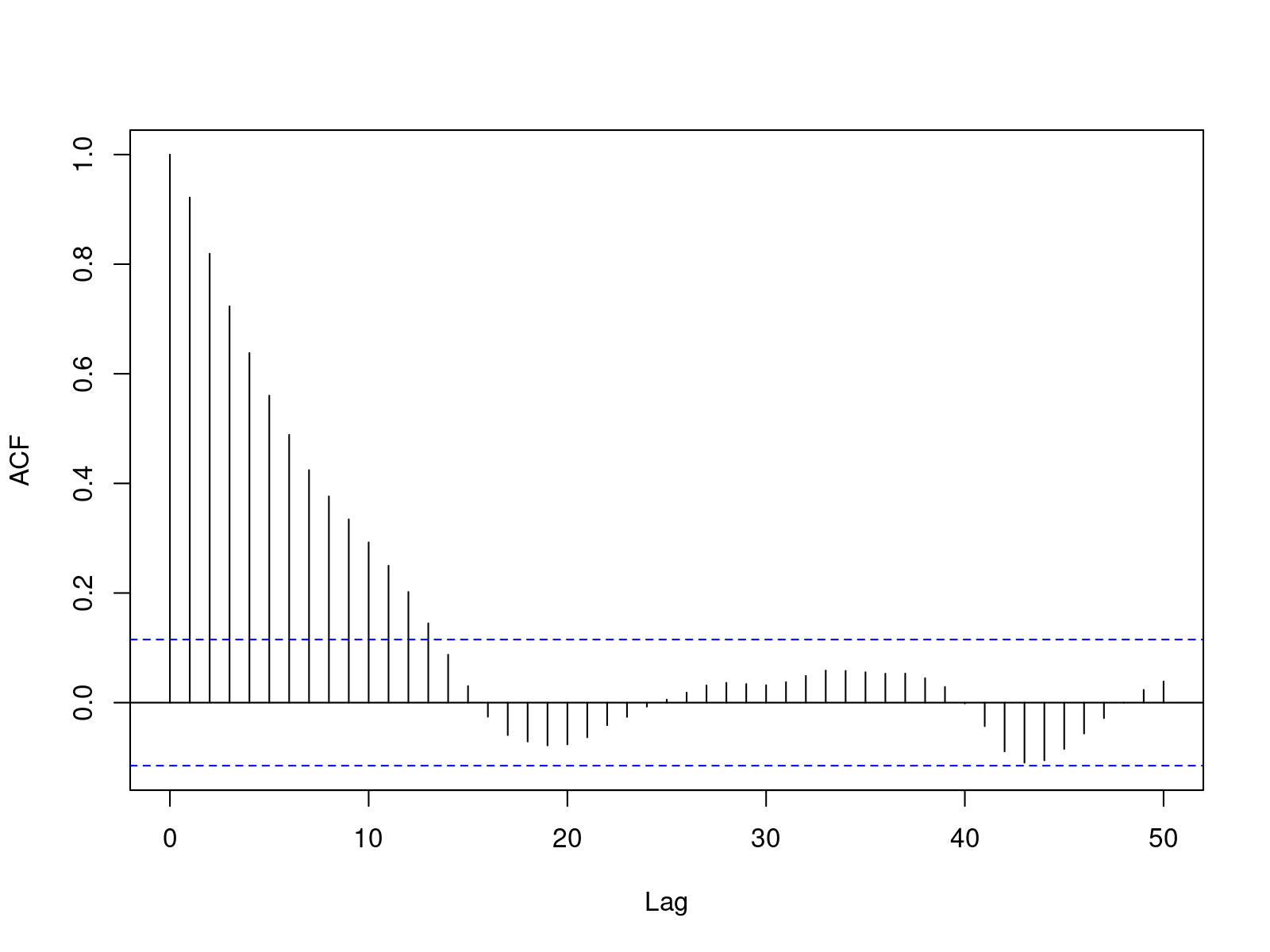}
        \caption{ACF from AVG}
        \label{fig:acf_avg_linear}
    \end{subfigure}
     \begin{subfigure}[b]{0.32\textwidth}
        \centering
        \includegraphics[width=0.8\linewidth]{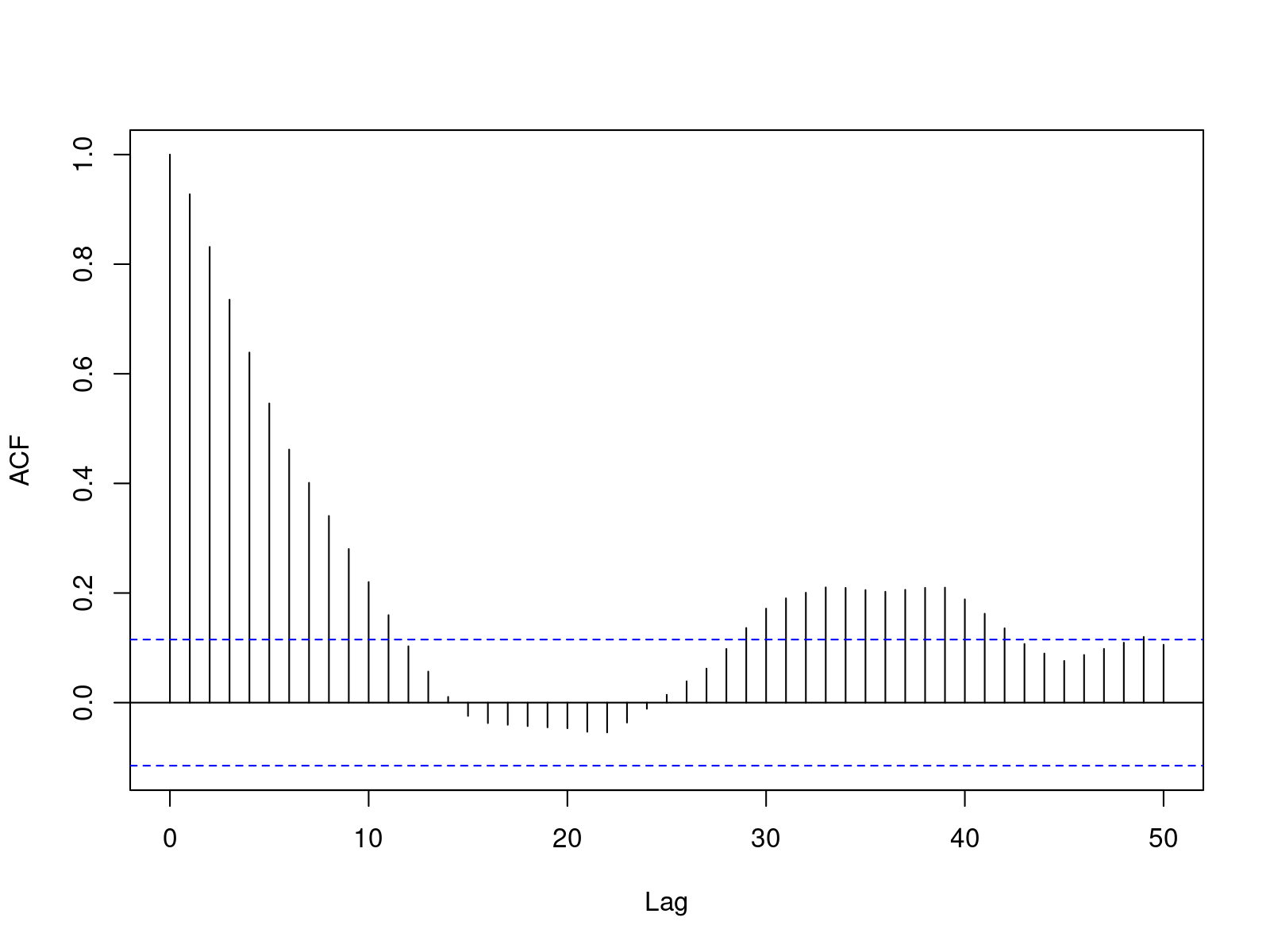}
        \caption{ACF from V-DHAMS}
        \label{fig:acf_Hams_linear}
    \end{subfigure}
     \begin{subfigure}[b]{0.32\textwidth}
        \centering
        \includegraphics[width=0.8\linewidth]{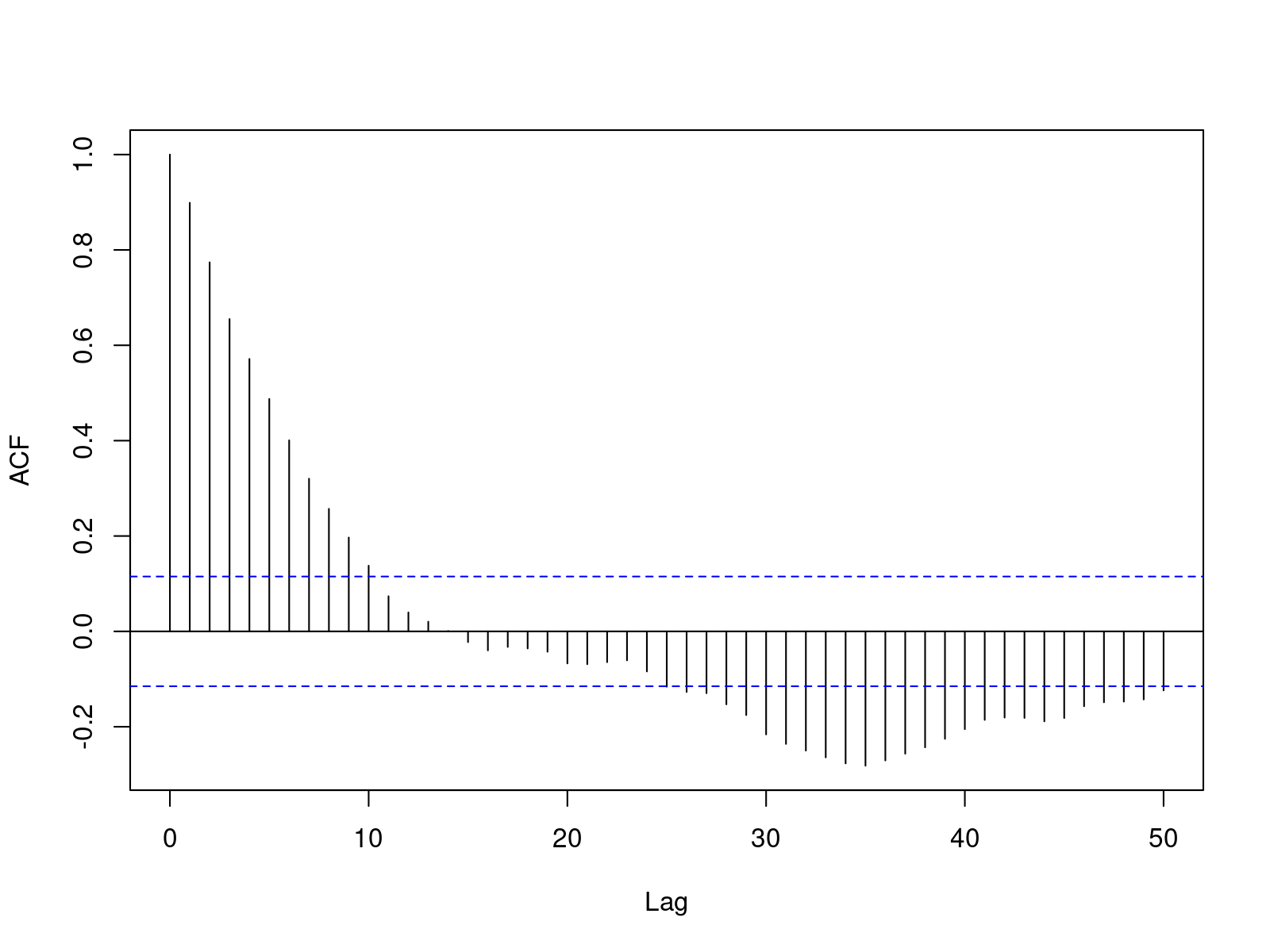}
        \caption{ACF from O-DHAMS}
        \label{fig:acf_overhams_linear}
    \end{subfigure}
\caption{ACF plots for $s_1$ in Bayesian linear regression}
\label{fig:acf_plots_linear}
\end{figure}

\begin{table}[H]
\centering
\begin{tabular}{|c|c|c|c|c|}
\hline
    Sampler & Parameter & Acceptance Rate\\
    \hline
    NCG & $\delta=0.136$ & 0.36\\
        AVG & $\delta=0.159$ & 0.28 \\
  V-DHAMS & $\epsilon=0.9, \delta=0.283, \phi=0.0, \beta=0.1$ & 0.29\\
    O-DHAMS & $\epsilon=0.9, \delta=0.260, \phi=0.0, \beta=0.1$ & 0.32\\ \hline
\end{tabular}
\caption{Parameters for high-dimensional Bayesian linear regression}
\label{tab:parma_highdim}
\end{table}
We also present ACF results for $s_1$ in Figure~\ref{fig:acf_plots_linear}, computed after thinning the first chain by a factor of 10 for each sampler. Both Discrete-HAMS samplers outperform AVG, and significantly outperform NCG, in terms of ACF decay. This suggests more frequent switching between the identical pair $s_1$ and $s_{601}$, indicating better mixing performance.
\section{Bayesian Linear Regression}
\subsection{Posterior Distribution Derivation}\label{sec:ridge_prior}
We derive the posterior distribution $\pi(s | y,X)$ stated in \eqref{eq:full_posterior}.
First, from \eqref{eq:sprior1} and \eqref{eq:sprior2}, the prior distribution for binary mask vector $s$ is
    \begin{align}
        \pi(s) & = \int \pi(s|\psi)\pi(\psi) d\psi \nonumber \\
             & = \int \prod_{i=1}^{d} \pi(s_{i}|\psi) \pi(\psi) d \psi \nonumber\\
             & \propto \int  \psi^{\alpha_{\psi}-1} (1-\psi) ^{\beta_{\psi}-1}\prod_{i=1}^{d} \psi^{s_{i}}(1-\psi)^{1-s_{i}}d\psi\nonumber \\
             & \propto  \Gamma(\sum\limits_{i=1}^{d}s_{i}+\alpha_{\psi}) \Gamma(d-\sum\limits_{i=1}^{d}s_{i}+\beta_{\psi}). \nonumber
    \end{align}
The corresponding posterior can be derived by integrating out $w_{s}$ and $\sigma^{2}$:
    \begin{align}
        &\quad \pi(s|y,X) \nonumber\\
        & =\int \pi(s, w_s, \sigma^{2}|y, X) dw_sd\sigma^{2} \nonumber\\
        & \propto \pi(s) \int \pi(\sigma^{2}) \pi(w_{s}|X,s,\sigma^2)\pi(y|X,s, w_{s}, \sigma^{2})  dw_{s} d\sigma^{2} \nonumber\\
        & \propto \pi(s)\frac{\sqrt{|X_{s}^{\T}X_{s}+\lambda I_{d_{s}}|}}{\sqrt{|(g+\kappa)X_{s}^{\T}X_{s}+\lambda I_{d}|}}  \nonumber\\
        & \quad \int (\sigma^{2})^{-\frac{n}{w}} \exp(-\frac{1}{2\sigma^{2}} [y^{\T}y-gy^{\T}X_{s}[(g+\kappa)X_{s}^{\T}X_{s}+\lambda I_{d_{s}}]^{-1}X_{s}^{\T}y]) \InvGamma (\sigma^{2}| \alpha_{\sigma}, \beta_{\sigma}) d\sigma^{2} \nonumber\\
        & \propto \pi(s)\frac{\sqrt{|X_{s}^{\T}X_{s}+\lambda I_{d_{s}}|}}{\sqrt{|(g+\kappa)X_{s}^{\T}X_{s}+\lambda I_{d}|}}(2\beta_{\sigma}+y^{\T}y-gy^{\T}X_{s}[(g+\kappa)X_{s}^{\T}X_{s}+\lambda I_{d_{s}}]^{-1}X_{s}^{\T}y)^{-\frac{2\alpha_{\sigma}+n}{2}}. \label{eq:s_posterior}
    \end{align}
We then calculate the negative potential function $f(s) \propto \log(\pi(s|X,y))$. For convenience, we define $\tilde{X}_{s} = X \diag(s)$. From \eqref{eq:s_posterior}, $f(s)$ can be calculated as
\begin{align}
    f(s) &= \log(\pi(s))+\frac{1}{2}\log(|\tilde{X}_{s}^{\T}\tilde{X}_{s}+\lambda I_{d}|)-\frac{1}{2}\log(|(g+\kappa)\tilde{X}_{s}^{\T}\tilde{X}_{s}+\lambda I_{d}|) \nonumber\\
    &-\frac{2\alpha_{\sigma}+N}{2} \log(2\beta_{\sigma}+y^{\T}y-gy^{\T}\tilde{X}_{s}[(g+\kappa)\tilde{X}_{s}^{\T}\tilde{X}_{s}+\lambda I_{d}]^{-1}\tilde{X}_{s}^{\T}y). \nonumber
\end{align}
Finally, we calculate the gradient of $f(s)$ as follows, which is required for implementing NCG, AVG and Discrete-HAMS:
\begin{align}
        \frac{\partial f(s)}{\partial s} & =  \DiGamma(\textbf{1}^{\T}_{d}s+\alpha_{\psi}) \textbf{1}_{d} - \DiGamma(d-\textbf{1}^{\T}_{d}s+\beta_{\psi})\textbf{1}_{d} \nonumber \\
        & +(\frac{\partial f}{\partial \tilde{X}_{s}} \circ X)^{\T}\textbf{1}_{n}, \nonumber
\end{align}
where $\circ$ denotes the Hadamard product.
By denoting $T = [(g+t)\tilde{X}_{s}^{\T}\tilde{X}_{s}+\lambda I_{d}]^{-1}$:
    \begin{align}
        \frac{\partial f}{\partial \tilde{X}_{s}} & = \frac{1}{2}\frac{\partial \log(|\tilde{X}_{s}^{\T}\tilde{X}_{s}+\lambda I_{d}|)}{\partial \tilde{X}_{s}}-\frac{1}{2} \frac{\partial \log(|(g+t)\tilde{X}_{s}^{\T}\tilde{X}_{s}+\lambda I_{d}|)}{\partial \tilde{X}_{s}} \nonumber \\
        &- (\alpha_{\sigma} +\frac{n}{2}) \frac{\partial \log(2\beta_{\sigma}+y^{\T}y-gy^{\T}\tilde{X}_{s}[(g+t)\tilde{X}_{s}^{\T}\tilde{X}_{s}+\lambda I_{d}]^{-1}\tilde{X}_{s}^{\T}y)}{\partial \tilde{X}_{s}} \nonumber\\
        & = \tilde{X}_{s}(\tilde{X}_{s}^{\T}\tilde{X}_{s}+\lambda I_{d})^{-1} -(g+t)\tilde{X}_{s}T \nonumber \\
        &+ (\alpha_{\sigma} +\frac{n}{2})\frac{2gyy^{\T}\tilde{X}_{s}T-2g(g+t)\tilde{X}_{s}T\tilde{X}_{s}^{\T}yy^{\T}\tilde{X}_{s}T}{2\beta_{\sigma}+yy^{\T}-gy^{\T}\tilde{X}_{s}T\tilde{X}_{s}^{\T}y}.    \label{eq:s_grad}
    \end{align}
To reduce computational complexity, we replace $\tilde{X}_{s}$ by $X_{s}$ in \eqref{eq:s_grad} and fill the columns of $ \frac{\partial f}{\partial \tilde{X}_{s}}$ that correspond to the entries of 1 in $s$, and set the other columns to zero.
\subsection{Hyper-parameter Choices for Ridge-type G-priors}\label{sec:ridge_prior_calib}

We discuss choices of hyper-parameters $(g, \kappa, \lambda)$ in the ridge-type g-prior \eqref{sec:ridge_prior}.
Consider the singular decompositions of $X_s$ given a binary mask vector  $s$:
    \begin{align}
        X_s &= U_sL_sW_s^{\T} \nonumber \\
        &= \sum\limits_{i=1}^{r}u_{s,i}l_{s,i}w_{s,i}^{\T}, \nonumber
    \end{align}
where $u_{s,i}$ is the $i$-th left singular vector, $l_{s,i}$ is the $i$-th singular value decreasing in $i$, $w_{s,i}$ is the $i$-th right singular vector and $r$ is the rank of matrix $X_{s}$.

First, by Borrell's inequality \citepSupp{Borell1975tailbound},
\begin{align}
    P(\|w_{s}\|_{2} > a | \sigma^{2}) & \leq \exp(-\frac{a^{2}}{2 \lambda_{max}(g\sigma^{2}(\kappa X_{s}^{\T}X_{s}+\lambda I_{d_{s}})^{-1})}) \nonumber \\
   &  \leq \exp(-\frac{a^{2}(\kappa l_{s,1}^{2} +\lambda)}{2g\sigma^{2}}), \nonumber
\end{align}
which implies that a larger value of $\lambda$ leads to a sharper prior on $w_{s}$.
In the absence of further information on $w_s$, a flatter prior is preferred and $\lambda$ is chosen with a small value close to $0$.
Second, the classical g-prior has an attractive feature that the covariance structure of the data is well-preserved (where $w_s$ is normal with zero mean and variance proportional to $(X_{s}^{\T}X_s)^{-1}$). To maintain this structure with the ridge-type g-prior as in \cite{Baragatti2012ridgegprior}, we calibrate the hyper-parameters  $\kappa$ and $\lambda$ such that the total variance of full data is maintained, by requiring
\begin{equation}
\tr(\frac{1}{g\sigma^{2}} (\kappa X^{\T}X+\lambda I_{d})) = \tr(\frac{1}{g\sigma^{2}} (X^{\T}X)), \nonumber
\end{equation}
which gives the following relation between $\lambda$ and $\kappa$:
\begin{equation}
        \lambda = \frac{(1-\kappa) \tr(X^{\T}X)}{d}.
    \label{eq:lambda_choice}
\end{equation}
From \eqref{eq:lambda_choice}, in order to keep $\lambda$ small for a flatter prior, we take $\kappa$ that is smaller than but close to 1.
Third, by \cite{Zellner1986gprior} and \citeSupp{Maruyama2011factorgprior}, $g = \mathcal{O}(n)$ is preferred for g-priors to achieve model selection consistency.
In our numerical studies, from the preceding considerations,
we choose $g=n$, $ \kappa=0.995$ and $\lambda$ using \eqref{eq:lambda_choice}.

\end{document}